\documentclass[aps,pra,reprint,amsmath,amssymb,groupedaddress,showpacs]{revtex4-1}
\usepackage{graphicx}
\usepackage{float}
\usepackage{dcolumn} 
\usepackage{bm}      
\usepackage[nomessages]{fp}
\usepackage[bookmarks=false]{hyperref} 
\usepackage[font={small}]{caption} 
\usepackage{enumitem}
\usepackage{textcomp}
\usepackage{pgfplots}              
\pgfplotsset{compat=1.9}           
\hypersetup{
    unicode=false,          
    pdftoolbar=true,        
    pdfmenubar=true,        
    pdffitwindow=true,      
    pdfstartview={},        
    pdftitle={Correlance and Discordance: Computable Measures of Nonlocal Correlation}, 
    pdfauthor={Samuel R. Hedemann},
    pdfsubject={Quantum Nonlocal Correlations},   
    pdfcreator={},          
    pdfproducer={},         
    pdfkeywords={Correlance,} {Discordance,} {Strong Discordance,} {Diagonal Correlance,} {Diagonal Discordance,} {Statance,} {Probablance,} {Entanglement,} {TGX States,} {Ent,} {Nonlocal Correlation,} {Quantum,} {Classicality,} {Strict Classicality,} {Strictly Classical,} {Quantum Discord}, 
    pdfnewwindow=true,      
    colorlinks=true,        
    linkcolor=black,        
    citecolor=black,        
    filecolor=black,        
    urlcolor=black          
}
\newcommand{\Sec}[1]{\hyperref[sec:#1]{Sec.{\kern 2pt}\ref*{sec:#1}}}
\newcommand{\Section}[1]{\hyperref[sec:#1]{Section~\ref*{sec:#1}}}
\newcommand{\Fig}[2][]{\hyperref[fig:#2]{Fig.{\kern 2pt}\ref*{fig:#2}#1}}
\newcommand{\Figure}[2][]{\hyperref[fig:#2]{Figure~\ref*{fig:#2}#1}}
\newcommand{\App}[1]{\hyperref[sec:App.#1]{App.{\kern 2pt}\ref*{sec:App.#1}}}
\newcommand{\Appendix}[1]{\hyperref[sec:App.#1]{Appendix~\ref*{sec:App.#1}}}
\newcommand{\EqLab}[1]{\\\noindent\smash{\raisebox{6pt}[0pt][0pt]{\hypertarget{eq:#1}{}}}\vspace{-11pt}}
\newcommand{\Eq}[1]{\protect\hyperlink{eq:#1}{(\ref*{eq.#1})}}
\newcommand{\Eqs}[2]{\protect\hyperlink{eq:#1}{(\ref*{eq.#1}--\ref*{eq.#2})}}
\newenvironment{Equation}[1]{\EqLab{#1}\begin{equation}\label{eq.#1}}{\end{equation}\par\noindent\ignorespacesafterend}
\newcommand{\Table}[2][]{\hyperref[tab:#2]{Table~\ref*{tab:#2}#1}}
\newcommand{\ThmLab}[1]{\noindent\smash{\raisebox{11pt}[0pt][0pt]{\hypertarget{Thm:#1}{}}}\textbf{Theorem{\kern 3pt}#1:~~}}
\newcommand{\Thm}[1]{\protect\hyperlink{Thm:#1}{Theorem{\kern 3pt}#1}}
\newcommand{\Thms}[2]{\protect\hyperlink{Thm:#1}{Theorems{\kern 3pt}#1--#2}}
\newcommand{\tr}[0]{\text{tr}}
\newcommand{\redx}[2]{{#1}{\kern -5.3pt}{{\textnormal{\raisebox{-1.2pt}{\scalebox{1.2}{\textasciicaron}}}}}{\kern -4.2pt}{~}^{(#2)}}
\newcommand{\redxprime}[2]{{#1}{\kern -5.3pt}{{\textnormal{\raisebox{-1.2pt}{\scalebox{1.2}{\textasciicaron}}}}}{\kern -4.2pt}{~}^{(#2)\prime}}
\newcommand{\redxpreprime}[2]{{#1}{\kern -5.3pt}{{\textnormal{\raisebox{-1.2pt}{\scalebox{1.2}{\textasciicaron}}}}}{\kern -4.2pt}{~}^{{\prime}{\kern 1pt}(#2)}}
\newcommand{\redxs}[3]{{#1}_{#2}^{{\kern -5.5pt}{\textnormal{\raisebox{-5.9pt}[0pt][0pt]{\scalebox{1.6}{\textasciicaron}}}}{\kern -1pt}#3}}
\newcommand{\redX}[2]{{#1}{\kern -6.9pt}{{\textnormal{\raisebox{1.2pt}{\scalebox{1.2}{\textasciicaron}}}}}{\kern -3.1pt}{~}^{(#2)}}
\newcommand{\mbar}[0]{{\mathop {m}\limits^{{\kern -3.5pt}~_{\overline{{\kern 7pt}}}}}}
\newcommand{\mbarsub}[0]{{{\kern 0.0pt}\mathop {m}\limits^{{\kern -0.3pt}{\overline{{\kern 5.5pt}}}}{\kern 0.0pt}}}
\newcommand{\nmaxnot}{n_{\,\overline{{\kern -1.8pt}\max^{~^{~^{~}}}\!\!\!\!\!\!\!\!\!\!}}\,}
\newcommand{\shiftmath}[2]{\textnormal{\raisebox{#1}[#1][#1]{$#2$}}}
\newcommand{\redxshift}[4]{{#1}{\kern -5.3pt}{{\textnormal{\raisebox{-1.2pt}{\scalebox{1.2}{\textasciicaron}}}}}{\kern -4.2pt}{~}^{{\kern #4}\shiftmath{#3}{(#2)}}}
\newcommand{\scalemath}[2]{\textnormal{\scalebox{#1}{$#2$}}}
\newcommand{\hsp}[1]{{\kern #1pt}}
\newcommand{\Correlance}[0]{\mathcal{X}}
\newcommand{\DiagonalCorrelance}[0]{\mathcal{X}_{D}}
\newcommand{\Probablance}[0]{\mathcal{P}}
\newcommand{\Statance}[0]{\mathcal{S}}
\newcommand{\StrongDiscordance}[0]{\mathcal{D}_{S}}
\newcommand{\Nondiagonality}[0]{\nu}
\newcommand{\Discordance}[0]{\mathcal{D}}
\newcommand{\DiagonalDiscordance}[0]{\mathcal{D}_{D}}

\newcommand{\RawStrongDiscordance}[0]{\widetilde{\mathcal{D}}_{S}}
\newcommand{\RawDiagonalStrongDiscordance}[0]{\widetilde{\mathcal{D}}_{DS}}
\newcommand{\Multicorrelance}[0]{\mathcal{M}_\Correlance}
\newcommand{\miniMcal}[0]{\scalemath{0.70}{\mathcal{M}}}
\newcommand{\vertsp}[1]{#1_{~_{~_{~_{~}}}}^{~^{~^{~^{~}}}}\!\!\!\!\!\!\!\!\!}
\newcommand{\Pearson}[0]{{r_{\text{P}}}}
\renewcommand{\geq}[0]{\geqslant}
\renewcommand{\ge}[0]{\geqslant}
\renewcommand{\leq}[0]{\leqslant}
\renewcommand{\le}[0]{\leqslant}
\begin{document}
\title{Correlance and Discordance: Computable Measures of Nonlocal Correlation}
\author{Samuel R. Hedemann}
\affiliation{Department of Physics, New York Institute of Technology, Old Westbury, New York 11568}
\date{January 09, 2020}
\begin{abstract}
We present six new measures of nonlocal correlation for discrete multipartite quantum systems; correlance, statance, probablance, strong discordance, discordance, and diagonal discordance.  The correlance measures all nonlocal correlation (even bound entanglement), and is exactly computable for all pure and mixed states. Statance and probablance are not yet computable, but motivate the strong discordance (for nonlocal correlation beyond that achievable by a strictly classical state), discordance (a measure of all nonlocal correlation in distinguishably quantum states), and diagonal discordance (for nonlocal correlation in diagonal states), all of which are exactly computable for all states. We discuss types of correlation and notions of classicality, and compare correlance, strong discordance, and discordance to quantum discord. We also define diagonal correlance to handle strictly classical probability distributions, providing a powerful tool with wide-ranging applications.%
\end{abstract}%
\maketitle
\section{\label{sec:I}Introduction}%
\vspace{-8pt}%
The strong nonlocal correlations achievable in quantum systems can cause novel physical effects that are impossible in classical systems, and have led to an intense worldwide search for applications of these properties, particularly in quantum computing \cite[]{Feyn,DiVi}, quantum cryptography \cite[]{BB84,B92,Eker}, and quantum communications \cite[]{BBCJ,BPM1,BPM2,HedA}.

While entanglement has been the star of this show, \textit{other} types of quantum nonlocal correlation exist that may also be useful. Thus, there is a need to quantify such nonlocal correlation to asses which states are most useful for a given application.  One such measure is \textit{quantum discord}, which we now briefly summarize.

In the case of bipartite mixed states \smash{$\rho\equiv\rho^{(1,2)}$}, where parenthetical superscripts denote labels of subsystems (modes) of a Hilbert space \smash{$\mathcal{H}\equiv\mathcal{H}^{(1,2)}\equiv\mathcal{H}^{(1)}\otimes\mathcal{H}^{(2)}$} of dimension $n\!\equiv\!\dim(\rho)\!=\! n_{1}n_{2}$, where \smash{$n_{m}$} is the dimension of mode $m$, the \textit{quantum mutual information} of \smash{$\rho$} is%
\begin{Equation}                      {1}
\mathcal{I}(\rho ) \equiv S(\redx{\rho}{1}) + S(\redx{\rho}{2}) - S(\rho ),
\end{Equation}
where \smash{$\redx{\rho}{m}$} is the reduced state for mode $m$ (see \App{A}), and\hsp{-1} \smash{$S(\rho ) \!\equiv \! - \tr[\rho \log _{n} (\rho )]$}\hsp{-1} is\hsp{-1} the\hsp{-1} von\hsp{-1} Neumann\hsp{-1} entropy\hsp{-1} \cite[]{NiCh}. It has been suggested that \smash{$\mathcal{I}(\rho )$} can be partitioned as%
\begin{Equation}                      {2}
\mathcal{I}(\rho )=\mathcal{C}(\rho)+\mathcal{Q}(\rho),
\end{Equation}
where \smash{$\mathcal{C}\equiv\mathcal{C}(\rho)$} is a measure of distinctly classical nonlocal correlation, and \smash{$\mathcal{Q}\hsp{-1}\equiv\hsp{-1}\mathcal{Q}(\rho)$} is a measure of all quantum (nonclassical) correlations called \textit{quantum discord} (see \App{B} for a full definition of quantum discord) \cite[]{OlZu,HeVe,AlRA}. 

It has been shown that \smash{$\mathcal{Q}$} is not necessarily only due to entanglement, nor is it only due to nonentanglement quantum correlations (so-called \textit{quantum nonlocality without entanglement}) \cite[]{BDFM,HHHO,NiCe}.

However, quantum discord is not the only way to quantify nonclassical correlations.  In this paper, we identify several mechanisms of nonlocal correlations (not necessarily all distinct), and propose six measures to quantify them.  The most general of these, \textit{correlance}, is computable for all possible input states and provides a single measure that can detect all forms of nonlocal correlation, while \textit{discordance} offers an alternative to quantum discord, yet is also computable for all states.
\subsection{\label{sec:I.A}General Mechanisms of Nonlocal Correlation}%
\vspace{-8pt}%
Here we define the condition of no nonlocal correlation, and three general mechanisms of nonlocal correlation.
\begin{itemize}[leftmargin=*,labelindent=4pt]\setlength\itemsep{0pt}
\item[\textbf{1.}]\hypertarget{NonlocMech:1}{}\textbf{Absence of Nonlocal Correlation:~~}An $N$-mode state $\rho$ (mixed or pure) has no nonlocal correlation if and only if (iff) it can be decomposed in \textit{product form},
\begin{Equation}                      {3}
\rho  = \mathop  \otimes \limits_{m = 1}^N \rho ^{(m)}  = \rho ^{(1)}  \otimes  \cdots  \otimes \rho ^{(N)} ,
\end{Equation}
\vspace{-10pt}\\%
where each \smash{$\rho ^{(m)}$} is a possibly mixed state in mode $m$. See \App{C} for details about purity.
\item[\textbf{2.}]\hypertarget{NonlocMech:2}{}\textbf{Entanglement Correlation:~~}An $N$-mode state $\rho \equiv \rho ^{(1, \ldots ,N)}$ is \textit{separable} (fully $N$-partite separable) iff it has a set of decomposition probabilities and pure decomposition states \smash{$\{p_{j},\rho_{j}\}$} such that
\begin{Equation}                      {4}
\rho  = \sum\nolimits_j {p_j \rho _j^{(1)}  \otimes  \cdots  \otimes \rho _j^{(N)} },
\end{Equation}
\vspace{-10pt}\\%
otherwise it is \textit{entangled} (fully $N$-partite entangled).\hsp{-1} See\hsp{-1} \App{D}\hsp{-1} for\hsp{-1} details\hsp{-1} and\hsp{-1} basic\hsp{-1} examples.
\item[\textbf{3.}]\hypertarget{NonlocMech:4}{}\textbf{Decomposition-State Correlation:~~}A necessary condition for $\rho$ to have product form is that it has a decomposition with pure decomposition states $\{\rho_{j}\}$ such that, for each $j$, the $\rho_{j}$ are \textit{pure product states},
\begin{Equation}                      {5}
\rho_{j}= \mathop  \otimes \limits_{m = 1}^N \rho _j^{(m)}  = \rho _j^{(1)}  \otimes  \cdots  \otimes \rho _j^{(N)} \;\;\forall j,
\end{Equation}
\vspace{-10pt}\\%
where each \smash{$\rho _j^{(m)}$} also has \textit{mode independence},
\begin{Equation}                      {6}
\rho _j^{(m)}= \rho _{(j_1 , \ldots ,j_N )}^{(m)}  = \rho _{j_m }^{(m)} \;\;\forall m \in 1, \ldots ,N.
\end{Equation}
\vspace{-10pt}\\%
If \Eqs{5}{6} are not satisfied, then $\rho$ must have some nonlocal correlation, even if it is separable, and we say there is \textit{decomposition-state correlation}. See \App{E} for an example illustrating mode independence.
\item[\textbf{4.}]\hypertarget{NonlocMech:3}{}\textbf{Probability Correlation:~~}
A necessary condition for $\rho$ to have product form is that it has a decomposition with probabilities \smash{$\{p_{j}\}$} with \textit{$N$ factors},
\begin{Equation}                      {7}
\begin{array}{*{20}l}
   {p_{j}} &\!\! {=\prod\nolimits_{m = 1}^N {p_j^{(m)} }  = p_j^{(1)}  \cdots p_j^{(N)} \;\;\forall j,}  \\
\end{array}
\end{Equation}
and also with \textit{mode independence}, $\forall j$,
\begin{Equation}                      {8}
\hsp{5}\begin{array}{*{20}l}
   {p_j^{(m)}\hsp{-2}} &\!\! {=\hsp{-1} p_{(j_1 , \ldots ,j_N )}^{(m)} \hsp{-2} =\hsp{-1} p_{j_m }^{(m)};\;\;\sum\nolimits_{j_m}p_{j_m}^{(m)}\hsp{-2}=\hsp{-2}1; \;\;\forall m \hsp{-2}\in\hsp{-2} 1 {\scriptstyle,\ldots,}\hsp{0.5} N .}  \\
\end{array}\hsp{-5}
\end{Equation}%
\vspace{-14pt}\\%
Any state with \textit{no} decomposition-state-correlation-minimizing decomposition (see \Sec{IV}) able to fulfill \Eqs{7}{8} must have some nonlocal correlation (even if separable), a feature we call \textit{probability correlation}.
\end{itemize}
\vspace{1pt}%
\ThmLab{1}Together, \Eqs{5}{8} form a set of necessary and sufficient conditions for $\rho$ to have product form. 
\noindent\textit{Proof:~~}Convert between general $\rho$ and \Eq{3}, as in \App{F}.\\
\vspace{-5pt}%

\begin{figure}[H]
\centering
\includegraphics[width=1.00\linewidth]{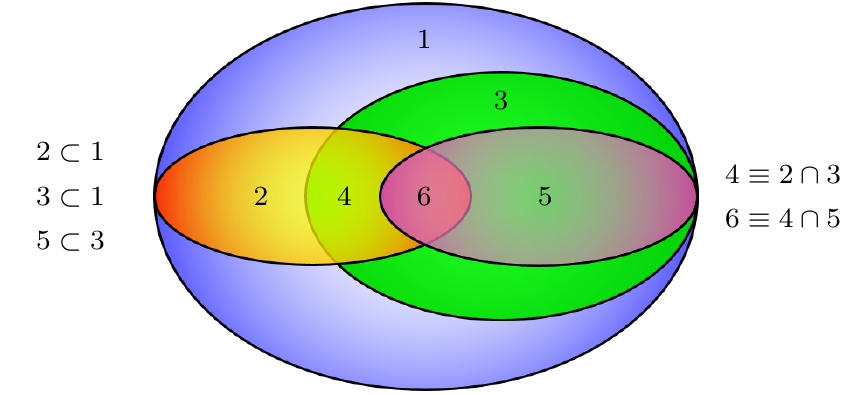}%
\vspace{-6pt}
\caption[]{(color online) Depiction of the six families of nonlocal correlation from \Table{1}. Not all set relations are written here; for example $5\subset 1$ and $6\subset 4$ are implied, etc. Not shown are \textit{Quasi-Families} $2|3$, $2|5$, and $4|5$, which consist of states with \textit{different} decompositions in different distinct families, but with no decomposition in their intersection; see \App{G.3} for details. For multipartite generalizations, see \Sec{VII}.}
\label{fig:1}
\end{figure}

The combinations of \Eqs{5}{8} yield the \textit{six families of nonlocal correlation} in \Table{1}, visualized in \Fig{1}. These families are not mutually exclusive and represent only the most general mechanisms of nonlocal correlation; more specific mechanisms are merely subsets of these, such as entanglement which is a subset of Families 1 and 2. Note that since \smash{$p_j \hsp{-1} =\hsp{-1} p_j ^{(1)} \hsp{-1} \cdots p_j ^{(N)} \;\forall j$} is \textit{always} possible, that does\hsp{0.5} not\hsp{0.5} generate\hsp{0.5} more\hsp{0.5} families, as proved in \App{G}.%
\vspace{-4pt}%
\begin{table}[H]
\caption{\label{tab:1}Six families of nonlocal correlation, meaning states that can be decomposed satisfying \protect\Eqs{5}{8} in these various ways. In all cases, $j\equiv j_{1},\ldots, j_{N}$. We abbreviate ``family'' as F, ``product-form'' as PF, and ``mode-independent'' as MI.}
\vspace{-4pt}%
\begin{ruledtabular}
\begin{tabular}{|l|p{223pt}|}
\;F\; & {\hspace{\stretch{1}}Description and State Form\hspace{\stretch{1}}} \\[0.5ex]
\hline
{\;\hsp{0.7}1} & {\rule{0pt}{12pt}{Generally\hsp{3}Nonlocally\hsp{3}Correlated:}}\\[-0.2ex]
{}  & {\hspace{\stretch{1}}$\rho  = \sum\nolimits_j {p_j \rho _j }$\hspace{\stretch{1}}} \\[0.2ex]
{\;\hsp{0.7}2} & {\rule{0pt}{6pt}{MI Probabilities:}}\\[-0.5ex]
{}  & {\hspace{62pt}{$\rho  = \sum\nolimits_j {(p_{j_1 } ^{(1)}  \cdots p_{j_N } ^{(N)} )\rho _j }$}}\\[0.2ex]
{\;\hsp{0.7}3} & {\rule{0pt}{12pt}{PF Decomposition States:}}\\[-0.5ex]
{}  & {\hspace{56pt}\rule{0pt}{12pt}{$\rho  = \sum\nolimits_j {p_j } \rho _j ^{(1)}  \otimes  \cdots  \otimes \rho _j ^{(N)}$}}\\[0.2ex]
{\;\hsp{0.7}4} & {\rule{0pt}{12pt}{MI Probabilities, PF Decomposition States:}}\\[-0.5ex]
{}  & {\hspace{35pt}\rule{0pt}{12pt}{$\rho  = \sum\nolimits_j {(p_{j_1 } ^{(1)}  \cdots p_{j_N } ^{(N)} )\rho _j ^{(1)}  \otimes  \cdots  \otimes \rho _j ^{(N)} } $}}\\[0.2ex]
{\;\hsp{0.7}5} & {\rule{0pt}{12pt}{MI Decomposition States:}}\\[-0.5ex]
{}  & {\hspace{56pt}{$\rho  = \sum\nolimits_j {p_j } \rho _{j_1 } ^{(1)}  \otimes  \cdots  \otimes \rho _{j_N } ^{(N)} $}}\\[0.2ex]
{\;\hsp{0.7}6} & {\rule{0pt}{12pt}{MI Probabilities, MI Decomposition States:}}\\[-0.5ex]
{}  & {\hspace{0pt}\rule{0pt}{12pt}{$\rho\hsp{-1}=\hsp{-2} \sum\nolimits_j {\hsp{-1}(p_{j_{1}} ^{(1)}  \cdots p_{j_{N}} ^{(N)} )} \rho _{j_1 } ^{(1)}  \hsp{-1}\otimes  \cdots  \otimes \rho _{j_N } ^{(N)}  \hsp{-1}=\hsp{-1} \rho ^{(1)}  \hsp{-1}\otimes  \cdots  \otimes \rho ^{(N)}$}}\\[1.2ex]
\end{tabular}
\end{ruledtabular}
\end{table}
\vspace{-16pt}%
\subsection{\label{sec:I.B}\hsp{-7}Correlance:\hsp{-1.3} A\hsp{-1.3} Measure\hsp{-1.3} of\hsp{-1.3} All\hsp{-1.3} Nonlocal\hsp{-1.3} Correlation}
\vspace{-1pt}%
Given any $N$-mode state $\rho\in\mathcal{H}\equiv\mathcal{H}^{(1)}\otimes\cdots\otimes\mathcal{H}^{(N)}$, a measure of all possible nonlocal $N$-mode correlation (full $N$-partite correlation) is the \textit{correlance}, given by
\begin{Equation}                      {9}
\Correlance(\rho ) \equiv \frac{{\widetilde{\Correlance}(\rho )}}{{\mathcal{N}_{\Correlance} }},
\end{Equation}
\vspace{-1pt}%
where the \textit{raw correlance }\!\! (unnormalized correlance) is
\begin{Equation}                      {10}
\widetilde{\Correlance}(\rho ) \equiv \tr[(\rho  - \varsigma )^2 ],
\end{Equation}
\vspace{-1pt}%
where we define the \textit{reduction product},
\begin{Equation}                      {11}
\varsigma  \equiv \varsigma (\rho ) \equiv \mathop  \otimes \limits_{m = 1}^N \redx{\rho}{m}  = \redx{\rho}{1}  \otimes  \cdots  \otimes \redx{\rho}{N},
\end{Equation}
\vspace{-1pt}%
where $\redx{\rho}{m}$ is the mode-$m$ reduction of $\rho$ (see \App{A}), and the normalization factor is
\begin{Equation}                      {12}
\mathcal{N}_{\Correlance}  \equiv \mathop {\max }\limits_{\forall \rho ' \in \mathcal{H}} [\widetilde{\Correlance}(\rho ')]= \widetilde{\Correlance}(\rho _{\text{ME}} ) = \widetilde{\Correlance}(\rho _{\text{ME}_{\text{TGX}} } ),
\end{Equation}
\vspace{-1pt}%
where \smash{$\rho _{\text{ME}}$} is any maximally entangled (ME) state in \smash{$\mathcal{H}$} and \smash{$\rho _{\text{ME}_{\text{TGX}}}$} is any maximally entangled true-generalized X (TGX) state where the TGX states, first defined in \cite[]{HedX} are generalizations of the Bell states (see \App{H}). The 13-step algorithm of \cite[]{HedE} makes ME TGX states methodically to compute \smash{$\mathcal{N}_{\Correlance}$}, while \Sec{II.B} gives exact forms.%
\vspace{-2pt}%
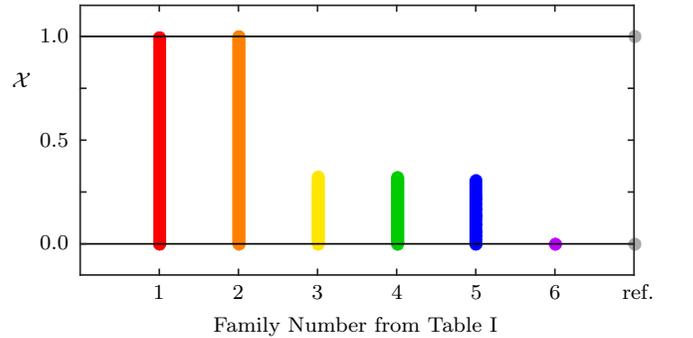
\begin{figure}[H]
\centering
\includegraphics[width=1.00\linewidth]{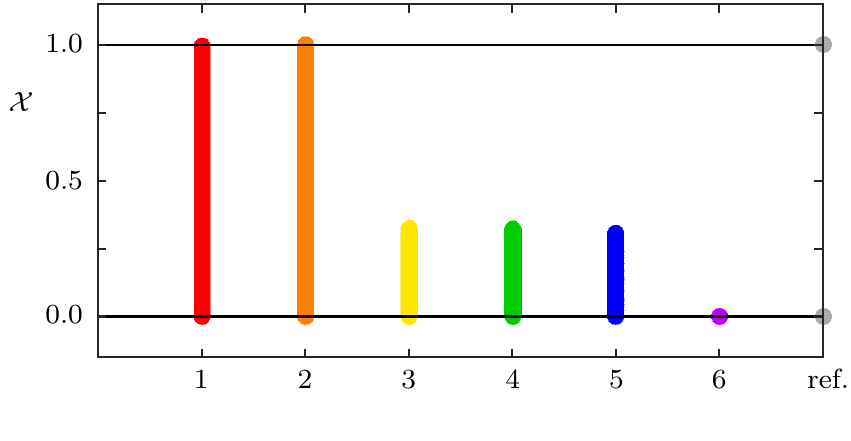}%
\vspace{-13pt}
\setlength{\unitlength}{0.01\linewidth}
\begin{picture}(100,0)
\put(32,0){\footnotesize Family Number from \Table{1}}
\end{picture}
\vspace{-12pt}%
\caption[]{(color online) Correlance of arbitrary two-qubit mixed states for each of the six families of nonlocal correlation from \Table{1}, $10^5$ states each (colors not related to \Fig{1}). This gives a sense of the upper limits of each family, and shows that merely having mode independence in either probabilities or decomposition states alone is not enough to get independent modes; both must have mode independence to get product form and $\Correlance=0$. Families 1 and 2 are generally entangled, Families 3--5 are generally nonlocally correlated without entanglement, and Family 6 has no nonlocal correlation at all.}
\label{fig:2}
\end{figure}
\vspace{-8pt}%

The correlance $\Correlance(\rho)$ is $0$ iff $\rho$ is expressible in product form as $\rho\hsp{-1}=\hsp{-1}\varsigma\hsp{-1}=\hsp{-1}\redx{\rho}{1} \hsp{-1} \otimes \hsp{-1} \cdots  \hsp{-1}\otimes \redx{\rho}{N}$,\hsp{-0.5} meaning\hsp{-0.5} that\hsp{-0.5} all\hsp{-0.5} modes\hsp{-0.5} are\hsp{-0.5} uncorrelated.\hsp{-0.5}  Furthermore,\hsp{-0.5} $\Correlance(\rho)\hsp{-2}=\hsp{-2}1$ iff $\rho$ is\hsp{-0.5} maximally\hsp{-0.5} entangled\hsp{-0.5} (meaning\hsp{-0.5} fully\hsp{-0.5} $N$-partite\hsp{-0.5} entangled\hsp{-0.5} here;\hsp{-0.5} see\hsp{-0.5} \Sec{VII}\hsp{-0.5} for\hsp{-0.5} generalizations),\hsp{-0.5} meaning\hsp{-0.5} that\hsp{-0.5} the\hsp{-0.5} $N$\hsp{-0.5} modes\hsp{-0.5} share\hsp{-0.5} the\hsp{-0.5} most\hsp{-0.5} nonlocal\hsp{-0.5} correlation\hsp{-0.5} possible.\hsp{-0.5} \Figure{2}\hsp{-0.5} explores\hsp{-0.5} the\hsp{-0.5} correlance\hsp{-0.5} for\hsp{-0.5} the\hsp{-0.5} families\hsp{-0.5} of\hsp{-0.5} \Table{1}.

Correlance $\Correlance$ measures all forms of nonlocal correlation; entanglement, decomposition-state correlation (includes entanglement), and  probability correlation.  See \Sec{II} for details, including proofs and numerical tests.
\section{\label{sec:II}TESTS AND PROOFS OF CORRELANCE}
Here, we test the correlance to check its performance and give details on its normalization. For \textit{proof} that correlance is necessary and sufficient for measuring all nonlocal correlation, see \App{I}.%
\vspace{-4pt}%
\subsection{\label{sec:II.A}Normalization Tests of Correlance}
\begin{figure}[H]
\centering
\includegraphics[width=1.00\linewidth]{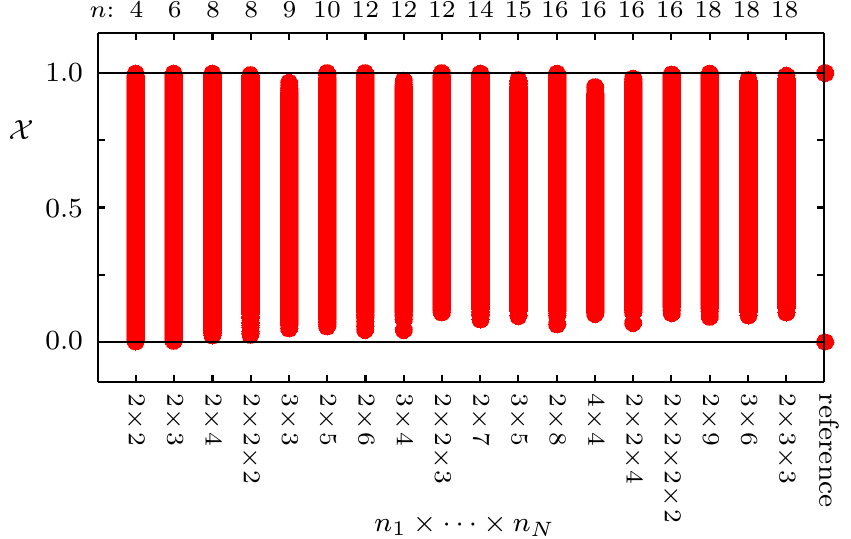}%
\vspace{-5pt}%
\caption[]{(color online) Normalization test of \Eq{9}, the correlance $\Correlance$ of arbitrary general quantum mixed states $\rho$ for all discrete multipartite systems up to $n=18$ levels, $30,\!000$ states for each system. This shows $540,\!000$ \textit{consecutive} examples that do not produce $\Correlance$ higher than $1$, providing strong evidence that $\Correlance$ is not improperly normalized. For \textit{proof} that \Eq{12} is the correct normalization factor, see \App{J}.}
\label{fig:3}
\end{figure}

\Figure{3} shows the results of a necessary test of the normalization of the correlance $\Correlance$, which is that if it is properly normalized, then we should not be able to find any states with a value of $\Correlance$ that exceeds $1$.  The \textit{proof} that $\Correlance$ is properly normalized is given in \App{J}.
\vspace{-16pt}%
\subsection{\label{sec:II.B}Exact Calculation of Correlance Normalization}
The correlance normalization factor for all systems can be computed without using any ME states as
\begin{Equation}                      {13}
\mathcal{N}_{\Correlance} = 1 - \prod\limits_{m = 1}^N {P_{\text{MP}}^{(m)} (L_* )},
\end{Equation}
as proved in \App{K}, where \smash{$P_{\text{MP}}^{(m)}(L_* )$} is the minimum physical reduction purity of mode $m$ given a pure maximally entangled parent state of $L_*$ levels of equal nonzero probability such that the \textit{combination} of all \smash{$P_{\text{MP}}^{(m)}(L_* )$} is minimized, the calculation of which is given in \App{K}.

For systems where more than one mode has the largest mode size \smash{$n_{\max}\equiv\max(\mathbf{n})=\max\{n_{1},\ldots,n_{N}\}$},
\begin{Equation}                      {14}
\mathcal{N}_{\Correlance} = 1 - \frac{1}{n}; \hsp{10}\text{(multiple modes with $n_{\max}$)},
\end{Equation}
for systems such as $2\times 3\times 3$ or $N$-qudit systems like $2\times 2$ or $3\times 3\times 3$. However for systems with \textit{exactly one} largest mode, like $2\times 3$ or $2\times 2\times 3$, we must use \Eq{13} or \Eq{12}.
\begin{table}[H]
\caption{\label{tab:2}Examples of normalization factors $\mathcal{N}_{\Correlance}$ for the correlance of \Eq{9} as computed by \Eq{13}, all of which were checked and found correct by using the 13-step algorithm $\mathcal{A}_{13}$ of \cite[]{HedE} to generate ME TGX states $\rho_{\text{ME}_{\text{TGX}}}$ for use in \Eq{12}.}
\begin{ruledtabular}
\begin{tabular}{|c|c|c|c|@{${\kern -4pt}$}c|c|c|}
$n$ & $n_{1}\times\cdots\times n_{N}$ & $\mathcal{N}_{\Correlance}$ & &{\kern 2pt}$n$ & $n_{1}\times\cdots\times n_{N}$ & $\mathcal{N}_{\Correlance}$ \\[0.5mm]
\cline{1-3} \cline{5-7}
$\vertsp{4}$ & $2\times 2$ & $3/4$ & &{\kern 2pt}$20$ & $2\times 10$ & $3/4$\\[0.5mm]
\cline{1-3} \cline{5-7}
$\vertsp{6}$ & $2\times 3$ & $3/4$ & &{\kern 2pt}$20$ & $4\times 5$ & $15/16$\\[0.5mm]
\cline{1-3} \cline{5-7}
$\vertsp{8}$ & $2\times 4$ & $3/4$ & &{\kern 2pt}$20$ & $2\times 2\times 5$ & $15/16$\\[0.5mm]
\cline{1-3} \cline{5-7}
$\vertsp{8}$ & $2\times 2\times 2$ &$7/8$ & &{\kern 2pt}$21$ & $3\times 7$ & $8/9$\\[0.5mm]
\cline{1-3} \cline{5-7}
$\vertsp{9}$ & $3\times 3$ & $8/9$ & &{\kern 2pt}$22$ & $2\times 11$ & $3/4$\\[0.5mm]
\cline{1-3} \cline{5-7}
$\vertsp{10}$ & $2\times 5$ & $3/4$ & &{\kern 2pt}$24$ & $2\times 12$ & $3/4$\\[0.5mm]
\cline{1-3} \cline{5-7}
$\vertsp{12}$ & $2\times 6$ & $3/4$ & &{\kern 2pt}$24$ & $3\times 8$ & $8/9$\\[0.5mm]
\cline{1-3} \cline{5-7}
$\vertsp{12}$ & $3\times 4$ & $8/9$ & &{\kern 2pt}$24$ & $4\times 6$ & $15/16$\\[0.5mm]
\cline{1-3} \cline{5-7}
$\vertsp{12}$ & $2\times 2\times 3$ & $29/32$ & &{\kern 2pt}$24$ & $2\times 2\times 6$ & $15/16$\\[0.5mm]
\cline{1-3} \cline{5-7}
$\vertsp{14}$ & $2\times 7$ & $3/4$ & &{\kern 2pt}$24$ & $2\times 3\times 4$ & $103/108$\\[0.5mm]
\cline{1-3} \cline{5-7}
$\vertsp{15}$ & $3\times 5$ & $8/9$ & &{\kern 2pt}$24$ & $2\times 2\times 2\times 3$ & $23/24$\\[0.5mm]
\cline{1-3} \cline{5-7}
$\vertsp{16}$ & $2\times 8$ & $3/4$ & &{\kern 2pt}$25$ & $5\times 5$ & $24/25$\\[0.5mm]
\cline{1-3} \cline{5-7}
$\vertsp{16}$ & $4\times 4$ & $15/16$ & &{\kern 2pt}$26$ & $2\times 13$ & $3/4$\\[0.5mm]
\cline{1-3} \cline{5-7}
$\vertsp{16}$ & $2\times 2\times 4$ & $15/16$ & &{\kern 2pt}$27$ & $3\times 9$ & $8/9$\\[0.5mm]
\cline{1-3} \cline{5-7}
$\vertsp{16}$ & $2\times 2\times 2\times 2$ & $15/16$ & &{\kern 2pt}$27$ & $3\times 3\times 3$ & $26/27$\\[0.5mm]
\cline{1-3} \cline{5-7}
$\vertsp{18}$ & $2\times 9$ & $3/4$ & &{\kern 2pt}$28$ & $2\times 14$ & $3/4$\\[0.5mm]
\cline{1-3} \cline{5-7}
$\vertsp{18}$ & $3\times 6$ & $8/9$ & &{\kern 2pt}$28$ & $4\times 7$ & $15/16$\\[0.5mm]
\cline{1-3} \cline{5-7}
$\vertsp{18}$ & $2\times 3\times 3$ & $17/18$ & &{\kern 2pt}$28$ & $2\times 2\times 7$ & $15/16$\\[0.5mm]
\end{tabular}
\end{ruledtabular}
\end{table}
\vspace{-15pt}%
\subsection{\label{sec:II.C}Correlance Adapted for Diagonal-Only Input}
\vspace{-5pt}%
If the input states $\rho$ are definitely always diagonal, we can define the \textit{diagonal correlance} as
\begin{Equation}                      {15}
\Correlance_D (\rho ) \equiv \frac{{\widetilde{\Correlance}(\rho )}}{{\mathcal{N}_{\Correlance_D } }},
\end{Equation}
where \smash{$\widetilde{\Correlance}(\rho )$} is from \Eq{10}, but the normalization is
\begin{Equation}                      {16}
\mathcal{N}_{\Correlance_D }\equiv\mathop {\max }\limits_{\{\rho_D '\} } [\widetilde{\Correlance}(\rho _D ')] = \widetilde{\Correlance}(\rho _{D_{\text{max}}} ),
\end{Equation}
where \smash{$\{\rho_D\}$} is the set of all diagonal states, and \smash{$\rho _{D_{\text{max}}} $} is any diagonal state that maximizes \smash{$\widetilde{\Correlance}(\rho )$}.  

\begin{figure}[H]
\centering
\includegraphics[width=1.00\linewidth]{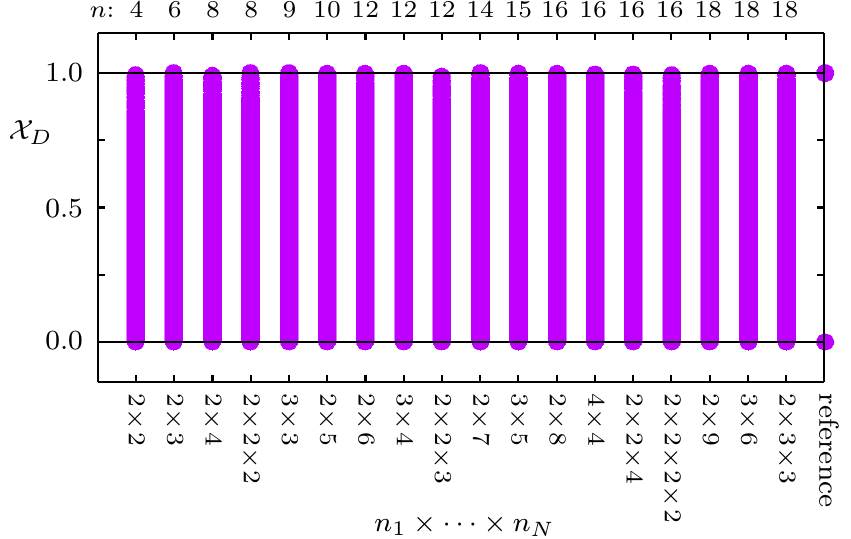}%
\vspace{-5pt}%
\caption[]{(color online) Normalization test of \Eq{15}, the diagonal correlance $\Correlance_D$ of arbitrary general diagonal mixed states $\rho_D$ for all discrete multipartite systems up to $n=18$ levels, $30,\!000$ states for each system. This shows $540,\!000$ \textit{consecutive} examples that do not produce $\Correlance_D$ higher than $1$, providing strong evidence that $\Correlance_D$ is not improperly normalized, where $\mathcal{N}_{\Correlance_D}$ of \Eq{18} was used. For a \textit{sketched proof} that \Eq{17} is a valid \smash{$\rho _{D_{\text{max}}} $}, and a derivation of \smash{$\mathcal{N}_{\Correlance_D }$} in \Eq{18}, see \App{L}.}
\label{fig:4}
\end{figure}

Surprisingly, the states \smash{$\rho _{D_{\text{max}}} $} that maximize \smash{$\Correlance_D$} are \textit{not} merely maximally dephased maximally entangled states in general. Instead, one simple example that maximizes \smash{$\Correlance_D$} in all systems is 
\begin{Equation}                      {17}
\rho _{D_{\max } }  = \frac{1}{2}(|1\rangle \langle 1| + |n\rangle \langle n|),
\end{Equation}
yielding the normalization factor, valid for all systems,
\begin{Equation}                      {18}
\mathcal{N}_{\Correlance_D }=\widetilde{\Correlance}(\rho _{D_{\text{max}}} )=\frac{1}{2}-\frac{1}{2^{N}}.
\end{Equation}
\Figure{4} gives a necessary test showing strong evidence that \Eq{18} is not wrong, while \App{L} gives a \textit{sketched proof} that \Eq{17} is a valid \smash{$\rho _{D_{\max } }$}, and uses it to derive \Eq{18}.

$\Correlance_D$ can be used for strictly classical probability distributions, for quantum states with no coherence, or for diagonal states exhibiting combinations of both classical and quantum features (see \Sec{III}). $\Correlance_D$ is useful because in strictly classical situations, entanglement is not possible, so ME states would not be a reasonable standard to use for normalization.
\vspace{-26pt}%
\subsection{\label{sec:II.D}Diagonal Correlance as a Measure of General Nonlocal Correlation of Strictly Classical Data}
\vspace{-4pt}%
For $N$ classical random variables (RVs), meaning $N$-tuples of variables  \smash{$\mathbf{x}\equiv(x^{(1)},\ldots,x^{(N)})$}, where the data consists of $n_{S}$ samples of $N$-tuple data points $X\equiv\{ \mathbf{x}_j \}  \equiv \{ (x^{(1)}_{j} , \ldots ,x^{(N)}_{j}) \} |_{j = 1}^{n_S }$, we can model the variables as a discrete system of $N$ modes, where mode $m$ represents RV $x^{(m)}$. The size of mode $m$ is $n_m$ (the number of discrete values $x^{(m)}$ can have or its number of histogram bins when viewed as a quantized continuous variable), so the total $N$-mode system has size $n=n_{1}\cdots n_{N}$.

\begin{figure}[H]
\centering
\includegraphics[width=1.00\linewidth]{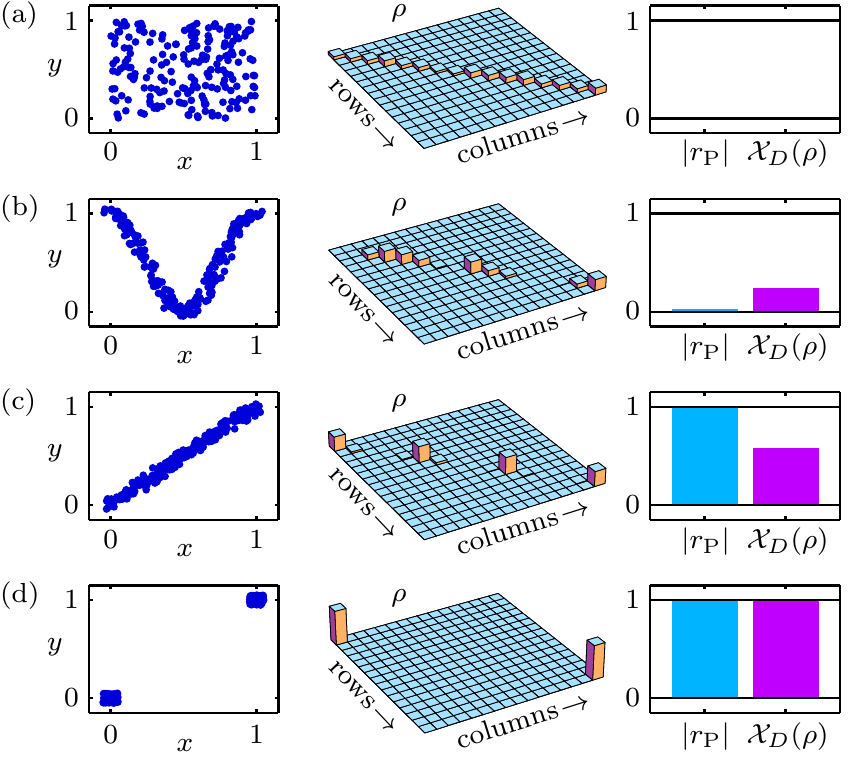}%
\vspace{-2pt}%
\caption[]{(color online) The left column of plots show four bivariate dimensionless data sets of $200$ points for random variables (RVs) $x$ and $y$, each quantized to four outcomes (not shown) before further calculation. Each RV has a uniformly distributed random deviation of $0.05$ about ideal functions: (a) $(x,y)=(\text{rand},\text{rand})$, (b) $(x,y)=(t,\frac{1}{2}[1+\cos(2\pi t)])$, (c) $(x,y)=(t,t)$, and (d) $(x,y)=(\text{round}[t],\text{round}[t])$, where $\text{rand}$ is a random number on $[0,1]$, and $t\in[0,1]$. The center column of plots are the density matrices $\rho$ for the quantized $2$-mode data in single-index form, and the right column of plots shows the magnitude of the Pearson correlation coefficient \smash{$|\Pearson|$} of \App{N} and the diagonal correlance $\Correlance_D$ from \Eq{15}.}
\label{fig:5}
\end{figure}

For simplicity, we will refer to the algorithm for constructing a density matrix $\rho$ from strictly classical data $X$ as $\mathcal{A}_{\rho}\equiv\mathcal{A}_{\rho}(\mathbf{n},X)$, and its output is
\begin{Equation}                      {19}
\rho=\mathcal{A}_{\rho}(\mathbf{n},X);
\end{Equation}
\App{M} shows how to implement $\mathcal{A}_{\rho}$. \Figure{5} shows four examples of bivariate data treated by this method, and compares the traditional Pearson correlation coefficient $\Pearson$ (see \App{N}) \cite[]{Galt,Pear,Devo} to the diagonal correlance $\Correlance_D$.

As seen in \Fig{5}, \Fig[a]{5} is where RVs $x$ and $y$ are completely independent, and both $|\Pearson|\approx 0.01$ and $\Correlance_{D}\approx 0.01$, which is appropriately near zero for a finite sample of uncorrelated variables. In \Fig[b]{5}, the data has some nonlinear correlation, and $|\Pearson|\approx 0.03$, while $\Correlance_{D}\approx 0.24$, showing that $\Correlance_{D}$ correctly detects the presence of nonlocal correlation, while $|\Pearson|$ has trouble detecting it because $|\Pearson|$ is merely a measure of \textit{linear} correlation.  \Figure[c]{5} shows the linear focus of $|\Pearson|$ since the linear data causes $|\Pearson|\approx 0.99$, close to its maximum value of $1$, while $\Correlance_{D}\approx 0.57$, showing that $\Correlance_{D}$ acknowledges this correlation but does not consider it to be maximal.
\vspace{-6pt}%
Finally, \Fig[d]{5} shows data that is indeed more correlated than linear since both $x$ and $y$ tend to have matching values while \textit{also} producing a $\rho$ that does not factor into a product state, giving $|\Pearson|\approx 1.00$ and $\Correlance_{D}\approx 1.00$, showing that $|\Pearson|$ cannot distinguish the data of \Fig[d]{5} from that of \Fig[c]{5}, while $\Correlance_{D}$ correctly detects the difference and values the data of \Fig[d]{5} as being maximally correlated for a diagonal state.

\textit{Important:} States such as $\rho$ in \Fig[d]{5} or \Eq{17} in general only have maximal diagonal correlance for the \textit{set of all diagonal states}, and are not generally maximally correlated in the context of all quantum states, as maximally entangled states are. Thus, if the data comes from a system with novel quantum correlations, application of $\Correlance_{D}$ could yield values greater than unity, even though this is guaranteed not to happen for strictly classical data.

Thus, the diagonal correlance $\Correlance_{D}$ provides a more sensitive measure of correlations in data with strictly diagonal density matrices than the Pearson correlation coefficient, and is not limited to two RVs.
\section{\label{sec:III}CLASSICAL VS. QUANTUM}
Before we can develop other measures capable of distinguishing the types of correlation from \Sec{I.A}, we must first confront the issue of classicality that is raised by the diagonal correlance. Our main questions here are \textit{What does it mean for a state to be diagonal in the context of classicality?} and \textit{Is there a precise and unambiguous definition for classicality?}  

These questions and their answers are extremely relevant in any discussion on quantum correlations that includes quantum discord, since its definition requires that we identify the part of a state that represents ``distinctly classical nonlocal correlation.''  Therefore here, we will make a series of observations, and from these deduce some concrete answers to these questions that will then allow us to construct the desired measures.
\subsection{\label{sec:III.A}Diagonal States Are Not Necessarily Classical}
Quantum superposition is an inherently nonclassical, \textit{distinctly quantum} phenomenon. The idea that an object's state could simultaneously have multiple exclusive outcomes at once is nonsense in any classical model.

For \textit{pure} quantum states, superposition appears in density matrices as \textit{nonzero off-diagonal elements}. For \textit{mixed} quantum states, having nonzero off-diagonals is called \textit{coherence}.\hsp{-1.5}  However, as shown in App.\hsp{2.5}H of \cite[]{HedC}, we can also have \textit{superposition without coherence}.

For example, given a general two-qubit pure parent state expanded as {$|\psi \rangle  \equiv a_1 |1\rangle \! +\! a_2 |2\rangle  \!+\! a_3 |3\rangle \! +\! a_4 |4\rangle$}, where \smash{$|1\rangle  \equiv |1,1\rangle $}, \smash{$|2\rangle  \equiv |1,2\rangle $}, \smash{$|3\rangle  \equiv |2,1\rangle $}, \smash{$|4\rangle  \equiv |2,2\rangle $}, if that state is the Bell state \smash{$|\psi\rangle=\frac{1}{\sqrt{2}}(|1\rangle+|4\rangle)$}, where \smash{$a_1 =\!a_4 \!=\frac{1}{\sqrt{2}}$} and \smash{$a_2 \!=\!a_3 \!=\!0$}, then its density matrix is
\begin{Equation}                      {20}
\begin{array}{*{20}l}
   \rho  &\!\! { =\!|\psi\rangle\langle\psi|\!= \!\!\left( {\begin{array}{*{20}c}
   {a_1 a_1 ^* } &  \cdot  &  \cdot  & {a_1 a_4 ^* }  \\
    \cdot  &  \cdot  &  \cdot  &  \cdot   \\
    \cdot  &  \cdot  &  \cdot  &  \cdot   \\
   {a_4 a_1 ^* } &  \cdot  &  \cdot  & {a_4 a_4 ^* }  \\
\end{array}} \right)\!\!} &\!\! { = \!\!\left( {\begin{array}{*{20}c}
   {\frac{1}{2}} &  \cdot  &  \cdot  & {\frac{1}{2}}  \\
    \cdot  &  \cdot  &  \cdot  &  \cdot   \\
    \cdot  &  \cdot  &  \cdot  &  \cdot   \\
   {\frac{1}{2}} &  \cdot  &  \cdot  & {\frac{1}{2}}  \\
\end{array}} \right)\!,}  \\
\end{array}
\end{Equation}
and therefore its mode-$1$ reduction is \textit{diagonal},
\begin{Equation}                      {21}
\redx{\rho}{1} \!=\!\left(\!\! {\begin{array}{*{20}c}
   {\rho _{1,1}  \!+\! \rho _{2,2} } &\, {\rho _{1,3}  \!+\! \rho _{2,4} }  \\
   {\rho _{3,1}  \!+\! \rho _{4,2} } &\, {\rho _{3,3}  \!+\! \rho _{4,4} }  \\
\end{array}}\!\! \right)\!{\kern -2pt}={\kern -2pt}\!\left(\!\! {\begin{array}{*{20}c}
   {a_1 a_1^* } &\!\! 0  \\
   0 &\!\! {a_4 a_4^* }  \\
\end{array}}\!\! \right)\! = \!\left(\! {\begin{array}{*{20}c}
   {\frac{1}{2}} & \cdot  \\
   \cdot & {\frac{1}{2}}  \\
\end{array}}\! \right)\!,
\end{Equation}
but its \textit{probabilities are directly inherited from the wave-function overlaps of the superposition amplitudes} of its Bell parent state as \smash{$p_1^{(1)} \equiv\redx{\rho}{1}_{1,1}=\rho_{1,1}=a_1 a_1^*$} and \smash{$p_2^{(1)} \equiv\redx{\rho}{1}_{2,2}=\rho_{4,4}=a_4 a_4^*$}, where \smash{$a_k  \equiv \langle k|\psi \rangle$} are wave-function overlaps of basis states \smash{$|k\rangle$} with the pure parent state \smash{$|\psi \rangle$}, which has matrix elements \smash{$\rho_{j,k}=a_j a_k^*$}. Thus, \Eq{21} is an example of \textit{superposition without coherence}.

In contrast, the probabilities of \textit{classical} discrete states are estimators of which outcome to expect from a step function of pure computational basis states on average.  Therefore, in principle, there is a sample-time window which could reveal the true step-function behavior of any truly classical mixed state.

For diagonal \textit{quantum} states such as \Eq{21}, no such time window exists, because \textit{the probabilities inherit the instantaneous nature of superposition from the wave-function overlaps of the pure parent state}, as in \Eq{21}.

A popular misconception is that Fock states \cite[]{Dira} (photon number states $|n\rangle$, not to be confused with our generic basis states used in the rest of this paper) are \textit{nonclassical}. This comes from the preconception that coherent states \cite[]{Gla1,Gla2} are somehow classical simply because they fulfill some necessary conditions for what we think the word ``classical'' should mean, such as having all quantum coherence functions equal $1$ \textit{which only measures similarity to coherent states} \cite[]{Gla1,Gla2,TiGl,GeKn} (see \App{O}).

The problem of misdiagnosing coherent states as classical and then mistakenly using them as standards of classicality is really a \textit{language problem}, as we explain next.%
\vspace{-10pt}%
\subsection{\label{sec:III.B}The Need for a New Term: \textit{Strictly Classical}}%
\vspace{-5pt}%
The main reason for confusion about classicality is that ``classical'' is already a colloquial word in everyday speech.  Similarly, ``work'' has an everyday meaning that has nothing to do with physics, but it also has a well-defined physics-meaning (though it is aptly named).

In physics, all models predating quantum theory were unofficially lumped into a category \textit{colloquially} referred to as ``classical,'' where its use is just an \textit{adjective} describing those models as part of the  pre-quantum mindset.

However, the colloquial use of the word ``classical'' is often \textit{incorrectly interpreted} as a formal physics definition, and this has led to some ridiculous conclusions, such as considering coherent states to be classical even though they have quantum superposition in the Fock basis.

Therefore, since improper use of the word ``classical'' is already ubiquitous, \textit{we need a new term with a clear, formal definition that lists the rules that a physical model must obey in a hypothetical world that is truly without quantum-mechanical properties}.
 
\begin{figure}[H]
\centering
\includegraphics[width=1.00\linewidth]{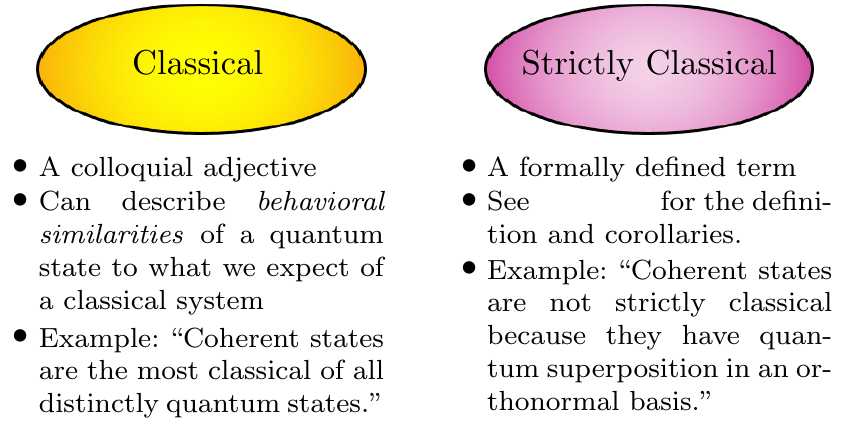}%
\vspace{-12pt}
\setlength{\unitlength}{0.01\linewidth}
\begin{picture}(100,0)
\put(63.1,24.4){\footnotesize \Sec{III.C}}
\end{picture}
\caption[]{(color online) Informal and formal ideas of classicality.}
\label{fig:6}
\end{figure}
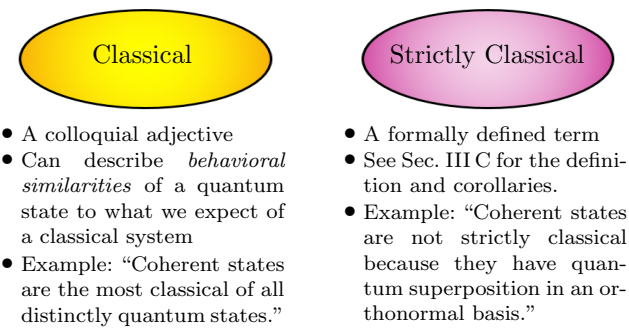

Thus, we propose the term \textit{strictly classical} as being the simplest, most transparent label for this idea that also lets existing literature have the word ``classical'' as either a colloquial adjective or a colloquial term improperly used as a formal physics term. \Figure{6} summarizes these terms, while \Sec{III.C} presents a formal definition of \textit{strict classicality}.
\subsection{\label{sec:III.C}Definition of Strictly Classical States}
Almost all (if not all) of the strange behavior of quantum mechanics comes from its distinctive feature of quantum superposition, the primary idea that caused so much reluctance to accept quantum mechanics, because it \textit{is} at such odds with classical physics.  Therefore we use \textit{the absence of superposition} as the core criterion for what constitutes strict classicality;
\\

{\noindent}\textbf{Strict Classicality:} Let a \textit{strictly classical system} be a hypothetical physical system for which a necessary condition is that quantum superposition plays no role, is impossible, and cannot be created. A system that is strictly classical can be said to have \textit{strict classicality}.
\\

While there may be other conditions we need to identify for strict classicality, the above definition immediately leads to the following rules;%
\begin{itemize}[leftmargin=*,labelindent=4pt]\setlength\itemsep{0pt}
\item[\textbf{1.}]\hypertarget{ClassicalReq:1}{}Strictly classical states must have no coherence (off-diagonals of the density matrix must be zero, so the density matrix is diagonal).
\item[\textbf{2.}]\hypertarget{ClassicalReq:2}{}Strictly classical states must have no diagonal superposition in any complete orthonormal pure product-state basis (no cases of \textit{superposition without coherence} as described in \Sec{III.A} and seen in \Eq{21}).
\item[\textbf{3.}]\hypertarget{ClassicalReq:3}{}The actual instantaneous state of any strictly classical system must be a pure computational basis state.
\item[\textbf{4.}]\hypertarget{ClassicalReq:4}{}\textit{Mixed} strictly classical states (having purity less than $1$) must be averages of step functions of computational basis states, and the lack of purity comes entirely from lack of human knowledge of the instantaneous state (which is always a pure computational basis state).
\item[\textbf{5.}]\hypertarget{ClassicalReq:5}{}(Corollary to \hyperlink{ClassicalReq:3}{Rule 3}) The only transformations possible for a discrete strictly classical system are permutation unitaries; no transformation can convert a strictly classical state to a state with quantum superposition. Thus, strictly classical states are strictly classical in all reference frames, and reference frames themselves are defined only in terms of strictly classical states.
\end{itemize}

Thus, by \hyperlink{ClassicalReq:2}{Rule 2} we must be careful when using the word\hsp{-1} ``classical.''\hsp{-2} A\hsp{-0.5} list\hsp{-0.5} of\hsp{-0.5} probabilities\hsp{-0.5} is\hsp{-0.5} not\hsp{-0.5} enough\hsp{-0.5} information\hsp{-0.5} to\hsp{-0.5} determine\hsp{-0.5} whether\hsp{-0.5} the\hsp{-0.5} system\hsp{-0.5} is\hsp{-0.5} strictly\hsp{-0.5} classical\hsp{-0.5} or\hsp{-0.5} quantum;\hsp{-0.5} we\hsp{-0.5} must\hsp{-0.5} also\hsp{-0.5} be\hsp{-0.5} told\hsp{-0.5} or\hsp{-0.5} find\hsp{-0.5} more\hsp{-0.5} information\hsp{-0.5} about\hsp{-0.5} the\hsp{-0.5} origins\hsp{-0.5} of\hsp{-0.5} the\hsp{-0.5} physical\hsp{-0.5} system\hsp{-0.5} itself.

A strictly classical situation often arises in the macroscopic world, where we happen to \textit{know} that outcomes are already in step functions of particular basis states, and we simply sample them over certain longer time windows and build estimators of the probabilities yielding a diagonal statistical mixture.  In such cases, we can use the diagonal correlance $\Correlance_D$ exclusively, but our physical interpretation of ``nonlocal correlations'' must be due to distinctly nonquantum features, meaning they have nothing to do with superposition and its related effects such as entanglement or relative quantum phase.

\begin{figure}[H]
\centering
\includegraphics[width=1.00\linewidth]{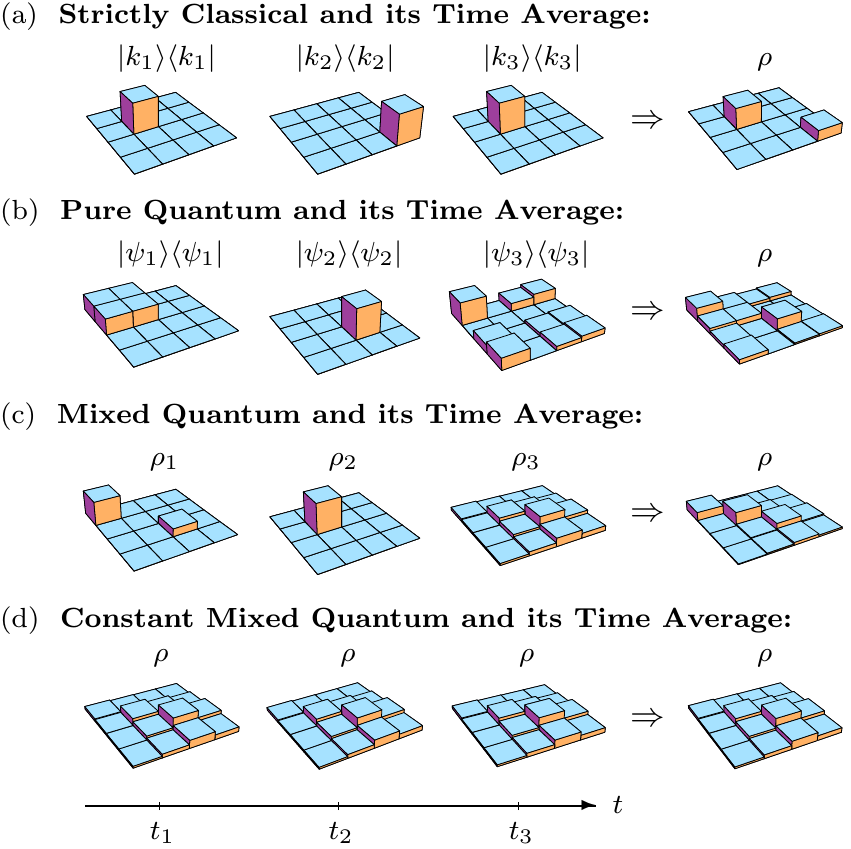}%
\vspace{-6pt}%
\caption[]{(color online) Depiction of density matrices of some types of physical systems (omitting quantum phase for simplicity). The first three columns of plots show the instantaneous state as a step function in time, and the fourth column shows the density matrix as it would be constructed from ideal tomography after using measurement time windows that were much longer than the duration of each step (to show the effect of human-induced mixing). The state in \Eq{21} is a special case of scenario (d) where the constant mixed quantum state is diagonal. In contrast, a strictly classical state can never be instantaneously mixed, as seen in (a). In these examples, the instantaneous states are \textit{ontic}, while the time-averaged states are \textit{epistemic} \cite[]{Spek,PuBR,HaSp}, as defined and discussed in \App{O}.}
\label{fig:7}
\end{figure}

If both superposition \textit{and} classical probability exist, such as a step function of quantum states sampled with a long time window giving a mixture containing both wave-function overlap probabilities and ignorance-based probabilities, our interpretation of $\Correlance$ or $\Correlance_D$ must acknowledge that \textit{both} probability mechanisms are part of the model.  \Figure{7} depicts some types of hybrid quantum/classical states.
\vspace{-4pt}%
\vspace{-10pt}%
At this point, a keen reader might object to nonlocal correlations in strictly classical systems.  However, the concept of \textit{state}, which belongs to both classical and quantum physics, is inherently \textit{global} and therefore nonlocal, so nonlocality is an inherent part of classical physics.  For example, even in a purely classical world, if you sneeze here on Earth, the state of the entire universe changes instantaneously.  Similarly, solutions to the classical heat equation propagate with \textit{infinite} velocity because the \textit{state} is global and therefore changes everywhere at once. Thus, nonlocal correlations are perfectly permissible in \textit{mixed} strictly classical states, because nonlocality is built-in to the concept of state.

Next, we define a few more specific measures which are not practical to calculate, but help to justify several fully computable measures similar to quantum discord.
\section{\label{sec:IV}STATANCE AND PROBABLANCE}%
\vspace{-6pt}%
Here we propose two measures of nonlocal correlation, \textit{statance} and \textit{probablance}, that are unlikely to be practical to calculate in general, but which nevertheless give us sufficient insight into the mechanisms of nonlocal correlation to define three more measures, \textit{strong discordance}, \textit{discordance}, and \textit{diagaonal discordance}, that \textit{are} exactly computable for all mixed and pure states and may be preferable to quantum discord.%
\vspace{-14pt}%
\subsection{\label{sec:IV.A}Statance}
\vspace{-8pt}%
The \textit{statance} $\hat{\Statance}(\rho)$ for any $N$-mode quantum state $\rho$ is 
\begin{Equation}                      {22}
\hat{\Statance}(\rho ) \equiv \frac{1}{{\mathcal{N}_{\Statance} }}\mathop {\min }\limits_{\{ U\} } [\Statance(\rho )],
\end{Equation}
over all unitary matrices $U$ of \smash{$D\in r,\ldots, r^2$} levels where \smash{$r\equiv\text{rank}(\rho)$}, \smash{$\mathcal{N}_{\Statance}$} normalizes \smash{$\min_{\{U\}}[\Statance(\rho )]$} over all $\rho$, and $\Statance$ is the \textit{unoptimized statance},
\begin{Equation}                      {23}
\Statance(\rho ) \equiv\hsp{-26} \sum\limits_{\hsp{26}j_1 , \ldots ,j_N =1,\ldots,1}^{D_{1},\ldots,D_{N}} {\hsp{-26}\hsp{-2}\tr [(\rho _{(j_1 , \ldots ,j_N )}  - \mu _{(j_1 , \ldots ,j_N )} )^2 ]}\hsp{0.5} ,
\end{Equation}
where \smash{$D_{m}\hsp{-0.5}\in\hsp{-0.5} 1,\ldots,n_{m}^2\text{~s.t.~}\hsp{-1.5}\prod\nolimits_{m=1}^{N}D_{m}\hsp{-0.5}=\hsp{-0.5}D$} (see \App{P}), and \smash{$\mu _j  \equiv \mu _{(j_1 , \ldots ,j_N )}$} are
\begin{Equation}                      {24}
\mu _j \equiv\hsp{-1} \mathop  \otimes \limits_{m = 1}^N \hsp{-1}\tr_{\mbarsub} \hsp{-2}\left(\hsp{-1.5} {\frac{1}{{D_{\mbarsub} }}\hsp{-1}\hsp{-26}\sum\limits_{\hsp{26}j_1^{\{ m\} } , \ldots ,j_N^{\{ m\} } =1,\ldots,1}^{D_{1},\ldots,D_{N}} {\hsp{-26}\hsp{-3}\delta _{j_m^{\{ m\} } ,j_m } \rho _{(j_1^{\{ m\} } , \ldots ,j_N^{\{ m\} } )} } }\hsp{-1.5} \right)\hsp{-2} ,
\end{Equation}
where \smash{$D_{\mbarsub}\equiv\frac{D}{D_{m}}$}, \smash{$\tr_{\mbarsub}(A)$} is the partial trace of $A$ over all modes that are \textit{not} mode $m$ (see \App{A}), superscripts $\{m\}$ in \smash{$j^{\{ m\} }\equiv(j_1^{\{ m\} } , \ldots ,j_N^{\{ m\} } )$} distinguish indices inside each factor of the tensor product, and the pure decomposition states \smash{$\rho _j  \equiv \rho _{(j_1 , \ldots ,j_N )} $} are
\begin{Equation}                      {25}
\rho _j  \equiv \frac{1}{{p_j }}\sum\limits_{k,l = 1,1}^{r,r} {\! U_{j,k} U_{j,l}^* \sqrt {\lambda _k \lambda _l } |e_k \rangle \langle e_l |},
\end{Equation}
where \smash{$\{\lambda_{k},|e_k \rangle\}$} are the eigenvalues and eigenstates of $\rho$ \smash{$\text{s.t.~}\lambda_{1}\geqslant\cdots\geqslant\lambda_{r}>0$}, and \smash{$ \sum\nolimits_{k  = 1}^{r } { |U_{j,k} |^2 }\neq 0\;\forall j$} as explained in \App{P}, and \smash{$p_j  \equiv \sum\nolimits_{k  = 1}^{r } {\lambda _{k} |U_{j,k} |^2 }$}. See \cite[]{HedU} for a useful parameterization of unitary matrices.

Basically, $\mu_j$ uses the $\rho_j$ to form states [inside the partial trace in \Eq{24}] constructed to have mode-independent (MI) \textit{probabilities} (due here to the fact that they are all equal) so that the only way the tensor product of the mode-$m$ reductions of those states could also be MI would be if all the $\rho_j$ were MI (see \App{Q}).

The statance \smash{$\hat{\Statance}(\rho)=0$} iff $\rho$ has a decomposition where all of the decomposition states \smash{$\rho_j$} have MI product form as \smash{$\rho _j  =  \otimes _{m = 1}^N \rho _{j_m }^{(m)} = $} \smash{$\rho _{j_1 }^{(1)}  \otimes  \cdots  \otimes \rho _{j_N }^{(N)}\;\;\forall j$}, so \smash{$\hat{\Statance}(\rho)=0$} is guaranteed in Families $5$ and $6$ of \Table{1}.\hsp{-2} If \smash{$\hat{\Statance}(\rho)\hsp{-1}>\hsp{-1}0$}, then $\rho$ has decomposition-state correlation, meaning that it has no decomposition for which all \smash{$\rho_j$} have both product form and mode independence, so its optimal decomposition has at least one \smash{$\rho_j$} that violates either \Eq{5} and \Eq{6} or violates just \Eq{6}.  Thus, statance \smash{$\hat{\Statance}(\rho)$} is a necessary and sufficient measure of decomposition-state correlation (see \App{Q} for \textit{proof}, and \App{R} for a special example).
\vspace{-12pt}%
\subsection{\label{sec:IV.B}Probablance}
\vspace{-8pt}%
Our goal here is to define a measure of probability correlation, but there are several fine points to consider.
\begin{itemize}[leftmargin=*,labelindent=4pt]\setlength\itemsep{0pt}
\item[\textbf{1.}]\hypertarget{ProbablanceFinePoints:1}{}Recalling that Family 5 of \Table{1}, with defining form \smash{$\rho=\sum\nolimits_j p_{j}\rho_{j_{1}}^{(1)}\otimes\cdots\otimes\rho_{j_{N}}^{(N)}$}, has the ability to yield $\Correlance(\rho)>0$ in \Fig{2}, this shows that there are indeed some states whose \textit{only} type of correlation is probability correlation (since those states were \textit{constructed} with MI decomposition states so they have no decomposition-state correlation).  Moreover, that means that for Family 5 only, $\Correlance(\rho)>0$ \textit{exclusively indicates probability correlation}.
\item[\textbf{2.}]\hypertarget{ProbablanceFinePoints:2}{}\textit{Probability correlation cannot be judged by itself over all decompositions}, because it is always possible to make a decomposition with equal probabilities which are therefore MI and have no probability correlation, as we will show in the next few paragraphs.
\item[\textbf{3.}]\hypertarget{ProbablanceFinePoints:3}{}The fact that Family 5 has MI decomposition states and yet can have $\Correlance(\rho)>0$ means that to get the defining optimal form of that family with respect to probability correlation, \textit{we must first find decompositions that minimize the decomposition-state correlation}, and \textit{then} minimize over the probability sets of \textit{those} decompositions to judge probability correlation. Thus, for Family 5, since this restriction causes us to only consider decompositions with MI decomposition states when calculating the probabilities, then by \hyperlink{ProbablanceFinePoints:1}{Fine Point 1}, if $\Correlance(\rho)>0$, we are \textit{guaranteed} to be \textit{unable} to find MI probabilities from that set of decompositions.
\end{itemize}
\vspace{-2pt}%

Keeping the above in mind, we define the \textit{probablance} for any $N$-mode quantum state $\rho$ as
\vspace{-2pt}%
\begin{Equation}                      {26}
\hat{\Probablance}(\rho )\equiv \frac{1}{{\mathcal{N}_\Probablance }}\mathop {\min }\limits_{\{ U^{\prime}\} } [\Probablance(\rho )];\hsp{6}\{ U^{\prime}\}\equiv\text{arg}\left({\mathop {\min }\limits_{\{ U\} } [\Statance(\rho )]}\right),
\end{Equation}
where \smash{$\{ U^{\prime}\}$} is the set of all unitaries that minimize $\Statance(\rho )$ from \Eq{23}, $\mathcal{N}_\Probablance$ normalizes over all $\rho$, and $\Probablance(\rho )$ is the \textit{unoptimized probablance},
\vspace{-4pt}%
\begin{Equation}                      {27}
\Probablance(\rho ) \equiv \sum\limits_{j=1}^{D} {\left| {p_{j}^{\prime}  - q_{j}^{\prime} } \right|^2 } ,
\end{Equation}
where $D$ and $j$ are as in \Sec{IV.A}, and $q_{j}^{\prime}\equiv q_{(j_1 , \ldots ,j_N )}^{\prime}$ is%
\begin{Equation}                      {28}
\begin{array}{*{20}l}
   {q_{(j_1 , \ldots ,j_N )}^{\prime}} &\!\! {\equiv\prod\limits_{m = 1}^N \redxpreprime{\varrho}{m}_{j_{m},j_{m}};\hsp{10}\varrho^{\prime}\equiv\sum\limits_{l=1}^{D}p_{l}^{\prime}|l\rangle \langle l|}  \\
   {} &\!\! {=\prod\limits_{m = 1}^N {\hsp{-3}\left( {\rule{0pt}{20pt}\hsp{-1}} \right.\hsp{-29}\sum\limits_{\hsp{29}\rule{0pt}{8pt}j_1^{\{ m\} } , \ldots ,\,j_N^{\{ m\} } =1,\ldots,1}^{D_{1},\ldots,D_{N}} {\hsp{-29}\delta _{j_m^{\{ m\} }\hsp{-1} ,\hsp{1}j_m } p_{(j_1^{\{ m\} } , \ldots ,j_N^{\{ m\} } )}^{\prime} } \left. {\rule{0pt}{20pt}\hsp{-3}} \right)},}  \\
\end{array}
\end{Equation}
where $|l\rangle\equiv|l_{1}\rangle\otimes\cdots\otimes|l_{N}\rangle$ are computational basis states in a Hilbert space of mode sizes  $\mathbf{n}^{\prime}\equiv(D_{1},\ldots,D_{N})$, and
\begin{Equation}                      {29}
p_j^{\prime}\equiv p_{(j_1 , \ldots ,j_N )}^{\prime}  \equiv \sum\limits_{k= 1}^{r } {\lambda _{k} |U^{\prime}_{(j_1 , \ldots ,j_N ),k} |^2 }\hsp{-1} ,
\end{Equation}
where $U^{\prime}$ is defined in \Eq{26}, based on $U$ from \Sec{IV.A}.

To see how \smash{$\hat\Probablance(\rho)$} works, $\varrho'$ in \Eq{28} creates a $D$-level diagonal state that automatically has $D$ MI decomposition states so that the only way $\varrho'$ can have product form is if its diagonal elements are MI, and those are the probabilities of a given decomposition specified by \smash{$U^{\prime}$}. The quantity $q_j^{\prime}$ then exploits this fact by making a product that  equals $p_j^{\prime}$ iff the set of $p_j^{\prime}$ is MI (see \App{S} for \textit{proof}).

The reason we cannot simply ignore statance in this definition is that if we \textit{did} ignore it, we could just choose $U$ to be an $r$-dimensional Fourier matrix, and then \Eq{29} would yield $p_j^{\prime} =\frac{1}{r}$ for $j\in 1,\ldots,r$ (so $D=r$), and since any set of equal probabilities is automatically MI, that would yield $\Probablance(\rho)=0$ for all $\rho$, which cannot be true for any valid measure of probability correlation, because the nonzero correlance of some states in Family 5 demonstrates that probability correlation exists.

Therefore, we must first find the particular $\{U^{\prime}\}$ that minimize $\Statance(\rho)$ from \Eq{23} and \textit{then} find a \textit{particular} $U^{\prime}$ that minimizes $\Probablance(\rho)$ to define the $p_j^{\prime}$ for $\hat\Probablance(\rho)$.

Probablance \smash{$\hat\Probablance (\rho)$} measures how far the decomposition probabilities are from having mode-independent factorizability for a decomposition that minimizes the decomposition-state correlation.  For a given $\rho$ with decompositions \smash{$\rho  = \sum\nolimits_j {p_j } \rho _j $}, we get \smash{$\hat{\Probablance}(\rho)=0$} iff there exists a decomposition-state-correlation-minimizing decomposition for which \smash{$p_j  = \prod\nolimits_{m = 1}^N {p_{j_{m}}^{(m)}  = p_{j_{1}}^{(1)}  \cdots p_{j_{N}}^{(N)} }\;\forall j$}, so \smash{$\hat{\Probablance}(\rho)=0$} is guaranteed for Families $2$, $4$, and $6$ of \Table{1}. If \smash{$\hat{\Probablance}(\rho)>0$}, then there is no decomposition-state-correlation-minimizing decomposition of $\rho$ for which all $p_j$ have mode-independent product form.  Thus, \smash{$\hat\Probablance (\rho)$} is a valid measure of probability correlation as defined in \Eqs{7}{8} (see \App{S} for \textit{proof}, and \App{T} for an example).

\textit{Caution:} While it is tempting to call $\hat\Probablance(\rho )$ the ``classical correlation,'' that would be inaccurate because although strictly classical states must have no superposition and are therefore diagonal in the computational basis, there are \textit{quantum} states of the same form with inherently different meaning, as explained in \Sec{III}.  Therefore, the correlation measured by $\hat\Probablance(\rho )$ can either be classical, quantum, or a composite of both depending on the physical system described by $\rho$, so we just call this correlation the ``probability correlation,'' since that is the most transparent term, and makes no extra assumptions.

For \textit{strictly classical} states, since they are step functions of computational basis states according to \hyperlink{ClassicalReq:3}{Rule 3} in \Sec{III.C}, \textit{only decompositions for which all decomposition states have no superposition in the computational basis are allowed}, so the only permissible decompositions have $D=r$, $U$ is an $r$-level \textit{permutation unitary}, and $\rho$'s eigenstates are required to be the computational basis sates. The reason $D=r$ is because $U$ is limited to a permutation unitary (because superposition is not allowed), which would lead to some $p_j =0$ if $D>r$, and that would cause our measures to include $\rho_j$ that do not contribute to $\rho$. Thus, for \textit{strictly classical states}, statance \smash{$\hat\Statance(\rho)$} only considers the finite set of $r!$ permutation unitaries of dimension $r$ as \smash{$\{U\}$} and probablance \smash{$\hat\Probablance(\rho)$} only considers the \smash{$\{U^{\prime}\}$} from that set that minimize statance.
\subsection{\label{sec:IV.C}Observations about Statance and Probablance}
\vspace{-8pt}%
Here, we use statance and probablance to make simple observations and do some special-case examples to lay the groundwork for a new measure that is comparable to quantum discord, but with several advantages, thereby showing statance and probablance to be worthwhile measures despite the difficulties in their calculation.
\vspace{7pt}\\
\ThmLab{2}Entanglement is merely sufficient for causing \smash{$\hat\Statance(\rho)>0$}, but is not necessary for that [i.e., states with \smash{$\hat\Statance(\rho)>0$} do not all have entanglement, but all states with entanglement have \smash{$\hat\Statance(\rho)>0$}]. \textit{Proof:} By definition, decomposition states of an entangled state cannot all achieve product form, so they also cannot achieve mode independence as in \Eq{6}, and therefore entangled states can never have \smash{$\hat\Statance(\rho)=0$}, which proves that entanglement is sufficient to cause \smash{$\hat\Statance(\rho)>0$}. The fact that there exist separable states \smash{$\rho  = \sum\nolimits_j {(p_{j_{1}}^{(1)}  \cdots p_{j_{N}}^{(N)}) \rho _j^{(1)}  \otimes  \cdots  \otimes \rho _j^{(N)} } $} which are not entangled by definition, but lack any decomposition states with mode independence (so they have \smash{$\hat\Statance(\rho)>0$} as in \Fig{2} for Family 4 [for which \smash{$\Correlance(\rho)>0$} implies \smash{$\hat\Statance(\rho)>0$}]), proves that entanglement is not necessary to achieve \smash{$\hat\Statance(\rho)>0$}. (Thus, entanglement is merely a special \textit{type} of decomposition-state correlation.)
\vspace{7pt}\\
\ThmLab{3}Entanglement is the only kind of nonlocal correlation that pure states can have. \textit{Proof:} If $\rho$ is \textit{pure}, then \smash{$\hat\Statance(\rho)$} would only measure entanglement because all pure states have no probability correlation (since its only decomposition is it itself with probability $1$), and any decomposition-state correlation (due to mode-independence violation) in a pure state is the same as violation of product form since its optimal decomposition is itself up to global phase, and violation of product form is the definition of entanglement in a pure state.
\vspace{7pt}\\
\indent All of this prompts the question: \textit{Can a strictly classical state violate mode independence with its decomposition states alone?  In other words, are there (diagonal) strictly classical states with \smash{$\hat\Probablance = 0$} but \smash{$\hat\Statance>0$}?}

The answer is \textit{yes}, but it depends on \textit{which} basis states correspond to nonzero eigenvalues, and is limited by $r$. 

For a strictly classical two-qubit state $\rho$, the case of $r=1$ always involves a single pure computational basis state, which always has product form, so \smash{$\hat\Statance=0$} \textit{and} \smash{$\hat\Probablance = 0$}.  For $r=4$, the complete set of computational basis states will always form a mode-independent (MI) set for some of the permutation unitaries $U$, so \smash{$\hat\Statance=0$} while \smash{$\hat\Probablance \geqslant 0$}.

For $r=2$ and $r=3$, the strictly-classical limitation of $D=r$ means that the probabilities will always appear MI, since for example a general set of $\{p_1, p_2 ,p_3\}$ can be interpreted as \smash{$\{p_1^{(1)}p_1^{(2)} ,p_1^{(1)}p_2^{(2)} , p_1^{(1)}p_3^{(2)}\}$}, where \smash{$p_1^{(1)}=1$} and \smash{$\sum_{j_2 =1}^{3}p_{j_2}^{(2)}=1$} where $\mathbf{D}=(1,3)$ since $D_1 D_2 =D$.  Therefore for $r=2$ and $r=3$ we get \smash{$\hat\Probablance = 0$} always.  

But for \textit{statance}, in $r=3$ no group of three computational basis states can form an MI set and all $3$-level permutation unitaries yield the same result. For example, suppose the decomposition states involved are
\begin{Equation}                      {30}
\begin{array}{*{20}l}
   {\rho _1 } &\!\! { \equiv \rho _{(1,1)} } &\!\! { = |1\rangle \langle 1| \otimes |1\rangle \langle 1|}  \\
   {\rho _2 } &\!\! { \equiv \rho _{(1,2)} } &\!\! { = |1\rangle \langle 1| \otimes |2\rangle \langle 2|}  \\
   {\rho _3 } &\!\! { \equiv \rho _{(1,3)} } &\!\! { = |2\rangle \langle 2| \otimes |1\rangle \langle 1|.}  \\
\end{array}
\end{Equation}
\vspace{-8pt}\\%
Then, putting \Eq{30} into \Eq{24} gives (for the 3-level \smash{$U=I$}),
\begin{Equation}                      {31}
\begin{array}{*{20}l}
   {\mu _1 } &\!\! { = \frac{2}{3}|1\rangle \langle 1| \otimes |1\rangle \langle 1| +\! \frac{1}{3}|2\rangle \langle 2| \otimes |1\rangle \langle 1|}  \\
   {\mu _2 } &\!\! { = \frac{2}{3}|1\rangle \langle 1| \otimes |2\rangle \langle 2| +\! \frac{1}{3}|2\rangle \langle 2| \otimes |2\rangle \langle 2|}  \\
   {\mu _3 } &\!\! { = \frac{2}{3}|1\rangle \langle 1| \otimes |1\rangle \langle 1| +\! \frac{1}{3}|2\rangle \langle 2| \otimes |1\rangle \langle 1|,}  \\
\end{array}
\end{Equation}
\vspace{-4pt}%
which, when put into \Eq{23} and \Eq{22} yields%
\vspace{-2pt}%
\begin{Equation}                      {32}
\hat \Statance(\rho ) \equiv \frac{1}{{\mathcal{N}_{\Statance} }}\mathop {\min }\limits_{\{ U\} } [\Statance(\rho )] = \frac{4}{3{\mathcal{N}_{\Statance} }} > 0,
\end{Equation}
\vspace{-6pt}\\%
since \smash{$\Statance(\rho)=\frac{4}{3}$} here for all $3$-level permutation unitaries, and thus we get another powerful result:%
\vspace{5pt}\\%
\ThmLab{4}Strictly classical states \textit{can} have nonzero statance even though all of the decomposition states are computational basis states.
\vspace{5pt}\\
\indent For\hsp{-1} $r\!=\!2$,\hsp{-1} the\hsp{-1} factorizability\hsp{-1} of\hsp{-1} the\hsp{-1} set\hsp{-1} of\hsp{-1} basis\hsp{-1} states\hsp{-1} belonging\hsp{-1} to\hsp{-1} nonzero\hsp{-1} eigenvalues\hsp{-1} determines\hsp{-1} the\hsp{-1} statance.\hsp{-1} For example if $\rho\!=\!p_{1}|1\rangle \langle 1| \otimes |1\rangle \langle 1| \!+\!p_{2}|1\rangle \langle 1| \otimes |2\rangle \langle 2|$, then \smash{$\hat\Statance\!=\!0$} since the mode-$1$ state is a common factor, but if $\rho\!=\!p_{1}|1\rangle \langle 1| \otimes |1\rangle \langle 1| \!+\!p_{2}|2\rangle \langle 2| \otimes |2\rangle \langle 2|$, then \smash{$\hat\Statance\!>\!0$}.

However, it is important to acknowledge that \textit{mixed} strictly classical states are by necessity \textit{epistemic}, so that the probabilities strictly represent observer ignorance of the actual \textit{ontic} state.  Since the ontic state of a strictly classical system is always pure, it can have no probability correlation, and since it is also a computational basis state, then it has no entanglement, and by \Thm{3} it also has no decomposition-state correlation.  Therefore we can summarize all of this as:
\vspace{5pt}\\
\ThmLab{5}All ontic strictly classical states have no nonlocal correlation of any kind. (Where again, \textit{ontic} means the actual state, which is instantaneously pure as defined in \Sec{III.C}.) \textit{Proof:} See the above text.
\vspace{5pt}\\
\ThmLab{6}Mixed strictly classical states (which are epistemic by necessity) \textit{can} have nonlocal correlation as any combination of decomposition-state correlation and probability correlation, but can never have entanglement correlation, and all of this nonlocal correlation is strictly observer-induced. (Again, \textit{epistemic} means that this is a state of observer knowledge, so any mixture of a strictly classical state is induced by the observer's ignorance of the actual state.) \textit{Proof:} See the text before \Thm{5}. Note that since strict classicality forbids superposition, reduction cannot induce mixture from \textit{pure} strictly classical states; it can only remix the pre-existing observer-induced mixture of a \textit{mixed} strictly classical parent state.%
\vspace{5pt}\\
\indent \Thm{5} is exactly what we should expect of the ontic state of a strictly classical system; nonlocality is impossible.  \Thm{6} is significant because it shows that nonlocality can arise in \textit{mixed} strictly classical states.

However, while these theorems are true for strictly classical states (which are all diagonal) they do not generally also apply to diagonal quantum states (for which general decomposition unitaries are allowed in statance and probablance calculations, not merely permutation unitaries).

So far, we have developed one general measure of nonlocal correlation (the correlance $\Correlance$), and two measures of specific kinds of nonlocal correlation (statance \smash{$\hat\Statance$} for decomposition-state correlation, and probablance \smash{$\hat\Probablance$} for probability correlation). For entanglement correlation, we can use any valid entanglement measure, such as the ent $\Upsilon$ \cite[]{HedE} or its generalization for distinguishing distinctly different types of multipartite entanglement as the ent-concurrence \smash{$\Upsilon_C$} \cite[]{HedC}. Thus, we have measures for each of the main mechanisms of nonlocal correlation from \Sec{I.A}, although among these measures, only the correlance $\Correlance$ is computable for mixed states at this time.

Now we return to the essential question that prompted the idea of quantum discord $\mathcal{Q}$; \textit{is it possible to construct a measure of nonlocal correlation that can distinguish between quantum and classical nonlocal correlation?}
\vspace{-12pt}%
\section{\label{sec:V}STRONG DISCORDANCE, DISCORDANCE, AND DIAGONAL DISCORDANCE}
\vspace{-8pt}%
Before we can answer the question at the end of \Sec{IV.C}, we need to define what we mean by \textit{classical correlations}.  Generalizing from Werner \cite[]{Wern}, any fully $N$-partite separable state in the form of \Eq{4} is ``classically correlated,'' however we \textit{cannot} accept this as accurate terminology here because separable quantum states in general fail the requirements for being strictly classical states in \Sec{III.C}, even when such states are diagonal, as explained in \Sec{III}.  Instead, we must recognize that \textit{strictly classical states are statistical mixtures built up from estimators of probabilities generated from using measurement time windows that are too long to show the true nature of the classical system as a step function of pure computational basis states, and contain no quantum superposition in any reference frame.} In other words, \textit{the most general truly classical state is represented by a strictly classical epistemic mixed diagonal density matrix.}

However, since ontic \textit{quantum} states can have the same form as epistemic strictly classical mixed states (with the key difference that \textit{no} choice of measurement time window could reveal a quantum state to be a step function of pure computational basis states unless that were the reality of the system) then we get another theorem:
\vspace{7pt}\\%
\ThmLab{7}\hsp{-2.5}Decomposition-state correlation and probability correlation are \textit{not exclusively} strictly-classical phenomena. \textit{Proof:} There exist ontic quantum states of the same form as epistemic strictly classical mixed states but with inherently different meaning that nevertheless have nonentanglement-decomposition-state correlation and/or probability correlation as their only source of nonlocal correlation, such as the diagonal two-qubit reductions of some fully $N$-partite entangled $N$-qubit state [similar to \Eq{21} but where the reduction is a two-qubit system]. In other words; \textit{reductions of ontic quantum states can lead to ontic quantum mixed states of the same form as epistemic strictly classical mixed states}.
\subsection{\label{sec:V.A}Strong Discordance: A Measure of Nonlocal Correlation Beyond that Achievable by a Strictly Classical State} 
\vspace{-8pt}%
Despite the fact that \Thm{4}, \Thm{6}, and \Thm{7} seem to indicate that quantum states cannot be so easily distinguished from strictly classical states, we can still use these theorems as motivation to create a measure that reveals whether a quantum state has correlations \textit{beyond those of a strictly classical strength}, which we will refer to as \textit{distinctly non-strictly-classical nonlocal correlation}, or more compactly, \textit{distinctly nonclassical nonlocal correlation}, or just \textit{distinctly quantum nonlocal correlation}. (Essentially, this means that we acknowledge \Thm{7} but will still consider \textit{any} nonlocal correlation in quantum states that is not greater than the correlation achievable by a strictly classical state to have classical \textit{strength} regardless of the inherently quantum physical origins that cause it.)

Therefore we define the \textit{strong discordance} of $\rho$ as a measure of  nonlocal correlation beyond that achievable by a strictly classical state, given by%
\vspace{-2pt}%
\begin{Equation}                      {33}
\StrongDiscordance(\rho ) \equiv \frac{1}{{\mathcal{N}_{\StrongDiscordance} }}\max \{ 0,\Correlance (\rho ) - \Correlance (\rho _{D_{\max } } )\} ,
\end{Equation}
\vspace{-8pt}\\%
where $\Correlance(\rho )$ is the correlance from \Eq{9}, with normalization%
\vspace{-2pt}%
\begin{Equation}                      {34}
\mathcal{N}_{\StrongDiscordance}  \equiv \mathop {\max }\limits_{\{ \rho '\} } \{ \Correlance (\rho ') - \Correlance (\rho _{D_{\max } } )\}  = 1 - \frac{2^{N-1}-1}{2^{N}\mathcal{N}_{\Correlance}},
\end{Equation}
\vspace{-8pt}\\%
where $\rho _{D_{\max } }$ are any \textit{diagonal} states that maximize $\DiagonalCorrelance(\rho )$ such as \Eq{17}, and \smash{$\Correlance(\rho _{D_{\max } } ) = \widetilde{\Correlance}(\rho _{D_{\max } } )/\mathcal{N}_{\Correlance}$}, where \smash{$\widetilde{\Correlance}(\rho _{D_{\max } } )=\mathcal{N}_{\Correlance _D }  = \frac{1}{2} - \frac{1}{{2^N }}$} from \Eq{18}, and $\mathcal{N}_{\Correlance}$ is from \Eq{12}, \Eq{13}, or \Eq{K.57}, so then \Eq{33} becomes%
\vspace{-2pt}%
\begin{Equation}                      {35}
\StrongDiscordance(\rho ) = \frac{2^{N}\mathcal{N}_{\Correlance}}{{2^{N}\mathcal{N}_{\Correlance}-2^{N - 1}  + 1}}\max\! \left\{ {0,\Correlance (\rho ) - \frac{2^{N-1}-1}{2^{N}\mathcal{N}_{\Correlance}}} \right\}\!.
\end{Equation}
Thus, \Eq{35} gives strong discordance in exactly computable form for all pure and mixed $N$-partite states. 

The definition of $\StrongDiscordance(\rho)$ is such that the only states that can attain $\StrongDiscordance(\rho)>0$ are those with correlance $\Correlance(\rho)$ that \textit{exceeds} that of the correlance-maximizing diagonal states $\rho _{D_{\max } }$ (since those have the form of the most correlated strictly classical states). Thus, strong discordance $\StrongDiscordance(\rho)$ is a measure of how much nonlocal correlation exists in $\rho$ \textit{beyond} that achievable by a mixed strictly classical state.

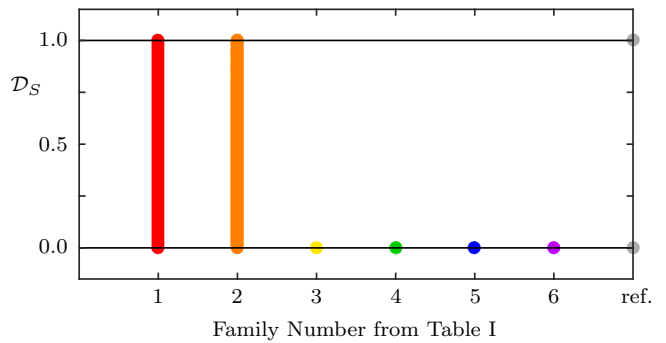
\begin{figure}[H]
\centering
\includegraphics[width=1.00\linewidth]{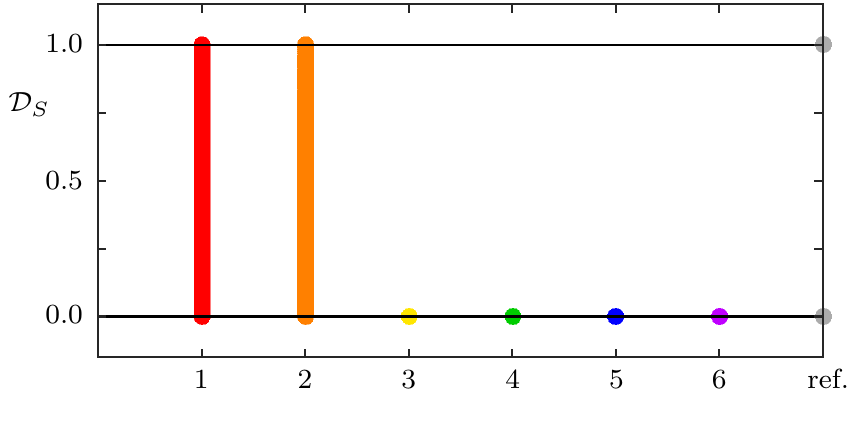}%
\vspace{-13pt}
\setlength{\unitlength}{0.01\linewidth}
\begin{picture}(100,0)
\put(32,0){\footnotesize Family Number from \Table{1}}
\end{picture}
\vspace{-12pt}%
\caption[]{(color online) Strong discordance of arbitrary two-qubit mixed states for each of the six families of nonlocal correlation of \Table{1}, $10^5$ states each (colors not related to \Fig{1}). This suggests that Families 3--6 (all separable) have $\StrongDiscordance=0$, and only if we allow entanglement, as in Families 1 and 2, can states achieve the full range of strong discordance up to $\StrongDiscordance=1$. Thus, Werner's observation is \textit{valid}, but should be restated as ``nonlocal correlation in all separable states is not stronger than that of \textit{strictly classical mixed states}.''}
\label{fig:8}
\end{figure}

$\StrongDiscordance(\rho)=0$ iff there exists a mixed strictly classical state that can achieve the same amount of nonlocal correlation as $\rho$.  Similarly, if $\StrongDiscordance(\rho)>0$, then there is definitely some nonlocal correlation that is distinctly quantum and not achievable by a strictly classical state, and $\StrongDiscordance(\rho)=1$ iff $\rho$ is maximally entangled. However, having $\StrongDiscordance(\rho)=0$ does \textit{not} guarantee that there is no nonlocal correlation, and it does not necessarily mean that the nonlocal correlation has nonquantum origins; as stated earlier, states with $\StrongDiscordance(\rho)=0$ can still have nonlocal correlation arising from distinctly quantum mechanisms such as entanglement, but the key point is that for these states, the correlation is weak enough that the same amount could be generated by a mixed strictly classical state, and that is why we consider it to be ``not distinctly quantum.'' Also as mentioned earlier, states with $\StrongDiscordance(\rho)>0$ do not necessarily have \textit{exclusively} quantum origins of nonlocal correlation; there may be multiple mechanisms, some of which can be nonquantum in nature (such as probability correlation arising from an epistemic mixture of nondiagonal quantum states).
\Figure{8} tests strong discordance $\StrongDiscordance(\rho)$ for the same families as \Fig{2}.%
\vspace{-3pt}%
\vspace{-10pt}%
In summary, a few caveats for strong discordance are:
\begin{itemize}[leftmargin=*,labelindent=4pt]\setlength\itemsep{0pt}
\item[\textbf{1.}]\hypertarget{StrongDiscordanceCaveat:1}{}The definition of $\StrongDiscordance(\rho)$ in \Eqs{33}{35} implies that \textit{$\StrongDiscordance(\rho)>0$ guarantees some distinctly nonclassical nonlocal correlation, but it does not necessarily mean there are no sources of strictly classical nonlocal correlation contributing to the total correlation}.
\item[\textbf{2.}]\hypertarget{StrongDiscordanceCaveat:2}{}A state for which $\StrongDiscordance(\rho)=0$ is \textit{not necessarily a strictly classical state or even a diagonal state}, but rather it is a state with no nonlocal correlations stronger than those of a strictly classical state, meaning that it may have any combination of probability correlation and decomposition-state correlation (\textit{including entanglement}), but that there exists a strictly classical state that could achieve the same amount of nonlocal correlation using a combination of only probability correlation and nonentanglement-decomposition-state correlation.
\item[\textbf{3.}]\hypertarget{StrongDiscordanceCaveat:3}{}A value of $\StrongDiscordance(\rho)>0$ guarantees some nondiagonality of $\rho$, but some nondiagonal states can have $\StrongDiscordance(\rho)=0$ (so $\StrongDiscordance(\rho)=0$ does not imply diagonality). \textit{Proof:} The maximum correlance of all diagonal states is the threshold for $\StrongDiscordance(\rho)=0$, so any states with $\StrongDiscordance(\rho)>0$ must have more correlance than the most correlated diagonal states, and therefore must not be diagonal. For the second claim, a proof by example is that $\StrongDiscordance(\rho)=0$ for a product state of nondiagonal mode states such as $\rho=\frac{1}{2}(\hsp{-2}~_{1}^{1}\hsp{1}~_{1}^{1})\otimes\frac{1}{2}(\hsp{-2}~_{-1}^{\phantom{-}1}\hsp{1}~_{\phantom{-}1}^{-1})$.
\item[\textbf{4.}]\hypertarget{StrongDiscordanceCaveat:4}{}The evidence that $\StrongDiscordance(\rho)=0$ for all separable states in \Fig{8} strongly supports the requirement that strictly classical states must be diagonal, because their value of $\Correlance(\rho_{D_{\max}})$ defines the threshold of distinctly quantum correlation strength in \Eq{33}.
\end{itemize}
Therefore, strong discordance is a \textit{more selective} measure than quantum discord, but because it rejects some weakly entangled states, it is \textit{too strong} to be a workable alternative to quantum discord. Therefore, next we develop a more inclusive measure that is closer in spirit to quantum discord, but more conceptually consistent.%
\vspace{-10pt}%
\subsection{\label{sec:V.B}Discordance: A Measure of Nonlocal Correlation in a Distinguishably Quantum State}%
\vspace{-5pt}%
Here, we define a more inclusive measure of nonlocal correlation that can never report a zero for distinctly quantum states such as entangled states, even if strictly classical states exist that have the same correlance.

To achieve this, we need a way to distinguish strictly classical states from nonlocally correlated quantum states with novel quantum features.  A practical feature for this purpose is \textit{coherence} (here, coherence means that $\rho$ has some nonzero off-diagonal elements in the computational basis). Even though superposition and coherence do not imply nonlocal correlation, they \textit{do} imply the presence of novel quantum effects, and are \textit{necessary} for some nonlocal correlations such as entanglement.  One caveat here is that, as we showed in \Eq{21}, diagonal quantum states can have ``superposition without coherence'' (such as reductions of maximally entangled states), so if we use coherence as a criterion for being a distinguishably quantum state, we would \textit{still} be rejecting \textit{diagonal quantum states} with nonzero correlance.

However, as discussed in \Sec{III.C}, it is not generally possible to determine whether a diagonal quantum state is quantum or strictly classical without doing a potentially impossible set of experiments or having more information beyond the state, such as its parent state before a reduction.  Therefore, for the purpose of measuring nonlocal correlation in a state that is definitely quantum using the state alone, \textit{the best we can do is use coherence as a criterion to reject strictly classical states at the expense of also rejecting diagonal quantum states that may have some nonlocal correlation}.

As a preliminary step, we define \textit{raw nondiagonality} as
\begin{Equation}                      {36}
\widetilde{\Nondiagonality}(\rho ) \equiv \tr[(\rho  - \Delta )^2 ],
\end{Equation}
where $\Delta$ is the maximally dephased input state,
\begin{Equation}                      {37}
\begin{array}{*{20}l}
   {\Delta} &\!\! {\equiv \Delta (\rho ) \equiv \sum\limits_{a = 1}^n {\rho _{a,a} |a\rangle \langle a|} ,}  \\
\end{array}
\end{Equation}
so that $\widetilde{\Nondiagonality}(\rho )=0$ iff $\rho$ is diagonal, and $\widetilde{\Nondiagonality}(\rho )>0$ iff $\rho$ is not diagonal. We could normalize $\widetilde{\Nondiagonality}(\rho )$ either over all states [by dividing $\widetilde{\Nondiagonality}(\rho )$ by $1-\frac{1}{n}$] or to only the subspace for which the diagonal elements are nonzero, but for our purposes, we are only interested in the sign of $\widetilde{\Nondiagonality}(\rho )$, so we do not need to normalize it.

Then, we can define a measure of nonlocal correlation in a distinguishably quantum state as the \textit{discordance},
\begin{Equation}                      {38}
\Discordance(\rho ) \equiv \text{sgn}[\widetilde{\Nondiagonality}(\rho )]\Correlance(\rho ),
\end{Equation}
where $\widetilde{\Nondiagonality}(\rho )$ is from \Eq{36} and $\Correlance(\rho )$ is the correlance from \Eq{9}. Thus, only states with some nondiagonality can have nonzero discordance, which ensures that all entangled states are recognized has having discordance, and only diagonal quantum states with nonzero correlance have their nonlocal correlation unrecognized (with the justifiability for that being that such states have density matrices with identical form to a strictly classical state, meaning that the state alone is not sufficient to distinguish them from strictly classical systems).

\begin{figure}[H]
\centering
\includegraphics[width=1.00\linewidth]{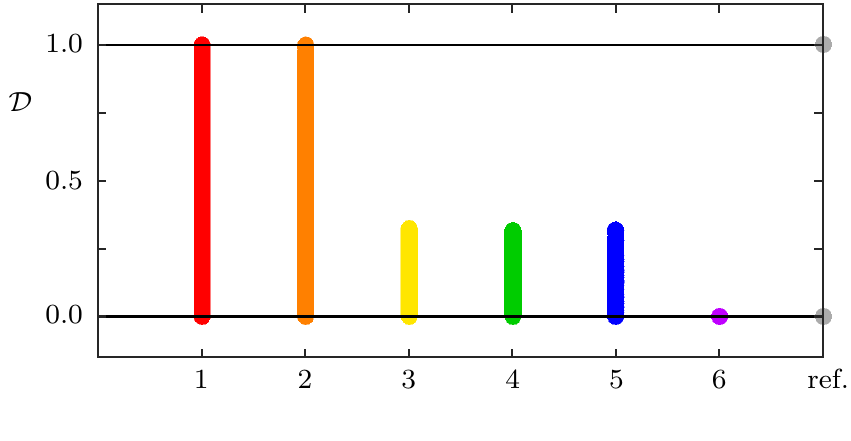}%
\vspace{-13pt}
\setlength{\unitlength}{0.01\linewidth}
\begin{picture}(100,0)
\put(32,0){\footnotesize Family Number from \Table{1}}
\end{picture}
\vspace{-12pt}
\caption[]{(color online) Discordance $\Discordance$ of arbitrary two-qubit mixed states for each of the six families of nonlocal correlation of \Table{1}, $10^5$ states each (colors not related to \Fig{1}). This shows that only Family 6 (product states, pure and mixed) has $\Discordance=0$ always, while Families 3--5 (all separable) can have $\Discordance>0$ (indicating some instances of \textit{quantum nonlocality without entanglement} \cite[]{BDFM,HHHO,NiCe}). Essentially, discordance is correlance recognized for nondiagonal states only. Again, only if entanglement is allowed, as in Families 1 and 2, can states achieve the full range of discordance up to $\Discordance=1$.}
\label{fig:9}
\end{figure}
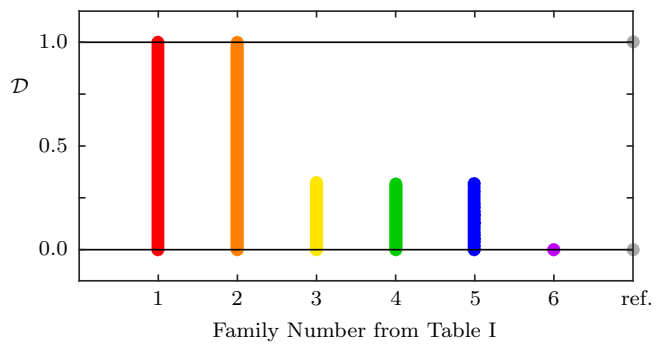

$\Discordance(\rho )=0$ if $\rho$ has no nonlocal correlation \textit{or} if the state is diagonal (whether it has nonlocal correlation or not), and $\Discordance(\rho )=1$ iff $\rho$ is maximally entangled. \Figure{9} plots the discordance for the families of \Table{1}.%
\vspace{-2pt}%
\vspace{-8pt}%
Thus, we now have \textit{three} main computable measures on all mixed and pure states to measure different degrees of general nonlocal correlation;
\begin{itemize}[leftmargin=*,labelindent=4pt]\setlength\itemsep{0pt}
\item[\textbf{1.}]\hypertarget{MainComputableMeasuresSummary:1}{}Strong discordance $\StrongDiscordance(\rho)$ (nonzero for states with more nonlocal correlation [as measured by correlance] than what is achievable by any strictly classical state).
\item[\textbf{2.}]\hypertarget{MainComputableMeasuresSummary:2}{}Discordance $\Discordance(\rho)$ (nonzero for quantum states with nonlocal correlation [as measured by correlance] that have the distinguishably quantum feature of coherence and are thus nondiagonal).
\item[\textbf{3.}]\hypertarget{MainComputableMeasuresSummary:3}{}Correlance $\Correlance(\rho)$ (nonzero for any nonlocally correlated state, whether quantum or strictly classical).
\end{itemize}
Ironically, as far as we know, \textit{all evidence suggests that there is really no such thing as a strictly classical state in the real world}, and that all states are truly quantum.  Therefore, correlance is really all we ever need to measure the presence of any nonlocal correlation in a state.

However, for the purpose of distinguishing quantum from strictly classical scenarios, the above measures give us powerful tools with different degrees of specificity. In fact, we will show that they obey a similar relationship to \Eq{2}, but first we will do some simple examples to compare these measures to quantum discord.%
\vspace{-12pt}%
\subsection{\label{sec:V.C}Comparisons of Discordance Measures with Quantum Discord}%
\vspace{-10pt}%
Here, we briefly compare correlance, discordance, and strong discordance with quantum discord, and also concurrence for reference.  Keep in mind that none of these new measures is meant to calculate quantum discord, but rather they are intended as different measures of the same intended qualitative features, with various restrictions that might be useful in particular applications.

\begin{figure}[H]
\centering
\includegraphics[width=1.00\linewidth]{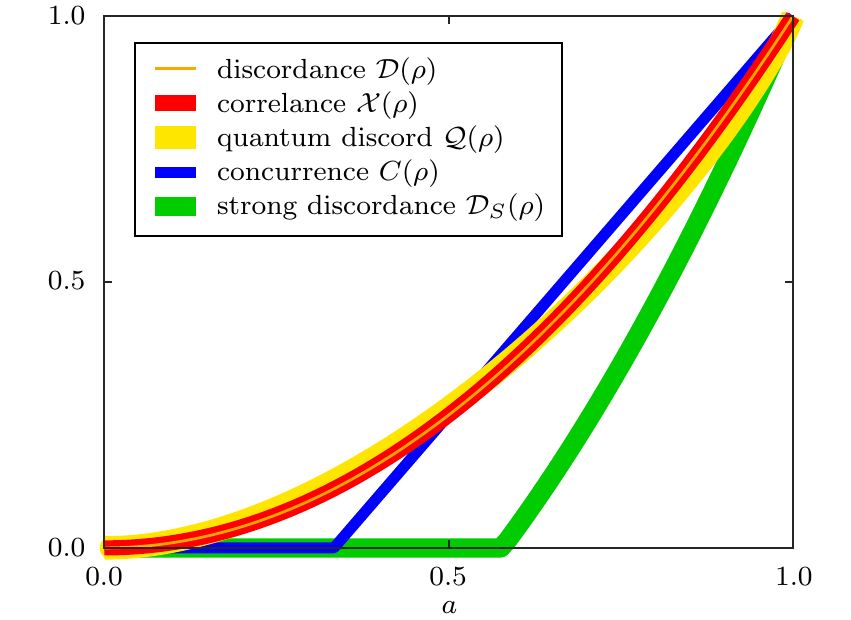}%
\vspace{-6pt}%
\caption[]{(color online) Comparison of strong discordance $\StrongDiscordance$, discordance $\Discordance$, and correlance $\Correlance$ with quantum discord $\mathcal{Q}$, and concurrence $C$ shown for reference for the Werner states of \Eq{39}. Note that $\StrongDiscordance$ is the most exclusive, rejecting entangled states weak enough that some strictly classical states have the same correlance ($\StrongDiscordance=0$ when $C>0$, proving \hyperlink{StrongDiscordanceCaveat:2}{Caveat 2} by example), while $\Discordance$ and $\Correlance$ both monotonically grow similarly to $\mathcal{Q}$ (their exact values relative to $\mathcal{Q}$ do not matter since they are not meant to calculate it; only their similar nonzero behavior is relevant). Here, $\Discordance=\Correlance$ since the state is nondiagonal except at $a=0$ where both are $0$ anyway.}
\label{fig:10}
\end{figure}

\begin{figure}[H]
\centering
\includegraphics[width=1.00\linewidth]{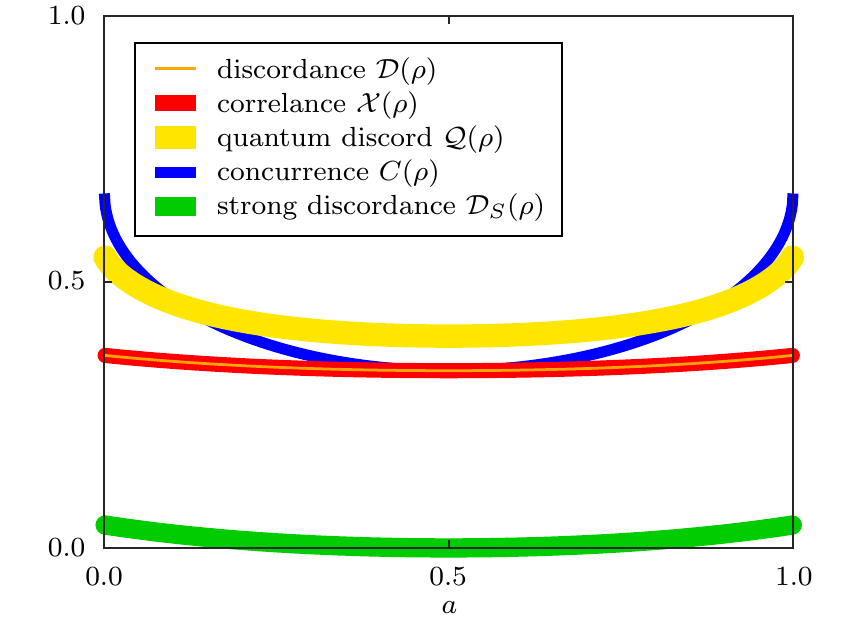}%
\vspace{-6pt}%
\caption[]{(color online) Comparison of strong discordance $\StrongDiscordance$, discordance $\Discordance$, and correlance $\Correlance$ with quantum discord $\mathcal{Q}$, and concurrence $C$ (for reference) for the states of \Eq{40}. Notice that $\Discordance$ and $\Correlance$ both monotonically grow symmetrically about $a=0.5$ similarly to $\mathcal{Q}$ (again, their exact values relative to $\mathcal{Q}$ are not important; only their similar nonzero behavior matters). Here, $\Discordance=\Correlance$ since the state is always nondiagonal.}
\label{fig:11}
\end{figure}

First, we consider two-qubit Werner states \cite{Wern},%
\vspace{-2pt}%
\begin{Equation}                      {39}
\rho  = a|\Psi ^ -  \rangle \langle \Psi ^ -  | + \frac{{1 - a}}{4}I,
\end{Equation}
\vspace{-8pt}\\%
where $a \in [0,1]$, \smash{$|\Psi ^ -  \rangle  \equiv \frac{1}{{\sqrt 2 }}(|1,2\rangle  - |2,1\rangle )$} is the (maximally entangled) singlet state (where the generic basis for each qubit is $\{|1\rangle,|2\rangle\}$), and \smash{$\frac{1}{4}I$} is the maximally mixed state. Using the quantum discord from \cite{AlRA}, \Fig{10} compares the various measures, showing that discordance and correlance behave similarly to quantum discord.%
\vspace{-4pt}%
\vspace{-8pt}%
Next, consider the mixture (from \cite{AlRA}),%
\vspace{-2pt}%
\begin{Equation}                      {40}
\rho  = \frac{1}{3}[(1 - a)|1,1\rangle \langle 1,1| + 2|\Psi ^ +  \rangle \langle \Psi ^ +  | + a|2,2\rangle \langle 2,2|],
\end{Equation}
\vspace{-10pt}\\%
where $a \in [0,1]$ and \smash{$|\Psi ^ +  \rangle  \equiv \frac{1}{{\sqrt 2 }}(|1,2\rangle  + |2,1\rangle )$}. Using its $\mathcal{Q}$ from \cite{AlRA}, \Fig{11} compares it to our various measures, again showing that $\Discordance$ and $\Correlance$ have similar behavior to $\mathcal{Q}$.%
\vspace{-6pt}%
\vspace{-10pt}%
Thus, these (limited) tests in \Fig{10} and \Fig{11} agree with our earlier reasoning that for the purpose of finding a substitute for quantum discord $\mathcal{Q}$, strong discordance $\StrongDiscordance$ is \textit{too exclusive} while correlance $\Correlance$ is \textit{too inclusive} (since it recognizes nonlocal correlation in diagonal states, which is not encountered in these examples). However, \textit{discordance} $\Discordance$ seems to be the most inclusive it is possible to be without having more information beyond the state itself, making it the most appropriate substitute for quantum discord.  In fact, all of these measures are related in an elegant way that formally parallels the theory of quantum discord, as we will show next.%
\vspace{-10pt}%
\subsection{\label{sec:V.D}Analogous Quantities to Quantum Mutual Information, Classical Correlation, and Quantum Discord}%
\vspace{-8pt}%
Since discordance $\Discordance$ only recognizes the nonlocal correlation of \textit{nondiagonal} states, here we define a measure for only \textit{diagonal} states as the \textit{diagonal discordance},%
\vspace{-2pt}%
\begin{Equation}                      {41}
\DiagonalDiscordance(\rho ) \equiv (1 - \text{sgn}[\widetilde{\Nondiagonality}(\rho )])\Correlance(\rho ),
\end{Equation}
\vspace{-12pt}\\%
which is just the correlance $\Correlance(\rho )$ of \Eq{9} with a sifting factor to map all nondiagonal states to $0$, where $\widetilde{\Nondiagonality}(\rho )$ is the raw nondiagonality from \Eq{36}. [We could define \Eq{41} in terms of $\DiagonalCorrelance$ from \Eq{15}, but as we show next, it is more useful to leave it in this ``unnormalized'' form. Also, $\DiagonalCorrelance$ does not have the sifting factor $(1 - \text{sgn}[\widetilde{\Nondiagonality}(\rho )])$ because $\DiagonalCorrelance$ is a stand-alone measure on diagonal states. The sifting factors in $\DiagonalDiscordance$ and $\Discordance$ are justified because they are meant to be used \textit{together}, as shown next.] Thus, diagonal discordance $\DiagonalDiscordance$ measures the nonlocal correlation in diagonal states, be they strictly classical or diagonal quantum, and is \textit{loosely} the analog of ``classical correlation'' $\mathcal{C}$ from the quantum discord theory.

Interestingly, comparing \Eq{38} and \Eq{41}, we see that
\begin{Equation}                      {42}
\Correlance(\rho ) = \DiagonalDiscordance(\rho ) + \Discordance(\rho ),
\end{Equation}
in formal analogy to \Eq{2} which was $\mathcal{I}(\rho)=\mathcal{C}(\rho)+\mathcal{Q}(\rho)$ where $\mathcal{I}$ is the quantum mutual information, $\mathcal{C}$ is the classical correlation, and $\mathcal{Q}$ is the quantum discord.

However, the similarity of \Eq{42} to \Eq{2} is a \textit{deceptive parallel} for several reasons;
\begin{itemize}[leftmargin=*,labelindent=4pt]\setlength\itemsep{0pt}
\item[\textbf{1.}]\hypertarget{ReasonsForDeceptiveParallel:1}{}Classical correlation $\mathcal{C}$ inherently contains a \textit{conceptual flaw}: the rank-$1$ projection operators (of the von Neumann measurements involved in its definition) are generally allowed to have \textit{superposition}, which sabotages the goal of measuring distinctly classical correlation by allowing it to depend on novel quantum properties.  In contrast, diagonal discordance $\DiagonalDiscordance$ only registers as nonzero for \textit{diagonal} states, which have no coherence, making them indistinguishable from strictly classical states, so $\DiagonalDiscordance$ is a much more appropriate measure of classical correlation (with the caveat that diagonal quantum states can have ``superposition without coherence'' as discussed in \Sec{III}).
\item[\textbf{2.}]\hypertarget{ReasonsForDeceptiveParallel:2}{}Since quantum discord $\mathcal{Q}$ is \textit{defined} as $\mathcal{Q}\equiv\mathcal{I}-\mathcal{C}$, it inherits the conceptual flaw of $\mathcal{C}$ from \hyperlink{ReasonsForDeceptiveParallel:1}{Reason 1}. In contrast, $\Discordance$ of \Eq{38} is designed to be the most general and inclusive measure of nonlocal correlation in distinguishably quantum states through its rejection of diagonal states, since, barring further information beyond the state, \textit{nondiagonality is the defining feature separating quantum from classical}. Again, we acknowledge the existence of diagonal quantum states, but they are indistinguishable from strictly classical states unless further information is known.
\item[\textbf{3.}]\hypertarget{ReasonsForDeceptiveParallel:3}{}Classical correlation $\mathcal{C}$ and quantum discord $\mathcal{Q}$ are \textit{not mutually exclusive} since they can be simultaneously nonzero. While this is not necessarily a problem since one can imagine simultaneous influences of classical and quantum mechanisms of correlation, the conceptual flaw mentioned in \hyperlink{ReasonsForDeceptiveParallel:1}{Reason 1} raises the question of whether there could be some overlap of what these two measures detect. Meanwhile, $\DiagonalDiscordance$ and $\Discordance$ are \textit{mutually exclusive}, since they depend on whether or not the state has coherence (with the value also depending on $\Correlance$). However, as mentioned earlier, since all states really are fundamentally quantum, it makes more sense to divide based on the \textit{novel quantum feature of coherence (nondiagonality)} than it does to define classicality based on measurements allowing projectors with superposition as in the definition of $\mathcal{C}$ and thus $\mathcal{Q}$.
\item[\textbf{4.}]\hypertarget{ReasonsForDeceptiveParallel:4}{}Quantum mutual information $\mathcal{I}$ is only defined for \textit{bipartite} systems, whereas correlance $\Correlance$ can handle \textit{$N$-partite} systems (and is extended to further multipartite generalizations in \Sec{VII}).
\end{itemize}

Therefore, while \Eq{42} does not exactly parallel the quantum-discord relation of \Eq{2}, it does have a solid grounding in the well-defined notion of nonlocal correlation in terms of actual quantum states and their distance from achieving mode-independence, which is the basis of correlance $\Correlance$.  It makes sense to use diagonality as the criterion for classicality since that is the form of strictly classical states of \Sec{III} which are carefully defined to avoid superposition both internally and through transformation.  

Furthermore, since we already showed in \Thm{4}, \Thm{6}, and \Thm{7} that statance $\hat{\Statance}$ and probablance $\hat{\Probablance}$ can arise in both diagonal and nondiagonal states, it makes sense that the general measure of correlance $\Correlance$ is used to calculate the actual value of correlation for both diagonal states in $\DiagonalDiscordance$ and nondiagonal states in $\Discordance$, since $\Correlance$ measures \textit{all} forms of nonlocal correlation whether or not the state is quantum or classical.

Thus, our definitions divide correlations in a much more conceptually consistent way than quantum discord simply by focusing on the state itself and carefully acknowledging the distinctly quantum feature of coherence.%
\vspace{-10pt}%
\section{\label{sec:VI}Hidden Variables Are Not Another Kind of Nonlocal Correlation}
\vspace{-7pt}%
Throughout this paper, we use the term \textit{nonlocal correlation} to mean a state's inability to be factored into a mode-independent product form as defined in \Sec{I.A}. However, there is a more subtle way to achieve nonlocal correlation that we must consider.

It is often said that ``correlation does not imply causation'' (meaning that correlation does not imply variable-dependence), but that does \textit{not} mean that causation cannot give rise to correlation (meaning that variable-dependence \textit{can} lead to correlation, just not always). 

This is essentially the idea of hidden-variable (HV) theories \cite[]{BaVa,EPR,Eins,HaSp}, where all states of quantum theory can be recast for each observable as diagonal states with variable-dependent probabilities that yield the same mean values of that observable as quantum theory, implying all possible correlations, including those perceived as entanglement.  Furthermore, HV models produce \textit{deterministic} theories, in which the future is exactly predictable if one knows the initial conditions of some set of ``hidden variables,'' whose trajectories, when sampled, yield the variable-dependent probabilities. Hidden-variable theories are generally regarded with skepticism for many reasons \cite[]{Bell,CHSH,AsDR}, and a full treatment of this subject is beyond the scope of this paper. 

However, since correlance measures nonlocal correlation as a state's distance from mode-independent product form, then \textit{the variable-dependence of an implied quantum state $\rho$ constructed from a complete HV theory only affects nonlocal correlation insofar as it affects the state's closeness to mode-independent product form}.  In other words, HV theories do \textit{not} constitute a new mechanism of nonlocal correlation, so all measures in this paper \textit{also} apply to states of HV theories.
\section{\label{sec:VII}Multipartite Generalizations Beyond $N$-Mode Correlations}
So far, correlance, diagonal correlance, statance, probablance, strong discordance, discordance, and diagonal discordance only measure nonlocal correlations \textit{over all $N$ modes of an $N$-mode coincidence system} (see \cite[]{HedE}, App.{\kern 2.5pt}A). For instance, the correlance $\Correlance(\rho)$ of \Eq{9} measures how far $\rho$ is from being a product state of $N$ modes as its distance from its own reduction product (since only product states are their own reduction products).

But in general, nonlocal correlation can arise in multipartite systems in many different ways, and they are not all equivalent. For example, from \cite[]{HedC}, a 4-qubit GHZ state \cite[]{GHZ,GHSZ,Merm} and a Bell-product state can have the same full $N$-partite entanglement by some necessary and sufficient measure of full $N$-partite entanglement, but the separability of the two Bell states in the Bell-product state shows that it has a distinctly different \textit{kind} of multipartite entanglement than the GHZ state for which no partitions of the $N$ modes are separable.  

Furthermore, we may only be interested in determining a correlation property for a \textit{subset} of the total modes, or we may want to view \textit{groups} of modes as if they were single modes and determine that property between those groups, or between several groups and a single mode.

In \cite[]{HedC}, a preliminary study was done on how best to adapt the \textit{ent} from an $N$-mode entanglement measure to a more general measure called the \textit{ent-concurrence} to measure \textit{distinctly multipartite entanglement}. The ent-concurrence (equal to the concurrence \cite[]{HiWo,Woot} for the case of two qubits) was shown to have the ability to distinguish between distinctly different types of multipartite entanglement for states that may have the same amount of full $N$-partite entanglement via the ent, such as a 4-qubit GHZ state and a Bell-product state.

This section explains how to generalize the correlance to measure the more general distinctly multipartite features described above, based on the formalism of \cite[]{HedC}. Generalizations of diagonal correlance, statance, probablance, strong discordance, discordance, and diagonal discordance follow that of correlance by direct analogy, and \textit{are all computable on all states}, except for generalizations of statance and probablance, for which computable expressions are not yet known.%
\\\vspace{-20pt}%
\subsection{\label{sec:VII.A}Multicorrelance}
The \textit{multicorrelance} for any \textit{pure or mixed} $\rho$ is
\begin{Equation}                      {43}
\Multicorrelance  (\rho ) \equiv \frac{1}{\mathcal{N}_{\Multicorrelance}  }\sum\limits_{k = 2}^N {\mathcal{M}_{\Correlance} ^{(\mathbf{N}^{(\mathbf{k})} )} (\rho )} ,
\end{Equation}
where \smash{$\mathcal{N}_{\Multicorrelance}$} is a normalization factor from maximizing the unnormalized $\Multicorrelance (\rho )$ over all possible input states, and we define the \textit{$N$-mode $k$-partitional multicorrelance},
\begin{Equation}                      {44}
\mathcal{M}_{\Correlance} ^{(\mathbf{N}^{(\mathbf{k})} )} (\rho ) \equiv \frac{1}{{\mathcal{N}_{\mathcal{M}_{\Correlance_k} } }}\sum\nolimits_{h = 1}^{\{ _k^N \} } {\sqrt {{\Correlance} ^{(\mathbf{N}_h^{(\mathbf{k})} )} (\rho )} } ,
\end{Equation}
where \smash{$\mathcal{N}_{\mathcal{M}_{\Correlance_k} }$} is a normalization factor again over all possible input states, \smash{$\{ _k^N \}  \equiv \frac{1}{{k!}}\sum\nolimits_{j = 0}^k {( - 1)^{k - j} (_j^k )j^N } $} are Stirling numbers of the second kind where \smash{$(_j^k ) \equiv \frac{{k!}}{{j!(k - j)!}}$}, and \smash{$\{ \Correlance ^{(\mathbf{N}_h^{(\mathbf{k})} )} (\rho )\} $}\rule{0pt}{10pt} is the set of all \textit{$N$-mode $k$-partitional correlances} (defined below), where $h$ is an index arbitrarily assigning a label to a given unique partition of modes.

To define the \textit{$N$-mode $k$-partitional correlances}, first we need a more explicit definition of \textit{correlance} as
\begin{Equation}                      {45}
\Correlance (\rho ) \equiv \Correlance (\rho ,\mathbf{n}) \equiv \frac{1}{{\mathcal{N}_{\Correlance} (\mathbf{n})}}\tr\left( {[\rho  - \varsigma (\rho ,\mathbf{n})]^2 } \right),
\end{Equation}
where \smash{$\mathcal{N}_{\Correlance} (\mathbf{n}) \equiv \max _{\forall \rho ' \in \mathcal{H}} \{ \tr([\rho ' - \varsigma (\rho ',\mathbf{n})]^2 )\} $}, where the maximizing states \smash{$\rho '_{\max }  \in \{ \{ \rho _{\text{ME}} \} ,\{ \rho _{\text{ME}_{\text{TGX}} } \} \} $} are maximally full-$N$-partite-entangled states of an $N$-mode $n$-level system where $\mathbf{n} \equiv (n_1 , \ldots ,n_N )$, mode $m$ has $n_m$ levels, $n \equiv n_1  \cdots n_N $, and the reduction product is
\begin{Equation}                      {46}
\varsigma (\rho ,\mathbf{n}) \equiv \mathop  \otimes \limits_{m = 1}^N \redx{\rho}{m} .
\end{Equation}
Then, recalling that \smash{$\redx{\rho}{m}$} is the $n_m$-level single-mode reduction of $\rho$ for mode $m$, and that \smash{$\redx{\rho}{\mathbf{m}}$} is the multimode reduction where $\mathbf{m} \equiv (m_1 , \ldots ,m_S )$ for $S\in 1,\ldots,N$, we see that the \textit{indices} of argument $\mathbf{n}$ are the mode labels $\mathbf{m}$ (with $S=N$) that label the reductions in $\varsigma (\rho ,\mathbf{n})$, so that $\mathbf{n}$ implicitly governs the tensor product in \Eq{46}.

Given the above, let the \textit{partitional correlance} be
\begin{Equation}                      {47}
\begin{array}{*{20}l}
   {\Correlance ^{(\mathbf{m}^{(\mathbf{T})} )} (\rho )} &\!\!\hsp{-1.0} { \equiv \Correlance (\redx{\rho}{\mathbf{m}} ,\mathbf{n}^{(\mathbf{m}^{(\mathbf{T})} )} )}  \\
   {} &\!\!\hsp{-1.0} { \equiv \frac{1}{{\mathcal{N}_{\Correlance} (\mathbf{n}^{(\mathbf{m}^{(\mathbf{T})} )} )}}\tr\hsp{-1.5}\left( {[\redx{\rho}{\mathbf{m}}\hsp{-1.7}  -\hsp{-1.0} \varsigma (\redx{\rho}{\mathbf{m}} ,\mathbf{n}^{(\mathbf{m}^{(\mathbf{T})} )} )]^2 } \right)\hsp{-2.0},}  \\
\end{array}
\end{Equation}
which lets us specify as input any reduction to mode group $\mathbf{m}$ of the input state with any repartitioning of its mode structure into new mode \mbox{groups as \smash{$\mathbf{m}^{(\mathbf{T})} \hsp{-1.2}\equiv\hsp{-1.2} (\mathbf{m}^{(1)} | \ldots |\mathbf{m}^{(T)} )$}, with levels \smash{$\mathbf{n}^{(\mathbf{m}^{(\mathbf{T})} )} \hsp{-1.2}\equiv$}\rule{0pt}{10pt}} \mbox{$ (n_{\mathbf{m}^{(1)} } , \ldots ,n_{\mathbf{m}^{(T)} } )$ where $T \hsp{-1.2}\in\hsp{-1.2} 1, \ldots ,S$,\hsp{2} and\hsp{2} \smash{$\mathcal{N}_{\Correlance} (\mathbf{n}^{(\mathbf{m}^{(\mathbf{T})} )} ) $}\rule{0pt}{10pt}} \smash{$\equiv\hsp{-1.2} \max _{\forall \rho ^{(\mathbf{m})\prime}  \in \mathcal{H}^{(\mathbf{m})} } \{ \tr([\rho ^{(\mathbf{m})\prime}  \hsp{-1.5}-\hsp{-1.5} \varsigma (\rho ^{(\mathbf{m})\prime} ,\mathbf{n}^{(\mathbf{m}^{(\mathbf{T})} )} )]^2 )\} $}\rule{0pt}{9pt} where the maximizing states \smash{$\rho _{\max }^{(\mathbf{m})\prime}  \in \{ \{ \rho _{\text{ME}}^{(\mathbf{m})} \} ,\{ \rho _{\text{ME}_{\text{TGX}} }^{(\mathbf{m})} \} \} $}\rule{0pt}{12pt} are maximally $T$-partite entangled states of a $T$-mode\rule{0pt}{10pt} \smash{$n_{\mathbf{m}^{(\mathbf{T})} }  \equiv n_{\mathbf{m}^{(1)} }  \cdots n_{\mathbf{m}^{(T)} } $}-level system of mode-structure \smash{$\mathbf{n}^{(\mathbf{m}^{(\mathbf{T})} )}  \equiv (n_{\mathbf{m}^{(1)} } , \ldots ,n_{\mathbf{m}^{(T)} } )$}\rule{0pt}{10pt}, where the input states can be specified in terms of the general reduction structure and underlying original modes as \smash{$\redx{\rho}{\mathbf{m}}$} where $\mathbf{m} \equiv (m_1 , \ldots ,m_S )$ since it is true by definition that \smash{$n_{\mathbf{m}}=n^{(\mathbf{m}^{(\mathbf{T})} )}$}\rule{0pt}{10pt} where \smash{$\mathbf{n}^{(\mathbf{m})}  \equiv (n_{m_1 } , \ldots ,n_{m_S } )$} and \smash{$n_{\mathbf{m}}  \equiv n_{m_1 }  \cdots n_{m_S } $}. See (\cite[]{HedC}, App.{\kern 2.5pt}C) for more details about the notation. We do not use the reduction symbol over density matrices in the normalization because this maximization is over all states in the Hilbert space \smash{$\mathcal{H}^{(\mathbf{m})}$} of the reduced system, not merely reductions from the parent state.  The reduction product in \Eq{47} is then
\begin{Equation}                      {48}
\varsigma (\redx{\rho}{\mathbf{m}} ,\mathbf{n}^{(\mathbf{m}^{(\mathbf{T})} )} ) \equiv \mathop  \otimes \limits_{q = 1}^T \redx{\rho}{\mathbf{m}^{(q)}} ,
\end{Equation}
where each multimode reduction \smash{$\redx{\rho}{\mathbf{m}^{(q)}}$} has \textit{internal} mode structure \smash{$\mathbf{m}^{(q)} \! \equiv\! (m_1^{(q)} , \ldots ,m_{G^{(q)} }^{(q)} )$} where $G^{(q)}  \in 1, \ldots ,S$\rule{0pt}{10pt}, in terms of the original indivisible modes $m_j$ such that all of them appear exactly once among all new mode groups $\mathbf{m}^{(q)}$ for $q \in 1, \ldots ,T$. Thus, in the case of the nonreduction ($S=N$), the input state to the partitional correlance is the full state $\redx{\rho}{\mathbf{N}}=\redx{\rho}{1,\ldots,N}=\rho$, and the new mode-group vector is (in this particular case) any element of the set \smash{$\mathbf{m}^{(\mathbf{T})}  \in \{ \mathbf{N}_h^{(\mathbf{k})} \}  \equiv$}\rule{0pt}{10pt} \smash{$ \{ (\mathbf{N}_h^{(1)} | \ldots |\mathbf{N}_h^{(k)} )\} |_{h = 1}^{\{ _k^N \} } $}\rule{0pt}{10pt}, meaning it is the set of all $\{ _k^N \} $ unique partitions of $N$ objects partitioned into $k$ groups.

As an example of how the notation works, for a tripartite system, $N=3$ and $\mathbf{N}\equiv(1,\ldots,N)=(1,2,3)$, so the possible $N$-mode partitions into $k=2$ groups are%
\vspace{-2pt}%
\begin{Equation}                      {49}
\begin{array}{*{20}l}
   {\mathbf{m}^{(\mathbf{T})} } &\!\! { \in \{ (1|2,3),(2|1,3),(3|1,2)\} }  \\
   {} &\!\! { \equiv \{ (\mathbf{N}_1^{(1)} |\mathbf{N}_1^{(2)} ),(\mathbf{N}_2^{(1)} |\mathbf{N}_2^{(2)} ),(\mathbf{N}_3^{(1)} |\mathbf{N}_3^{(2)} )\} }  \\
   {} &\!\! { \equiv \{ \mathbf{N}_h^{(\mathbf{2})} \} |_{h = 1}^{\{ _2^3 \} }  \equiv \{ \mathbf{N}_h^{(\mathbf{k})} \}, }  \\
\end{array}
\end{Equation}
\vspace{-6pt}\\%
where notice that reordering is not considered unique here, so for instance we do not list $(3|2,1)$ as a unique option since we have already listed $(3|1,2)$. 

Therefore, specifying $\mathbf{m}^{(\mathbf{T})}  \in \{ \mathbf{N}_h^{(\mathbf{k})} \} $ in the partitional correlance gives the \textit{$N$-mode $k$-partitional correlances} as%
\vspace{-2pt}%
\begin{Equation}                      {50}
\begin{array}{*{20}l}
   {\Correlance ^{(\mathbf{N}_h^{(\mathbf{k})} )} (\rho )} &\!\! { \equiv \Correlance (\redx{\rho}{\mathbf{N}} ,\mathbf{n}^{(\mathbf{N}_h^{(\mathbf{k})} )} ) = \Correlance (\rho ,\mathbf{n}^{(\mathbf{N}_h^{(\mathbf{k})} )} )}  \\
   {} &\!\! { \equiv \frac{1}{{\mathcal{N}_{\Correlance} (\mathbf{n}^{(\mathbf{N}_h^{(\mathbf{k})} )} )}}\tr\left( {[\rho  - \varsigma (\rho ,\mathbf{n}^{(\mathbf{N}_h^{(\mathbf{k})} )} )]^2 } \right),}  \\
\end{array}
\end{Equation}
\vspace{-6pt}\\%
with \textit{$N$-mode $k$-partitional reduction products},%
\vspace{-2pt}%
\begin{Equation}                      {51}
\varsigma (\rho ,\mathbf{n}^{(\mathbf{N}_h^{(\mathbf{k})} )} ) \equiv \mathop  \otimes \limits_{q = 1}^k \redx{\rho}{\mathbf{N}_h^{(q)}} .
\end{Equation}
\vspace{-6pt}\\%
Therefore, \Eq{50} appears in the terms of \Eq{44}. The multicorrelance $\mathcal{M}_{\Correlance}(\rho)$ measures the simultaneous amount of all \textit{$N$-mode $k$-partitional multicorrelances} \smash{$\mathcal{M}_{\Correlance}^{(\mathbf{N}^{(\mathbf{k})})}(\rho)$}\rule{0pt}{10pt}, while each individual \smash{$\mathcal{M}_{\Correlance}^{(\mathbf{N}^{(\mathbf{k})})}(\rho)$}\rule{0pt}{10pt} measures not only the combination of all possible $N$-mode $k$-partitional correlances, but it also measures how equally distributed they are by use of the square root. Alternatively, we may also define the $h$th \textit{$N$-mode $k$-partitional root-correlance} as%
\vspace{-1pt}%
\begin{Equation}                      {52}
\mathcal{M}_{\Correlance} ^{(\mathbf{N}_h^{(\mathbf{k})} )} (\rho ) \equiv \sqrt {\Correlance ^{(\mathbf{N}_h^{(\mathbf{k})} )} (\rho )} ,
\end{Equation}
\vspace{-12pt}\\%
which simplifies the notation for later concepts. 
Note that in this nomenclature, the prefix \textit{multi} implies a sum over all varieties of something, whereas the absence of \textit{multi} means no sum over varieties. For example, the $N$-mode $k$-partitional multicorrelance implies a sum over all $N$-mode $k$-partitional root-correlances which are each single-term quantities for a specific $N$-mode $k$-partition.

The multicorrelance $\Multicorrelance(\rho)$ is built to reveal whether there is \textit{any} nonlocal correlation at all in a given state at its $N$-mode scale, even including different perspectives created by grouping those modes together in a way that the total size of the input state remains the same.

As an example, for a $4$-partite system, the (unnormalized) multicorrelance is, from \Eq{43},
\begin{Equation}                      {53}
\widetilde{\mathcal{M}}_{\Correlance} (\rho ) = {\mathcal{M}}_{\Correlance} ^{(\mathbf{N}^{(\mathbf{2})} )} (\rho ) + {\mathcal{M}}_{\Correlance} ^{(\mathbf{N}^{(\mathbf{3})} )} (\rho ) + {\mathcal{M}}_{\Correlance} ^{(\mathbf{N}^{(\mathbf{4})} )} (\rho ),
\end{Equation}
where \smash{$\mathcal{M}_{\Correlance} ^{(\mathbf{N}_h^{(\mathbf{k})} )} (\rho )\equiv\widetilde{\mathcal{M}}_{\Correlance} ^{(\mathbf{N}^{(\mathbf{k})} )} (\rho )/\mathcal{N}_{\mathcal{M}_{\Correlance_k} }$}\rule{0pt}{10pt} are the $N$-mode $k$-partitional multicorrelances from \Eq{44}, given in unnormalized form (with correlance inputs suppressed) by
\vspace{-2pt}%
\begin{Equation}                      {54}
\begin{array}{*{20}l}
   {\widetilde{\mathcal{M}}_{\Correlance} ^{(\mathbf{N}^{(\mathbf{2})} )} (\rho ) = } &\!\! {\sqrt {\Correlance ^{(1|2,3,4)} }  + \sqrt {\Correlance ^{(2|1,3,4)} }  + \sqrt {\Correlance ^{(3|1,2,4)} }  }  \\
   {} &\!\! {+ \sqrt {\Correlance ^{(4|1,2,3)} } + \sqrt {\Correlance ^{(1,2|3,4)} }  + \sqrt {\Correlance ^{(1,3|2,4)} }  }  \\
   {} &\!\! { + \sqrt {\Correlance ^{(1,4|2,3)} } }  \\
   {\widetilde{\mathcal{M}}_{\Correlance} ^{(\mathbf{N}^{(\mathbf{3})} )} (\rho ) = } &\!\! {\sqrt {\Correlance ^{(1|2|3,4)} }  + \sqrt {\Correlance ^{(1|3|2,4)} }  + \sqrt {\Correlance ^{(1|4|2,3)} } }  \\
   {} &\!\! { + \sqrt {\Correlance ^{(2|3|1,4)} }  + \sqrt {\Correlance ^{(2|4|1,3)} }  + \sqrt {\Correlance ^{(3|4|1,2)} } }  \\
   {\widetilde{\mathcal{M}}_{\Correlance} ^{(\mathbf{N}^{(\mathbf{4})} )} (\rho ) = } &\!\! {\sqrt {\Correlance ^{(1|2|3|4)} }  = \sqrt {\Correlance (\rho )} ,}  \\
\end{array}
\end{Equation}
where the radicands above are obtained from \Eq{50}.

The specification of the structure vectors for each term then determines how each one is specifically calculated by controlling which effective modes are recognized for each term.  The result is that if \textit{any} nonlocal correlation exists between any partitioning of the $N$ modes, the multicorrelance will report the presence of that correlation as a value relative to some maximum over all states.

The square root of each partitional correlance is used here in analogy with the definition of ent-concurrence from \cite[]{HedC}, where it was shown to be able to distinguish different types of distinctly multipartite entanglement, whereas without the square root, states like the $4$-qubit GHZ state and Bell-product states did not appear to have different multipartite entanglement. 
In this context, tests of the $4$-qubit tier-1, tier-2, and tier-3 maximally entangled states of \cite[]{HedC} have shown that the square root is necessary to distinguish the $N$-mode 3-partitional multicorrelances of the tier-2 and tier-3 states (GHZ state and Bell-product states) which are identical if the square root is omitted. See \cite[]{HedC} for mathematical details about why the square root is appropriate for the ent-concurrence; the same argument applies here as well.

The purpose of the detailed notation is just to keep track of partitions for the purpose of telling the correlance function over which groups of modes we want to measure nonlocal correlations.  As we will soon see, this degree of specificity will allow us to extract information about \textit{all} possible nonlocal correlations (with the limitation that the correlance cannot distinguish between particular types of nonlocal correlation such as free entanglement and bound entanglement \cite[]{Hor1,Hor3,Hor4}).  For a more in-depth exposition of partitions in this same notation, see (\cite[]{HedC}, App.{\kern 2.5pt}C), and for further details on multipartite reductions see (\cite[]{HedE}, App.{\kern 2.5pt}B).%
\vspace{-10pt}%
\subsection{\label{sec:VII.B}$N$-mode Partitional Multicorrelance Vector}%
\vspace{-8pt}%
For a finer-grained picture of nonlocal correlation that tells us \textit{between which mode groups} correlation exists, we define the \textit{$N$-mode partitional multicorrelance vector},%
\vspace{-2pt}%
\begin{Equation}                      {55}
\Xi _{\mathcal{M}_{\Correlance}  }^{(\mathbf{N})} (\rho ) \equiv \left( {\begin{array}{*{20}c}
   {\{ \mathcal{M}_{\Correlance} ^{(\mathbf{N}_h^{(\mathbf{2})} )} (\rho )\} }  \\
    \vdots   \\
   {\{ \mathcal{M}_{\Correlance} ^{(\mathbf{N}_h^{(\mathbf{N})} )} (\rho )\} }  \\
\end{array}} \right)\hsp{-1.5},
\end{Equation}
where \smash{$\{ \mathcal{M}_{\Correlance} ^{(\mathbf{N}_h^{(\mathbf{k})} )} (\rho )\}$} is the set of all $N$-mode $k$-partitional root-correlances, and is valid for both pure and mixed states $\rho$. For example, in a $4$-mode system (suppressing input arguments),
\begin{Equation}                      {56}
\scalebox{0.89}{$\begin{array}{l}
 \Xi_{\mathcal{M}_\Correlance  }^{(\mathbf{4})}  \equiv \Xi_{\mathcal{M}_\Correlance  }^{(1,2,3,4)}=  \\ 
 \rule{0pt}{28.5pt}\!\!\left(\!\!{\kern -0.6pt} {\begin{array}{*{20}c}
   {\miniMcal_\Correlance ^{(1|2,\!3,\!4)} {\kern 1.0pt}\miniMcal_\Correlance ^{(2|1,\!3,\!4)} {\kern 1.0pt}\miniMcal_\Correlance ^{(3|1,\!2,\!4)} {\kern 1.0pt}\miniMcal_\Correlance ^{(4|1,\!2,\!3)} {\kern 1.0pt}\miniMcal_\Correlance ^{(1,\!2|3,\!4)} {\kern 1.0pt}\miniMcal_\Correlance ^{(1,\!3|2,\!4)} {\kern 1.0pt}\miniMcal_\Correlance ^{(1,\!4|2,\!3)} }  \\
   {\rule{0pt}{12.5pt}\!\miniMcal_\Correlance ^{(1|2|3,\!4)} {\kern 1.0pt}\miniMcal_\Correlance ^{(1|3|2,\!4)} {\kern 1.0pt}\miniMcal_\Correlance ^{(1|4|2,\!3)} {\kern 1.0pt}\miniMcal_\Correlance ^{(2|3|1,\!4)} {\kern 1.0pt}\miniMcal_\Correlance ^{(2|4|1,\!3)} {\kern 1.0pt}\miniMcal_\Correlance ^{(3|4|1,\!2)} }  \\
   {\rule{0pt}{12.5pt}\!\miniMcal_\Correlance ^{(1|2|3|4)} }  \\
\end{array}}\!\! \right)\!{\kern -1pt}, \\ 
 \end{array}$}
\end{Equation}
where each particular \smash{$\mathcal{M}_{\Correlance} ^{(\mathbf{N}_h^{(\mathbf{k})} )} (\rho )$} is given by \Eq{52}. The top row gives each $N$-mode $2$-partitional root-correlance, and each row farther down treats increasing partitions until the bottom row gives the $N$-mode $N$-partitional root-correlance which is just the square root of the actual correlance from \Eq{9}. Thus, \smash{$\Xi_{\mathcal{M}_\Correlance  }^{(\mathbf{N})}$}\rule{0pt}{10pt} gives a fine-grained, location-specific view of nonlocal correlation.

For a more \textit{intermediate} picture of nonlocal correlation, the $N$-mode $k$-partitional multicorrelances from \Eq{44}, expressible as [with the help of \Eq{52}],
\begin{Equation}                      {57}
\Multicorrelance ^{(\mathbf{N}^{(\mathbf{k})} )} (\rho ) \equiv \frac{1}{{\mathcal{N}_{\mathcal{M}_{\Correlance_k} } }}\sum\nolimits_{h = 1}^{\{ _k^N \} } {\Multicorrelance ^{(\mathbf{N}_h^{(\mathbf{k})} )} (\rho )} ,
\end{Equation}
give a measure of the total possible nonlocal correlation over all partitions of a certain $k$-value for an $N$-mode system, such as all possible bipartitions.

Note that the more aggregated measures such as multicorrelance and $N$-mode $k$-partitional multicorrelance do not necessarily imply that all of these nonlocal correlation resources are \textit{available for use simultaneously}; rather they indicate that such correlations are present, and typically only \textit{some} of them may be used simultaneously.
\subsection{\label{sec:VII.C}Absolute Multicorrelance}
Again following \cite[]{HedC}, while the multicorrelance \smash{$\Multicorrelance(\rho)$} and its more specific $N$-mode partitional multicorrelance vector \smash{$\Xi_{\mathcal{M}_\Correlance  }^{(\mathbf{N})}(\rho)$} and the even more specific $N$-mode $k$-partitional root-correlances \smash{$\mathcal{M}_{\Correlance} ^{(\mathbf{N}_h^{(\mathbf{k})} )} (\rho )$}\rule{0pt}{10pt} give us a sense of the nonlocal correlation present in the \textit{full input state within its full Hilbert space}, we can get an even more in-depth picture of the resources available in a state by evaluating the nonlocal correlations \textit{within reductions of the input state}.

To this end, rather than look at all $N$ modes, we focus on a subset of modes denoted by mode group $\mathbf{m}\equiv(m_{1},\ldots,m_{S})$ for $S\in 2,\ldots,N$ (we exclude $S = 1$ here since nonlocal correlation is correlation between at least two modes), and thus we now want to look for multipartite nonlocal correlations of some $S$-mode \textit{reduction} of $N$-mode state $\rho$. Thus, we define the \textit{$S$-mode partitional multicorrelance vector} as
\begin{Equation}                      {58}
\Xi _{\Multicorrelance  }^{(\mathbf{m})} (\rho ) \equiv \left( {\begin{array}{*{20}c}
   {\{ \Multicorrelance ^{(\mathbf{m}_h^{(\mathbf{2})} )} (\rho )\} }  \\
    \vdots   \\
   {\{ \Multicorrelance ^{(\mathbf{m}_h^{(\mathbf{S})} )} (\rho )\} }  \\
\end{array}} \right),
\end{Equation}
where \smash{$\{ \Multicorrelance ^{(\mathbf{m}_h^{(\mathbf{k})} )} (\rho )\}$} is the set of all \textit{$S$-mode $T$-partitional root-correlances} of a given reduction $\redx{\rho}{\mathbf{m}}$, where $T\in 2,\ldots,S$ here and each particular partitioning is labeled by $h$, and each \textit{$S$-mode $T$-partitional root-correlance} is%
\begin{Equation}                      {59}
\Multicorrelance ^{(\mathbf{m}_h^{(\mathbf{T})} )} (\rho ) \equiv \sqrt {\Correlance ^{(\mathbf{m}_h^{(\mathbf{T})} )} (\rho )},
\end{Equation}
where the $h$th \textit{$S$-mode $T$-partitional correlance} is
\begin{Equation}                      {60}
\begin{array}{*{20}l}
   {\Correlance ^{(\mathbf{m}_h^{(\mathbf{T})} )} (\rho )}\hsp{-2.0} &\!\! { \equiv\hsp{-1.5} \Correlance (\redx{\rho}{\mathbf{m}} ,\mathbf{n}^{(\mathbf{m}_h^{(\mathbf{T})} )} )}  \\
   {}\hsp{-2.0} &\!\! { \equiv\hsp{-1.5} \frac{1}{{\mathcal{N}_\Correlance (\mathbf{n}^{(\mathbf{m}_h^{(\mathbf{T})} )} )}}\tr\hsp{-1.5}\left( {[\redx{\rho}{\mathbf{m}}  \hsp{-1.5}-\hsp{-1.5} \varsigma (\redx{\rho}{\mathbf{m}} ,\mathbf{n}^{(\mathbf{m}_h^{(\mathbf{T})} )} )]^2 } \right)\hsp{-1.5},}  \\
\end{array}
\end{Equation}
with \textit{$S$-mode $T$-partitional reduction products},
\begin{Equation}                      {61}
\varsigma (\rho ^{(\mathbf{m})} ,\mathbf{n}^{(\mathbf{m}_h^{(\mathbf{T})} )} ) \equiv \mathop  \otimes \limits_{q = 1}^T \redx{\rho}{\mathbf{m}_h^{(q)}} ,
\end{Equation}
where \Eq{61} is written in terms of a general input state $\rho ^{(\mathbf{m})}$ in the Hilbert space of the reduced system to help show that the \textit{reductions of this state} must be taken to calculate this quantity.  Therefore, in \Eq{60} we are taking \textit{reductions of reductions} to calculate \smash{$\varsigma (\redx{\rho}{\mathbf{m}} ,\mathbf{n}^{(\mathbf{m}_h^{(\mathbf{T})} )} )$}. Again, since correlance can handle mixed states, all of these definitions work for general states both pure and mixed.  Thus, in \Eq{58}, row $T-1$ is a list of all $S$-mode $T$-partitional root-correlances of $\redx{\rho}{\mathbf{m}}$ which is the mode-$\mathbf{m}$ reduction of $\rho$.

Then, to account for the fact that we can make \smash{$\Xi _{\Multicorrelance  }^{(\mathbf{m})} (\rho )$} for \textit{many} different reduction mode-groups $\mathbf{m}$ for a given parent state $\rho$ , we can collect them all in a larger object called the \textit{multicorrelance array} $\widetilde{\nabla} _{\Multicorrelance } $ (not a gradient), the elements of which are all possible $S$-mode partitional multicorrelance vectors,
\begin{Equation}                      {62}
(\widetilde{\nabla}_{\Multicorrelance  } )_{k,l} (\rho ) \equiv \Xi _{\Multicorrelance  }^{([\text{nCk}(\mathbf{N},k)]_{l, \cdots } )} (\rho ),
\end{Equation}
where again $\mathbf{N}\equiv(1,\ldots,N)$, and $k\in 2,\ldots,N$, and $l \in 1, \ldots ,(_k^N )$, where $(_k^N ) \equiv \frac{{N!}}{{k!(N - k)!}}$, and $\text{nCk}(\mathbf{N},k)$ is the vectorized ``$n$-choose-$k$'' function yielding the matrix whose rows are each unique combinations of the elements of $\mathbf{N}$ chosen $k$ at a time, and we use the notation $A_{l,\cdots}$ to mean the $l$th row of matrix $A$. Note that the tilde in \Eq{62}, in keeping with earlier notation, implies that this quantity has not been normalized over all states.  As we will soon see, this is actually \textit{preferable} in this case, since each smallest element of the resulting object is already normalized, and therefore it is most useful to just see the actual root-correlance values as they are.

Continuing our example from earlier, a $4$-partite multicorrelance array has the form (suppressing inputs)
\begin{Equation}                      {63}
\widetilde{\nabla} _{\Multicorrelance  } (\rho )\hsp{-1} =\hsp{-4} \left(\hsp{-5} {\begin{array}{*{20}c}
   {\begin{array}{*{20}c}
   {\Xi _{\Multicorrelance  }^{(1,2)} } & {\Xi _{\Multicorrelance  }^{(1,3)} }  & {\Xi _{\Multicorrelance  }^{(1,4)} } & {\Xi _{\Multicorrelance  }^{(2,3)} } \hfill & {\Xi _{\Multicorrelance  }^{(2,4)} } & {\Xi _{\Multicorrelance  }^{(3,4)} }   \\
\end{array}}  \\
   {\begin{array}{*{20}c}
   {\Xi _{\Multicorrelance  }^{(1,2,3)} } & {\Xi _{\Multicorrelance  }^{(1,2,4)} } & {\Xi _{\Multicorrelance  }^{(1,3,4)} } & {\Xi _{\Multicorrelance  }^{(2,3,4)} }  \\
\end{array}}  \\
   {\Xi _{\Multicorrelance  }^{(1,2,3,4)} }  \\
\end{array}}\hsp{-5} \right)\hsp{-3.0},
\end{Equation}
where the $4$-mode partitional multicorrelance vector \smash{$\Xi _{\Multicorrelance  }^{(1,2,3,4)}$} was given in \Eq{56}, while the $3$-mode partitional multicorrelance vectors \smash{$\Xi _{\Multicorrelance  }^{(a,b,c)}$} have the form
\begin{Equation}                      {64}
\Xi _{\Multicorrelance  }^{(a,b,c)}  = \left( {\begin{array}{*{20}c}
   {\begin{array}{*{20}c}
   {\Multicorrelance ^{(a|b,c)} } & {\Multicorrelance ^{(b|a,c)} } & {\Multicorrelance ^{(c|a,b)} }  \\
\end{array}}  \\
   {\Multicorrelance ^{(a|b|c)} }  \\
\end{array}} \right)\hsp{-1.5},
\end{Equation}
and the $2$-mode partitional multicorrelance vectors all have just one element of the form
\begin{Equation}                      {65}
\Xi _{\Multicorrelance  }^{(a,b)}  = \Multicorrelance ^{(a|b)}  = \Multicorrelance ^{(a,b)} .
\end{Equation}
The multicorrelance array \smash{$\widetilde{\nabla} _{\Multicorrelance  } (\rho )$} gives the \textit{most detailed picture} of the nonlocal correlations available in $\rho$, showing \textit{in which reductions} such correlations exist, as well as \textit{where} (between which mode groups) in those reductions they exist.  Again, it is unlikely that all correlations within a state are available to be exploited simultaneously, but this treatment may give us a powerful way to categorize states for their potential resources of nonlocal correlation.

Lastly, for the most aggregated measure of the total potential nonlocal correlation resources within a state, we can define the \textit{absolute multicorrelance} as
\begin{Equation}                      {66}
\mathcal{M}_{\Correlance _{\text{abs}} } (\rho ) \equiv \frac{{||\widetilde{\nabla} _{\Multicorrelance  } (\rho )||_1 }}{{\max _{\forall \rho ' \in \mathcal{H}} \{ ||\widetilde{\nabla} _{\Multicorrelance  } (\rho ')||_1 \} }},
\end{Equation}
which is the normalized $1$-norm of the multicorrelance array, where the $1$-norm of vector $\mathbf{v}$ is  $||\mathbf{v}||_1  \equiv \sum\nolimits_{k = 1}^{\dim (\mathbf{v})} {|v_k |} $, and the $1$-norm is taken over the smallest scalar elements of $\widetilde{\nabla} _{\Multicorrelance  } (\rho )$, meaning for example that even though $\widetilde{\nabla} _{\Multicorrelance  } (\rho )$ in \Eq{63} has 11 ``elements,'' the fact that each of them is an object containing several scalar elements means that the total number of scalar-element terms in the $1$-norm in $\mathcal{M}_{\Correlance _{\text{abs}} } (\rho )$ for a $4$-partite system is $36$.

Note that all of this multipartite generalization follows the formalism of \cite[]{HedC} and \cite[]{HedE} closely, with the exception that here there is no difficulty in handling mixed states and our measure considers all nonlocal correlation, not just entanglement.

Keep in mind that all state measures rate states based on different criteria, and in general, the decision for which measure to use depends on the application. For example, if we only intend on using a particular state $\rho$ in its full $N$-mode form for the correlations that exist between partitions of its $N$ modes, then none of the measures in \Sec{VII.C} are directly relevant for that application, and the measures of \Sec{VII.A} and \Sec{VII.B} are more appropriate.  However, even then, if we do not care about the total available correlations among the $N$ modes and instead want it only between a specific partition of mode groups, then we would only need a \textit{particular} $N$-mode $k$-partitional root-correlance as our measure of choice to compare candidate states for our application.  Therefore, care should be taken to precisely define the needs of a given application before jumping in and applying one of these measures; otherwise the value of ``goodness'' they give on a scale of $0$ to $1$ may mean something unsuited to a given application.

Finally, all of the definitions in \Sec{VII} can be applied to diagonal correlance, strong discordance, discordance, and diagonal discordance as well (and also statance and probablance although they are not yet generally computable); to do so in these equations simply replace all occurrences of $\Correlance$ with $\mathcal{L}$ where $\mathcal{L}$ is the label for that particular measure such as $\Discordance$ for discordance or $\DiagonalDiscordance$ for diagonal discordance etc. For the nomenclature, just replace all occurrences of ``correlance'' (including within compound words) with the name of new measure, such as in ``multidiscordance'' or ``multi-diagonal-discordance,'' etc.
\section{\label{sec:VIII}Conclusions}
In this paper, we have introduced several measures of nonlocal correlation for discrete $N$-partite systems, with explicit generalizations of them for multipartite correlations \textit{beyond} merely $N$-partite correlations in an $N$-mode system. Most of these measures are \textit{computable} for all states both mixed and pure, giving them distinct advantages over other measures such as quantum discord.

To facilitate this goal, in \Sec{I.A} we first defined several mechanisms of nonlocal correlation, not all of which are mutually exclusive, but all of which are important to acknowledge. The two main kinds identified are called \textit{decomposition-state correlation} and \textit{probability correlation}. We also identified \textit{entanglement correlation} (which we later showed to be a special kind of decomposition-state correlation), and we defined general mixed product states as the \textit{absence of nonlocal correlation}.

From the above general mechanisms of nonlocal correlation, we were able to identify the \textit{Six Families of Nonlocal Correlation} in \Table{1} as a set of forms that density matrices can take that yield nonlocal correlation. While these families are not all mutually exclusive, identifying them was an extremely helpful conceptual aid to investigating nonlocal correlations in multipartite systems.

From these initial observations, we immediately constructed the main measure of interest as the \textit{correlance} $\Correlance$ in \Sec{I.B}, which measures all possible $N$-mode nonlocal correlation by gauging a state's distance from having mode-independent product form, meaning its distance from its own reduction product (since only product states are their own reduction products), the conditions for which are established in \Sec{I.A}. The only drawback of $\Correlance$ is that it cannot distinguish between \textit{different kinds} of correlation. Its advantages are that it is computable for all states, and it is capable of detecting \textit{all} nonlocal correlation (including, for instance, the notoriously difficult-to-detect bound entanglement \cite[]{Hor1,Hor3,Hor4}).

In \Sec{II}, we proved the validity of and derived the normalization factor of $\Correlance$ (contained in various appendices), and did extensive numerical tests in a wide variety of multipartite systems as a set of necessary checks against the proofs. Furthermore, we constructed a related measure as the \textit{diagonal correlance} $\DiagonalCorrelance$ intended for only diagonal states and we sketched a similar proof of its normalization from which we derived its normalization factor, and numerically tested its normalization as well.  We also compared $\DiagonalCorrelance$ to the well-known Pearson correlation and gave extensive instructions for how to handle classical data for use with $\DiagonalCorrelance$ in \App{M}.  The diagonal correlance $\DiagonalCorrelance$ is intended for classical probability distributions only, since those must always be diagonal since superposition and thus coherence (nonzero off-diagonal elements of a density matrix) are not possible in classical physics. (This also provided a nice segue into a discussion of what we really mean by classical states; a crucial topic for any paper on nonlocal correlations including quantum discord and notions of classicality.)

\Section{III} delves deeply into the distinction between classical and quantum, and ultimately proposes the new term \textit{strictly classical} to provide a well-defined and conceptually consistent meaning for true classicality beyond the often carelessly and ambiguously used term ``classical.'' Ultimately, we conclude that the main defining feature for strict classicality must be the \textit{absence of quanutm superposition}, both intrinsically and through transformation.  This notion is a departure from the popular belief that coherent states are the most appropriate standards of classicality despite having quantum superposition.  This topic is discussed in depth in \App{O} which describes the consistency of strict classicality with established physics and also explains why using coherent states as standards of classicality is not conceptually consistent. 

\Section{III} also makes the important observation that there exist diagonal states that are quantum, and identifies the property of a state having ``superposition without coherence,'' meaning that in cases such as when a diagonal state is a reduction of a larger maximally entangled parent state, the probabilities of the reduction are directly inherited from the superposition coefficients of the parent state, which are themselves overlaps of the parent state with the basis states. Thus we can have diagonal states that still have superposition in an inherited sense.  All of these ideas relating classicality to the edge of distinctly quantum properties form powerful conceptual tools for the remaining discussions.

In \Sec{IV}, we present the hard-to-compute measures \textit{statance} $\hat{\Statance}(\rho)$ and \textit{probablance} $\hat{\Probablance}(\rho)$, which measure the decomposition-state correlation and probability correlation respectively. From these definitions, we learn several very useful new facts, one of the most interesting of which is that \textit{strictly classical states} (which are diagonal) can have nonzero $\hat{\Statance}(\rho)$ and $\hat{\Probablance}(\rho)$, just as general quantum states can. Furthermore, we showed that these types of correlation can exist independently of each other in strictly classical states as well.  This taught us that correlations of a classical kind cannot be purely ascribed to one mechanism such as probability correlation or decomposition-state correlation, because in the cases of diagonal quantum states where we have ``superposition without coherence,'' both of these types of nonlocal correlation are \textit{entirely quantum} in origin, having come from the superposition of the parent state before the reduction.  Therefore, since there is no fundamental way to distinguish strictly classical states from diagonal quantum states without having more information beyond the state, this established that \textit{diagonal quantum states are not distinguishable from strictly classical states}.

\Section{V} then began by defining the \textit{strong discordance} $\StrongDiscordance$ as a measure of nonlocal correlation \textit{beyond} that achievable by a strictly classical state, as a reasonable first step towards constructing a measure sensitive to uniquely quantum correlations.  However, despite being exactly computable, we soon found that $\StrongDiscordance$ is \textit{too exclusive}, since it rejects weak quantum correlation such as entanglement that produces a value of correlance $\Correlance$ that is also achievable by some strictly classical states.

Therefore, since we desired a measure that can detect all novel quantum correlation such as entanglement, and since we established that diagonality is the fundamental defining feature of strictly classical states but also the point at which diagonal quantum states cannot be distinguished from strictly classical states, then this prompted the definition of \textit{discordance} $\Discordance$ in \Sec{V.B} as a measure of nonlocal correlation in a distinguishably quantum state.  The fully computable measure of discordance $\Discordance$ uses nondiagonality as the threshold for recognizing nonlocal correlation, so that it is guaranteed to report ``zero discordance'' for strictly classical states (which are all diagonal), while it is also guaranteed to correctly report the net correlation for all entangled states as well as for all quantum states that are distinguishably quantum by being nondiagonal.  Its only weakness is that $\Discordance$ rejects the nonlocal correlation of  \textit{diagonal quantum} states as being ``not quantum enough'' to qualify as discordance.

However, that prompted the definition in \Sec{V.D} of another computable measure, the \textit{diagonal discordance} $\DiagonalDiscordance$ which only recognizes the nonlocal correlation in \textit{diagonal} states, and rejects all distinguishably quantum states (nondiagonal states) as being ``too quantum.''

Together, the three computable measures of correlance $\Correlance$, diagonal discordance $\DiagonalDiscordance$, and discordance $\Discordance$ were shown to obey the relation $\Correlance=\DiagonalDiscordance+\Discordance$ in \Eq{42}, which formally parallels the quantum-discord relation from \Eq{2}, $\mathcal{I}=\mathcal{C}+\mathcal{Q}$, where $\mathcal{I}$ is the quantum mutual information, $\mathcal{C}$ is the classical correlation, and $\mathcal{Q}$ is the quantum discord.

We then showed that this similarity is only one of form, but not meaning, since we identified the main conceptual flaw of quantum discord $\mathcal{Q}$ is that it is defined in terms of classical correlation $\mathcal{C}$ which is based on von Neumann projectors that are allowed to have \textit{quantum superposition}, something that is completely at odds with classical physics.  Therefore, it may be that $\mathcal{C}$ and $\mathcal{Q}$ have some conceptually inconsistent overlaps in what they measure, making them \textit{conceptually} less preferable, \textit{beyond} the fact that they are difficult calculate.

Alternatively, we could treat the \textit{raw strong discordance} \smash{$\RawStrongDiscordance(\rho)\equiv\max\{0,\Correlance(\rho)-\Correlance(\rho_{D_{\max}})\}$} as an analog of \smash{$\mathcal{Q}$}, and then define an analog to \smash{$\mathcal{C}$} as the \textit{raw diagonal strong discordance} \smash{$\RawDiagonalStrongDiscordance(\rho)\equiv\Correlance(\rho)-\RawStrongDiscordance(\rho)=$} \smash{$\min\{\Correlance(\rho),\Correlance(\rho_{D_{\max}})\}$}.  Then \smash{$\Correlance=\RawDiagonalStrongDiscordance+\RawStrongDiscordance$}\rule{0pt}{10pt} would parallel the original idea of \smash{$\mathcal{I}=\mathcal{C}+\mathcal{Q}$} more closely since strong-discordance states (those with \smash{$\StrongDiscordance>0$}) would simultaneously have nonzero values of both \smash{$\RawDiagonalStrongDiscordance$} and \smash{$\RawStrongDiscordance$} (all \smash{$\StrongDiscordance>0$} states would have maximal \smash{$\RawDiagonalStrongDiscordance$}).  However, the \smash{$\RawStrongDiscordance$} part would still behave discontinuously, yielding zero for weakly entangled states until the maximally-correlated diagonal-state threshold, as for \smash{$\StrongDiscordance$} in \Fig{10}. Thus, discordance \smash{$\Discordance$} is a better replacement for quantum discord since \smash{$\Discordance$} never ignores entanglement, no matter how weak. Nevertheless, this alternative trio of measures \smash{$\{\Correlance,\RawDiagonalStrongDiscordance,\RawStrongDiscordance\}$} may be useful in some applications.

Some possible extensions and improvements of these ideas would be their generalization to systems of infinite levels, and then to continuous systems. Beyond that, hybrid systems such as those where some modes are discrete and some modes are continuous would be another interesting area to explore. It would also be nice to have a more explicit proof of the normalization of diagonal correlance $\DiagonalCorrelance$ rather than the sketch given in \App{L}, even though the numerical tests seem to vindicate it to excellent precision. However, no such difficulty arose for correlance $\Correlance$, the normalization of which was explicitly proved in \App{J}, and that is the more important measure since it covers all states regardless of diagonality.

Note that while the explicit normalization factors given are exact, as system size gets larger (either in number of levels $n$ or number of modes $N$ or both), the accumulated numerical errors in typical computers may lead to increasingly large errors in normalization. Therefore, in those cases it may be beneficial to find a maximally correlated state of the desired type to act as a normalization standard, which may help compensate for the errors since then the standard state must undergo the same pre-normalized computation sequence as the input state.

In closing, this paper provides several useful measures of nonlocal correlation that are carefully motivated by the possible mechanisms of nonlocal correlation and also the distinctly quantum feature of superposition and its related feature of coherence.  The main measures of correlance $\Correlance$, diagonal discordance $\DiagonalDiscordance$, and discordance $\Discordance$ have the advantages of being conceptually consistent ways to distinguish quantum correlations from classical correlations through their acknowledgment that quantum superposition cannot be allowed in any reasonable definition of classicality, which is the bedrock of the notion of strict classicality.  Since these measures are also exactly computable for both pure and mixed states and work in all multipartite discrete systems, they provide us with extremely powerful tools for studying nonlocal correlation.
\begin{appendix}
\section{\label{sec:App.A}Brief Review of Reduced States}
Let the \textit{multimode reduction} from $N$-mode \textit{parent state} $\rho$ to a composite subsystem of $S\hsp{-1}\in\hsp{-1} 1,\ldots,N$ possibly noncontiguous and reordered modes $\mathbf{m} \equiv (m_1 , \ldots ,m_S)$ be
\begin{Equation}                      {A.1}
\redx{\rho}{\mathbf{m}}  \equiv \tr_{\mathbf{\mbarsub}} (\rho ),
\end{Equation}
where the $\raisebox{-4pt}{\scalebox{1.2}{\textasciicaron}}$ symbol in $\redx{\rho}{\mathbf{m}}$ indicates that it is a \textit{reduction} of $\rho$ (and not merely an isolated system of the same size as mode group $\mathbf{m}$), and the bar in $\mathbf{\mbar}$ means ``not $\mathbf{m}$,'' meaning that we trace over all modes whose labels are \textit{not} in $\mathbf{m}$. See App.{\kern 2.5pt}B of \cite[]{HedE} for details.
\section{\label{sec:App.B}Quantum Discord Definition}
A suggested definition for quantum discord \cite[]{OlZu,AlRA} is
\begin{Equation}                      {B.1}
\mathcal{Q}(\rho)\equiv\mathcal{I}(\rho )-\mathcal{C}(\rho),
\end{Equation}
with $\mathcal{I}(\rho )$ from \Eq{1}, where the \textit{classical correlation} is
\begin{Equation}                      {B.2}
\mathcal{C}(\rho)\equiv\mathop {\max }\limits_{\{ P_k^{(2)} \} } \left[ {\mathcal{I}(\rho |\{ P_k^{(2)} \} )} \right],
\end{Equation}
where we define 
\begin{Equation}                      {B.3}
\mathcal{I}(\rho |\{ P_k^{(2)} \} )\equiv S(\redx{\rho}{1}) - S(\rho |\{ P_k^{(2)} \} )
\end{Equation}
as the quantum mutual information of the von Neumann measurement with projection operators $\{ P_k^{(2)} \} $ for mode $2$, where
\begin{Equation}                      {B.4}
S(\rho |\{ P_k^{(2)} \} ) \equiv \sum\nolimits_k {p_k } S(\rho _k )
\end{Equation}
is the conditional von Neumann entropy given {this\hsp{-1} measurement,\hsp{-1} where\hsp{-1} \smash{$p_k  \hsp{-1}\equiv\hsp{-1} \tr[(I^{(1)} \hsp{-1} \otimes\hsp{-1} P_k^{(2)} )\rho (I^{(1)} \hsp{-1} \otimes\hsp{-1} P_k^{(2)} )^ \dag  ]$}} are the probabilities of conditional measurement-outcome states \smash{$\rho _k  \equiv \frac{1}{{p_k }}(I^{(1)}  \otimes P_k^{(2)} )\rho (I^{(1)}  \otimes P_k^{(2)} )^ \dag  $}.
\section{\label{sec:App.C}Purity}
Recall that the \textit{purity} of $\rho$ is the trace of its square,
\begin{Equation}                      {C.1}
P \equiv P(\rho ) \equiv \tr(\rho ^2 ),
\end{Equation}
with range $P \in [\frac{1}{n},1]$ for an isolated system, and $\rho$ is \textit{pure} iff $P=1$. Otherwise, if $P<1$ then $\rho$ is \textit{mixed}.  Thus, for mode $m$, $P(\rho ^{(m)} ) \in [\frac{1}{{n_m }},1]$ for an isolated mode $m$. 

\textit{Beware:} if the parent state $\rho$ is not a product state, then its reductions $\redx{\rho}{m}$ can have constraints on the limits of their purity, as proved in \cite[]{HedE}. The reason \Eq{3} has no nonlocal correlation is that its tensor product ensures that each reduction (marginal state) is $\redx{\rho}{m}=\rho ^{(m)} $, and therefore has no dependence on any other modes.
\section{\label{sec:App.D}Some Basic Details about Entanglement}
In separable states, as in \Eq{4}, notice that the probabilities $p_j$ do not need any special structure for \Eq{4} to be satisfied, and that the decomposition states $\rho_j$ merely need \textit{product form}, but not mode independence.  Therefore, product form of the full state is not necessary for separability in general. However, for \textit{pure} states, product form is both necessary and sufficient for separability. 

For example, in a two-qubit system where each qubit has a generic basis with labels starting on $1$ (our convention in this paper) as $\{ |1\rangle ,|2\rangle \} $, the pure state $|\psi \rangle  = ac|1\rangle  \otimes |1\rangle  + ad|1\rangle  \otimes |2\rangle  + bc|2\rangle  \otimes |1\rangle  + bd|2\rangle  \otimes |2\rangle$ is separable because it can be factored as $|\psi \rangle  = (a|1\rangle  + b|2\rangle ) \otimes (c|1\rangle  + d|2\rangle )$, so that $\rho  = |\psi \rangle \langle \psi | = \rho ^{(1)}  \otimes \rho ^{(2)}$ where $\rho ^{(m)}  = |\psi ^{(m)} \rangle \langle \psi ^{(m)} |$, and $|\psi ^{(1)} \rangle  = a|1\rangle  + b|2\rangle$ and $|\psi ^{(2)} \rangle  = c|1\rangle  + d|2\rangle $.  In contrast, the pure state \smash{$|\Phi ^ +  \rangle  = \frac{1}{{\sqrt 2 }}(|1\rangle  \otimes |1\rangle  + |2\rangle  \otimes |2\rangle )$} is not separable so it is entangled (and happens to be maximally entangled) because it cannot be factored into a product form (since this example is for pure states).
\section{\label{sec:App.E}Example of Product Form}
To get product form for a two-qubit state $\rho  = \sum\nolimits_j {p_j \rho_j }$ (and to demonstrate mode independence), \Thm{1} requires \Eqs{5}{8} by setting \smash{$\rho_j \!\equiv\! \rho _j^{(1)}  \!\otimes\! \rho _j^{(2)}$} and
\begin{Equation}                      {E.1}
\!\!\scalebox{0.95}{
$\begin{array}{*{20}l}
   {p_1 } &\!\!\! { =\! p_{(1,1)} } &\!\!\! { \equiv\! p_1^{(1)} p_1^{(2)} ,\;\;} &\!\!\! {\rho _1^{(1)} } &\!\!\! { =\! \rho _{(1,1)}^{(1)} } &\!\!\! { \equiv\! \rho _{(1)}^{(1)} ,\;\;} &\!\!\! {\rho _1^{(2)} } &\!\!\! { =\! \rho _{(1,1)}^{(2)} } &\!\!\! { \equiv\! \rho _{(1)}^{(2)} ,} \\
   {p_2 } &\!\!\! { =\! p_{(1,2)} } &\!\!\! { \equiv\! p_1^{(1)} p_2^{(2)} ,\;\;} &\!\!\! {\rho _2^{(1)} } &\!\!\! { =\! \rho _{(1,2)}^{(1)} } &\!\!\! { \equiv\! \rho _{(1)}^{(1)} ,\;\;} &\!\!\! {\rho _2^{(2)} } &\!\!\! { =\! \rho _{(1,2)}^{(2)} } &\!\!\! { \equiv\! \rho _{(2)}^{(2)} ,} \\
   {p_3 } &\!\!\! { =\! p_{(2,1)} } &\!\!\! { \equiv\! p_2^{(1)} p_1^{(2)} ,\;\;} &\!\!\! {\rho _3^{(1)} } &\!\!\! { =\! \rho _{(2,1)}^{(1)} } &\!\!\! { \equiv\! \rho _{(2)}^{(1)} ,\;\;} &\!\!\! {\rho _3^{(2)} } &\!\!\! { =\! \rho _{(2,1)}^{(2)} } &\!\!\! { \equiv\! \rho _{(1)}^{(2)} ,} \\
   {p_4 } &\!\!\! { =\! p_{(2,2)} } &\!\!\! { \equiv\! p_2^{(1)} p_2^{(2)} ,\;\;} &\!\!\! {\rho _4^{(1)} } &\!\!\! { =\! \rho _{(2,2)}^{(1)} } &\!\!\! { \equiv\! \rho _{(2)}^{(1)} ,\;\;} &\!\!\! {\rho _4^{(2)} } &\!\!\! { =\! \rho _{(2,2)}^{(2)} } &\!\!\! { \equiv\! \rho _{(2)}^{(2)} ,} \\
\end{array}$}
\end{Equation}
where all states in \Eq{E.1} are pure. Then $\rho$ becomes
\begin{Equation}                      {E.2}
\begin{array}{*{20}l}
   \rho  &\!\!  =  &\!\! {p_1^{(1)} p_1^{(2)} \rho _{(1)}^{(1)}  \otimes \rho _{(1)}^{(2)}  + p_1^{(1)} p_2^{(2)} \rho _{(1)}^{(1)}  \otimes \rho _{(2)}^{(2)} }  \\
   {} &\!\! {} &\!\! {\!+ p_2^{(1)} p_1^{(2)} \rho _{(2)}^{(1)}  \otimes \rho _{(1)}^{(2)}  + p_2^{(1)} p_2^{(2)} \rho _{(2)}^{(1)}  \otimes \rho _{(2)}^{(2)} }  \\
   {} &\!\!  =  &\!\! {p_1^{(1)} \rho _{(1)}^{(1)}  \otimes (p_1^{(2)} \rho _{(1)}^{(2)}  + p_2^{(2)} \rho _{(2)}^{(2)} )}  \\
   {} &\!\! {} &\!\! {\!+ p_2^{(1)} \rho _{(2)}^{(1)}  \otimes (p_1^{(2)} \rho _{(1)}^{(2)}  + p_2^{(2)} \rho _{(2)}^{(2)} )}  \\
   {} &\!\!  =  &\!\! {(p_1^{(1)} \rho _{(1)}^{(1)}  + p_2^{(1)} \rho _{(2)}^{(1)} ) \otimes (p_1^{(2)} \rho _{(1)}^{(2)}  + p_2^{(2)} \rho _{(2)}^{(2)} )}  \\
   {} &\!\!  =  &\!\! {\rho ^{(1)}  \otimes \rho ^{(2)}, }  \\
\end{array}
\end{Equation}
with\hsp{-0.5} \smash{$\rho ^{(1)} \hsp{-1.5} \equiv\hsp{-1} p_1^{(1)} \rho _{(1)}^{(1)}  \hsp{-1.5}+\hsp{-1} p_2^{(1)} \rho _{(2)}^{(1)} $} and \smash{$\rho ^{(2)}  \hsp{-1.5}\equiv\hsp{-1} p_1^{(2)} \rho _{(1)}^{(2)}  \hsp{-1.5}+\hsp{-1} p_2^{(2)} \rho _{(2)}^{(2)} $}. Notice the redundant probabilities and decomposition states in \Eq{E.1}; this is a consequence of applying \Thm{1}. Also, if either or both mode reductions have lower rank, some of the probabilities in \Eq{E.1} will be zero.

Thus, the smallest maximum number of decomposition states required for any mixed product state is
\begin{Equation}                      {E.3}
r_{1}\cdots r_{N}=r,
\end{Equation}
(which is also their minimum number of decomposition states) where the ranks of each single-mode reduction of the product state are $r_{m}\in 1,\ldots,n_{m}$, and $r\equiv\text{rank}(\rho)$, which is just an application the well-known property for tensor products that $\text{rank}(A\otimes B)=\text{rank}(A)\text{rank}(B)$. In fact, by making mode-independent (MI) tensor products of the eigenstates of the reductions, and forming the corresponding probabilities from MI products of the eigenvalues of the reductions, we can always construct an $r$-member MI decomposition of any product state.
\section{\label{sec:App.F}Proof of \Thm{1}}
Starting with a general mixed state,
\begin{Equation}                      {F.1}
\rho  = \sum\nolimits_j {p_j \rho _j },
\end{Equation}
applying \Eq{5} gives a separable state,
\begin{Equation}                      {F.2}
\rho  = \sum\nolimits_j {p_j \rho _j^{(1)}  \otimes  \cdots }  \otimes \rho _j^{(N)},
\end{Equation}
which is not enough to definitely get product form. If we then also use \Eq{7}, we get
\begin{Equation}                      {F.3}
\begin{array}{*{20}l}
   {\rho} &\!\! {=\sum\nolimits_j {(p_j^{(1)}  \cdots p_j^{(N)} )\rho _j^{(1)}  \otimes  \cdots  \otimes \rho _j^{(N)} }}  \\
   {} &\!\! {= \sum\nolimits_j {p_j^{(1)} \rho _j^{(1)}  \otimes  \cdots  \otimes p_j^{(N)} \rho _j^{(N)} },}  \\
\end{array}
\end{Equation}
which is \textit{still} not enough for product form because the indices are ``locked'' across the modes, indicating nonlocal correlation.  Therefore, vectorizing the indices as
\begin{Equation}                      {F.4}
\rho =\hsp{-4.5} \sum\limits_{j_1 , \ldots ,j_N } {}\hsp{-4.5} p_{(j_1 , \ldots ,j_N )}^{(1)} \rho _{(j_1 , \ldots ,j_N )}^{(1)} \hsp{-1} \otimes  \cdots  \otimes p_{(j_1 , \ldots ,j_N ) }^{(N)} \rho _{(j_1 , \ldots ,j_N )}^{(N)},
\end{Equation}
and then applying both \Eq{8} and \Eq{6}, we get
\begin{Equation}                      {F.5}
\begin{array}{*{20}l}
   {\rho} &\!\! {= \sum\nolimits_{j_1 } { \cdots \sum\nolimits_{j_N } {p_{j_1 }^{(1)} \rho _{j_1 }^{(1)}  \otimes  \cdots  \otimes p_{j_N }^{(N)} \rho _{j_N }^{(N)} } }}  \\
   {} &\!\! {= (\sum\nolimits_{j_1 } {p_{j_1 }^{(1)} \rho _{j_1 }^{(1)} } ) \otimes  \cdots  \otimes (\sum\nolimits_{j_N } {p_{j_N }^{(N)} \rho _{j_N }^{(N)} } )}  \\
   {} &\!\! {= \rho ^{(1)}  \otimes  \cdots  \otimes \rho ^{(N)}}  \\
\end{array}
\end{Equation}
where \smash{$\rho ^{(m)}  \equiv \sum\nolimits_{j_m } {p_{j_m }^{(m)} \rho _{j_m }^{(m)} } $}, which proves the sufficiency of \Eqs{5}{8} for product form, and thus the absence of all nonlocal correlation.  

To prove the necessity of \Eqs{5}{8} for product form, we simply start from the product-form definition in \Eq{3},
\begin{Equation}                      {F.6}
\rho  = \rho ^{(1)}  \otimes  \cdots  \otimes \rho ^{(N)} .
\end{Equation}
Then, since each mode's state can be generally mixed as 
\begin{Equation}                      {F.7}
\rho ^{(m)}  = \sum\nolimits_{j_m } {p_{j_m }^{(m)} \rho _{j_m }^{(m)} } ,
\end{Equation}
we can put \Eq{F.7} into \Eq{F.6} to get 
\begin{Equation}                      {F.8}
\rho  = (\sum\nolimits_{j_1 } {p_{j_1 }^{(1)} \rho _{j_1 }^{(1)} } ) \otimes  \cdots  \otimes (\sum\nolimits_{j_N } {p_{j_N }^{(N)} \rho _{j_N }^{(N)} } ).
\end{Equation}
Rearranging shows that product states must have form
\begin{Equation}                      {F.9}
\begin{array}{*{20}l}
   {\rho} &\!\! {= \sum\nolimits_{j_1 } { \cdots \sum\nolimits_{j_N } {p_{j_1 }^{(1)} \rho _{j_1 }^{(1)}  \otimes  \cdots  \otimes p_{j_N }^{(N)} \rho _{j_N }^{(N)} } }}  \\
   {} &\!\! {= \sum\limits_{j_1 , \ldots ,j_N } {(p_{j_1 }^{(1)}  \cdots p_{j_N }^{(N)} )\rho _{j_1 }^{(1)}  \otimes  \cdots  \otimes \rho _{j_N }^{(N)} }.}  \\
\end{array}
\end{Equation}
Thus, line 2 of \Eq{F.9} verifies that this form is equivalent to product-form.  Then, to show how \Eq{F.9} relates to the notation of a general mixed state, we can unify these all with a vector index only if each mode has redundantly defined quantities of the forms \smash{$p_{j_m }^{(m)}  = p_{(j_1 , \ldots ,j_N )}^{(m)} $} and \smash{$\rho _{j_m }^{(m)}  = \rho _{(j_1 , \ldots ,j_N )}^{(m)} $}, regardless of the values of the indices of labels other than $m$, so then \Eq{F.9} becomes
\begin{Equation}                      {F.10}
\begin{array}{*{20}l}
   {\rho} &\!\! {= \hsp{-20}\sum\limits_{\hsp{17}j_1 , \ldots ,j_N \rule{0pt}{6.5pt}} {\hsp{-18}(p_{(j_1 , \ldots ,j_N )}^{(1)}  \hsp{-3}\cdots\hsp{-2} p_{(j_1 , \ldots ,j_N )}^{(N)} )\rho _{(j_1 , \ldots ,j_N )}^{(1)}  \hsp{-2}\otimes\hsp{-3}  \cdots\hsp{-3}  \otimes\hsp{-2} \rho _{(j_1 , \ldots ,j_N )}^{(N)} }}  \\
   {} &\!\! {= \sum\nolimits_j {(p_j^{(1)}  \cdots p_j^{(N)} )} \rho _j^{(1)}  \otimes  \cdots  \otimes \rho _j^{(N)},}  \\
\end{array}
\end{Equation}
and then if we abbreviate \smash{$p_j=p_j^{(1)}  \cdots p_j^{(N)}$} and $\rho _j =$ \smash{$ \rho _j^{(1)}  \otimes  \cdots  \otimes \rho _j^{(N)}$}, then \Eq{F.10} takes the general form
\begin{Equation}                      {F.11}
\rho  = \sum\nolimits_j {p_j \rho _j } ,
\end{Equation}
which shows that since we started from product states in \Eq{F.6} and used \Eqs{5}{8} to recover the form of a general state in \Eq{F.11}, then \Eqs{5}{8} are necessary for product form.  Thus, we have now proven that \Eqs{5}{8} are necessary and sufficient conditions for a state to have product form.
\section{\label{sec:App.G}Proofs and Caveats for the Six Families of Nonlocal Correlation}
Here we develop some important facts about the six families of nonlocal correlation from \Table{1}.
\subsection{\label{sec:App.G.1}Proof that All Decomposition Probabilities are Expressible in Product Form Without Mode Normalization}
Given a set of normalized decomposition probabilities $\{ p_j \}$, we can \textit{always} express them in product form as
\begin{Equation}                      {G.1}
p_j  = \prod\nolimits_{m = 1}^N {p_j ^{(m)} }  = p_j ^{(1)}  \cdots p_j ^{(N)}\;\; \forall j,
\end{Equation}
by using the parameterization
\begin{Equation}                      {G.2}
p_j ^{(m)}  = p_j ^{x_m^2 (\bm{\theta} ^{\{j\}} )}\;\; \forall j;\;\;m \in 1, \ldots ,N,
\end{Equation}
where for each $j$, the \smash{$x_m (\bm{\theta} ^{\{j\}} )$} for $m \in 1, \ldots ,N$ are $N$-dimensional unit-hyperspherical coordinates \cite[]{HedU} with $N-1$ angles \smash{$\bm{\theta} ^{\{j\}}\equiv (\theta _1^{\{j\}} , \ldots ,\theta _{N - 1}^{\{j\}} )$} which can be restricted to
\smash{$\theta _k^{\{j\}}  \in [0,\frac{\pi}{2}]$} for this application, with a different set for each $j$.  These coordinates have the property that \smash{$\sum\nolimits_{m = 1}^N {x_m^2 (\bm{\theta} ^{\{j\}} )}  = 1\;\;\forall j$}. Notice that in \Eq{G.2}, we are setting each mode-specific factor \smash{$p_j ^{(m)}$} equal to the full probability \smash{$p_j$} raised to the power \smash{$x_m^2 (\bm{\theta} ^{\{j\}} )$}.  Thus,
\begin{Equation}                      {G.3}
\begin{array}{*{20}l}
   {p_j } &\!\! {= p_j ^1  = p_j ^{x_1^2 (\bm{\theta} ^{\{j\}} ) +  \cdots  + x_N^2 (\bm{\theta} ^{\{j\}} )}  = p_j ^{x_1^2 (\bm{\theta} ^{\{j\}} )}  \cdots p_j ^{x_N^2 (\bm{\theta} ^{\{j\}} )}}  \\
   {} &\!\! {= p_j ^{(1)}  \cdots p_j ^{(N)}\;\;\forall j,}  \\
\end{array}
\end{Equation}
which also works right to left, given \Eq{G.2}. Therefore, we have proven that all decomposition probabilities are always expressible in product form as in \Eq{G.1}, where we do \textit{not} require mode normalization over $j$ as \smash{$\sum\nolimits_j {p_j ^{(m)} }  \hsp{-2}=\hsp{-2} 1\;\forall m$}, which is actually \textit{impossible} as we prove in \App{G.2}.

The consequence of \Eq{G.3} is that we do not need to define separate sets of families based on whether they have decompositions with product-form probabilities as \smash{$p_j = p_j ^{(1)}  \cdots p_j ^{(N)}$}, since such families would be redundant to other families with general probabilities $p_j$ and the same decomposition-state forms. For example, defining a family as \smash{$\rho = \sum\nolimits_j (p_j ^{(1)}  \cdots p_j ^{(N)})\rho_j$} would be exactly the same as Family 1, which is \smash{$\rho = \sum\nolimits_j p_{j}\rho_j$} of \Table{1}.
\subsection{\label{sec:App.G.2}Proof that Product-Form Probabilities with Mode Normalization on the Full Decomposition Index Are Impossible Except Trivially}
Suppose we require that in addition to having product form (PF), the decomposition probabilities $p_j$ must also have \textit{mode normalization} over the full decomposition index $j$, so that
\begin{Equation}                      {G.4}
p_j  = \prod\nolimits_{m = 1}^N {p_j ^{(m)} }  = p_j ^{(1)}  \cdots p_j ^{(N)} \;\;\forall j;\;\;\sum\nolimits_j {p_j ^{(m)} }  = 1\;\;\forall m,
\end{Equation}
which is \textit{not} the same thing as mode independence (MI) from \Eq{8} due to the sum over the full $j$ rather than mode-specific index $j_m$ where $j \equiv (j_1 , \ldots ,j_N )$.

First, for the PF part, we can use the parameterization of \App{G.1}, which, together with the mode normalization over $j$, requires that 
\begin{Equation}                      {G.5}
\sum\nolimits_j {p_j ^{(m)} }  = \sum\nolimits_j {p_j ^{x_m^2 (\bm{\theta} ^{\{ j\} } )} }  = 1\;\;\forall m.
\end{Equation}
For decompositions with $D$ states, this requires that
\begin{Equation}                      {G.6}
p_1 ^{x_m^2 (\bm{\theta} ^{\{ 1\} } )}  +  \cdots  + p_D ^{x_m^2 (\bm{\theta} ^{\{ D\} } )}  = 1\;\;\forall m.
\end{Equation}

Now, given that any collection of $D$ real numbers on $(0,1]$ such as $\{p_{j}\}$ can be viewed as eigenvalues of some physical $D$-level state, then they are always exactly determined by a set of $D$ equations for their power sums as
\begin{Equation}                      {G.7}
p_1^k  +  \cdots  + p_D^k  = \sum\nolimits_{j=1}^{D} {p_j^k }  \equiv P_k ,
\end{Equation}
for integers $k \in 1, \ldots ,D$. Thus, \Eq{G.6} constitutes a set of generally \textit{additional constraints} beyond the main set in \Eq{G.7}, so together \Eq{G.6} and \Eq{G.7} form a generally \textit{overdetermined set of nonlinear equations}.

It turns out that this overdetermined set \textit{does} have solutions, but only for cases that are irrelevant for the purpose of defining nontrivial families of decompositions for probability correlation, as we now briefly explain.

One way to get a solution to an overdetermined set is to find conditions for which the additional constraints simplify to the main constraints \Eq{G.7}. Since the powers of $p_j$ in \Eq{G.6} are squared unit-hyperspherical coordinates and therefore are each no greater than $1$, then for a given $m$ in \Eq{G.6}, we can never achieve the condition of having both all powers being equal and all being integers, without preventing that for all other $m$. Specifically, for a particular $m$, the only way to involve integer powers is to set \smash{$x_m^2 (\bm{\theta} ^{\{ j\} } ) = 1\;\forall j$} and thus satisfy \Eq{G.6} for that $m$, but then \textit{for all other $m$}, we would have \smash{$x_{\mbarsub}^2 (\bm{\theta} ^{\{ j\} } ) = 0\;\forall j$}, so that we would have $N-1$ unsatisfied equations of the form of \Eq{G.6}. 

There are only two ways that could work. It could work for a \textit{unipartite} system (one with $N=1$ mode, meaning no physical coincidence behavior, which is a system that can never have nonlocal correlation of any kind).  Alternatively, it could work with multiple modes if $D=1$, but that means the state is \textit{pure}, in which case the question of achieving a special new kind of decomposition probability correlation through \Eq{G.4} is irrelevant since there is only one unique decomposition state with probability $1$.

Therefore, we have outlined the proof that expressing decomposition probabilities $p_j$ as products of $N$ factors where each is separately normalized over full decomposition index $j$ is \textit{impossible} in all cases except pure states or unipartite states, neither of which can have probability correlation. (An alternative proof would be to use $p$-norms to achieve $DN$ separate inequalities, which lead to the same trivial exceptions.)
\subsection{\label{sec:App.G.3}Quasi-Families}
Since state decompositions are generally not unique, it may happen that for some states, they belong to \textit{multiple partially-intersecting families} from \Table{1}, but only through \textit{different decompositions}, so that they do not belong to the intersection of those families since they do not have a single decomposition in that intersection.

For example, since Family 2 is \smash{$\rho  = \sum\nolimits_j {(p_{j_1 }^{(1)}  \cdots p_{j_N }^{(N)} )\rho _j } $}, while\hsp{-1.5} Family\hsp{-1.5} 3\hsp{-1.5} is\hsp{-1.5} \smash{$\rho \hsp{-1} =\hsp{-3} \sum\nolimits_j {p_j \rho _j^{(1)}  \hsp{-2.5}\otimes  \hsp{-1.5}\cdots \hsp{-1.5} \otimes\hsp{-1} \rho _j^{(N)} } $}\rule{0pt}{11pt}, we see that the probabilities of Family 2 are a strict subset of Family 3, but the decomposition states of Family 2 are a strict \textit{superset} of Family 3, meaning that \textit{neither Family 2 nor Family 3 can be strict subsets of each other, although\hsp{-0.5} they\hsp{-0.5} can\hsp{-0.5} have\hsp{-0.5} an\hsp{-0.5} intersection,\hsp{-0.5} which\hsp{-0.5} is\hsp{-0.5} Family\hsp{-0.5} 4} \smash{$\rho  = \sum\nolimits_j {(p_{j_1 }^{(1)}  \cdots p_{j_N }^{(N)} )\rho _j^{(1)}  \otimes  \cdots  \otimes \rho _j^{(N)} } $}.  While any state with a decomposition in Family 4 is in both Family 2 and Family 3, there may exist some states that have \textit{different decompositions} in Family 2 and Family 3 separately, but no decomposition in Family 4.

Thus we define a \textit{quasi-family} as being a set of states with the property of belonging to multiple partially intersecting families without belonging to their intersection.

To identify all quasi-families, we have to check every possible set of every possible number of families, keeping in mind that if any family is a proper subset of another family, then that pair does \textit{not} constitute a quasi-family, since then membership in the subset guarantees membership in the superset for the same decomposition.

Checking all family combinations in \Table{1} shows that there are only three quasi-families in this group,
\begin{Equation}                      {G.8}
\begin{array}{*{20}l}
   {\text{Quasi-Family}} & {2|3,}  \\
   {\text{Quasi-Family}} & {2|5,}  \\
   {\text{Quasi-Family}} & {4|5,}  \\
\end{array}
\end{Equation}
where Quasi-Family $x_{1}|\cdots|x_{M}$ is a quasi-family involving $M$ families; Family $x_{1}$ through Family $x_{M}$. Note that due the various subset memberships in this small set of families, there are no quasi-families between more than two families in \Table{1}.

The quasi-families may not be as important as the main families. In particular, being able to satisfy multiple family definitions with a single decomposition is what leads to such extreme behavior as achieving product form, whereas a state that only achieved  mode-independent (MI) probabilities and MI decomposition states with separate decompositions alone would not achieve product form.  Therefore, \textit{the fact that a state can have membership to multiple families but not their intersection does not diminish the significance of the families regarding their roles in producing nonlocal correlation}.  

Ultimately, since a state needs to have at least a \textit{single decomposition} that satisfies a given family definition for it to have the nonlocal correlation achievable by that family, then quasi-families are not as important as families regarding nonlocal correlation.

Furthermore, due to the difficulty of constructing states that have different decompositions of particular forms, we do not have examples of quasi-family states at this time (keeping in mind that states within the intersection of the two families involved in a quasi-family are \textit{not} part of the quasi-family, so we cannot use such intersections to generate examples of quasi-family members).  It may be that no states exist in any quasi-families, in which case they can be ignored as physically irrelevant.

Nevertheless, we mention quasi-families here in case they turn out to be physically meaningful in some way, and thus this is a possible area for further research.%
\vspace{-8pt}%
\section{\label{sec:App.H}True-Generalized X (TGX) States}%
\vspace{-6pt}%
True-generalized X (TGX) states are defined as a special family of states that are conjectured to be related to all general states (both pure and mixed) by an entanglement-preserving unitary (EPU) transformation, so that the TGX state and the general state connected by such an EPU have the same entanglement, a property called \textit{EPU equivalence}.  

The name TGX means ``the true generalization of X states with respect to entanglement for all systems as big as or larger than two qubits,'' meaning that, just as the X states are EPU equivalent to general states for $2\times 2$ systems (which is now proven by two independent methods as detailed below), TGX states (if they exist) are generally the larger-system analog of that two-qubit family, having the defining property of EPU equivalence with the set of general states.

The leading candidates for TGX states in all systems are called \textit{simple} states, defined as those states for which their $N$ single-mode reductions are all diagonal in the computational basis, such that all of the off-diagonal parent-state matrix elements appearing in the formal off-diagonals of those reductions are identically zero (meaning that those parent elements do not merely add to zero, but are each themselves zero).

As a simple nontrivial example of TGX states, in $2\times 3$, since the single-mode reductions are
\begin{Equation}                      {H.1}
\begin{array}{*{20}l}
   {\redx{\rho}{1}\! } &\!\! { =\! \left(\! {\begin{array}{*{20}c}
   {\rho _{1,1}  + \rho _{2,2}  + \rho _{3,3} } & {\rho _{1,4}  + \rho _{2,5}  + \rho _{3,6} }  \\
   {\rho _{4,1}  + \rho _{5,2}  + \rho _{6,3} } & {\rho _{4,4}  + \rho _{5,5}  + \rho _{6,6} }  \\
\end{array}}\! \right)}  \\
   {\redx{\rho}{2}\! } &\!\! { =\! \left(\! {\begin{array}{*{20}c}
   {\rho _{1,1}  + \rho _{4,4} } & {\rho _{1,2}  + \rho _{4,5} } & {\rho _{1,3}  + \rho _{4,6} }  \\
   {\rho _{2,1}  + \rho _{5,4} } & {\rho _{2,2}  + \rho _{5,5} } & {\rho _{2,3}  + \rho _{5,6} }  \\
   {\rho _{3,1}  + \rho _{6,4} } & {\rho _{3,2}  + \rho _{6,5} } & {\rho _{3,3}  + \rho _{6,6} }  \\
\end{array}}\! \right)\!,}  \\
\end{array}
\end{Equation}
then the parent elements contributing to the off-diagonals of these reductions are $\rho_{4,1}$, $\rho_{5,2}$, $\rho_{6,3}$,  $\rho_{2,1}$, $\rho_{5,4}$, $\rho_{3,1}$, $\rho_{6,4}$, $\rho_{3,2}$, $\rho_{6,5}$, and their index-swapped counterparts, so setting all of these to zero not only makes the reductions diagonal, but defines a \textit{simple} parent state as
\begin{Equation}                      {H.2}
\rho  =\! \left( {\begin{array}{*{20}c}
   {\rho _{1,1} } &  \cdot  &  \cdot  &  \cdot  & {\rho _{1,5} } & {\rho _{1,6} }  \\
    \cdot  & {\rho _{2,2} } &  \cdot  & {\rho _{2,4} } &  \cdot  & {\rho _{2,6} }  \\
    \cdot  &  \cdot  & {\rho _{3,3} } & {\rho _{3,4} } & {\rho _{3,5} } &  \cdot   \\
    \cdot  & {\rho _{4,2} } & {\rho _{4,3} } & {\rho _{4,4} } &  \cdot  &  \cdot   \\
   {\rho _{5,1} } &  \cdot  & {\rho _{5,3} } &  \cdot  & {\rho _{5,5} } &  \cdot   \\
   {\rho _{6,1} } & {\rho _{6,2} } &  \cdot  &  \cdot  &  \cdot  & {\rho _{6,6} }  \\
\end{array}} \right)\!,
\end{Equation}
which we take as a working hypothesis to be the family of $2\times 3$ TGX states (where dots represent zeros to help show its form). Note that in all of the work on TGX states so far, all evidence strongly supports the hypothesis that simple states are TGX states, so the two terms are often used interchangeably. However, if simple states are ever proved not to have EPU equivalence, the idea of TGX states can then be reserved for EPU equivalent states if they exist.  See \cite[]{HedX} for many other examples.
\\
\\
\noindent\textit{A Brief History of TGX States:}\vspace{-4pt}
\begin{itemize}[leftmargin=*,labelindent=4pt]\setlength\itemsep{0pt}
\item[\textbf{1.}]\hypertarget{CorrelanceTGXHistory:1}{}(2013) \cite[]{HedX} gave the first definition of TGX states, and the general form was conjectured to be that of \textit{simple} states.  The idea of EPU equivalence was also introduced, and strong numerical evidence was shown that simple TGX states are EPU equivalent to all states for $2\times 2$ and $2\times 3$ systems, and this property was conjectured to hold for TGX states in all quantum systems. Numerical evidence was also given showing that literal X states cannot in general be EPU equivalent to general states, with respect to negativity. The maximally entangled-basis (MEB) theorem was conjectured and shown to be fulfilled by TGX states for several example systems.
\item[\textbf{2.}]\hypertarget{CorrelanceTGXHistory:2}{}(2014) \cite[]{HedD} presented the Bloch-vector form of simple candidates for TGX states.
\item[\textbf{3.}]\hypertarget{CorrelanceTGXHistory:3}{}(2014) \cite[]{MeMG} proved the conjecture of \cite[]{HedX} for the $2\times 2$ case by showing the implicit existence of an EPU connecting all general states to X states (which are TGX states in $2\times 2$ systems).
\item[\textbf{4.}]\hypertarget{CorrelanceTGXHistory:4}{}(2016) \cite[]{HedE} presented the multipartite entanglement measure \textit{the ent}.  TGX states (simple candidates) were used to prove the MEB theorem for all discrete quantum systems. It was proved that ME TGX states have the special property of having balanced superposition. Furthermore, ME TGX states were shown to yield indexing patterns that can function as a multipartite Schmidt decomposition state for full $N$-partite entanglement. This also presented the 13-step algorithm as a method for deterministically constructing all possible ME TGX states in all discrete quantum systems.
\item[\textbf{5.}]\hypertarget{CorrelanceTGXHistory:5}{}(2017) \cite[]{MeMH} proved that in $2\times 3$ systems, literal X states definitely cannot achieve EPU equivalence to general states with respect to negativity, which also proves that literal X states cannot have EPU equivalence in general systems if negativity is a valid measure of entanglement in $2\times 2$ and $2\times 3$.  This study also added further numerical evidence agreeing with that of \cite[]{HedX} suggesting that the TGX states \textit{may} indeed achieve EPU equivalence in $2\times 3$ systems.
\item[\textbf{6.}]\hypertarget{CorrelanceTGXHistory:6}{}(2018) \cite[]{HeXU} presented an explicit family of $2\times 2$ X states parameterized by concurrence and spectrum and proved it to be EPU-equivalent to the set of all states, providing an explicit proof of the original conjecture of \cite[]{HedX}, and proving the existence of an explicit formula for the EPU of the transformation, as well as yielding an explicit ready-to-use EPU-equivalent state family.
\end{itemize}
\vspace{-23pt}%
\section{\label{sec:App.I}Proof that Correlance is a Necessary and Sufficient Measure of All Nonlocal Correlation}%
\vspace{-9pt}%
First, from the definition of correlance $\Correlance$ in \Eqs{9}{12},
\begin{Equation}                      {I.1}
\Correlance(\rho ) = 0\;\;\text{iff}\;\;\rho  = \varsigma (\rho ),
\end{Equation}
since \smash{$\Correlance(\rho )=\tr[(\rho  - \varsigma )^2 ]/\mathcal{N}_{\Correlance}$} is proportional to the square magnitude of the difference of the Bloch vectors of $\rho$ and $\varsigma (\rho )$, which is zero iff $\rho=\varsigma (\rho )$.  Next, by \Thm{1}, proven in \App{F},
\begin{Equation}                      {I.2}
\rho  = \varsigma (\rho )\;\;\text{iff}\;\;\rho \;\;\text{has product form}.
\end{Equation}
\vspace{-12pt}\\%
Furthermore, from \Eq{3},%
\vspace{-4pt}%
\begin{Equation}                      {I.3}
\mbox{$\rho$ has no nonlocal correlation \,iff\, $\rho$ has product form,}
\end{Equation}
which is true since discarding any modes by partial tracing leaves the states of the remaining modes unchanged.  Then, putting \Eq{I.3} into \Eq{I.2} and the result into \Eq{I.1} gives%
\vspace{-1pt}%
\begin{Equation}                      {I.4}
\mbox{$\Correlance(\rho )=0$ \,iff\, $\rho$ has no nonlocal correlation,}
\end{Equation}
\vspace{-12pt}\\%
which is what we set out to prove. This means we can use $\Correlance$ to detect any and all kinds of nonlocal correlation. The only drawback is that it cannot tell us which kind(s) of correlation is(are) present. Nevertheless, if $\Correlance(\rho)>0$, we are guaranteed that $\rho$ has some nonlocal correlation.
\vspace{-25pt}%
\section{\label{sec:App.J}Proof that Correlance is Properly Normalized}%
\vspace{-9pt}%
Here we prove that it is valid to normalize correlance $\Correlance$ with maximally entangled (ME) states (see \App{K} for a derivation of the normalization factor). For ease of display, the proof's steps are given as numbered facts.
\begin{itemize}[leftmargin=*,labelindent=4pt]\setlength\itemsep{0pt}
\item[\textbf{1.}]\hypertarget{CorrelanceNormProof:1}{}
$\Correlance$ measures \textit{nonlocal correlation}, as proved in \App{I}. (``Nonlocal correlation'' abbreviates ``full $N$-partite nonlocal correlation'' here; see \Sec{VII} for generalizations to distinctly multipartite nonlocal correlation).
\item[\textbf{2.}]\hypertarget{CorrelanceNormProof:2}{}From \Eq{10}, the raw (unnormalized) correlance \smash{$\widetilde{\Correlance}(\rho)\equiv$} $\tr[(\rho-\varsigma)^{2}]=\tr(\rho^{2})-2\tr(\rho\varsigma)+\tr(\varsigma^{2})$ is a function of input state $\rho$ and its reduction product \smash{$\varsigma\equiv\varsigma(\rho)\equiv\redx{\rho}{1}\otimes\cdots\otimes\redx{\rho}{N}$}.  As we will see later, \smash{$\widetilde{\Correlance}(\rho)$} \textit{is simply a function of the squared Euclidean distance between the Bloch vectors of} $\rho$ \textit{and} $\varsigma(\rho)$. Thus \textit{correlance} \smash{$\Correlance(\rho)\equiv\widetilde{\Correlance}(\rho)/{\mathcal{N}_\Correlance}$} \textit{is a measure of distance between} $\rho$ \textit{and its reduction product} $\varsigma$.
\item[\textbf{3.}]\hypertarget{CorrelanceNormProof:3}{}\textit{Pure} states can be used to maximize $\Correlance(\rho)$. \textit{Proof:} $\Correlance(\rho)$ is proportional to a squared distance between Bloch vectors (BVs) as \smash{$\Correlance(\rho ) = \frac{1}{\mathcal{N}_{\Correlance}}\frac{n - 1}{n}|\bm{\Gamma}  - \bm{\Gamma}_\varsigma  |^2 $} (see \App{K} for details about our BV\rule{0pt}{10pt} notation) where $\bm{\Gamma}$ is the BV of $\rho$ and $\bm{\Gamma}_\varsigma$ is the BV of $\varsigma\equiv\varsigma(\rho)$, both of which have real components in a Hermitian operator basis. $\Correlance(\rho)$ can be adapted for BV input as $\Correlance^{\prime}(\bm{\Gamma})\equiv\Correlance(\rho)$, and since both terms in the \textit{reduction difference vector} $\bm{\Delta}\equiv\bm{\Delta}(\bm{\Gamma})\equiv\bm{\Gamma}-\bm{\Gamma}_\varsigma$ depend on the input state, we can rewrite it as \smash{$\Correlance^{\prime\prime}(\bm{\Delta})\equiv\Correlance^{\prime}(\bm{\Gamma})\equiv\Correlance(\rho)=\frac{1}{\mathcal{N}_{\Correlance}}\frac{n - 1}{n}|\bm{\Delta}|^2$}. Then, recall that $f(x)$ is \textit{strongly convex} iff\rule{0pt}{10pt} for all $x$ and $y$ in its domain and $p\geq 0$, there exists some scalar $m\geq 0$ such that $f(px + (1 - p)y) \le pf(x) + (1 - p)f(y) - \frac{1}{2}mp(1 - p)||x - y||_2^2 $, and that any strongly convex function is also \textit{convex}. Thus, given reduction difference vectors $\mathbf{X}$ and $\mathbf{Y}$, since the quantities $\Correlance^{\prime\prime}(p\mathbf{X}+(1-p)\mathbf{Y})$ and $p\Correlance^{\prime\prime}(\mathbf{X}) + (1 - p)\Correlance^{\prime\prime}(\mathbf{Y}) - \frac{1}{2}mp(1 - p)|\mathbf{X} - \mathbf{Y}|^2 $ are \textit{equal} if $m=\frac{2}{\mathcal{N}_{\Correlance}}\frac{n - 1}{n}$, then $\Correlance^{\prime\prime}(\bm{\Delta})$ and thus $\Correlance(\rho)$ are both strongly convex and convex. Then, recalling Jensen's inequality for convex $f(\mathbf{x})$, that \smash{$f(\sum\nolimits_{j = 1}^{\hsp{1}D} { p_j \mathbf{x}_j } ) \le \sum\nolimits_{j = 1}^D { p_j f(\mathbf{x}_j )}$}, to which a corollary is \smash{$f(\sum\nolimits_{j = 1}^{\hsp{1}D} { p_j \mathbf{x}_j } ) \le \max \{ f(\mathbf{x}_1 ), \ldots ,f(\mathbf{x}_D )\}$}\rule{0pt}{10pt}, then for any mixed input $\rho$, $\Correlance(\rho ) \le \max \{ \Correlance(\rho _1) , \ldots ,\Correlance(\rho _D ) \}$ where $\rho _j$ are \textit{pure} decomposition states of $\rho$. Thus maximizers of $\Correlance(\rho )$ over all $\rho$ are \textit{pure}.
\item[\textbf{4.}]\hypertarget{CorrelanceNormProof:4}{}Since pure states have the trivial decomposition probability of $p_{1}=1$, with only one pure decomposition state up to global phase, then \textit{the only mechanism for nonlocal correlation in pure states is the nonfactorizability of the state itself, meaning its entanglement}.
\item[\textbf{5.}]\hypertarget{CorrelanceNormProof:5}{}Since entanglement is the only nonlocal correlation possible for pure states by \hyperlink{CorrelanceNormProof:4}{Fact 4}, the pure states of highest $\Correlance$ are those of highest entanglement. (Since \App{I} proved that nonlocal correlation is what $\Correlance$ measures, this also means that for pure states, the states of highest nonlocal correlation are the states of highest entanglement.)
\item[\textbf{6.}]\hypertarget{CorrelanceNormProof:6}{}The pure states with the highest entanglement are any $\rho$ for which all of its single-mode reductions $\redx{\rho}{m}$ have the lowest simultaneous purities possible for them to have, given their pure parent state $\rho$. We define these states as \textit{maximally full-$N$-partite-entangled} states, or just ``maximally entangled'' (ME) states here, and represent them as $\rho_{\text{ME}}$.  (This definition led to the derivation of the automatically normalized entanglement measure the \textit{ent} $\Upsilon(\rho)$ in \cite[]{HedE}.)
\item[\textbf{7.}]\hypertarget{CorrelanceNormProof:7}{}Therefore, by \hyperlink{CorrelanceNormProof:5}{Fact 5} and \hyperlink{CorrelanceNormProof:6}{Fact 6}, the highest $\Correlance$ for pure states is achieved by pure ME states $\rho_{\text{ME}}$, so then by \hyperlink{CorrelanceNormProof:3}{Fact 3}, the pure ME states maximize $\Correlance$ over \textit{all} states (mixed and pure), which proves the part of \Eq{12} that says $\mathcal{N}_{\Correlance}=\widetilde{\Correlance}(\rho_{\text{ME}})$.
\item[\textbf{8.}]\hypertarget{CorrelanceNormProof:8}{}In \cite[]{HedE} it was shown that the \textit{simplest} maximally full-$N$-partite-entangled states are ME TGX states $\rho_{\text{ME}_{\text{TGX}}}$, since they achieve the minimum simultaneous single-mode purities while also having equal superposition coefficients, with not all levels being nonzero.  Therefore, since ME TGX states have the same entanglement as general ME states, then by \hyperlink{CorrelanceNormProof:7}{Fact 7}, this proves the part of \Eq{12} that says $\mathcal{N}_{\Correlance}=\widetilde{\Correlance}(\rho_{\text{ME}_{\text{TGX}}})$.
\end{itemize}

Thus, we have proven that any pure ME state can be used to normalize $\Correlance$ over all states, both mixed and pure. The reason for using ME TGX states is that they are generally simpler than general ME states, since it was proved in \cite[]{HedE} that ME TGX states always have balanced superposition and not all levels are nonzero.  Furthermore, the ME TGX states can be methodically generated, using the 13-step algorithm of \cite[]{HedE}. Regarding the distance interpretation of $\Correlance$ from \hyperlink{CorrelanceNormProof:2}{Fact 2}, see \cite[]{StKB} for the intimate connection between distance measures of entanglement and convex-roof extensions of entanglement monotones.%
\vspace{-8pt}%
\section{\label{sec:App.K}Proof and Calculation of Explicit Correlance Normalization Factors}%
\vspace{-6pt}%
This proof makes extensive use of multipartite Bloch vectors \cite[]{Blch,HedD}, and therefore this appendix has two parts; \App{K.1} reviews multipartite-Bloch-vector formalism, and \App{K.2} derives the normalization factors.%
\vspace{-6pt}%
\subsection{\label{sec:App.K.1}Review of Bloch-Vector Quantities}%
\vspace{-6pt}%
The concept of a Bloch vector is simply to use a set of operators that is somehow \textit{complete} in that it allows us to expand any operator as a linear combination of that set of operators.  The Bloch vector is then the list of scalar coefficients of that expansion.

The idea of Bloch vectors actually originated as \textit{Stokes parameters} in 1852 \cite[]{Stok}, which were used as an operational method of describing \textit{classical} light.  However, the quantum-mechanical density matrix was not invented until 1927, by von Neumann \cite[]{VNeu}, and the modern idea of Bloch vectors for quantum states came from Felix Bloch's 1946 treatment of mixed-state qubits \cite[]{Blch}, which were soon-after connected with the density matrix.  The 1961 development of the Gell-Mann (GM) matrices by Ne'eman and Gell-Mann \cite[]{Neem,Gell} paved the way for describing states larger than a qubit, but it was not until 1981 that a unipartite $n$-level Bloch vector was devised, by Hioe and Eberly \cite[]{HiEb}.  Soon after, many multipartite descriptions were attempted and many works treat simple cases of these.  Therefore, here we present a brief \textit{general} treatment of multipartite Bloch vectors, from the more complete work in \cite[]{HedD} from 2014.

Consider a multipartite system of $N$ subsystems (modes), with Hilbert space $\mathcal{H}^{(1)}\otimes\mathcal{H}^{(2)}\otimes\cdots\otimes\mathcal{H}^{(N)}$, where $\mathcal{H}^{(m)}$ is the Hilbert space of mode $m$, where $\mathcal{H}$ has $n$ total levels such that $n = n_1  \cdots n_N $, where $n_m$ is the number of levels of mode $m$.

Now let \smash{$\{ \nu _k ^{(m)} \}$} be a complete basis of $n_m^2$ operators for mode $m$, such that all operators in \smash{$\mathcal{H}^{(m)}$} can be expanded as linear combinations of \smash{$\{ \nu _k ^{(m)} \}$}, and where \smash{$\nu _0 ^{(m)}  \equiv I^{(m)} $} is the identity for \smash{$\mathcal{H}^{(m)}$}.  Furthermore, under the Hilbert-Schmidt (HS) inner product $A\cdot B\equiv\text{tr}(A^{\dag}B)$, suppose that \smash{$\{ \nu _k ^{(m)} \}$} has \textit{uniform} orthogonality,
\begin{Equation}                      {K.1}
\tr(\nu _{j_{m}}^{(m)\dag} \nu _{k_{m}}^{(m)} ) = n_{m} \delta _{{j_{m}},{k_{m}}},
\end{Equation}
for all ${j_{m}},{k_{m}}\in 0,\ldots,n_{m}^{2}-1$, where ``uniform'' means that only one case is needed to cover all indices, which allows the simplest transition to a multipartite basis.  Thus, we can define a multipartite basis as
\begin{Equation}                      {K.2}
\nu_{\mathbf{k}}\equiv \nu _{k_1 , \ldots ,k_N }^{(1,\ldots ,N)} \equiv \nu _{k_1 }^{(1)}  \otimes  \cdots  \otimes \nu _{k_N }^{(N)}
\end{Equation}
for \smash{$k_{m}\in 0,\ldots,n_{m}^{2}-1\;\,\forall m\in\{1,\ldots,N\}$}, where the vector-index subscript indicates the multipartite nature of the basis, where the number of elements in the vector is the number of modes $N$.  The set $\{\nu_{\mathbf{k}}\}$ inherits the uniform orthogonality of its modes as
\begin{Equation}                      {K.3}
\tr(\nu_{\mathbf{j}}^{\dag} \nu_{\mathbf{k}} ) = n\delta _{\mathbf{j},\mathbf{k}},
\end{Equation}
valid for all \smash{$k_{m}\in 0,\ldots,n_{m}^{2}-1\;\,\forall m\in\{1,\ldots,N\}$}, and where \smash{$\delta _{\mathbf{j},\mathbf{k}}\equiv\delta _{j_1 ,k_1 }  \cdots \delta _{j_N ,k_N }$}.

The HS completeness of \smash{$\{\nu_{\mathbf{k}}\}$} lets us express all density operators as \smash{$\rho  =\sum\nolimits_{\mathbf{k}}{c_{\mathbf{k}}\nu_{\mathbf{k}} }  = \frac{1}{n}\! ( {\nu _{\mathbf{0}}  + A \sum\nolimits_{\mathbf{k}\neq\mathbf{0}} {\Gamma _{\mathbf{k}} \nu_{\mathbf{k}} } } )$}, where \smash{$c_{\mathbf{k}}$} are generally complex scalars, and \smash{$\Gamma_{\mathbf{k}}\equiv\Gamma_{\mathbf{k}\neq \mathbf{0}}\equiv\Gamma_{k_{1},\ldots,k_{N}}\equiv\frac{1}{A}\frac{c_{k_{1},\ldots,k_{N}}}{c_{0,\ldots,0}}$} are the $n^2 -1$ scalars known as Bloch components that constitute a Bloch vector $\bm{\Gamma}$, and we choose \smash{$\Gamma_{0,\ldots,0}\equiv\frac{c_{0,\ldots,0}}{c_{0,\ldots,0}}=1$} by convention.  Then, computing the purity \smash{$P\equiv\tr(\rho^2)$}\rule{0pt}{10pt} and applying the \textit{unitization} condition that $|\bm{\Gamma}|=1$ for pure states $\rho$, then \smash{$A=\sqrt{n-1}$}, and we obtain the multipartite Bloch-vector expansion of $\rho$ as
\begin{Equation}                      {K.4}
\begin{array}{*{20}c}
   {\rho} &\!\! {=\frac{1}{n}(\nu_{\mathbf{0}}  + \sqrt{n - 1} \bm{\Gamma}\cdot\bm{\nu} ),}  \\
\end{array}
\end{Equation}
where $\bm{\Gamma}$ is the list of scalars $\{\Gamma _{\mathbf{k}  \neq \mathbf{0}}\}$, and $\bm{\nu}$ is the list of operators $\{\nu_{\mathbf{k}\neq \mathbf{0}}\}$ in the same order as $\{\Gamma _{\mathbf{k}  \neq \mathbf{0}}\}$, and the dot product in the context of \Eq{K.4} is just an abbreviation for $\sum\nolimits_{\mathbf{k}\neq\mathbf{0}} {\Gamma _{\mathbf{k}} \nu_{\mathbf{k}} }$. Then, applying \Eq{K.3} to \Eq{K.4} gives the multipartite Bloch components as
\begin{Equation}                      {K.5}
\begin{array}{*{20}c}
   {\Gamma _{\mathbf{k}  \neq \mathbf{0}}} &\!\! {= \frac{1}{\sqrt {n - 1}}\tr(\rho \nu_{\mathbf{k}}^{\dag} ).}  \\
\end{array}
\end{Equation}
Note that if $\{\nu_{\mathbf{k}}\}$ consists entirely of Hermitian operators, then the $\Gamma _{\mathbf{k}  \neq \mathbf{0}}$ will all be real.  The Bloch vector $\bm{\Gamma}$ contains all of the same information as $\rho$ and can be used as an alternative method of representing any physical state.

The purity of $\rho$ is $P \equiv \tr(\rho ^2 ) = \frac{1}{n}(1 + (n - 1)|\bm{\Gamma} |^2 )$, so we can define the \textit{Bloch purity} as
\begin{Equation}                      {K.6}
\begin{array}{*{20}c}
   {P_{\text{B}}} &\!\! {\equiv |\bm{\Gamma} |^2 = \sum\limits_{\mathbf{k} \ne \mathbf{0}} {|\Gamma _{\mathbf{k}} |^2 } = \frac{n\tr(\rho ^2 ) - 1}{n - 1},}  \\
\end{array}
\end{Equation}
which obeys $0\leq P_{\text{B}}\leq 1$, such that $P_{\text{B}}=1$ for pure states, $0\leq  P_{\text{B}}<1$ for general strictly mixed states, and $P_{\text{B}}=0$ for the maximally mixed state.

So far, we have merely specified properties of $\{\nu_{\mathbf{k}}\}$ without explaining how to make it. One simple way to construct it is, for each $m\in\{1,\ldots,N\}$, let
\begin{Equation}                      {K.7}
\begin{array}{*{20}c}
   {\nu_0^{(m)}  \equiv \lambda _0^{(m)}  \equiv I^{(m)}} & {\text{and}} & {\nu _{k_{m}}^{(m)}  \equiv \sqrt {\frac{n_{m}}{2}} \lambda _{k_{m}}^{(m)},}  \\
\end{array}
\end{Equation}
where \smash{${k_{m}}\in 1,\ldots,n_{m}^{2}-1$}, and the \smash{$\lambda _{k_{m}}^{(m)}$} are generalized Gell-Mann (GM) matrices in mode $m$ of $n_m$ levels, given by the implicit equations
\begin{Equation}                      {K.8}
\begin{array}{*{20}l}
   {\lambda_{\scriptstyle \alpha ^2  - 2(\alpha -\beta ) - 1 \hfill \atop 
  \scriptstyle \alpha ^2  - 2(\alpha -\beta ) \hfill}^{(m)}  \equiv \begin{array}{*{20}c}
   {\; E_{(\alpha,\beta)}^{[n_m]}+E_{(\beta,\alpha)}^{[n_m]} }  \\
   {\! i\!\left({E_{(\alpha,\beta)}^{[n_m]}-E_{(\beta,\alpha)}^{[n_m]} }\right) }  \\
\end{array}\!\! ; \;\alpha  > \beta ,}  \\
   {\lambda_{\alpha ^2  - 1}^{(m)}  \equiv \sqrt{\frac{2}{\alpha (\alpha  - 1)}} \left(\! { - (\alpha  - 1)E_{(\alpha,\alpha)}^{[n_m]}  +\! \sum\limits_{q = 1}^{\alpha  - 1} {E_{(q,q)}^{[n_m]} } }\! \right)\! ,}  \\
\end{array}
\end{Equation}
where $2 \le \alpha \le n_m$ and $1 \le \beta \le \alpha-1$, and $E_{(\alpha,\beta)}^{[n_m]}\equiv|\alpha\rangle\langle\beta|$ is the $n_m\times n_m$ matrix with a $1$ in the row-$\alpha$, column-$\beta$ entry and $0$ elsewhere, given that the top row is row $1$, and the left column is column $1$, and \smash{$\lambda_{0}^{(m)}\equiv I^{[n_m]}$} is the $n_m\times n_m$ identity matrix for mode $m$.  A given pair of integers $\alpha,\beta$ determines the conventional label for each GM matrix.

Note that the GM matrices are not preferable as a basis for general multipartite Bloch vectors (though they are often used for that), because their orthogonality relations require two cases to include the identity, resulting in $2^N$ cases for the orthogonality of a general $N$-partite system.  Thus, putting \Eq{K.8} into \Eq{K.7} and using that in \Eq{K.2}, we obtain a realization for the multipartite basis $\{\nu_{\mathbf{k}}\}$, which has only \textit{one} orthogonality case, which is given by \Eq{K.3}.

We now develop some formalism that will be useful in establishing the results we will use to prove the normalization factors of the correlance.

For multipartite systems, the implicit-basis representation of $\bm{\Gamma}$ is not intuitive, since any vector or matrix representation uses relative positions on a page to encode the basis, so listing out components with vector indices is not helpful. Therefore, we will simply keep basis-explicit notation using a vector-indexed basis, and to that end we introduce a \textit{uniform standard basis} (USB) as 
\begin{Equation}                      {K.9}
\mathbf{b}_{\mathbf{k}}  \equiv \mathbf{b}_{k_1 , \ldots ,k_N }^{(1, \ldots ,N)}  \equiv \frac{1}{{\sqrt n }}\nu _{\mathbf{k}}  = \frac{1}{{\sqrt n }}\nu _{k_1 }^{(1)}  \otimes  \cdots  \otimes \nu _{k_N }^{(N)} .
\end{Equation}
If we then define the Hilbert-Schmidt inner product between Bloch vectors $\mathbf{A}$ and $\mathbf{B}$ in the USB as
\begin{Equation}                      {K.10}
\mathbf{A} \cdot \mathbf{B} \equiv \tr(\mathbf{A}^{\dag}  \mathbf{B}),
\end{Equation}
then any USB as defined above has orthonormality
\begin{Equation}                      {K.11}
\mathbf{b}_\mathbf{j}  \cdot \mathbf{b}_\mathbf{k}  = \delta _{\mathbf{j},\mathbf{k}} .
\end{Equation}
Then, in the USB, Bloch vectors have explicit-basis form,
\begin{Equation}                      {K.12}
\bm{\Gamma}  = \sum\limits_{\mathbf{k} \ne \mathbf{0}}^\mathbf{K} {\Gamma _\mathbf{k} \mathbf{b}_\mathbf{k} },
\end{Equation}
where $K_m \equiv n_{m}^{2}-1$ and $k_m \in 0,\ldots,K_m$, so only the case of all $k_m =0$ is excluded from the sum.  Thus, in the USB, $\bm{\Gamma}$ is a matrix, and $\Gamma _\mathbf{k}$ have the same values as in \Eq{K.5} but are now given by
\begin{Equation}                      {K.13}
\Gamma _{\mathbf{k} \ne \mathbf{0}}  = \bm{\Gamma}  \cdot \mathbf{b}_\mathbf{k}  = \tr(\bm{\Gamma} \mathbf{b}_\mathbf{k} ^{\dag}  ) = \sqrt {\frac{n}{{n - 1}}} \tr(\rho \mathbf{b}_\mathbf{k} ^{\dag}  ),
\end{Equation}
and the density matrix can be expanded more simply as
\begin{Equation}                      {K.14}
\rho  = \frac{1}{n}(I + \sqrt {n(n - 1)} \bm{\Gamma} ),
\end{Equation}
where note that we use boldness to distinguish Bloch-vector objects from density matrices, despite both being represented as matrices here.

The (informal) overlap of any two states $\rho_A$ and $\rho_B$ is then (using their Hermiticity in the HS inner product), 
\begin{Equation}                      {K.15}
\tr(\rho _A \rho _B ) = \frac{1}{n}\left( {1 + (n - 1)\bm{\Gamma} _A  \cdot \bm{\Gamma} _B } \right),
\end{Equation}
where, since all $\bm{\Gamma}$ are also Hermitian, using \Eq{K.12} and supposing that the $\mathbf{b}_\mathbf{k}$ are Hermitian as well so that all Bloch components are real, then
\begin{Equation}                      {K.16}
\bm{\Gamma} _A  \cdot \bm{\Gamma} _B  = \tr(\bm{\Gamma} _A \bm{\Gamma} _B ) = \sum\limits_{\mathbf{k} \ne \mathbf{0}}^\mathbf{K} {(\Gamma _A )_\mathbf{k} (\Gamma _B )_\mathbf{k} } ,
\end{Equation}
and so the Bloch purity of a Bloch vector in the USB is
\begin{Equation}                      {K.17}
P_\text{B}  \equiv |\bm{\Gamma} |^2  = \bm{\Gamma}  \cdot \bm{\Gamma}  = \tr(\bm{\Gamma}^2 ) = \sum\limits_{\mathbf{k} \ne \mathbf{0}}^\mathbf{K} {|\Gamma _\mathbf{k} |^2 } .
\end{Equation}
For reference, a useful way to obtain $\bm{\Gamma}$ in the USB is 
\begin{Equation}                      {K.18}
\bm{\Gamma}  = \frac{1}{{\sqrt {n(n - 1)} }}(n\rho  - I).
\end{Equation}

The multipartite \textit{reduction} to $S\in 1,\ldots, N$ modes notated by mode-label vector $\mathbf{m}\equiv(m_{1},\ldots,m_{S})$ is given in density-matrix form expanded by its USB reduced Bloch vector \smash{$\redX{\bm{\Gamma}}{\mathbf{m}}$} as
\begin{Equation}                      {K.19}
\redx{\rho}{\mathbf{m}}  = \frac{1}{{n_{\mathbf{m}} }}(I^{(\mathbf{m})}  + \sqrt {n_{\mathbf{m}} (n_{\mathbf{m}}  - 1)} \hsp{1}\redX{\bm{\Gamma}}{\mathbf{m}} ),
\end{Equation}
where $n_{\mathbf{m}}\equiv n_{m_1 }  \cdots n_{m_S }$, and \smash{$\redX{\bm{\Gamma}}{\mathbf{m}}$} has components
\begin{Equation}                      {K.20}
\redX{\Gamma}{\mathbf{m}}_{\mathbf{k}_{\mathbf{m}}  \ne \mathbf{0}_\mathbf{m}}=\redX{\bm{\Gamma}}{\mathbf{m}}\cdot\mathbf{b}_{\mathbf{k}_\mathbf{m} }^{(\mathbf{m})} ,
\end{Equation}
where $\mathbf{k}_{\mathbf{m}}\equiv(k_{m_1 } , \ldots ,k_{m_S } )$, $\mathbf{0}_{\mathbf{m}}\equiv(0_{m_1 } , \ldots ,0_{m_S } )$, and the USB of the reduced system is
\begin{Equation}                      {K.21}
\mathbf{b}_{\mathbf{k}_\mathbf{m} }^{(\mathbf{m})}\equiv\frac{1}{{\sqrt {n_\mathbf{m} } }}\nu _{k_{m_1 } }^{(m_1 )}  \otimes  \cdots  \otimes \nu _{k_{m_S } }^{(m_S )} ,
\end{Equation}
where the mode-specific basis operators \smash{$\nu _{k_{m } }^{(m )}$} are defined in \Eq{K.7}.  Note that modes in $\mathbf{m}$ are not necessarily contiguous or ordered, but if order is changed, permutation unitaries are needed; see (\cite[]{HedE}, App.{\kern 2.5pt}B).

Here is where our formalism will start to show benefits (with more to follow below). First, note that these definitions cause \smash{$\redX{\bm{\Gamma}}{\mathbf{m}}$} to be automatically unitized, meaning that \smash{$|\redX{\bm{\Gamma}}{\mathbf{m}}|=1$} iff \smash{$\redX{\bm{\Gamma}}{\mathbf{m}}$} is a pure state and \smash{$|\redX{\bm{\Gamma}}{\mathbf{m}}|=0$} iff \smash{$\redX{\bm{\Gamma}}{\mathbf{m}}$} is ideally maximally mixed for a general system of mode structure $\mathbf{m}$.

In particular, we get a simple relationship that connects the Bloch components of the reduction to the full Bloch-vector components by a common factor as
\begin{Equation}                      {K.22}
\begin{array}{*{20}l}
   {\redX{\Gamma}{\mathbf{m}}_{\mathbf{k}_\mathbf{m}\ne \mathbf{0}_\mathbf{m}} } &\!\! {=\sqrt {\frac{{n - 1}}{{n_\mathbf{m}  - 1}}} \Gamma _{\mathbf{k}\delta _{\mathbf{k}_{\mathbf{\mbarsub}} ,\mathbf{0}_{\mathbf{\mbarsub}} } }}  \\
   {} &\!\! {= \sqrt {\frac{{n - 1}}{{n_\mathbf{m}  - 1}}} \Gamma _{0, \ldots ,0,k_{m_1 } ,0, \ldots ,0,k_{m_S } ,0, \ldots ,0} }  \\
\end{array},
\end{Equation}
where $\mathbf{\mbar}$ is the ordered set of all full-system mode labels \textit{not} in $\mathbf{m}$, and where $\mathbf{k}\equiv(k_1 , \ldots ,k_N )$, $\mathbf{k}_{\mathbf{\mbarsub}}\equiv(k_{\mbarsub_1 } , \ldots ,k_{\mbarsub_{N - S} } )$, and $\delta _{\mathbf{k}_{\mathbf{\mbarsub}} ,\mathbf{0}_{\mathbf{\mbarsub}} }\equiv\delta _{k_{\mbarsub_1 } ,0}  \cdots \delta _{k_{\mbarsub_{N - S} } ,0}$, where $\mbar_j$ is the $j$th mode label in $\mathbf{\mbar}$. Thus, the explicit-basis form of a multipartite reduced Bloch vector is
\begin{Equation}                      {K.23}
\redX{\bm{\Gamma}}{\mathbf{m}} \hsp{-1}=\hsp{-4} \sum\limits_{\mathbf{k}_\mathbf{m}  \ne \mathbf{0}_\mathbf{m} }^{\mathbf{K}_\mathbf{m} } {\hsp{-7}\redX{\Gamma}{\mathbf{m}}_{\mathbf{k}_{\mathbf{m}}} \mathbf{b}_{\mathbf{k}_\mathbf{m} }^{(\mathbf{m})} } \hsp{-1} =\hsp{-1} \sqrt {\hsp{-2}\frac{{n - 1}}{{n_\mathbf{m}  - 1}}}\hsp{-2} \sum\limits_{\mathbf{k}_\mathbf{m}  \ne \mathbf{0}_\mathbf{m} }^{\mathbf{K}_\mathbf{m} } {\hsp{-7}\Gamma _{\mathbf{k}\delta _{\mathbf{k}_{\mathbf{\mbarsub}} ,\mathbf{0}_{\mathbf{\mbarsub}} } } \mathbf{b}_{\mathbf{k}_\mathbf{m} }^{(\mathbf{m})} } ,
\end{Equation}
where $\mathbf{K}_{\mathbf{m}}\equiv(K_{m_{1}},\ldots,K_{m_{S}})$, where $K_{m_{j}}\equiv n_{m_{j}}^2 -1$. The Bloch purity of a reduction in a Hermitian USB is
\begin{Equation}                      {K.24}
\redX{P}{\mathbf{m}}_\text{B}  \hsp{-2}\equiv\hsp{-1} |\redX{\bm{\Gamma}}{\mathbf{m}} |^2  \hsp{-2}=\hsp{-2} \redX{\bm{\Gamma}}{\mathbf{m}}  \cdot \redX{\bm{\Gamma}}{\mathbf{m}} \hsp{-2}=\hsp{-1} \tr({\redX{\bm{\Gamma}}{\mathbf{m}}}^{2}) \hsp{-1}=\hsp{-5}\sum\limits_{\mathbf{k}_\mathbf{m} \ne \mathbf{0}_\mathbf{m}}^{\mathbf{K}_\mathbf{m}} {\hsp{-7}|\redX{\Gamma}{\mathbf{m}} _{\mathbf{k}_\mathbf{m}} |^2 } \hsp{-1}.
\end{Equation}

It is often tidier to simply organize the full Bloch-vector components corresponding to certain reductions into groups by defining \textit{correlation vectors} (in the USB of the full Bloch vector) as
\begin{Equation}                      {K.25}
\mathbf{X}^{(\mathbf{m})}\equiv\sum\limits_{\mathbf{k}_\mathbf{m}  = \mathbf{1}_\mathbf{m} }^{\mathbf{K}_\mathbf{m} } {\hsp{-7}\Gamma _{\mathbf{k}\delta _{\mathbf{k}_{\mathbf{\mbarsub}} ,\mathbf{0}_{\mathbf{\mbarsub}} } }\mathbf{b}_{\mathbf{k}\delta _{\mathbf{k}_{\mathbf{\mbarsub}} ,\mathbf{0}_{\mathbf{\mbarsub}} } }},
\end{Equation}
where $\mathbf{1}_{\mathbf{m}}\equiv(1_{m_{1}},\ldots,1_{m_{S}})$, with the property that
\begin{Equation}                      {K.26}
\bm{\Gamma}  = \sum\limits_{k = 1}^N {\sum\limits_{l = 1}^{\binom{N}{k}} {\mathbf{X}^{((\text{nCk}[\mathbf{c},k])_{l, \cdots } )} } } ,
\end{Equation}
where $\mathbf{c}\equiv(1,\ldots,N)$, and $\text{nCk}[\mathbf{v},k]$ is the vectorized $n$-choose-$k$ function that gives a matrix whose rows are the unique combinations of the elements of $\mathbf{v}$ chosen $k$ at a time, and $A_{l,\cdots}$ is the $l$th row of a matrix $A$. So, for example, for a tripartite system (meaning $N=3$),
\begin{Equation}                      {K.27}
\begin{array}{*{20}l}
   {\bm{\Gamma} = } &\!\! {\mathbf{X}^{(1)}  + \mathbf{X}^{(2)}  + \mathbf{X}^{(3)} }  \\
   {} &\!\! { + \mathbf{X}^{(1,2)}  + \mathbf{X}^{(2,3)}  + \mathbf{X}^{(1,3)} }  \\
   {} &\!\! { + \mathbf{X}^{(1,2,3)}, }  \\
\end{array}
\end{Equation}
where, just expanding a few terms as examples,
\begin{Equation}                      {K.28}
\hsp{-4}\begin{array}{*{20}c}
   {\begin{array}{*{20}c}
   {\mathbf{X}^{(2)}  \hsp{-1}=\hsp{-5} \sum\limits_{k_2  = 1}^{K_2 } {\hsp{-3}\Gamma _{0,k_2 ,0} \mathbf{b}_{0,k_2 ,0} },} & {\mathbf{X}^{(1,3)} \hsp{-1} =\hsp{-5} \sum\limits_{k_1 ,k_3  = 1,1}^{K_1 ,K_3 } {\hsp{-7}\Gamma _{k_1 ,0,k_3 } \mathbf{b}_{k_1 ,0,k_3 } },}  \\
\end{array}}  \\
   {\mathbf{X}^{(1,2,3)} \hsp{-1} =\hsp{-5} \sum\limits_{k_1 ,k_2 ,k_3  = 1,1,1}^{K_1 ,K_2 ,K_3 } {\hsp{-9}\Gamma _{k_1 ,k_2 ,k_3 } \mathbf{b}_{k_1 ,k_2 ,k_3 } }.}  \\
\end{array}\hsp{-4}
\end{Equation}
We can even define \textit{unitized} correlation vectors as
\begin{Equation}                      {K.29}
\widetilde{\mathbf{X}}^{(\mathbf{m})}\equiv \sqrt{\frac{n-1}{n_{\mathbf{m}}-1}}\mathbf{X}^{(\mathbf{m})}=\sum\limits_{\mathbf{k}_\mathbf{m}  = \mathbf{1}_\mathbf{m} }^{\mathbf{K}_\mathbf{m} } {\hsp{-7}\redX{\Gamma}{\mathbf{m}}_{\mathbf{k}_\mathbf{m}}\mathbf{b}_{\mathbf{k}\delta _{\mathbf{k}_{\mathbf{\mbarsub}} ,\mathbf{0}_{\mathbf{\mbarsub}} } }},
\end{Equation}
with the property that the Bloch purity of $\widetilde{\mathbf{X}}^{(\mathbf{m})}$ matches that of $\redX{\bm{\Gamma}}{\mathbf{m}}$ so that $|\widetilde{\mathbf{X}}^{(\mathbf{m})}|^2 = |\redX{\bm{\Gamma}}{\mathbf{m}}|^2$, and the only difference between them is that the matrix basis of $\widetilde{\mathbf{X}}^{(\mathbf{m})}$ lives in the full space of parent state $\bm{\Gamma}$, whereas the matrix basis of $\redX{\bm{\Gamma}}{\mathbf{m}}$ lives in the space of mode group $\mathbf{m}$.

As proved in \cite[]{HedD}, we only need the \textit{single}-mode reductions to quantify full $N$-partite entanglement, and as such, it is useful to define the \textit{liaison vector},
\begin{Equation}                      {K.30}
\bm{\Lambda}  \equiv \bm{\Gamma} - \sum\limits_{m = 1}^N {\mathbf{X}^{(m)} }  = \bm{\Gamma} - \sum\limits_{m = 1}^N {\sqrt {\frac{{n_m  - 1}}{{n - 1}}} \widetilde{\mathbf{X}}^{(m)} } ,
\end{Equation}
which is the sum of all strictly multipartite reductions, so for example, in a tripartite system, 
\begin{Equation}                      {K.31}
\bm{\Lambda} = \mathbf{X}^{(1,2)}  + \mathbf{X}^{(2,3)}  + \mathbf{X}^{(1,3)}  + \mathbf{X}^{(1,2,3)}.
\end{Equation}
Since $\bm{\Lambda}$ is the group of all \textit{strictly multimode} Bloch components, the vector indices $\mathbf{k}$ of its components $\Lambda_{\mathbf{k}}$ always have \textit{at least two} nonzero indices $k_m$. For example, $\mathbf{X}^{(1,3)}$ and $\mathbf{X}^{(1,2,3)}$ of \Eq{K.28} both have two or more nonzero indices, as do all correlation vectors that make up $\bm{\Lambda}$.

Note that the Bloch purity is expressible as
\begin{Equation}                      {K.32}
|\bm{\Gamma}|^2  = |\bm{\Lambda}|^2  + \sum\limits_{m = 1}^N {|\mathbf{X}^{(m)} |^2 } ,
\end{Equation}
which we use in our correlance-normalization proof.  Another useful fact is that all \textit{separable} states obey
\begin{Equation}                      {K.33}
\Gamma _{\mathbf{k} \ne \mathbf{0}}  = \frac{{A_\mathbf{k} }}{{\sqrt {n - 1} }}\sum\nolimits_j {p_j \redX{\Gamma}{1}_{j|k_1 }  \cdots \redX{\Gamma}{N}_{j|k_N } } ,
\end{Equation}
where $\redX{\Gamma}{m}_{j|k_m }\equiv(\redX{\bm{\Gamma}}{m}_{j})_{k_m}$, and
\begin{Equation}                      {K.34}
A_{\mathbf{k}}\equiv A_{k_{1}}\cdots A_{k_{N}};\hsp{8}A_{k_{m}}\equiv(\sqrt{n_{m}-1}\hsp{1})^{\text{sgn}(k_{m})}.
\end{Equation}
For \textit{product-form} states, using \Eqs{5}{8} in \Eq{K.33} leads to
\begin{Equation}                      {K.35}
\Gamma _{\mathbf{k} \ne \mathbf{0}}  = \frac{{A_\mathbf{k} }}{{\sqrt {n - 1} }}\redX{\Gamma}{1}_{k_1 }  \cdots \redX{\Gamma}{N}_{k_N } ,
\end{Equation}
where we used the fact that $\redX{\Gamma}{m}_{k_m }\equiv\sum\nolimits_{j_m}{p_{j_m}\redX{\Gamma}{m}_{{j_m}|k_m }}$.

There are many useful applications of this formalism (and much more to say about it), which was introduced in the present form in \cite[]{HedD}, but this will suffice as a good working reference both here and for future research.
\subsection{\label{sec:App.K.2}Proof of Explicit Correlance Normalization Factors}
Expanding \Eq{12} in its simplest form in ME TGX states (since they have nice properties such as balanced superposition, multiple levels of zero probability, and diagonal reductions), we get
\begin{Equation}                      {K.36}
\mathcal{N}_{\Correlance}  = 1 - 2\tr(\rho _{\text{ME}_{\text{TGX}} } \varsigma _{\text{ME}_{\text{TGX}} } ) + \tr(\varsigma _{\text{ME}_{\text{TGX}} } ^2 ),
\end{Equation}
where $\varsigma _{\text{ME}_{\text{TGX}} }\equiv\varsigma(\rho_{\text{ME}_{\text{TGX}} })$ where $\varsigma(\rho)$ is the reduction product from \Eq{11}. Since the reduction product has product form, then its purity also has product form as
\begin{Equation}                      {K.37}
\begin{array}{*{20}l}
   {\tr(\varsigma _{\text{ME}_{\text{TGX}} } ^2 )} &\!\! { = P(\redx{\rho}{1}_{\text{ME}_{\text{TGX}} }) \cdots P(\redx{\rho}{N}_{\text{ME}_{\text{TGX}} })}  \\
   {} &\!\! { = \prod\nolimits_{m = 1}^N {P_{\text{MP}}^{(m)} (L_* )},\rule{0pt}{11pt} }  \\
\end{array}
\end{Equation}
where \smash{$P(\rho)\equiv\tr(\rho^{2})$}, and \smash{$P_{\text{MP}}^{(m)} (L_* )$} is the minimum physical reduction purity of mode $m$ given a pure maximally full-$N$-partite-entangled parent state with $L_*$ levels of equal nonzero probabilities.

To calculate each $P_{\text{MP}}^{(m)} (L_* )$, first make the more general function,
\begin{Equation}                      {K.38}
\begin{array}{*{20}l}
   {P_{\text{MP}}^{(m)} (L)\equiv} &\!\! { \bmod (L,n_m )\left( {\frac{{1 + \text{floor} (L/n_m )}}{L}} \right)^2 }  \\
   {} &\!\! { + (n_m  - \bmod (L,n_m ))\left( {\frac{{\text{floor}(L/n_m )}}{L}} \right)^2 ,}  \\
\end{array}
\end{Equation}
where $\bmod (a,b) \equiv a - \text{floor}(a/b)b$, and use that to define
\begin{Equation}                      {K.39}
M(L) \equiv 1 - \frac{1}{N}\sum\limits_{m = 1}^N {\frac{{n_m P_{\text{MP}}^{(m)} (L) - 1}}{{n_m  - 1}}} .
\end{Equation}
Then find the set $\mathbf{L}_{*}\equiv\{L_{*}\}$ which are the values of $L$ that satisfy
\begin{Equation}                      {K.40}
\mathop {\min }\limits_{L \in 2, \ldots ,\nmaxnot } [1 - M(L)],
\end{Equation}
where $\nmaxnot\!\equiv\! n/n_{\max}$ and $n_{\max}\!\equiv\!\max(\mathbf{n})\!=\!\{n_1,\ldots,n_N\}$. Then by convention let $L_{*}\equiv\min(\mathbf{L}_{*})$ and use that to compute each \smash{$P_{\text{MP}}^{(m)} (L_* )$}.  Note that we could simplify things slightly, but these quantities have physical significance in the context of the entanglement measure the \textit{ent} \cite[]{HedE}, so we use these forms for conceptual consistency.  Thus, putting \Eq{K.37} into \Eq{K.36} gives
\begin{Equation}                      {K.41}
\mathcal{N}_{\Correlance}  = 1 - 2\tr(\rho _{\text{ME}_{\text{TGX}} } \varsigma _{\text{ME}_{\text{TGX}} } ) + \prod\limits_{m = 1}^N {P_{\text{MP}}^{(m)} (L_* )}.
\end{Equation}

To get the overlap term in \Eq{K.41}, it is helpful to use a Bloch-vector formalism, such as that given in \App{K.1}. First, we note that Bloch components of $\varsigma$ have the product form of \Eq{K.35} as
\begin{Equation}                      {K.42}
\Gamma _{\mathbf{k} \ne \mathbf{0}}(\varsigma)  = \frac{{A_\mathbf{k} }}{{\sqrt {n - 1} }}\redX{\Gamma}{1}_{k_1 }(\varsigma)  \cdots \redX{\Gamma}{N}_{k_N }(\varsigma) ,
\end{Equation}
where $\mathbf{k}\equiv(k_{1},\ldots,k_{N})$ is a vector index for multipartite Bloch vectors where $k_{m}\in 0,\ldots,n_{m}^2 -1$ for $m\in 1,\ldots,N$, (so $\mathbf{k}\ne\mathbf{0}$ means only the case of \textit{all} $k_m =0$ is excluded), $A_\mathbf{k}$ is given in \Eq{K.34}, and \smash{$\redX{\Gamma}{m}_{k_m }(\varsigma)$} are Bloch components of the mode-$m$ reduction of $\bm{\Gamma}(\varsigma)$ which is the Bloch vector of $\varsigma$ (see \App{K.1} for more details).

Now, consider the following facts:
\begin{itemize}[leftmargin=*,labelindent=4pt]\setlength\itemsep{0pt}
\item[\textbf{1.}]\hypertarget{CorrelanceNormBlochProof:1}{}If the parent state is an ME TGX state and \textit{multiple} modes have size $n_{\max}$, then, as proved in (\cite[]{HedE}, App.{\kern 2.5pt}D.4.c), all mode-$m$ reductions are ideally maximally mixed, having purities $P(\redx{\rho}{m})=\frac{1}{n_{m}}\;\forall m\in 1,\ldots,N$.
\item[\textbf{2.}]\hypertarget{CorrelanceNormBlochProof:2}{}If the parent state is an ME TGX state and \textit{exactly one} mode (call it mode $N$) has size $n_{\max}$, then, as proved in (\cite[]{HedE}, App.{\kern 2.5pt}D.4.c), all nonlargest mode-$m$ reductions are ideally maximally mixed, with purities $P(\redx{\rho}{m})=\frac{1}{n_{m}}\;\forall m\in 1,\ldots,N-1$, while $\redx{\rho}{N}$ has purity \smash{$P_{\text{MP}}^{(m)}(L_{*})$} as given in \Eqs{K.38}{K.40}, which is larger than that of the ideal maximally mixed state for an isolated system of size $n_m$ levels, which is due to the purity and maximal entanglement of its parent state (see \cite[]{HedE}, App.{\kern 2.5pt}D for full explanations). Note that \smash{$P_{\text{MP}}^{(m)}(L_{*})$} simplifies to the correct value for all modes in this case, not just the largest mode.
\item[\textbf{3.}]\hypertarget{CorrelanceNormBlochProof:3}{}From \hyperlink{CorrelanceNormBlochProof:1}{Step 1} and \hyperlink{CorrelanceNormBlochProof:2}{Step 2}, \textit{all maximally full-$N$-partite entangled states have at least one single-mode reduction that is ideally maximally mixed}.
\item[\textbf{4.}]\hypertarget{CorrelanceNormBlochProof:4}{}As given in \App{K.1}, any mode-$m$ ideally maximally-mixed Bloch vector has magnitude \smash{$|\redX{\bm{\Gamma}}{m}|=0$}, and thus all its components are \smash{$\redX{\Gamma}{m}_{k_{m}>0}=0$}.
\item[\textbf{5.}]\hypertarget{CorrelanceNormBlochProof:5}{}From \hyperlink{CorrelanceNormBlochProof:1}{Steps 1}, \hyperlink{CorrelanceNormBlochProof:2}{2}, \hyperlink{CorrelanceNormBlochProof:3}{3}, and \hyperlink{CorrelanceNormBlochProof:4}{4}, \textit{for ME states, there always exist at least $N-1$ reduced Bloch vectors with} \smash{$|\redX{\bm{\Gamma}}{m}(\rho_{\text{ME}})|=0$} \textit{and thus} \smash{$\redX{\Gamma}{m}_{k_{m}>0}(\rho_{\text{ME}})=0$} \textit{for} $m\in 1,\ldots,N-1$. (Note, here and throughout we use the convention that the system is organized in increasing mode size, so that mode $N$ is always the largest even if there are multiple largest modes.)
\item[\textbf{6.}]\hypertarget{CorrelanceNormBlochProof:6}{}The liaison vector $\bm{\Lambda}$ from \Eq{K.30} is composed entirely of components of the full Bloch vector $\bm{\Gamma}$ that have two or more nonzero Bloch indices $k_m$.
\item[\textbf{7.}]\hypertarget{CorrelanceNormBlochProof:7}{}From \Eq{K.42} and \hyperlink{CorrelanceNormBlochProof:6}{Step 6}, the components of $\bm{\Lambda}(\varsigma_{\text{ME}})$ take the form
\begin{Equation}                      {K.43}
\Lambda _{\mathbf{k} \ne \mathbf{0}}(\varsigma_{\text{ME}})  = \frac{{A_\mathbf{k} }}{{\sqrt {n - 1} }}\redX{\Gamma}{1}_{k_1 }(\varsigma_{\text{ME}})  \cdots \redX{\Gamma}{N}_{k_N }(\varsigma_{\text{ME}}) =0 ,
\end{Equation}
where $\varsigma_{\text{ME}}\equiv\varsigma(\rho_{\text{ME}})$, because even though the largest mode can have nonzero components \smash{$\redX{\Gamma}{N}_{k_N >0 }(\varsigma_{\text{ME}})$} (if it is the only largest mode), all other modes have \smash{$\redX{\Gamma}{m<N}_{k_m >0 }(\varsigma_{\text{ME}})=0$}, so since $\mathbf{k}$ of $\Lambda_\mathbf{k}$ must always have at least two nonzero indices, \textit{at least one of the factors in} $\Lambda_\mathbf{k}(\varsigma_{\text{ME}})$ \textit{will always be zero}. In the case of multiple largest modes, at least \textit{two} factors in $\Lambda_\mathbf{k}(\varsigma_{\text{ME}})$ will always be zero, so \Eq{K.43} holds true for all systems. Therefore \Eq{K.43} means that
\begin{Equation}                      {K.44}
\bm{\Lambda}(\varsigma_{\text{ME}})=\mathbf{0}.
\end{Equation}
\end{itemize}
Now, putting \Eq{K.44} into \Eq{K.30}, we see that the full Bloch vector of $\varsigma_{\text{ME}}$ is a function of only single-mode reduced Bloch components as
\begin{Equation}                      {K.45}
\bm{\Gamma}(\varsigma _\text{ME} ) = \mathbf{X}^{(1)} (\varsigma _\text{ME} ) +  \cdots  + \mathbf{X}^{(N)} (\varsigma _\text{ME} ),
\end{Equation}
where from putting \Eq{K.22} into \Eq{K.25}, we see that, in general, $\mathbf{X}^{(m)}$ are Bloch vectors whose only nonzero components are proportional to single-mode reduction Bloch components, but whose basis matrices are in the space of the full Bloch vector, given by
\begin{Equation}                      {K.46}
\mathbf{X}^{(m)}\equiv\sqrt{\frac{n_{m}-1}{n-1}}\sum\limits_{k_m  = 1 }^{K_m } {\hsp{-2}\redX{\Gamma}{m}_{k_m}\mathbf{b}_{\mathbf{k}\delta _{\mathbf{k}_{\mbarsub} ,\mathbf{0}_{\mbarsub} } }},
\end{Equation}
where \smash{$\mathbf{b}_{\mathbf{k}\delta _{\mathbf{k}_{\mbarsub} ,\mathbf{0}_{\mbarsub} } }\equiv\mathbf{b}_{0_{1},\ldots,0_{m-1},k_{m},0_{m+1},\ldots,0_{N}}$} is part of a matrix basis for the full Bloch vector, as defined in \App{K.1}.  Thus, \Eq{K.45} becomes
\begin{Equation}                      {K.47}
\bm{\Gamma}(\varsigma _\text{ME} ) \hsp{-1}=\hsp{-2}\sum\limits_{m=1}^{N}{\hsp{-3}{\textstyle\sqrt{\frac{n_{m}-1}{n-1}}}\hsp{-2}\sum\limits_{k_m  = 1 }^{K_m } {\hsp{-3.5}\redX{\Gamma}{m}_{k_m}(\rho _\text{ME} )\mathbf{b}_{\mathbf{k}\delta _{\mathbf{k}_{\mbarsub} ,\mathbf{0}_{\mbarsub} } }}}\hsp{-2}.
\end{Equation}
where we used the fact that $\redX{\Gamma}{m}_{k_m}(\varsigma _\text{ME} )=\redX{\Gamma}{m}_{k_m}(\rho _\text{ME} )$ since the reduction product $\varsigma_{\text{ME}}$ is constructed from a tensor product of the reductions of $\rho_{\text{ME}}$ and therefore it has exactly the same reduced Bloch components as $\rho_{\text{ME}}$, with the main difference between the two states being that $\bm{\Lambda}(\rho_{\text{ME}})\ne\mathbf{0}$. Thus, we also have, using \Eq{K.30},
\begin{Equation}                      {K.48}
\bm{\Gamma}(\rho _\text{ME} ) \hsp{-1}=\hsp{-2}\bm{\Lambda}(\rho_{\text{ME}})+\hsp{-3}\sum\limits_{m=1}^{N}{\hsp{-3}{\textstyle\sqrt{\frac{n_{m}-1}{n-1}}}\hsp{-2}\sum\limits_{k_m  = 1 }^{K_m } {\hsp{-3.5}\redX{\Gamma}{m}_{k_m}(\rho _\text{ME} )\mathbf{b}_{\mathbf{k}\delta _{\mathbf{k}_{\mbarsub} ,\mathbf{0}_{\mbarsub} } }}}\hsp{-2}.
\end{Equation}
In the computation of the overlap of these two states, we will need to take the dot product of their Bloch vectors.  However, due to \Eq{K.44}, the liaison vector of $\rho_{\text{ME}}$ will have no contribution to the overlap, and furthermore, since all of the remaining terms in \Eq{K.47} and \Eq{K.48} are \textit{identical}, then the Bloch overlap of these states is simply
\begin{Equation}                      {K.49}
\bm{\Gamma}(\rho _{\text{ME}} ) \cdot \bm{\Gamma} (\varsigma _{\text{ME}} ) = \sum\limits_{m = 1}^N {\frac{{n_m  - 1}}{{n - 1}}|\redX{\bm{\Gamma}}{m} (\rho _{\text{ME}} )|^2 }.
\end{Equation}
However, this is the same as the Bloch purity of $\varsigma _{\text{ME}}$,
\begin{Equation}                      {K.50}
|\bm{\Gamma}(\varsigma _{\text{ME}} )|^2 =\bm{\Gamma}(\varsigma _{\text{ME}} ) \cdot \bm{\Gamma} (\varsigma _{\text{ME}} ) = \sum\limits_{m = 1}^N {\frac{{n_m  - 1}}{{n - 1}}|\redX{\bm{\Gamma}}{m} (\rho _{\text{ME}} )|^2 },
\end{Equation}
so we have
\begin{Equation}                      {K.51}
\bm{\Gamma}(\rho _{\text{ME}} ) \cdot \bm{\Gamma} (\varsigma _{\text{ME}} ) =\bm{\Gamma}(\varsigma _{\text{ME}} ) \cdot \bm{\Gamma} (\varsigma _{\text{ME}} ).
\end{Equation}
Then, putting \Eq{K.51} into the fact from \Eq{K.15} that $\tr(\rho _A \rho _B ) = \frac{1}{n}\left( {1 + (n - 1)\bm{\Gamma} _A  \cdot \bm{\Gamma} _B } \right)$, we obtain the important result that
\begin{Equation}                      {K.52}
\tr(\rho _{\text{ME}} \varsigma (\rho _{\text{ME}} )) = P(\varsigma (\rho _{\text{ME}} )).
\end{Equation}
Then, since this holds true for pure ME TGX states as well, then using \Eq{K.37} in \Eq{K.52} gives
\begin{Equation}                      {K.53}
\tr(\rho _{\text{ME}_{\text{TGX}}} \varsigma (\rho _{\text{ME}_{\text{TGX}}})) =\prod\limits_{m = 1}^N {P_{\text{MP}}^{(m)} (L_* )},
\end{Equation}
which, when put into \Eq{K.41} yields
\begin{Equation}                      {K.54}
\mathcal{N}_{\Correlance}  = 1 - \prod\limits_{m = 1}^N {P_{\text{MP}}^{(m)} (L_* )},
\end{Equation}
which proves the result from \Eq{13}, valid for all systems, where \smash{$P_{\text{MP}}^{(m)} (L_* )$} is given by \Eq{K.38} with $L=L_{*}$ where $L_{*}$ is determined from \Eq{K.40}.

For the special-case systems where more than one mode has size $n_{\max}$, we can use the fact, proved in (\cite[]{HedE}, App.{\kern 2.5pt}D.4.c), that for maximally entangled states, all reductions are ideally maximally mixed, so then \Eq{K.53} is
\begin{Equation}                      {K.55}
\begin{array}{*{20}l}
   {\prod\nolimits_{m = 1}^N {P_{\text{MP}}^{(m)} (L_* )}} &\!\! { = \tr[(\frac{1}{{n_1 }}I^{(1)}  \otimes  \cdots  \otimes \frac{1}{{n_N }}I^{(N)} )^2 ]}  \\
   {} &\!\! { = \frac{1}{{(n_1  \cdots n_N )^2 }}\tr[I^{(1)}  \otimes  \cdots  \otimes I^{(N)} ]\rule{0pt}{11pt}}  \\
   {} &\!\! { = \frac{1}{{n^2 }}n\rule{0pt}{11pt}}  \\
   {} &\!\! { = \frac{1}{n},\rule{0pt}{11pt}}  \\
\end{array}
\end{Equation}
which, when put into \Eq{K.54} gives
\begin{Equation}                      {K.56}
\mathcal{N}_{\Correlance}  = 1 - \frac{1}{n},
\end{Equation}
in agreement with \Eq{14}, but again, is only valid for systems with multiple modes of size $n_{\max}$. For systems with exactly one largest mode, we must use \Eq{K.54}.

Finally, since all cases of mode structures have the property that all modes except for the nominally largest mode (which we will call $m_{\max}$ here for generality, meaning the single mode with size $n_{\max}$ designated as the nominally largest mode regardless of whether other modes also have size $n_{\max}$) have minimal reductions that are ideally maximally mixed when the parent is maximally full-$N$-partite entangled, then \Eq{K.54} can be simplified to
\begin{Equation}                      {K.57}
\mathcal{N}_{\Correlance}  = 1 - \frac{P_{\text{MP}}^{(m_{\max})} (L_* )}{\nmaxnot},
\end{Equation}
which may be used as an alternative to \Eq{13} or \Eq{K.54}, where $\nmaxnot\equiv n/n_{\max}$.
\section{\label{sec:App.L}Proof-Sketch and Derivation of the Normalization of Diagonal Correlance}
Here we sketch a proof of why \Eq{17} is valid for finding the normalization of diagonal correlance $\Correlance_D$, and derive the explicit result in \Eq{18}.
\begin{itemize}[leftmargin=*,labelindent=4pt]\setlength\itemsep{0pt}
\item[\textbf{1.}]\hypertarget{DiagCorrelanceNormProof:1}{}Since all diagonal pure states are pure product states with no nonlocal correlation, that disqualifies pure states as maximizers of $\Correlance_D$.
\item[\textbf{2.}]\hypertarget{DiagCorrelanceNormProof:2}{}Therefore, by \hyperlink{DiagCorrelanceNormProof:1}{Step 1}, maximizers of $\Correlance_D$ (called $\rho_{D_{\max}}$) must have rank $2$ or higher.
\item[\textbf{3.}]\hypertarget{DiagCorrelanceNormProof:3}{}By definition, \smash{$\rho_{D_{\max}}$} maximize \smash{$\widetilde{\Correlance}_D$} as in \Eq{16}. Therefore, while $\rho_{D_{\max}}$ must have rank $r\geq 2$ by \hyperlink{DiagCorrelanceNormProof:2}{Step 2}, they must also \textit{not} have product form by definition, and their Bloch vectors must \textit{also} have the largest distance from their own reduction products $\varsigma_{D_{\max}}\equiv\varsigma(\rho_{D_{\max}})$ under the Hilbert-Schmidt inner product, out of all states compared to their own reduction products.  (Note: $\Correlance_D$ and $\Correlance$ are \textit{not} measures of distance to the set of all product states from the given input $\rho$; instead they are measures of the distance between $\rho$ and its \textit{own} reduction product.  Therefore, all $\rho_{D_{\max}}$ have the biggest distance between themselves and their own $\varsigma_{D_{\max}}$ out of all diagonal states $\rho_D$, but their $\varsigma_{D_{\max}}$ are generally neither the farthest nor the closest product states from themselves.)
\item[\textbf{4.}]\hypertarget{DiagCorrelanceNormProof:4}{}Since all \smash{$\rho_D$} are diagonal, then by the definition of partial trace, their reductions \smash{$\redx{\rho}{m}_{\!D}$} are also diagonal.
\item[\textbf{5.}]\hypertarget{DiagCorrelanceNormProof:5}{}From \Eq{16}, the function to be maximized by \smash{$\rho_{D_{\max}}$} is \smash{$\widetilde{\Correlance}(\rho _{D} )\!=\!P(\rho _{D})\!+\!P(\redx{\rho}{1}_{\!D})\cdots P(\redx{\rho}{N}_{\!D})\!-2\tr[\rho _{D}\varsigma_{D}]$}
over all $\rho _{D}$, where $P(\sigma)\equiv\tr(\sigma^2)$ is the purity of $\sigma$, and the overlap is \smash{$\tr[\rho _{D}\varsigma_{D}]=\tr[\rho _{D}(\redx{\rho}{1}_{\!D}\otimes\cdots\otimes\redx{\rho}{N}_{\!D})]$}.
\item[\textbf{6.}]\hypertarget{DiagCorrelanceNormProof:6}{}By \hyperlink{DiagCorrelanceNormProof:5}{Step 5}, maximizers of $\widetilde{\Correlance}(\rho _{D} )$ need to simultaneously fulfill the conditions of having highest purity, highest single-mode reduction purities, and lowest overlap of parent state and reduction product.
\item[\textbf{7.}]\hypertarget{DiagCorrelanceNormProof:7}{}By \hyperlink{DiagCorrelanceNormProof:6}{Step 6}, we need to minimize $\tr[\rho _{D}\varsigma_{D}]$ (while also keeping the purity terms as high as possible). So first, expanding it as $\tr[\rho _{D}\varsigma_{D}]=\sum\nolimits_{a = 1}^n {} (\rho_D )_{a,a} (\varsigma _D )_{a,a}=\sum\nolimits_{a_{1},\ldots,a_{N}=1,\ldots,1}^{n_{1},\ldots,n_{N}} {(\rho_D )_{\mathbf{a},\mathbf{a}}[(\redx{\rho}{1}_{\!D})_{a_{1},a_{1}}\cdots(\redx{\rho}{N}_{\!D})_{a_{N},a_{N}}]}$ where $\mathbf{a}\equiv(a_{1},\ldots,a_{N})$, shows that each term is a product of the probabilities of $\rho_D$ with the probabilities of each of the $N$ single-mode \textit{reductions} of $\rho_D$ (which are \textit{also} functions of probabilities of $\rho_D$), by \hyperlink{DiagCorrelanceNormProof:4}{Step 4}.  This suggests that the more diagonal elements of $\rho_D$ are zero, the lower this overlap function will be, which would \textit{also} increase the purity of $\rho_D$, but not necessarily the reduction purity product. Therefore this suggests that we \textit{reduce rank as much as possible} while also looking at the \textit{particular ways to choose nonzero elements to maximize} $\widetilde{\Correlance}(\rho _{D} )$. For example, in $2\times 3$, abbreviating with $\rho\equiv\rho_D$ and $\varsigma_{D}\equiv\varsigma(\rho_{D})$,%
\begin{Equation}                      {L.1}
\begin{array}{*{20}l}
   {(\varsigma _D )_{1,1} =\redx{\rho}{1}_{1,1}\redx{\rho}{2}_{1,1}} &\!\! { = (\rho _{1,1}  + \rho _{2,2}  + \rho _{3,3} )(\rho _{1,1}  + \rho _{4,4} )}  \\
   {(\varsigma _D )_{2,2} =\redx{\rho}{1}_{1,1}\redx{\rho}{2}_{2,2}} &\!\! { = (\rho _{1,1}  + \rho _{2,2}  + \rho _{3,3} )(\rho _{2,2}  + \rho _{5,5} )}  \\
   {(\varsigma _D )_{3,3} =\redx{\rho}{1}_{1,1}\redx{\rho}{2}_{3,3}} &\!\! { = (\rho _{1,1}  + \rho _{2,2}  + \rho _{3,3} )(\rho _{3,3}  + \rho _{6,6} )}  \\
   {(\varsigma _D )_{4,4} =\redx{\rho}{1}_{2,2}\redx{\rho}{2}_{1,1}} &\!\! { = (\rho _{4,4}  + \rho _{5,5}  + \rho _{6,6} )(\rho _{1,1}  + \rho _{4,4} )}  \\
   {(\varsigma _D )_{5,5} =\redx{\rho}{1}_{2,2}\redx{\rho}{2}_{2,2}} &\!\! { = (\rho _{4,4}  + \rho _{5,5}  + \rho _{6,6} )(\rho _{2,2}  + \rho _{5,5} )}  \\
   {(\varsigma _D )_{6,6} =\redx{\rho}{1}_{2,2}\redx{\rho}{2}_{3,3}} &\!\! { = (\rho _{4,4}  + \rho _{5,5}  + \rho _{6,6} )(\rho _{3,3}  + \rho _{6,6} ),}  \\
\end{array}
\end{Equation}
which shows that some choices of nonzero parent elements lead to more nonzero elements in $\varsigma_D$ than others. Since the rank of $\rho$ determines the number of terms in $\tr[\rho _{D}\varsigma_{D}]$ anyway, then to minimize it for a given rank, we need to choose the nonzero parent elements in a way that \textit{maximizes the number of nonzero terms in $\varsigma_D$} so that their values are lower due to normalization, which allows a lower value of $\tr[\rho _{D}\varsigma_{D}]$.
\item[\textbf{8.}]\hypertarget{DiagCorrelanceNormProof:8}{}Considering rank and choice of nonzero elements, notice that for \textit{all states}, $\rho_{1,1}$ always appears in \smash{$\redx{\rho}{m}_{1,1}$} and $\rho_{n,n}$ always appears in \smash{$\redx{\rho}{m}_{{n_m},{n_m}}\;\forall m\in 1,\ldots,N$}. Thus for any number of $N$ modes, if a rank-2 parent state only has $\rho_{1,1}\neq 0$ and $\rho_{n,n}\neq 0$, then $\varsigma_{1,1}$ and $\varsigma_{n,n}$ will always be nonzero \textit{and} several other elements of $\varsigma$ may still be nonzero as well, as in \Eq{L.1} which has \textit{four} nonzero elements in this case. This is important because since the parent state has all other $n-2$ diagonal elements as $0$, then $n-2$ terms of the negative overlap term in \hyperlink{DiagCorrelanceNormProof:5}{Step 5} vanish as seen by \hyperlink{DiagCorrelanceNormProof:7}{Step 7}, reducing its ability to lessen the objective function, while the factors from $\varsigma$ are lower than they would be for other choices of nonzero parent elements, since there are more than two nonzero elements of $\varsigma$. Also, the parent purity has its largest minimum since $P(\rho)\in[\frac{1}{2},1)$, while $P(\varsigma)\in[\frac{1}{n},1)$ with possibly a larger minimum due to the parent state's correlation (as shown for reduction products of maximally entangled parent states in \cite[]{HedE}), where we used $\rho\equiv\rho_D$ and $\varsigma\equiv\varsigma_D$.
\item[\textbf{9.}]\hypertarget{DiagCorrelanceNormProof:9}{}By \hyperlink{DiagCorrelanceNormProof:7}{Step 7}, since every term of the objective function $\widetilde{\Correlance}_D$ (for any rank) is a function of parent elements, their normalization and nonnegativity makes $\widetilde{\Correlance}_D$ a linear combination of products of squares of $r$ unit-hyperspherical coordinates $\{x_k\}$.  It is well-known that the \textit{sum} of even powers of $\{x_k\}$ is minimized when \smash{$x_k=\frac{1}{\sqrt{r}}\;\forall k\in 1,\ldots,r$} (see \cite[]{HedE}, App.{\kern 2.5pt}D.2).  Here, this same solution maximizes the rank-$2$ case of $\widetilde{\Correlance}_D$, \textit{and for ranks $r\geq 3$, neither balanced probabilities nor any other combination of values of hyperspherical coordinates can cause $\widetilde{\Correlance}_D$ to get as large as the rank-$2$ case, because they generally cause a smaller parent-state purity and a smaller reduction-purity product than the rank-$2$ case}. The overlap term behaves less consistently, however its \textit{combination} with the other terms in $\widetilde{\Correlance}_D$ is such that the rank-$2$ value of $\widetilde{\Correlance}_D$ is always the largest when maximized over all combinations of nonzero-levels for the equal-probabilities case.  This was confirmed by brute-force combinatorial comparison of all ranks and nonzero-element combinations for several multipartite systems in equal-probability states, while \Fig{4} provides strong numerical evidence that it is true over all $\rho_D$. However, a rigorous proof of this is still lacking.
\item[\textbf{10.}]\hypertarget{DiagCorrelanceNormProof:10}{}Therefore, \hyperlink{DiagCorrelanceNormProof:9}{Step 9} gives evidence that $\rho _{D_{\max } }$ must have rank $2$ \textit{and equal probabilities} (note that if either probability is larger, the state would be closer to being a pure product state, so it would have nonmaximal correlation).  Furthermore, \hyperlink{DiagCorrelanceNormProof:8}{Step 8} suggests one way to chose particular nonzero elements that will minimize the overlap the most out of all possible pairs, since it causes the most terms in the overlap to be zero, while causing the most terms of $\varsigma$ to be nonzero which lowers the values of the surviving overlap terms. Therefore, our candidate maximizer states are \smash{$\rho _{D_{\max } }  = \text{diag}\{ \frac{1}{2},0, \ldots ,0,\frac{1}{2}\} =\frac{1}{2}(|1\rangle \langle 1| + |n\rangle \langle n|)$}.
\item[\textbf{11.}]\hypertarget{DiagCorrelanceNormProof:11}{}To see that the candidate $\rho _{D_{\max } }$ from \hyperlink{DiagCorrelanceNormProof:10}{Step 10} \textit{also} satisfies the requirement from \hyperlink{DiagCorrelanceNormProof:3}{Step 3} that it not have product form, note that in any diagonal state of equal nonzero probabilities, the factorizability of basis elements of the nonzero terms completely determines whether it has product form. A rank-$2$ diagonal state of equal nonzero elements that does not have product form in all multipartite systems is $\rho_{D_{\max}}=\frac{1}{2}(|1\rangle \langle 1| + |n\rangle \langle n|)$ of \Eq{17}, since its basis states of nonzero probabilities yield the coincidence form $\rho_{D_{\max}}=\frac{1}{2}(|1^{(1)}\rangle\langle 1^{(1)}|\otimes\cdots\otimes|1^{(N)}\rangle\langle 1^{(N)}|+|n_{1}^{(1)}\rangle\langle n_{1}^{(1)}|\otimes\cdots\otimes|n_{N}^{(N)}\rangle\langle n_{N}^{(N)}|)$, \textit{which is unfactorizable because the projectors in each mode are different in each term, preventing any single-mode projectors from being factored out}. Using the first and last basis element guarantees that the projectors will always have different labels in each mode between the two terms, which is the simplest way to guarantee this property for all systems.  Thus, since this is exactly the candidate state of \hyperlink{DiagCorrelanceNormProof:10}{Step 10}, we have justified why it is a prototypical $\rho _{D_{\max } }$.
\end{itemize}

Now that we have a motivation for why $\rho_{D_{\max}}$ from \hyperlink{DiagCorrelanceNormProof:11}{Step 11} is a nonfactorizable diagonal state furthest from its own reduction product, we need to calculate its raw diagonal correlance to get the normalization factor $\mathcal{N}_{\Correlance_D}$ of \Eq{18}. Therefore, we start with the raw correlance of $\rho_{D_{\max}}=\frac{1}{2}(|1\rangle \langle 1| + |n\rangle \langle n|)$ from \hyperlink{DiagCorrelanceNormProof:11}{Step 11} above, as
\begin{Equation}                      {L.2}
\begin{array}{*{20}l}
   {\mathcal{N}_{\Correlance_D } } &\!\! { \equiv \widetilde{\Correlance}(\rho _{D_{\max } } ) = \tr[(\rho _{D_{\max } }  - \varsigma _{D_{\max } } )^2 ]}  \\
   {} &\!\! { = P(\rho _{D_{\max } } ) - 2\tr[\rho _{D_{\max } } \varsigma _{D_{\max } } ] + P(\varsigma _{D_{\max } } ),}  \\
\end{array}
\end{Equation}
where $\varsigma _{D_{\max } }  \equiv \varsigma (\rho _{D_{\max } } )$, and 
\begin{Equation}                      {L.3}
\rho _{D_{\max } }  = \text{diag}\{ \frac{1}{2}_1 ,0_2 , \ldots ,0_{n - 1} ,\frac{1}{2}_n \} ,
\end{Equation}
(where we use subscripts to help indicate which basis element corresponds to each matrix element), and thus
\begin{Equation}                      {L.4}
P(\rho _{D_{\max } } ) = \frac{1}{2}.
\end{Equation}
Then, using the fact from \hyperlink{DiagCorrelanceNormProof:8}{Step 8} that writing any single-mode reduction in terms of the matrix elements of the parent state always features the first and last diagonal parent matrix elements in \textit{separate} diagonal matrix elements of the reduced states (specifically the first and last element of each, where the multipartite basis is ordered by the standard register-counting convention), then each mode-$m$ reduction mirrors the parent state as
\begin{Equation}                      {L.5}
\redx{\rho}{m}_{D_{\max } } = \text{diag}\{ \frac{1}{2}_1 ,0_2 , \ldots ,0_{n_m  - 1} ,\frac{1}{2}_{n_m } \},
\end{Equation}
which means that the reduction product is
\begin{Equation}                      {L.6}
\varsigma _{D_{\max } }  \equiv \mathop  \otimes \limits_{m = 1}^N \redx{\rho}{m}_{D_{\max } }  = \mathop  \otimes \limits_{m = 1}^N \text{diag}\{ \frac{1}{2}_1 ,0_2 , \ldots ,0_{n_m  - 1} ,\frac{1}{2}_{n_m } \},
\end{Equation}
and therefore has purity
\begin{Equation}                      {L.7}
P(\varsigma _{D_{\max } } ) = \prod\limits_{m = 1}^N {\frac{1}{2}}  = \frac{1}{{2^N }}.
\end{Equation}
For the overlap in \Eq{L.2}, since all but the first and last elements of $\rho _{D_{\max } }$ are zero, only the first and last elements of $\varsigma _{D_{\max } }$ affect the overlap, and since those elements are both $\frac{1}{{2^N }}$ by expanding \Eq{L.6}, then
\begin{Equation}                      {L.8}
\tr(\rho _{D_{\max } } \varsigma _{D_{\max } } ) = \tr(\text{diag}\{ \frac{1}{2} \cdot \frac{1}{{2^N }},0 , \ldots ,0 ,\frac{1}{2} \cdot \frac{1}{{2^N }}\} ) = \frac{1}{{2^N }}.
\end{Equation}
Thus, putting \Eq{L.4}, \Eq{L.7}, and \Eq{L.8} into \Eq{L.2} gives
\begin{Equation}                      {L.9}
\mathcal{N}_{\Correlance_D }= \frac{1}{2} - 2\frac{1}{{2^N }} + \frac{1}{{2^N }} = \frac{1}{2} - \frac{1}{{2^N }},
\end{Equation}
which is the result in \Eq{18}.

Note that any given diagonal state $\rho_D$ can be converted to another diagonal state of equal $\Correlance_D$ by a local-permutation unitary (LPU) (which can generally be complex), meaning a tensor-product of permutation unitaries of each mode, since these operators cause neither superpositions between modes nor superpositions within the modes, preserving both locality and diagonality.  Local unitaries (LU) also preserve $\Correlance_D$, but not diagonality, so they cannot be used to reach all states of a given $\Correlance_D$ from a single $\rho_D$. Thus, since $\rho_{D_{\max}}$ only has rank 2 and balanced probabilities, the set of \textit{all} $\rho_{D_{\max}}$ is found by applying any LPU to the prototypical $\rho_{D_{\max}}$ of \Eq{17}.  
\section{\label{sec:App.M}Algorithm $\mathcal{A}_{\rho}$ to Form a Multipartite Density Matrix from Strictly Classical Data}
The purpose of this algorithm is to generate a density-matrix estimator $\rho$ from the data set of measurements of a strictly classical multivariable system, so that we can then measure the general nonlocal correlation in the data using the diagonal correlance $\Correlance_D (\rho)$ of \Eq{15}.

Continuing from the description leading up to \Eq{19}, the main idea of this algorithm is that the data for each of the $N$ random variables (RVs) $\{x^{(m)}\}$ is first quantized separately, and then combined as a set of $n_{S}$ quantized $N$-tuple data points $\{ \mathbf{x}_j ' \}$.  During the quantization, we make a list of the center values of every bin of each mode as \textit{representative bin values} $\widetilde{\mathbf{x}}^{(m)}\equiv(\widetilde{x}^{(m)}_{1},\ldots,\widetilde{x}^{(m)}_{n_{m}})^T$. Then, we create the density matrix $\rho$ for the quantized data by cycling through $N$-tuples of representative bin values in ``register format'' (for example, $\{(\widetilde{x}^{(1)}_{1},\widetilde{x}^{(2)}_{1}),(\widetilde{x}^{(1)}_{1},\widetilde{x}^{(2)}_{2}),(\widetilde{x}^{(1)}_{2},\widetilde{x}^{(2)}_{1}),(\widetilde{x}^{(1)}_{2},\widetilde{x}^{(2)}_{2})\}$ in the simplest nontrivial case), and counting the number of quantized data points that match each of them as an entire $N$-tuple, forming a \textit{multipartite histogram}, which is then normalized and placed in the main-diagonal elements of an $n\times n$ matrix, giving us $\rho$.

The following steps constitute algorithm $\mathcal{A}_{\rho}$:
\begin{itemize}[leftmargin=*,labelindent=4pt]\setlength\itemsep{0pt}
\item[\textbf{1.}]\hypertarget{ClassicalDensityMatrixAlg:1}{}Identify the $N$ RVs of a given data set as
\begin{Equation}                      {M.1}
\mathbf{x} \equiv (x^{(1)} , \ldots ,x^{(N)} ).
\end{Equation}
\item[\textbf{2.}]\hypertarget{ClassicalDensityMatrixAlg:2}{}Organize the data as the $n_S  \times N$ matrix,
\begin{Equation}                      {M.2}
X\hsp{-1.5} \equiv\hsp{-4} \left(\hsp{-3} {\begin{array}{*{20}c}
   {\mathbf{x}_1 }  \\
    \vdots   \\
   {\mathbf{x}_{n_S } }  \\
\end{array}}\hsp{-3} \right) \hsp{-4}=\hsp{-4} \left(\hsp{-2} {\begin{array}{*{20}c}
   {(x_1^{(1)} , \ldots ,x_1^{(N)} )}  \\
    \vdots   \\
   {(x_{n_S }^{(1)} , \ldots ,x_{n_S }^{(N)} )}  \\
\end{array}}\hsp{-2} \right) \hsp{-4}\equiv\hsp{-4} \left(\hsp{-3} {\begin{array}{*{20}c}
   {X_{1,1} } &\hsp{-7}  \cdots  &\hsp{-7} {X_{1,N} }  \\
   {} &\hsp{-7}  \vdots  &\hsp{-7} {}  \\
   {X_{n_S ,1} } &\hsp{-7}  \cdots  &\hsp{-7} {X_{n_S ,N} }  \\
\end{array}}\hsp{-4} \right)\hsp{-2.5},
\end{Equation}
where the $j$th $N$-dimensional data point $\mathbf{x}_j$ is an $N$-tuple of measured values, for a total of $n_S$ measurements of $N$ values each.  Thus, elements of $X$ are \smash{$X_{j,m}  \equiv x_j^{(m)} $} for \smash{$j \in 1, \ldots ,n_S $} and \smash{$m \in 1, \ldots ,N$}. 
\item[\textbf{3.}]\hypertarget{ClassicalDensityMatrixAlg:3}{}Choose quantization\hsp{-1} (bin)\hsp{-1} numbers\hsp{-1} $n_m$\hsp{-1} for\hsp{-1} each\hsp{-1} RV\hsp{-1} as
\begin{Equation}                      {M.3}
\mathbf{n} \equiv (n_1 , \ldots ,n_N ).
\end{Equation}
For discrete RVs, $n_m$ is the number of possible outcomes for RV $x^{(m)}$. For continuous RVs, $n_m$ is the finite number of \textit{recognized} possible values for RV $x^{(m)}$, and there are many techniques for choosing $n_m$ appropriately based on the data.  Choice of $n_m$ sets the bin number in a histogram of the data in column $m$ of $X$. Note: in most cases, the precision of the measuring devices used to obtain the data already imposes some quantization on the RVs.
\item[\textbf{4.}]\hypertarget{ClassicalDensityMatrixAlg:4}{}Define the domain of each RV as a bounded a pair of extreme values \smash{$[x_{\min }^{(m)},x_{\max }^{(m)}]$}. \textit{In some cases} it might be preferable to choose bounds that hug the extremes of the data itself, such as by using
\begin{Equation}                      {M.4}
\begin{array}{*{20}c}
   {\begin{array}{*{20}c}
   {x_{\min }^{(m)}  \equiv \min \{ X_{:,m} \} ,}  \\
   {x_{\max }^{(m)}  \equiv \max \{ X_{:,m} \} ,}  \\
\end{array}} & {X_{:,m}  \equiv \left(\hsp{-2} {\begin{array}{*{20}c}
   {X_{1,m} }  \\
    \vdots   \\
   {X_{n_S ,m} }  \\
\end{array}}\hsp{-2} \right)\!.}  \\
\end{array}
\end{Equation}
However, note that \textit{defining} \smash{$[x_{\min }^{(m)},x_{\max }^{(m)}]$} as the extreme values of what \textit{could} happen for each RV (regardless of whether data reaches those values) is technically the more correct method here, so these should generally be \textit{specified inputs} instead of using \Eq{M.4}.
\item[\textbf{5.}]\hypertarget{ClassicalDensityMatrixAlg:5}{}Get bin-edge lists $\mathbf{e}^{(m)} $ for each RV as
\begin{Equation}                      {M.5}
\mathbf{e}^{(m)}  \equiv (e_1^{(m)} , \ldots ,e_{n_m  + 1}^{(m)} )^T ,
\end{Equation}
where the $n_m +1$ bin edges for RV $x^{(m)}$ are
\begin{Equation}                      {M.6}
e_{k_m }^{(m)}  \equiv x_{\min }^{(m)}  + \frac{{x_{\max }^{(m)}  - x_{\min }^{(m)} }}{{n_m }}(k_m  - 1),
\end{Equation}
for $k_m  \in 1, \ldots ,n_m  + 1$.
\item[\textbf{6.}]\hypertarget{ClassicalDensityMatrixAlg:6}{}Get representative bin-value lists for each RV as
\begin{Equation}                      {M.7}
\widetilde{\mathbf{x}}^{(m)}  \equiv (\widetilde{x}_1^{(m)} , \ldots ,\widetilde{x}_{n_m }^{(m)} )^T ,
\end{Equation}
with elements
\begin{Equation}                      {M.8}
\widetilde{x}_q^{(m)}  \equiv \frac{{e_q^{(m)}  + e_{q + 1}^{(m)} }}{2};\quad q \in 1, \ldots ,n_m .
\end{Equation}
\item[\textbf{7.}]\hypertarget{ClassicalDensityMatrixAlg:7}{}Quantize the data as an $n_S  \times N$ matrix $X'$ using
\begin{Equation}                      {M.9}
\begin{array}{l}
 X' = 0^{[n_S  \times N]}  \\ 
\hsp{-2}\begin{array}{*{20}l}
   {\text{for}\;m \in 1, \ldots ,N} \hfill  \\
   {\hsp{8}\text{for}\;j \in 1, \ldots ,n_S } \hfill  \\
   {\hsp{16}\text{for}\;q \in 1, \ldots ,n_m } \hfill  \\
   {\hsp{24}\text{if}\;(q \hsp{-1}==\hsp{-1} 1)\;\&\; (X_{j,m}  \hsp{-1}<\hsp{-1} e_2^{(m)} )} \hfill  \\
   {\hsp{32}X'_{j,m}  = \widetilde{x}_1^{(m)} } \hfill  \\
   {\hsp{24}\text{elseif}\;[(q \hsp{-1}>\hsp{-1} 1)\;\&\; (q \hsp{-1}<\hsp{-1} n_m )} \hfill  \\
   {\hsp{52}\&\; (X_{j,m}  \hsp{-1}\ge\hsp{-1} e_q^{(m)} )\;\&\; (X_{j,m}  \hsp{-1}<\hsp{-1} e_{q + 1}^{(m)} )]} \hfill  \\
   {\hsp{32}X'_{j,m}  = \widetilde{x}_q^{(m)} } \hfill  \\
   {\hsp{24}\text{elseif}\;(q \hsp{-1}==\hsp{-1} n_m )\;\& \;(X_{j,m}  \hsp{-1}\ge\hsp{-1} e_{n_m }^{(m)} )} \hfill  \\
   {\hsp{32}X'_{j,m}  = \widetilde{x}_{n_m }^{(m)} } \hfill  \\
\end{array} \\ 
 \end{array}
\end{Equation}
where the notation \smash{$M^{[a\times b]}$} means an $a\times b$ matrix $M$. Notice that for bin $1$ and bin $n_m$, the conditions assign data outside the RV extremes to those end bins; that is to allow variations in measured values to go under or over the extremes, but again, this may not be applicable in all situations, so modify as appropriate.
\item[\textbf{8.}]\hypertarget{ClassicalDensityMatrixAlg:8}{}Generate the multipartite histogrammic $n\times n$ density matrix $\rho$ (where $n=n_{1}\cdots n_{N}$) using
\begin{Equation}                      {M.10}
\begin{array}{l}
 \rho  = 0^{[n\times n]}  \\ 
 \text{for}\;a \in 1, \ldots ,n \\ 
 \hsp{8}\mathbf{a} = \mathbf{a}_a^{\{ N,\mathbf{n}\} }  \\ 
 \hsp{8}\text{for}\;j \in 1, \ldots ,n_{S} \\ 
 \hsp{16}\widetilde{\mathbf{x}}_j  = X'_{j, \cdots }  \\ 
 \hsp{16}\mathbf{v} = 0^{[1 \times N]} \\
 \hsp{16}\text{for}\;m \in 1, \ldots ,N \\ 
 \hsp{24}v_{m}  = \widetilde{x}_{a_m }^{(m)}  \\ 
 \hsp{16}\text{if}\;\text{sum}(\widetilde{\mathbf{x}}_j  == \mathbf{v}) == N \\ 
 \hsp{24}\rho _{a,a}  = \rho _{a,a}  + 1 \\ 
 \rho  = \rho /\tr(\rho ) \\ 
 \end{array}
\end{Equation}
where $\mathbf{a}_a^{\{ N,\mathbf{n}\} }$ is the \textit{inverse register function} given in \App{U} that maps scalar index $a$ to vector index $\mathbf{a}\equiv(a_{1},\ldots,a_{N})$, and $\mathbf{v}\equiv(v_1,\ldots,v_N)$ is a temporary vector to hold the multipartite representative bin value corresponding to scalar index $a$.

This step counts all the occurrences of each multipartite representative bin value that arises in the quantized data $X'$ and tabulates it as a relative frequency in the main diagonal of a density matrix indexed with the corresponding scalar row-column indices for that value. Note that, being diagonal by design, this $\rho$ can be stored as a vector of length $n$, since it is just a classical discrete probability density function, but we keep it in matrix form here for conceptual continuity with the full quantum problem.
\end{itemize}

The result of this algorithm is $\rho$, an estimator of the strictly classical density matrix, which can then be used to calculate the diagonal correlance $\Correlance_D (\rho)$ in \Eq{15}.

The above steps may not necessarily be the most efficient implementation of $\mathcal{A}_{\rho}$; however in this form, the steps are at least conceptually clear.
\section{\label{sec:App.N}Pearson Correlation Coefficient}
Consider a system of $N$ classical random variables (RVs) $\mathbf{x} \equiv (x^{(1)} , \ldots ,x^{(N)} )$, for a dataset $X$ defined as a matrix of $N$ columns representing the RVs, and $n_S$ rows each consisting of an $N$-tuple sample of particular measured (observed) values $\mathbf{x}_j \equiv (x_j^{(1)} , \ldots ,x_j^{(N)} )$ for the $N$ RVs, for a total sample number of $n_S$ measured $N$-tuples [see \Eq{M.2}].  Thus, $X_{j,m}$ is the $j$th observed value for the $m$th RV $x^{(m)}$.  

The \textit{sample Pearson correlation coefficient} \cite[]{Galt,Pear,Devo}, which is defined only for the case of two variables, is given for any two RVs $x^{(a)}$ and $x^{(b)}$ by
\begin{Equation}                       {N.1}
r_{\text{P}}   \equiv r_{\text{P}} ^{(a,b)}  \equiv \frac{{V_S ^{(a,b)} }}{{s^{(a)} s^{(b)} }},
\end{Equation}
where \smash{$V_S ^{(a,b)}\equiv \text{cov}_{S}(x^{(a)},x^{(b)})$} is the (unbiased) \textit{sample covariance} between \smash{$x^{(a)}$} and \smash{$x^{(b)}$},
\begin{Equation}                       {N.2}
\begin{array}{*{20}l}
   {V_S ^{(a,b)} } &\!\! {\equiv\frac{1}{{n_S  - 1}}\sum\limits_{j = 1}^{n_S } {(x_j ^{(a)}  - \mu _S^{(a)} )} (x_j ^{(b)}  - \mu _S^{(b)} )}  \\
   {} &\!\! { = \frac{1}{{n_S  - 1}}\sum\limits_{j = 1}^{n_S } {(X_{j,a}  - \mu _S^{(a)} )} (X_{j,b}  - \mu _S^{(b)} ),}  \\
\end{array}
\end{Equation}
where the \textit{sample mean} of RV $x^{(m)}$ is
\begin{Equation}                       {N.3}
\mu _S^{(m)}  \equiv \overline{x}^{(m)}  \equiv \frac{1}{{n_S }}\sum\limits_{j = 1}^{n_S } {x_j^{(m)} }  = \frac{1}{{n_S }}\sum\limits_{j = 1}^{n_S } {X_{j,m} } ,
\end{Equation}
and the \textit{sample standard deviation} is
\begin{Equation}                       {N.4}
\begin{array}{*{20}l}
   {s^{(m)}  \equiv \sqrt {V_S ^{(m,m)} } } &\!\! { \equiv \sqrt {\frac{1}{{n_S  - 1}}\sum\limits_{j = 1}^{n_S } {(x_j ^{(m)}  - \mu _S^{(m)} )^2 } } }  \\
   {} &\!\! { = \sqrt {\frac{1}{{n_S  - 1}}\sum\limits_{j = 1}^{n_S } {(X_{j,m}  - \mu _S^{(m)} )^2 } } .}  \\
\end{array}
\end{Equation}

We can also form an $N$-by-$N$ \textit{sample covariance matrix} $V_S$ with elements \smash{$(V_S )_{a,b}  \equiv V_S ^{(a,b)} $} and an $N$-by-$N$ \textit{sample Pearson correlation matrix} $r_\text{P}$ with elements \smash{$(r_\text{P} )_{a,b}  \equiv r_\text{P}^{(a,b)} $}, each containing all pairwise covariances and Pearson correlations between the $N$ RVs. However, it is common practice to simply write \smash{$r_{\text{P}}   \equiv r_{\text{P}} ^{(a,b)}$}, as in \Eq{N.1}, to denote the scalar value of the sample Pearson correlation coefficient between two particular variables when those are the only two variables in the problem and no Pearson correlation matrix is used elsewhere in the analysis, which is the convention we use. Also, it is not standard to use the ``$\text{P}$'' subscript, but we use it to distinguish it from our use of $r$ to mean rank.

The Pearson correlation has range $[-1,1]$, and is merely a measure of \textit{linear} correlation between two RVs only.  Thus, two variables can have strong \textit{nonlinear} correlations and yet produce $r_{\text{P}}=0$, so it is not a measure of general correlation. Furthermore, the old adage that ``correlation does not imply causation'' means that having a nonzero $r_\text{P}$ does \textit{not} mean that the RVs share some functional dependence on a common parameter, but rather it means that the data of the two RVs have a linear correspondence whether they are functions of the same underlying variables or not.
\section{\label{sec:App.O}Consistency of Strict Classicality with Established Physics}
Here, we list a few additional reasons why the strictly classical states defined in \Sec{III.C} are compatible with existing ideas in physics, and are a much more appropriate standard for classicality than coherent states.

\begin{itemize}[leftmargin=*,labelindent=4pt]\setlength\itemsep{0pt}
\item[\textbf{1.}]\hypertarget{NiceStrictClassicalityFeatures:1}{}For multiple modes of the same size, the set of all distinct pure strictly classical states has the same form as the set of possible wave functions that yield Maxwell-Boltzmann statistics \cite[]{Saku,Reif}, which arise from a collection of \textit{identical but distinguishable} particles, and is a regime of statistics that would apply if the universe were not quantum-mechanical, i.e.\hsp{-2} in a strictly classical universe.\hsp{-2} For example, for two identical, distinguishable two-level particles, the possible wave functions are \smash{$\{ \psi _1 (x_1 ) \otimes \psi _1 (x_2 ),$} \smash{$\psi _1 (x_1 ) \otimes \psi _2 (x_2 ),$} \smash{$\psi _2 (x_1 ) \otimes \psi _1 (x_2 ),$} \smash{$\psi _2 (x_1 ) \otimes \psi _2 (x_2 )\} $}, where \smash{$\psi _a (x_m ) \equiv$} \smash{$ \langle x^{(m)} |\psi _a^{(m)} \rangle $}, \smash{$|\psi _a^{(m)} \rangle $} is the $a$th possible state of particle $m$, and \smash{$|x^{(m)} \rangle $} is the position eigenstate of particle $m$. These states form a set of pure computational basis states, which qualify (in form) as strictly classical states as defined in \Sec{III.C}.
\item[\textbf{2.}]\hypertarget{NiceStrictClassicalityFeatures:2}{}In regards to probability representing the lack of knowledge of the observer, there are really \textit{two} general kinds of states with different meanings \cite[]{Spek,PuBR,HaSp};
    \begin{itemize}[leftmargin=*,labelindent=4pt]\setlength\itemsep{0pt}
    \item[\textbf{a.}]\hypertarget{FundamentalStateTypes:a}{}\textit{Ontic states}: the \textit{actual} state of a system, part of fundamental reality.
    \item[\textbf{b.}]\hypertarget{FundamentalStateTypes:b}{}\textit{Epistemic states}: the state in the context of an observer, given limitations and lack of full knowledge; is usually a statistical mixture of ontic states each assigned probabilities based on observer ignorance.
    \end{itemize}
\textit{Pure} strictly classical states are ontic, since the state is definite. \textit{Mixed} strictly classical states are epistemic, since by the definition of strict classicality, any probability distributions arise from lack of specific knowledge (i.e., taking sample measurements with time windows too long to notice the fact that the system is instantaneously pure). This dichotomy fits nicely with the classical notion of probability; the \textit{only} kind of probability in a strictly classical system is the ignorance-induced kind, and there is no fundamental quantum probability such as that which arises in pure quantum states with superposition (we exclude the possibility of hidden variable theories until \Sec{VI}).  Furthermore, since pure strictly classical states are pure product states of computational basis states in all reference frames, \textit{nonlocal correlation is impossible in an ontic strictly classical state}. In \Sec{III}--\Sec{V} we discuss the more interesting case of mixed strictly classical nonlocal correlation in detail.
\item[\textbf{3.}]\hypertarget{NiceStrictClassicalityFeatures:3}{}Quantum coherence functions arise from finding mean values of quantities composed of electric field operators in analogy to classical coherence functions. What this leads to is a certification of whether or not the state of the electric field is a \textit{coherent state} (which is only a necessary and sufficient certification of being in a coherent state if the field has $m$th-order quantum coherence values of $|g ^{(m)} (x_1 , \ldots ,x_m )| = 1$ $\forall m \in \{ 1, \ldots ,\infty \} $, meaning that it is ``infinite-order coherent'' \cite[]{GeKn}, where here $x_m\equiv(\bm{\eta}_m,t_m)$ where $\bm{\eta}_m \equiv\mathbf{r}-\mathbf{r}_m$ is a separation vector from source point $\mathbf{r}_m$ to field point $\mathbf{r}$ and $t_m \equiv t-\frac{|{\bm{\eta}_m}|}{c}$ is the retarded time for light to reach $\mathbf{r}$ from $\mathbf{r}_m$). 

However, \textit{measuring how close a state is to infinite-order coherence does not prove anything about classicality; rather it is a test of how close the state is to being a coherent state} (with lower-order failures such as $|g ^{(2)} (x_1 ,x_2 )| \ne 1$ being sufficient to conclude that a state is not a coherent state, thereby making such a test practical to use).  

The fallacy of assuming that a state's classicality is determined by its similarity to a coherent state as quantified by quantum coherence functions is evident for several reasons. The main red flag indicating this fallacy is that a coherent state requires quantum superposition in the Fock basis, an impossibility in truly classical physics. Another red flag is that quantum-coherence-function results are often interpreted in conjunction with quantum-phase space probability distributions where nonclassicality is said to be evident by the appearance of \textit{negative probabilities}, which are not truly allowed in quantum mechanics; a density matrix with any negative diagonal elements is \textit{nonphysical}. In contrast, the definition of strict classicality forbids quantum superposition, a restriction that is a necessary requirement of classicality not satisfied by coherent states, and strict classicality requires no negative probabilities to certify it.
\end{itemize}

Regarding \hyperlink{NiceStrictClassicalityFeatures:3}{Reason 3} above, the classical coherence functions arise by defining \textit{abbreviations} for quantities not naturally expressible in terms of electric field intensities at single spacetime points alone.  The quantum coherence functions simply promote the classical field quantities to operators and promote the time average to a quantum ensemble average; thus its main claim to classicality is through a generalization of Ehrenfest's theorem \cite[]{Ehre,Grif}, which is merely an approximate condition for correspondence of quantum operators to classical quantities, but not sufficient (nor exactly necessary) to determine classicality since it is not a fundamental law of quantum mechanics.  In other words; even though construction of quantum coherence functions is permissible, the ability of certain quantum states (such as coherent states) to produce mean values that we expect in a classical theory is not sufficient to conclude that such states are truly classical; at best we could say that they \textit{exhibit classical behavior}, and as such, coherent states might be considered \textit{the most classical of all quantum states that possess superposition in a computational basis}.
\section{\label{sec:App.P}Limits of Decomposition Indices}
In \Sec{IV.A}, the decomposition unitary $U$ is limited to dimension $D$ of at least $r\equiv\text{rank}(\rho)$ (guaranteed by the existence of the spectral decomposition), and a minimax of at most $r^2$ (proved by P. Horodecki using the Caratheodory theorem justified by a Bloch-vector expansion \cite[]{Hor1,Cara}, which sets the maximum number of decomposition states needed to describe a separable state with mode-independent (MI) decomposition states [Family 5 from \Table{1}]), so that the number of pure decomposition states can always be limited to $D \in r, \ldots ,r^2 $, even though we are free to use $D \in r, \ldots ,\infty $. 

Note that in \cite[]{Hor1}, the term ``dimension'' is used to mean ``rank'' which can be seen in a later work by the same author \cite[]{Hor6} in which the upper decomposition limit is clearly identified as $r^2$. The fact that this limit should be $r^2$ rather than $n^2$ is clear when we consider the case of pure states, which have $r=1$ and only need one decomposition state (the pure state itself up to global phase), regardless of dimension.  However, if for some reason this interpretation is incorrect and the use of $\max(D)=r^2$ turns out to be too restrictive, then we would simply use $\max(D)=n^2$ instead. Since none of the main results in this paper rely on actually computing statance or probablance, this issue has no bearing on the main results.

Then, since $j \equiv (j_1 , \ldots ,j_N )$ is our index for decomposition states of an $N$-mode state, we need to determine the bounds of each $j_m$ such that we achieve a given total of $D$ decomposition states.  Therefore, if we let the range of each mode label be $j_m  \in 1, \ldots ,D_m $, we can define a vector \smash{$\mathbf{D}\equiv(D_{1},\ldots,D_{N})$} such that \smash{$D = \prod\nolimits_{m = 1}^N {D_m }  = D_1  \cdots D_N $} [due to the nested sums in quantities like \Eq{24}] so that as $j \equiv (j_1 , \ldots ,j_N )$ counts over all allowed values, it has exactly $D$ of them.

Now we just need to determine the value of a given $D_m$. By the Caratheodory theorem \cite[]{Hor1,Cara}, the upper necessary limit $D_m$ for each mode label for a Family-5 state is $n_{m}^2$, where $n_{m}$ is the number of levels of mode $m$. Thus, by this line of reasoning, we need to use 
\begin{Equation}                       {P.1}
D_m  \in 1, \ldots ,n_m ^2 \;\;\,\text{s.t.}\;\prod\limits_{m = 1}^N {D_m }  = D\;\;\text{for}\;\;D \in r, \ldots ,r^2 .
\end{Equation}

However, note that in some cases, the Caratheodory theorem's application by P. Horodecki may be too restrictive for the following reason.  Recall that in Wootters's full concurrence paper \cite[]{Woot}, he included an explicit method showing that it is always possible to decompose any separable two-qubit state with exactly $r$ separable decomposition states (he used $n$ to mean rank there) \textit{of the form of Family 3} from \Table{1}. Yet in P. Horodecki's paper \cite[]{Hor1}, he placed an upper bound on the number of necessary separable decomposition states of $r^2$, but he used states of the form of Family 5 from \Table{1}, which are \textit{only a small subset of all separable states}, the full set being Family 3.

The interesting thing about Wootters's decomposition is that \textit{it also applies to the Family-5 states}, meaning that even states for which statance is zero (since they have MI decomposition states) have a separable decomposition of the Family-5 form with only $r$ members!  But there is no contradiction here; the reason the Wootters decomposition works for all separable states is that it is based merely on minimizing average \textit{entanglement}, a special kind of decomposition-state correlation, which means that at least for two qubits, all states with an MI set of decomposition states (Families 5 and 6 from \Table{1}) are \textit{guaranteed} to have a \textit{separable} decomposition of only $r$ members.  But because statance measures \textit{all} decomposition-state correlation, of which \textit{entanglement is only a part}, the decomposition that minimizes statance is generally more restrictive and therefore requires more members, so that is why we must use the Caratheodory theorem to set the limits in the statance calculation.  Basically, P. Horodecki used a set of states that is \textit{only a subset} of separable states; his proof applies only to Family 5 and its subset, Family 6. However, his result is still correct; it is just too loose of an upper bound for more general separable states such as those of Families 3 and 4 that are not also in Families 5 and 6.

To see how \Eq{P.1} fits into the decomposition of zero-statance states (since they are the standard by which zero statance is achieved), \Table{3} shows all the possible decomposition vectors $\mathbf{D}$ allowable for two qubits for the purpose of keeping $D \in r,\ldots,r^2$.
\begin{table}[H]
\caption{\label{tab:3}Allowable minimally necessary numbers of two-qubit decomposition states $D$ given $r\hsp{-1}\equiv\hsp{-1}\text{rank}(\rho)$, and vectors of possible mode-independent single-mode index limits $\mathbf{D}\hsp{-1}\equiv\hsp{-1}(D_{1},D_{2})$ such that $\prod\nolimits_{m=1}^{N}D_{m}\hsp{-1}=\hsp{-1}D$ for $D\hsp{-1}\in\hsp{-1} r,\ldots,r^2$ as in \Eq{P.1}.}
\vspace{-4pt}%
\centering
\begin{tabular}{|c|c|p{149pt}|}
\hline 
\hline
{$r\rule{0pt}{9pt}$} & {$\{D\}$} & {$\hspace{\stretch{1}}\{\mathbf{D}\}\hspace{\stretch{1}}$} \\[0.5mm]
\hline 
{$1\rule{0pt}{9pt}$} & {$\{1\}$} & {$\hspace{\stretch{1}}\{(1,1)\}\hspace{\stretch{1}}$} \\
\hline
$\begin{array}{*{20}c}
   {{2}\rule{0pt}{9pt}}  \\
   {}  \\
\end{array}$ & $\begin{array}{*{20}c}
   {{\{2,3,4\}}\rule{0pt}{9pt}}  \\
   {}  \\
\end{array}$ & {$\hspace{\stretch{1}}\begin{array}{*{20}l}
   {{\{(1,2),(2,1),(1,3),(3,1),(1,4),}\rule{0pt}{9pt}}  \\
   {\hsp{4.7}(4,1),(2,2)\} }  \\
\end{array}\hspace{\stretch{1}}$} \\
\hline
$\begin{array}{*{20}c}
   {{3}\rule{0pt}{9pt}}  \\
   {}  \\
\end{array}$ & $\begin{array}{*{20}c}
   {{\{3,4,6,8,9\}}\rule{0pt}{9pt}}  \\
   {}  \\
\end{array}$ & {$\hspace{\stretch{1}}\begin{array}{*{20}l}
   {{\{(1,3),(3,1),(1,4),(4,1),(2,2),}\rule{0pt}{9pt}}  \\
   {\hsp{4.7}(2,3),(3,2),(2,4),(4,2),(3,3) }  \\
\end{array}\hspace{\stretch{1}}$} \\
\hline
$\begin{array}{*{20}c}
   {{4}\rule{0pt}{9pt}}  \\
   {}  \\
   {}  \\
\end{array}$ & $\begin{array}{*{20}c}
   {{\{4,6,8,9,12,16\}}\rule{0pt}{9pt}}  \\
   {}  \\
   {}  \\
\end{array}$ & {$\hspace{\stretch{1}}\begin{array}{*{20}l}
   {{\{(1,4),(4,1),(2,2),(2,3),(3,2),}\rule{0pt}{9pt}}  \\
   {\hsp{4.7}(2,4),(4,2),(3,3),(3,4),(4,3), }  \\
   {\hsp{4.7}(4,4)\}}  \\
\end{array}\hspace{\stretch{1}}$} \\
\hline
\hline
\end{tabular}
\end{table}
Notice that $D$ is never a prime number for $D> n$, and for the same reason, any numbers whose only factors include values that exceed any $n_{m}^2$ are also excluded, such as $D=10$ whose only factors are $1\cdot 10$ and $2\cdot 5$, since both $10$ and $5$ exceed $n_{2}^2=4$. We \textit{can} use decompositions such as that, but if P. Horodecki's proof is correct, then we are guaranteed to always be able to find a decomposition of member numbers from the set in \Table{3} instead.  

For example, if a decomposition for a rank-$3$ state is known to have $10$ MI states, we should always be able to find a different MI decomposition with only $D\in\{3,4,6,8,9\}$ MI states. However, in a rank-$4$ state, starting with a known decomposition of $10$ MI states may require that we increase to $D=12$ or $D=16$ states, depending on whether a different MI decomposition with fewer MI states than $10$ exists for the input state.  But if we allow things like $D=10$ here, then our single-mode limits would need to exceed the single-mode Caratheodory bounds. Therefore, the single-mode Caratheodory bounds of $n_{m}^2$ may actually lead to larger decompositions (or they may instead always allow a smaller decomposition), but we do not have a proof for the convertibility of such decompositions at this time. 

To see how $\mathbf{D}$ relates to a state with zero statance, consider a two-qubit state such as
\begin{Equation}                       {P.2}
\begin{array}{*{20}r}
   {\rho  = } &\!\!\! {p_1 \rho _1^{(1)}  \otimes \rho _1^{(2)} +} &\!\!\! {  p_6 \rho _2^{(1)}  \otimes \rho _1^{(2)} }  \\
   {} &\!\!\! { + p_2 \rho _1^{(1)}  \otimes \rho _2^{(2)} +} &\!\!\! {  p_7 \rho _2^{(1)}  \otimes \rho _2^{(2)} }  \\
   {} &\!\!\! { + p_3 \rho _1^{(1)}  \otimes \rho _3^{(2)} +} &\!\!\! {  p_8 \rho _2^{(1)}  \otimes \rho _3^{(2)} }  \\
   {} &\!\!\! { + p_4 \rho _1^{(1)}  \otimes \rho _4^{(2)} +} &\!\!\! {  p_9 \rho _2^{(1)}  \otimes \rho _4^{(2)} }  \\
   {} &\!\!\! { + p_5 \rho _1^{(1)}  \otimes \rho _5^{(2)} +} &\!\!\! { p_{10} \rho _2^{(1)}  \otimes \rho _5^{(2)} }  \\
\end{array}\hsp{-2}.
\end{Equation}
Since statance ignores the decomposition probabilities, then regardless of what they are, we can merely look at the decomposition states themselves, and here we notice that we can generate the same set from
\begin{Equation}                       {P.3}
\{ \rho _1^{(1)} ,\rho _2^{(1)} \}  \otimes \{ \rho _1^{(2)} ,\rho _2^{(2)} ,\rho _3^{(2)} ,\rho _4^{(2)} ,\rho _5^{(2)} \}  \equiv \{ \rho _{(j_1 ,j_2 )} \}, 
\end{Equation}
where the tensor product distributes over commas (if you prefer, the commas can be replaced by plus signs, and then the \textit{terms} of the resulting sum comprise \smash{$\{ \rho _{(j_1 ,j_2 )} \}$}. We can simplify \Eq{P.3} further using the notation
\begin{Equation}                       {P.4}
\begin{array}{*{20}l}
   {\{ 1,2\}  \otimes \{ 1,2,3,4,5\} } &\!\!\! { \equiv\! \{ (j_1 ,j_2 )\} }  \\
   {} &\!\!\! { =\! \{ (1,1),(1,2),(1,3),(1,4),(1,5),}  \\
   {} &\!\!\! {\phantom{=\!\{}(2,1),(2,2),(2,3),(2,4),(2,5)\} .}  \\
\end{array}\!\!
\end{Equation}
The decomposition vector $\mathbf{D}\equiv(D_{1},\ldots,D_{N})$  is a list of the maximal index labels in each mode that are needed to decompose a zero-statance state whose optimal decomposition has $D=D_{1}\cdots D_{N}$ MI decomposition states.  Thus, in this example, the decomposition in \Eq{P.2} has
\begin{Equation}                       {P.5}
\mathbf{D}\equiv(2,5),
\end{Equation}
[which are also the sizes of the mode sets in \Eq{P.4}].  According to P. Horodecki's decomposition limit, which applies to states such as this, $D_2 =5$ is larger than is needed for the mode-$2$ limit, so we should be able to find a \textit{different} decomposition with mode-2 limit $D_2 \leqslant 4$. If $r=3$, then \Table{3} shows that we should be guaranteed to find such an MI decomposition of fewer members where $D\in\{3,4,6,8,9\}$, but if $r=4$, then we may or may not be able to find a different MI decomposition with fewer members, as mentioned earlier, though we are guaranteed to find one that has at most $D=r^2 =16$ MI members.

Regarding the discrepancy between Wootters's decompositions and P. Horodecki's limits, we may hypothesize that Wootters's result is true for \textit{all} systems (not just two qubits); that all full $N$-partite separable states have a decomposition of exactly $r$ separable pure decomposition states in the form of Family 3, but that for states with zero statance, they require at least between $r$ and $r^2$ pure decomposition states to achieve a set of MI decomposition states (and furthermore, states belonging to Families 5 and 6 all have both types of decompositions; an $r$-member separable decomposition, \textit{and} an $(r,\ldots,r^2)$-member zero-statance decomposition).

Note that in cases where the spectral decomposition is also a statance-minimizing decomposition, it is not necessarily $U=I^{[r]}$ that achieves this (where square bracketed superscripts indicate matrix dimension), but rather it is the $r$-level \textit{permutation unitary} that minimizes statance. This is important when constructing examples using orthogonal decomposition states, since the descending-order convention (DOC) of eigenvalues used to define their labels does not always produce the statance-minimizing decomposition when $U=I^{[r]}$; therefore even when constructing such states, we still need to search the order of their labels to truly minimize the statance for that decomposition. Furthermore, since the eigenvalues in the probability formula \smash{$p_j  = \sum\nolimits_{k = 1}^r {\lambda _k |U_{j,k} |^2 } $} in the DOC prevent any columns of $U$ above column $r$ from affecting the probability, the factors $|U_{j,k} |^2$ must not all be zero for a given $j$ so that we consider only sets where $p_j  > 0\;\;\forall j$. A necessary and sufficient way to ensure this is to specify that
\smash{$\sum\nolimits_{k = 1}^r {|U_{j,k} |^2 }  \ne 0\;\;\forall j$}. Then, since there is always at least one nonzero eigenvalue, all probabilities are nonzero.
\section{\label{sec:App.Q}Proof that Statance is a Necessary and Sufficient Measure of Decomposition-State Correlation}
Here we prove that achieving zero statance \smash{$\hat\Statance(\rho)=0$} is necessary and sufficient for any mixed or pure state $\rho$ to have no decomposition-state correlation as defined in \Sec{I.A}, and that having \smash{$\hat\Statance(\rho)>0$} is necessary and sufficient for $\rho$ to have some decomposition-state correlation, proving that statance is a necessary and sufficient measure of decomposition-state correlation.

First, the unoptimized statance $\Statance(\rho)$ of \Eq{23} has form
\begin{Equation}                       {Q.1}
\Statance(\rho ) \equiv ||\mathbf{w}||_{1} = \sum\nolimits_{j = 1}^D {|w_j |} ,
\end{Equation}
which is just a $1$-norm of the vector $\mathbf{w}$ with components
\begin{Equation}                       {Q.2}
w_j  \equiv \tr[(\rho _j  - \mu _j )^2 ],
\end{Equation}
where the absolute value $|w_j |$ in \Eq{Q.1} is because the operator $\Delta_j\equiv \rho _j  - \mu _j$ is Hermitian and therefore has only real eigenvalues, so the trace of its square as $w_j  \equiv \tr[\Delta_j^2]$ in \Eq{Q.2} is just the sum of the squares of its real eigenvalues, which is always nonnegative. Thus, it is always true that $w_j =|w_j |$. Therefore, since the necessary and sufficient condition for any $1$-norm of a vector $\mathbf{w}$ to be zero is that $\mathbf{w}=\mathbf{0}$ which means that all of its components are zero $w_j =0\;\forall j$, then the necessary and sufficient condition for $\Statance(\rho)$ to be zero is that $w_j =0\;\forall j$, or equivalently, that $\rho _j =\mu _j\;\forall j$ (since that is the unique condition that causes $\rho _j  - \mu _j=0$ which means that $w_j  = \tr[(\rho _j  - \mu _j )^2] = 0\;\forall j$).

Then, due to the definition in \Eq{22} of statance \smash{$\hat\Statance(\rho)$} as being the minimum value of $\Statance(\rho)$ over all decompositions of $\rho$ , then the necessary and sufficient condition for \smash{$\hat\Statance(\rho)=0$} can be stated as
\begin{Equation}                       {Q.3}
\hat\Statance(\rho ) = 0\;\;\text{iff}\;\;\exists \{ p_j ,\rho _j \} \;\;\text{s.t.}\;\;\rho _j  = \mu _j \;\forall j,
\end{Equation}
where $\{ p_j ,\rho _j \}$ is some \textit{particular} decomposition set of $\rho$.

To show that statance \smash{$\hat\Statance(\rho)$} is a necessary and sufficient measure of decomposition-state correlation, we will first show that a sufficient condition for \smash{$\hat\Statance(\rho)=0$} is that $\rho$ has a decomposition with mode-independent decomposition states ($\text{MI}\{\rho_j\}$), which means that $\hat\Statance(\rho)=0$ is necessary for $\text{MI}\{\rho_j\}$. Then, we will show that \smash{$\hat\Statance(\rho)=0$} is also sufficient for $\text{MI}\{\rho_j\}$, which means that $\text{MI}\{\rho_j\}$ is necessary for \smash{$\hat\Statance(\rho)=0$}.

To prove that having $\text{MI}\{\rho_j\}$ is sufficient for \smash{$\hat\Statance(\rho)=0$}, we need to show that a $\rho$ with $\text{MI}\{\rho_j\}$ causes $\rho _j  = \mu _j \;\forall j$. Therefore, suppose that $\rho$ has an optimal decomposition with $\text{MI}\{\rho_j\}$. Then, by definition of MI, combining \Eq{5} and \Eq{6} (using a different dummy index to keep things distinct), the decomposition states take the special form
\begin{Equation}                       {Q.4}
\rho _j  \equiv \rho _{(j_1 , \ldots ,j_N )}  = \mathop  \otimes \limits_{q = 1}^N \rho _{j_q }^{(q)} \;\;\forall j.
\end{Equation}
For $\mu _j  \equiv \mu _{(j_1 , \ldots ,j_N )} $ in \Eq{24}, we need the quantity
\begin{Equation}                       {Q.5}
\rho _{(j_1 ^{\{ m\} } , \ldots ,j_N ^{\{ m\} } )}  = \mathop  \otimes \limits_{q = 1}^N \rho _{j_q ^{\{ m\} } }^{(q)} ,
\end{Equation}
so then, the quantity
\begin{Equation}                       {Q.6}
R_{j_m }^{(m)}  \equiv \hsp{-27}\sum\limits_{\hsp{27}j_1 ^{\{ m\} } , \ldots ,j_N ^{\{ m\} }  = 1, \ldots ,1}^{D_1 , \ldots ,D_N } {\hsp{-27}\delta _{j_m ^{\{ m\} } ,j_m } \rho _{(j_1 ^{\{ m\} } , \ldots ,j_N ^{\{ m\} } )} } 
\end{Equation}
becomes
\begin{Equation}                       {Q.7}
\begin{array}{*{20}l}
   {R_{j_m }^{(m)} }\! &\!\!\! { = \hsp{-27}\sum\limits_{\hsp{27}\rule{0pt}{9pt}j_1 ^{\{ m\} } , \ldots ,j_N ^{\{ m\} }  = 1, \ldots ,1}^{D_1 , \ldots ,D_N } {\hsp{-29}\delta _{j_m ^{\{ m\} } ,j_m } \mathop  \otimes \limits_{q = 1}^N \rho _{j_q ^{\{ m\} } }^{(q)} } }  \\
   {}\! &\!\!\! { =\rule{0pt}{22pt} \hsp{-134}\sum\limits_{\hsp{130}\rule{0pt}{13pt}j_1 ^{\{ m\} } , \ldots ,j_{m - 1} ^{\{ m\} } ,j_{m + 1} ^{\{ m\} } , \ldots ,j_N ^{\{ m\} }  = 1, \ldots ,1,1, \ldots ,1}^{\hsp{78}{D_1 , \ldots ,D_{m - 1} ,D_{m + 1} , \ldots ,D_N }_{\vphantom{\rule{0pt}{7pt}}}} {\hsp{-134}\textnormal{\scalebox{0.94}{$\rho _{j_1 ^{\{ m\} } }^{(1)} \hsp{-2} \otimes \hsp{-2} \cdots\hsp{-2}  \otimes\hsp{-1} \rho _{j_{m - 1} ^{\{ m\} } }^{(m - 1)} \hsp{-2} \otimes\hsp{-1} \rho _{j_m }^{(m)} \hsp{-2} \otimes\hsp{-1} \rho _{j_{m + 1} ^{\{ m\} } }^{(m + 1)} \hsp{-2} \otimes \hsp{-2}  \cdots \hsp{-2} \otimes \hsp{-1} \rho _{j_N ^{\{ m\} } }^{(N)} $}}} }  \\
   {}\! &\!\!\! { =\rule{0pt}{24pt}\hsp{-5}\left[ {\hsp{-3}\left( {\sum\limits_{j_1 ^{\{ m\} }  = 1}^{D_1 } {\hsp{-3}\rho _{j_1 ^{\{ m\} } }^{(1)} } } \right)\hsp{-4} \otimes\hsp{-2}  \cdots\hsp{-2}  \otimes\hsp{-4} \left( {\sum\limits_{j_{m - 1} ^{\{ m\} }  = 1}^{D_{m - 1} } {\hsp{-3}\rho _{j_{m - 1} ^{\{ m\} } }^{(m - 1)} } } \right)\hsp{-4} \otimes\hsp{-1} \rho _{j_m }^{(m)} } \right.}  \\
   {}\! &\!\!\! {\left. { \hsp{10}\otimes\hsp{-4} \left( {\sum\limits_{j_{m + 1} ^{\{ m\} }  = 1}^{D_{m + 1} } {\hsp{-3}\rho _{j_{m + 1} ^{\{ m\} } }^{(m + 1)} } } \right) \hsp{-4}\otimes  \cdots  \otimes \hsp{-4}\left( {\sum\limits_{j_N ^{\{ m\} }  = 1}^{D_N } {\hsp{-3}\rho _{j_N ^{\{ m\} } }^{(N)} } } \right)\hsp{-3}} \right]\hsp{-3},}  \\
\end{array}
\end{Equation}
and since \Eq{24} has the form
\begin{Equation}                       {Q.8}
\mu _j  \equiv \mathop  \otimes \limits_{m = 1}^N \tr_{\mbarsub} \left( {\frac{1}{{D_1  \cdots D_{m - 1} D_{m + 1}  \cdots D_N }}R_{j_m }^{(m)} } \right),
\end{Equation}
then putting \Eq{Q.7} into \Eq{Q.8} gives
\begin{Equation}                       {Q.9}
\begin{array}{*{20}l}
   {\mu _j } &\!\! { =\!\! \mathop  \otimes \limits_{m = 1}^N \hsp{-4}\left[ {\tr\hsp{-3}\left( {\hsp{-3}\frac{1}{{D_1 }}\hsp{-6}\sum\limits_{\rule{0pt}{9pt}j_1 ^{\{ m\} }  = 1}^{D_1 } {\hsp{-6}\rho _{j_1 ^{\{ m\} } }^{(1)} } \hsp{-3}} \right) \hsp{-4}\cdots\hsp{-1} \tr\hsp{-3}\left( {\hsp{-3}\frac{1}{{D_{m - 1} }}\hsp{-6}\sum\limits_{\rule{0pt}{9pt}j_{m - 1} ^{\{ m\} }  = 1}^{D_{m - 1} } {\hsp{-6}\rho _{j_{m - 1} ^{\{ m\} } }^{(m - 1)} } \hsp{-3}} \right)\hsp{-2}\rho _{j_m }^{(m)} } \right.}  \\
   {} &\!\!\! {\left. { \hsp{28}\times \hsp{-1}\tr\hsp{-3}\left( {\hsp{-3}\frac{1}{{D_{m + 1} }}\hsp{-6}\sum\limits_{\rule{0pt}{9pt}j_{m + 1} ^{\{ m\} }  = 1}^{D_{m + 1} } {\hsp{-6}\rho _{j_{m + 1} ^{\{ m\} } }^{(m + 1)} } \hsp{-3}} \right) \hsp{-4}\cdots\hsp{-1} \tr\hsp{-3}\left( {\hsp{-3}\frac{1}{{D_N }}\hsp{-6}\sum\limits_{\rule{0pt}{9pt}j_N ^{\{ m\} }  = 1}^{D_N } {\hsp{-6}\rho _{j_N ^{\{ m\} } }^{(N)} } \hsp{-3}} \right)\hsp{-4}} \right]}  \\
   {} &\!\! { =\!\! \mathop  \otimes \limits_{m = 1}^N \rho _{j_m }^{(m)} }  \\
\end{array}
\end{Equation}
and since \Eq{Q.4} holds for each $j$, then \Eq{Q.9} does as well, so then putting \Eq{Q.9} into \Eq{Q.4} we see that
\begin{Equation}                       {Q.10}
\rho _j  = \mu _j \;\forall j,
\end{Equation}
which, by \Eq{Q.3} proves that having $\text{MI}\{\rho_j\}$ is \textit{sufficient} to cause \smash{$\hat\Statance(\rho)=0$}.

To show that having $\text{MI}\{\rho_j\}$ is also \textit{necessary} to cause \smash{$\hat\Statance(\rho)=0$}, we will simply show that \smash{$\hat\Statance(\rho)=0$} is sufficient for $\rho$ to have $\text{MI}\{\rho_j\}$. Therefore, supposing that \smash{$\hat\Statance(\rho)=0$}, then by \Eq{Q.3}, $\rho _j  = \mu _j \;\forall j$, which by \Eq{24} is
\begin{Equation}                       {Q.11}
\rho _{(j_1 , \ldots ,j_N )}  = \mu _{(j_1 , \ldots ,j_N )}  = \mathop  \otimes \limits_{m = 1}^N \beta _{j_m }^{(m)} ,
\end{Equation}
where each \smash{$\beta _{j_m }^{(m)}$} is a physical state in mode $m$ given by
\begin{Equation}                       {Q.12}
\beta _{j_m }^{(m)}  \equiv \tr_{{\mbarsub}} \hsp{-3}\left( {\hsp{-2}\frac{1}{{D_{\mbarsub} }}\hsp{-130}\sum\limits_{\hsp{130}\rule{0pt}{11pt}j_1 ^{\{ m\} } , \ldots ,j_{m - 1} ^{\{ m\} } ,j_{m + 1} ^{\{ m\} } , \ldots ,j_N ^{\{ m\} }  = 1, \ldots ,1,1, \ldots ,1}^{\hsp{78}{D_1 , \ldots ,D_{m - 1} ,D_{m + 1} , \ldots ,D_N }_{\vphantom{\rule{0pt}{2pt}}}} {\hsp{-128}\rho _{(j_1 ^{\{ m\} } , \ldots ,j_{m - 1} ^{\{ m\} } ,j_m ,j_{m + 1} ^{\{ m\} } , \ldots ,j_N ^{\{ m\} } )} } \hsp{5}} \right)\hsp{-2},
\end{Equation}
where note that in \Eq{Q.12} we did not expand the decomposition states further because we are making no assumption about their form in this part of the proof.

Thus, \Eq{Q.11} shows that the condition of \smash{$\hat\Statance(\rho)=0$} automatically leads to a situation where the $\rho _j  \equiv \rho _{(j_1 , \ldots ,j_N )} $ have a (tensor) \textit{product form} of $N$ factors guaranteed for all $j$, and each factor depends on a different mode-specific index $j_m$, which is the definition of the \textit{form} of mode independence for decomposition states. Furthermore, since each \smash{$\beta _{j_m }^{(m)} $} is a convex sum of physical states (which are just the pure decomposition states with special mode-specific index arguments) with equal probabilities such that \smash{$\tr(\beta _{j_m }^{(m)} ) = 1$}, then each \smash{$\beta _{j_m }^{(m)} $} qualifies as a physical state, and we can simply rename them as states
\begin{Equation}                       {Q.13}
\beta _{j_m }^{(m)}  \equiv \rho _{j_m }^{(m)} ,
\end{Equation}
which, put into \Eq{Q.11}, is the \textit{definition} of $\text{MI}\{\rho_j\}$, and so we have proven that \smash{$\hat\Statance(\rho)=0$} is a sufficient condition for $\text{MI}\{\rho_j\}$, and thus $\text{MI}\{\rho_j\}$ is \textit{necessary} for \smash{$\hat\Statance(\rho)=0$}.

We have now shown that the condition of a state having a decomposition with mode-independent decomposition states ($\text{MI}\{\rho_j\}$) is both necessary and sufficient to cause \smash{$\hat\Statance(\rho)=0$}. Then, since a violation of $\text{MI}\{\rho_j\}$ implies \smash{$\hat\Statance(\rho)\neq 0$}, and by its definition this means that \smash{$\hat\Statance(\rho)>0$}, we have proven that statance \smash{$\hat\Statance(\rho)$} is a valid measure of decomposition-state correlation, since the violation of $\text{MI}\{\rho_j\}$ is the definition of decomposition-state correlation, and the achievement of $\text{MI}\{\rho_j\}$ is the necessary and sufficient condition for \smash{$\hat\Statance(\rho)=0$}.
\section{\label{sec:App.R}Statance Example}
Suppose we have a two-qubit state of the type in Family 5 of \Table{1}, with the statance-minimizing decomposition \smash{$\rho  = \sum\nolimits_j {p_j \rho _{j_1 }^{(1)}  \otimes \rho _{j_2 }^{(2)} } $}, where the probabilities \smash{$p_j  \equiv p_{(j_1 , j_2 )} $} in this example happen to \textit{not} have mode independence, while by definition the decomposition states \textit{do} have mode independence.  Therefore, in this particular example, the decomposition states are
\begin{Equation}                       {R.1}
\begin{array}{*{20}l}
   {\rho _1 } &\!\! { \equiv \rho _{(1,1)} } &\!\! { = \rho _1^{(1)}  \otimes \rho _1^{(2)} }  \\
   {\rho _2 } &\!\! { \equiv \rho _{(1,2)} } &\!\! { = \rho _1^{(1)}  \otimes \rho _2^{(2)} }  \\
   {\rho _3 } &\!\! { \equiv \rho _{(2,1)} } &\!\! { = \rho _2^{(1)}  \otimes \rho _1^{(2)} }  \\
   {\rho _4 } &\!\! { \equiv \rho _{(2,2)} } &\!\! { = \rho _2^{(1)}  \otimes \rho _2^{(2)}. }  \\
\end{array}
\end{Equation}
To find the statance of $\rho$, first write the general form of \smash{$\mu _j  \equiv \mu _{(j_1 ,j_2 )} $} from \Eq{24} with $\mathbf{D}=(2,2)$ as
\begin{Equation}                       {R.2}
\begin{array}{*{20}l}
   {\mu _j} &\!\! {\equiv \tr_{\overline{1}} \left( {\frac{1}{2}[\rho _{(j_1 ,1)}  \hsp{-1}+\hsp{-1} \rho _{(j_1 ,2)} ]} \right)\hsp{-2} \otimes \tr_{\overline{2}} \left( {\frac{1}{2}[\rho _{(1,j_2 )} \hsp{-1}+\hsp{-1} \rho _{(2,j_2 )} ]} \right)\hsp{-1.5}.}  \\
\end{array}
\end{Equation}
Writing these out for each index and then plugging-in the particular states in \Eq{R.1} gives
\begin{Equation}                       {R.3}
\begin{array}{*{20}l}
   {\mu _1 } &\!\! { = \tr_{\overline{1}} (\rho _1^{(1)}  \otimes \frac{1}{2}[\rho _1^{(2)}  \hsp{-1.5}+\hsp{-1} \rho _2^{(2)} ]) \otimes \tr_{\overline{2}} (\frac{1}{2}[\rho _1^{(1)}  \hsp{-1.5}+\hsp{-1} \rho _2^{(1)} ] \otimes \rho _1^{(2)} )}  \\
   {} &\!\! { = \rho _1^{(1)}  \otimes \rho _1^{(2)} }  \\
   {\mu _2 } &\!\! { = \tr_{\overline{1}} (\rho _1^{(1)}  \otimes \frac{1}{2}[\rho _1^{(2)}  \hsp{-1.5}+\hsp{-1} \rho _2^{(2)} ]) \otimes \tr_{\overline{2}} (\frac{1}{2}[\rho _1^{(1)}  \hsp{-1.5}+\hsp{-1} \rho _2^{(1)} ] \otimes \rho _2^{(2)} )}  \\
   {} &\!\! { = \rho _1^{(1)}  \otimes \rho _2^{(2)} }  \\
   {\mu _3 } &\!\! { = \tr_{\overline{1}} (\rho _2^{(1)}  \otimes \frac{1}{2}[\rho _1^{(2)}  \hsp{-1.5}+\hsp{-1} \rho _2^{(2)} ]) \otimes \tr_{\overline{2}} (\frac{1}{2}[\rho _1^{(1)}  \hsp{-1.5}+\hsp{-1} \rho _2^{(1)} ] \otimes \rho _1^{(2)} )}  \\
   {} &\!\! { = \rho _2^{(1)}  \otimes \rho _1^{(2)} }  \\
   {\mu _4 } &\!\! { = \tr_{\overline{1}} (\rho _2^{(1)}  \otimes \frac{1}{2}[\rho _1^{(2)}  \hsp{-1.5}+\hsp{-1} \rho _2^{(2)} ]) \otimes \tr_{\overline{2}} (\frac{1}{2}[\rho _1^{(1)}  \hsp{-1.5}+\hsp{-1} \rho _2^{(1)} ] \otimes \rho _2^{(2)} )}  \\
   {} &\!\! { = \rho _2^{(1)}  \otimes \rho _2^{(2)} .}  \\
\end{array}
\end{Equation}
Then, putting \Eq{R.1} and \Eq{R.3} into \Eq{23}, we get
\begin{Equation}                       {R.4}
\begin{array}{*{20}l}
   {\Statance(\rho )} &\!\! { \equiv \sum\limits_{j_1 ,j_2 =1,1}^{D_{1},D_{2}} {\hsp{-4}\tr[(\rho _{(j_1 ,j_2 )}  - \mu _{(j_1 ,j_2 )} )^2 ]} }  \\
   {} &\!\! { = \sum\limits_{j_1 ,j_2 =1,1}^{D_{1},D_{2}} {\hsp{-4}\tr[(\rho _{j_1 }^{(1)}  \otimes \rho _{j_2 }^{(2)}  - \rho _{j_1 }^{(1)}  \otimes \rho _{j_2 }^{(2)} )^2 ]}  = 0,}  \\
\end{array}
\end{Equation}
since for this example, each $\mu_j$ is exactly equal to $\rho_j$ regardless of what the probabilities were. Then, since \Eq{R.4} was calculated with an optimal decomposition [because it minimizes the unoptimized statance $\Statance(\rho )$], $\Statance(\rho )$ yields the statance from \Eq{22} as
\begin{Equation}                       {R.5}
\hat \Statance(\rho ) \equiv \frac{1}{{\mathcal{N}_{\Statance} }}\mathop {\min }\limits_{\{ U\} } [\Statance(\rho )] = 0.
\end{Equation}
Thus, this example shows that any states belonging to Families 5 or 6 from \Table{1} have zero statance regardless of their decomposition probabilities, because those states all have mode-independent sets of decomposition states for their statance-minimizing decompositions.
\section{\label{sec:App.S}Proof that Probablance is a Necessary and Sufficient Measure of Probability Correlation}
Here we prove that achieving zero probablance \smash{$\hat\Probablance(\rho)=0$} is necessary and sufficient for any mixed or pure state $\rho$ to have no probability correlation as defined in \Sec{I.A}, and that having \smash{$\hat\Probablance(\rho)>0$} is necessary and sufficient for the state to have some probability correlation, proving that probablance is a necessary and sufficient measure of probability correlation.

First, note that the unoptimized probablance $\Probablance(\rho)$ from \Eq{27} can be written as
\begin{Equation}                       {S.1}
\Probablance(\rho ) \equiv |\mathbf{v}|^2  = \sum\nolimits_j {|v_j |^2 } ,
\end{Equation}
which is just a nonzero scalar times a $2$-norm of the vector
\begin{Equation}                       {S.2}
\mathbf{v}\equiv\mathbf{p}^{\prime}-\mathbf{q}^{\prime},
\end{Equation}
where $\mathbf{p}^{\prime}$ has components $p_j^{\prime}  \equiv p_{(j_1 , \ldots ,j_N )}^{\prime} $ which is the set of decomposition probabilities of an arbitrary statance-minimizing (SM) decomposition of $\rho$ as given by \Eq{29}, and $\mathbf{q}^{\prime}$ has components $q_j^{\prime}  \equiv q_{(j_1 , \ldots ,j_N )}^{\prime} $ defined in \Eq{28}, so $\mathbf{v}$ has components $v_j  \equiv p_j^{\prime}  - q_j^{\prime} $.  Therefore, since the necessary and sufficient condition for any $2$-norm of a vector $\mathbf{v}$ to be zero is that $\mathbf{v}=\mathbf{0}$, which means that all of its components are zero $v_j  \equiv 0\;\forall j$, then the necessary and sufficient condition for $\Probablance(\rho)$ to be zero is that $v_j  \equiv 0\;\forall j$, or equivalently, that $p_j^{\prime}  = q_j^{\prime} \;\forall j$.  Then, due to the definition in \Eq{26} of \smash{$\hat\Probablance(\rho)$} as being the minimum value of $\Probablance(\rho)$ over all SM decompositions of $\rho$, the necessary and sufficient condition for \smash{$\hat\Probablance(\rho)=0$} can be stated as
\begin{Equation}                       {S.3}
\hat\Probablance(\rho ) \!=\! 0\hsp{5} \text{iff} \hsp{5} \exists \{ p_j^{\prime} ,\rho _j^{\prime} \} \hsp{5}\text{s.t.}\hsp{5}p_j^{\prime} \! =\! q_j^{\prime} \;\forall j\hsp{5}\text{and}\hsp{5}\hat\Statance(\rho)\!=\!\hat\Statance_{\min},
\end{Equation}
where $\{ p_j^{\prime} ,\rho _j^{\prime} \} $ is some \textit{particular} SM decomposition of $\rho$, \smash{$\hat\Statance(\rho)$} is the statance from \Sec{IV.A}, and \smash{$\hat\Statance_{\min}$} is the minimal \smash{$\hat\Statance(\rho)$} over \textit{all} decompositions of $\rho$. (The requirement that $\{ p_j^{\prime} ,\rho _j^{\prime} \} $ causes $\hat\Statance_{\min}$ is actually already part of the definition of $\{ p_j^{\prime} ,\rho _j^{\prime} \} $, so it is redundant in \Eq{S.3}, but stated anyway to emphasize its importance).

To show that \smash{$\hat\Probablance(\rho)$} is a necessary and sufficient measure of probability correlation, we will first show that a sufficient condition for \smash{$\hat\Probablance(\rho)=0$} is that $\rho$ has an SM decomposition with mode-independent probabilities (\smash{$\text{MI}\{p_{j}\}_{\text{SM}}\equiv\text{MI}\{p_{j}^{\prime}\}$}), meaning that \smash{$\hat\Probablance(\rho)=0$} is necessary for \smash{$\text{MI}\{p_{j}\}_{\text{SM}}$}. Then, we will show that \smash{$\hat\Probablance(\rho)=0$} is also sufficient for \smash{$\text{MI}\{p_{j}\}_{\text{SM}}$}, meaning that \smash{$\text{MI}\{p_{j}\}_{\text{SM}}$} is also necessary for \smash{$\hat\Probablance(\rho)=0$}.

To prove that having \smash{$\text{MI}\{p_{j}\}_{\text{SM}}$} is sufficient for \smash{$\hat\Probablance(\rho)=0$}, we need to show that a state $\rho$ with \smash{$\text{MI}\{p_{j}\}_{\text{SM}}$} causes $p_j^{\prime}  = q_j^{\prime} \;\forall j$. Therefore, suppose that $\rho$ has an SM decomposition with \smash{$\text{MI}\{p_{j}\}_{\text{SM}}$}. Then, by definition, combining \Eq{7} and \Eq{8} (using a different dummy index to keep things distinct),
\begin{Equation}                       {S.4}
p_j^{\prime}  \equiv p_{(j_1 , \ldots ,j_N )}^{\prime}  = \prod\nolimits_{q = 1}^N p_{j_q }^{(q)} \;\;\forall j,
\end{Equation}
where \smash{$\sum\nolimits_{j_q } {p_{j_q }^{(q)} }  = 1;\;\;\forall q \in 1, \ldots ,N$}.  For \smash{$q_j^{\prime}  \equiv q_{(j_1 , \ldots ,j_N )}^{\prime}$} in \Eq{28}, we need the quantity
\begin{Equation}                       {S.5}
p_{(j_1 ^{\{ m\} } , \ldots ,j_N ^{\{ m\} } )}^{\prime}  = \prod\nolimits_{q = 1}^N p_{j_q ^{\{ m\} } }^{(q)} ,
\end{Equation}
so then, the quantity 
\begin{Equation}                       {S.6}
Q_{j_m }^{(m)}  \equiv \hsp{-27}\sum\limits_{\hsp{27}j_1 ^{\{ m\} } , \ldots ,j_N ^{\{ m\} }= 1, \ldots ,1 }^{D_1 , \ldots ,D_N } {\hsp{-27}\delta _{j_m ^{\{ m\} } ,j_m } p_{(j_1 ^{\{ m\} } , \ldots ,j_N ^{\{ m\} } )}^{\prime} } 
\end{Equation}
takes on the value
\begin{Equation}                       {S.7}
\begin{array}{*{20}l}
   {Q_{j_m }^{(m)} } &\!\! { = \hsp{-27}\sum\limits_{\rule{0pt}{10pt}\hsp{27}j_1 ^{\{ m\} } , \ldots ,j_N ^{\{ m\} }= 1, \ldots ,1 }^{D_1 , \ldots ,D_N } {\hsp{-27}\delta _{j_m ^{\{ m\} } ,j_m } \prod\nolimits_{q = 1}^N {p_{j_q ^{\{ m\} } }^{(q)} } } }  \\
   {} &\!\! { = \rule{0pt}{22pt} \hsp{-134}\sum\limits_{\hsp{130}\rule{0pt}{14pt}j_1 ^{\{ m\} } , \ldots ,j_{m - 1} ^{\{ m\} } ,j_{m + 1} ^{\{ m\} } , \ldots ,j_N ^{\{ m\} }  = 1, \ldots ,1,1, \ldots ,1}^{\hsp{78}{D_1 , \ldots ,D_{m - 1} ,D_{m + 1} , \ldots ,D_N }_{\vphantom{\rule{0pt}{8pt}}}} {\hsp{-131}p_{j_1 ^{\{ m\} } }^{(1)}  \cdots p_{j_{m - 1} ^{\{ m\} } }^{(m - 1)} p_{j_m }^{(m)} p_{j_{m + 1} ^{\{ m\} } }^{(m + 1)}  \cdots p_{j_N ^{\{ m\} } }^{(N)} } }  \\
   {} &\!\! {
  = \left[ {\hsp{-3}\left({\sum\limits_{j_1 ^{\{ m\} } =1}^{D_{1}} {\hsp{-3}p_{j_1 ^{\{ m\} } }^{(1)} } }\right) \cdots \left({\sum\limits_{j_{m - 1} ^{\{ m\} } =1}^{D_{m-1}} {\hsp{-3}p_{j_{m - 1} ^{\{ m\} } }^{(m - 1)} } }\right) p_{j_m }^{(m)} } \right.} \\ 
 {} &\!\! {\left. { \hsp{13}\times \hsp{-5}\left({\sum\limits_{j_{m + 1} ^{\{ m\} } =1}^{D_{m+1}} {\hsp{-3}p_{j_{m + 1} ^{\{ m\} } }^{(m + 1)} } }\right) \cdots \left({\sum\limits_{j_N ^{\{ m\} } =1}^{D_{N}} {\hsp{-3}p_{j_N ^{\{ m\} } }^{(N)} } }\right)\hsp{-3}} \right]} \\ 
   {} &\!\! { = p_{j_m }^{(m)} ,}  \\
\end{array}
\end{Equation}
which, when put into \Eq{28}, gives
\begin{Equation}                       {S.8}
q_j^{\prime}  \equiv q_{(j_1 , \ldots ,j_N )}^{\prime}  = \prod\nolimits_{m = 1}^N p_{j_m }^{(m)} .
\end{Equation}
Since \Eq{S.4} holds for each $j$, then \Eq{S.8} does as well, so then putting \Eq{S.8} into \Eq{S.4}, we see that
\begin{Equation}                       {S.9}
p_j^{\prime}  = q_j^{\prime} \;\forall j,
\end{Equation}
which, by \Eq{S.3} proves that having \smash{$\text{MI}\{p_{j}\}_{\text{SM}}$} is \textit{sufficient} to cause \smash{$\hat\Probablance(\rho)=0$}.

To show that having \smash{$\text{MI}\{p_{j}\}_{\text{SM}}$} is also \textit{necessary} to cause \smash{$\hat\Probablance(\rho)=0$}, we will simply show that \smash{$\hat\Probablance(\rho)=0$} is sufficient for $\rho$ to have \smash{$\text{MI}\{p_{j}\}_{\text{SM}}$}. Therefore, supposing that \smash{$\hat\Probablance(\rho)=0$}, then by \Eq{S.1} and \Eq{S.2}, $p_j^{\prime}  = q_j^{\prime} \;\forall j$, which expands using \Eq{28} as
\begin{Equation}                       {S.10}
p_{(j_1 , \ldots ,j_N )}^{\prime}  = \prod\nolimits_{m = 1}^N {b_{j_m }^{(m)} } ,
\end{Equation}
where we define the quantity
\begin{Equation}                       {S.11}
b_{j_m }^{(m)}  \equiv \hsp{-132}\sum\limits_{\hsp{130}\rule{0pt}{11pt}j_1 ^{\{ m\} } , \ldots ,j_{m - 1} ^{\{ m\} } ,j_{m + 1} ^{\{ m\} } , \ldots ,j_N ^{\{ m\} }  = 1, \ldots ,1,1, \ldots ,1}^{\hsp{78}{D_1 , \ldots ,D_{m - 1} ,D_{m + 1} , \ldots ,D_N }_{\vphantom{\rule{0pt}{3pt}}}} {\hsp{-128}p_{(j_1 ^{\{ m\} } , \ldots ,j_{m - 1} ^{\{ m\} } ,j_m ,j_{m + 1} ^{\{ m\} } , \ldots ,j_N ^{\{ m\} } )}^{\prime} } ,
\end{Equation}
where we cannot simplify the probabilities further because we are making no assumptions about their form here.  Also, since the sum of all the $p_j^{\prime} \equiv p_{(j_1 , \ldots ,j_N )}^{\prime} $ is $1$, and the sum in \Eq{S.11} is \textit{missing} one of the mode sums that would complete that sum to $1$, and since the terms in \Eq{S.11} are all probabilities existing on $[0,1]$, then these quantities \smash{$b_{j_m }^{(m)}$} generally satisfy
\begin{Equation}                       {S.12}
0 \le b_{j_m }^{(m)}  \le 1.
\end{Equation}
Thus, \Eq{S.10} shows that the condition of \smash{$\hat\Probablance(\rho)=0$} automatically leads to a situation where each $p_j^{\prime} \equiv p_{(j_1 , \ldots ,j_N )}^{\prime} $ has a \textit{product form} of $N$ factors guaranteed for all $j$, and each factor depends on a different mode-specific index $j_m$, which is the definition of the \textit{form} of mode independence. Furthermore, since \smash{$b_{j_m }^{(m)} \in [0,1]$} by \Eq{S.12}, then by the normalization property of the $p_j^{\prime} \equiv p_{(j_1 , \ldots ,j_N )}^{\prime} $, the mode-independence leads to all the sums producing positive factors, seen by summing \Eq{S.10} as
\begin{Equation}                       {S.13}
\begin{array}{*{20}l}
   {\hsp{-27}\sum\limits_{\rule{0pt}{8pt}\hsp{27}j_1 , \ldots ,j_N = 1, \ldots ,1}^{D_1 , \ldots ,D_N } {\hsp{-31}p_{(j_1 , \ldots ,j_N )} } } &\!\! { = \hsp{-4}\left({\sum\limits_{j_1 =1}^{D_{1}} {b_{j_1 }^{(1)} } }\right) \cdots \left({\sum\limits_{j_N =1}^{D_{N}} {b_{j_N }^{(N)} } }\right)}  \\
   {\hsp{40}1} &\!\! { = B^{(1)}  \cdots B^{(N)} ,}  \\
\end{array}
\end{Equation}
where \smash{$B^{(m)}  \equiv \sum\nolimits_{j_m } {b_{j_m }^{(m)} } $} such that \smash{$B^{(m)} \in[0,1]$}. Then, the only way that the right-side product, with each factor on $[0,1]$, could equal the left is if \textit{all} of the factors are $1$, meaning \smash{$B^{(m)} =1\;\forall m$}, which means that each \smash{$b_{j_m }^{(m)}$} then satisfies all of the conditions for being a probability since \smash{$b_{j_m }^{(m)} \in [0,1]$} and \smash{$\sum\nolimits_{j_m } {b_{j_m }^{(m)} }=1$}. [Also, \smash{$\sum\nolimits_{j_m } {b_{j_m }^{(m)} }=1$} \textit{due to the definition} in \Eq{S.11} because it supplies the missing mode sum for the full probabilities to sum to $1$.] Therefore, we can rename them as mode-specific probabilities
\begin{Equation}                       {S.14}
b_{j_m }^{(m)}  \equiv p_{j_m }^{(m)} ,
\end{Equation}
obeying \smash{$\sum\nolimits_{j_m } {p_{j_m }^{(m)} }=1$} so that \Eq{S.10} becomes
\begin{Equation}                       {S.15}
p_{(j_1 , \ldots ,j_N )}^{\prime}  = \prod\nolimits_{m = 1}^N {p_{j_m }^{(m)} };\;\;\sum\nolimits_{j_m } {p_{j_m }^{(m)} }=1\;\;\forall m \;\forall j ,
\end{Equation}
which is the definition of \smash{$\text{MI}\{p_{j}\}_{\text{SM}}$}, and therefore we have proven that \smash{$\hat\Probablance(\rho)=0$} is a sufficient condition for \smash{$\text{MI}\{p_{j}\}_{\text{SM}}$}, and thus \smash{$\text{MI}\{p_{j}\}_{\text{SM}}$} is \textit{necessary} for \smash{$\hat\Probablance(\rho)=0$}.

We have now shown that the condition of a state having an SM decomposition with mode-independent decomposition probabilities (\smash{$\text{MI}\{p_{j}\}$}) is both necessary and sufficient to cause \smash{$\hat\Probablance(\rho)=0$}. Then, since a violation of \smash{$\text{MI}\{p_{j}\}_{\text{SM}}$} implies \smash{$\hat\Probablance(\rho)\neq 0$}, and by its definition this means that \smash{$\hat\Probablance(\rho)> 0$}, we have proven that probablance \smash{$\hat\Probablance(\rho)$} is a valid measure of probability correlation, since violation of \smash{$\text{MI}\{p_{j}\}_{\text{SM}}$} is the definition of probability correlation, and achievement of \smash{$\text{MI}\{p_{j}\}_{\text{SM}}$} is the necessary and sufficient condition for \smash{$\hat\Probablance(\rho)=0$}. See \Sec{IV.B} for the explanation of why we must use a statance-minimizing decomposition to compute probablance. 
\section{\label{sec:App.T}Probablance Example}
To see how probablance works for a state with mode-independent (MI) probabilities of a statance-minimizing (SM) decomposition, consider the special case of some \textit{product-form} two-qubit input mixed state $\rho$, which, due to its product form, has an optimal decomposition [one that minimizes the unoptimized probablance of \Eq{27}] with probabilities of the form
\begin{Equation}                       {T.1}
\begin{array}{*{20}l}
   {p_{(1,1)}^{\prime} } &\!\! { = p_1^{(1)} p_1^{(2)} }  \\
   {p_{(1,2)}^{\prime} } &\!\! { = p_1^{(1)} p_2^{(2)} }  \\
   {p_{(2,1)}^{\prime} } &\!\! { = p_2^{(1)} p_1^{(2)} }  \\
   {p_{(2,2)}^{\prime} } &\!\! { = p_2^{(1)} p_2^{(2)}, }  \\
\end{array}
\end{Equation}
where \smash{$p_1^{(m)}  + p_2^{(m)}  = 1$}, and these \smash{$p_{(j_1 , \ldots ,j_N )}^{\prime} $} are found from solving the optimization problem over all the unitaries $U^{\prime}$ of an SM decomposition that produce \smash{$p_{(j_1 , \ldots ,j_N )}^{\prime} $} from \Eq{29} to minimize $\mathcal{P}(\rho)$ in \Eq{27} to give the probablance \smash{$\hat{\mathcal{P}}(\rho )$} in \Eq{26}. (In practice, we would just get a list of numbers that have this structure, but we would not know it at this point in the calculation; it would only be clear when we finish computing the probablance.)

In \App{E} we proved that any mixed product-form state has an optimal decomposition of $r=\text{rank}(\rho)$ decomposition states, where the rank of each single-mode reduction is $r_m$ and $r_1  \cdots r_N  = r$. Using this fact, then $\mathbf{D}=(2,2)$, and the overall form of $q_{j}^{\prime}$ from \Eq{28} is
\begin{Equation}                       {T.2}
q_{(j_1 ,j_2 )}^{\prime}=(p_{(j_1 ,1)}^{\prime}  + p_{(j_1 ,2)}^{\prime} )(p_{(1,j_2 )}^{\prime}  + p_{(2,j_2 )}^{\prime} ).
\end{Equation}

Then, using the particular values of \Eq{T.1} in \Eq{T.2} gives
\begin{Equation}                       {T.3}
\begin{array}{*{20}l}
   {q_{(1,1)}^{\prime} } &\!\! { = (p_1^{(1)} p_1^{(2)}  + p_1^{(1)} p_2^{(2)} )(p_1^{(1)} p_1^{(2)}  + p_2^{(1)} p_1^{(2)} )}  \\
   {} &\!\! { = p_1^{(1)} (p_1^{(2)}  + p_2^{(2)} )(p_1^{(1)}  + p_2^{(1)} )p_1^{(2)} }  \\
   {} &\!\! { = p_1^{(1)} p_1^{(2)}, }  \\
\end{array}
\end{Equation}
and similarly,
\begin{Equation}                       {T.4}
\begin{array}{*{20}l}
   {q_{(1,2)}^{\prime} } &\!\! { = (p_1^{(1)} p_1^{(2)} \!\! +\! p_1^{(1)} p_2^{(2)} )(p_1^{(1)} p_2^{(2)} \!\! +\! p_2^{(1)} p_2^{(2)} )} &\!\! { = p_1^{(1)} p_2^{(2)} }  \\
   {q_{(2,1)}^{\prime} } &\!\! { = (p_2^{(1)} p_1^{(2)} \!\! +\! p_2^{(1)} p_2^{(2)} )(p_1^{(1)} p_1^{(2)} \!\! +\! p_2^{(1)} p_1^{(2)} )} &\!\! { = p_2^{(1)} p_1^{(2)} }  \\
   {q_{(2,2)}^{\prime} } &\!\! { = (p_2^{(1)} p_1^{(2)} \!\! +\! p_2^{(1)} p_2^{(2)} )(p_1^{(1)} p_2^{(2)} \!\! +\! p_2^{(1)} p_2^{(2)} )} &\!\! { = p_2^{(1)} p_2^{(2)} ,}  \\
\end{array}
\end{Equation}
which are exactly the same as the optimal decomposition probabilities themselves [which only happens when the optimal decomposition probabilities have mode independence as in \Eq{T.1}].  Thus, putting \Eqs{T.3}{T.4} into \Eq{27},
\begin{Equation}                       {T.5}
\mathcal{P}(\rho ) = \sum\limits_{j_1 ,j_2 =1,1}^{D_{1},D_{2}} {|p_{(j_1 ,j_2 )}^{\prime}  - q_{(j_1 ,j_2 )}^{\prime} |^2 }  = 0,
\end{Equation}
which also yields the fully optimized probablance as \smash{$\hat{\mathcal{P}}(\rho )=0$} here, since we specified that the $U^{\prime}$ we used to get the \smash{$p_{(j_1 ,j_2 )}^{\prime}$} produced an SM decomposition, and since $0$ is the lowest value that \smash{$\mathcal{P}(\rho )$} can have, then this decomposition minimizes the unoptimized probablance.

Thus, this example shows the mechanism by which states with SM decompositions of mode-independent product-form probabilities produce a probablance of $0$.
\section{\label{sec:App.U}Register and Inverse Register Functions}
\textit{Register counting} is any ordered set of sets of characters for which the rightmost character increases in its ordered set until it reaches the end, and then the character to its immediate left can increment by one member in \textit{its} ordered set, but the character to its right simultaneously resets and must go through another cycle before the character to its left can increment again.

For example, in a two-character word where each character has an alphabet of two letters (a system with structure $2\times 2$ such as two qubits), register counting goes as $\{(1,1),(1,2),(2,1),(2,2)\}$.  We can relabel this vector-index list with the scalar-index labels $\{1,2,3,4\}$.

For an $N$-mode discrete quantum system with structure $\mathbf{n}=(n_1 ,\ldots,n_{N})$, where mode $m$ has $n_m$ outcomes labeled in increasing order from $1$ to $n_m$, the map from a particular vector-index outcome labeled as $\mathbf{a}=(a_1 ,\ldots,a_{N})$, where $a_m  \in 1, \ldots ,n_m \;\forall m\in 1,\ldots,N$ to the scalar-index outcome in the set $1,\ldots,n$, where $n\equiv n_{1}\cdots n_N$, is given by the \textit{indical register function} \cite[]{HedD,HedE},
\begin{Equation}                       {U.1}
\begin{array}{*{20}l}
   {R_\mathbf{a}^{\{ N,\mathbf{n}\} } } &\!\! { \equiv 1 +\! \sum\limits_{m = 1}^N {\left( {(a_m  - 1)\!\!\!\!\prod\limits_{j = m + 1}^N \!\!\!\!{n_j } } \right)}}  \\
   {} &\!\! { = (a_1  - 1)(n_2  \cdots n_N )\! +\! (a_2  - 1)(n_3  \cdots n_N ) \!+  \cdots }  \\
   {} &\!\! {\;\;\;\, + (a_{N - 2}  - 1)(n_{N - 1} n_N )\! +\! (a_{N - 1}  - 1)n_N \! +\! a_N ,}  \\
\end{array}
\end{Equation}
which maps the vector index $\mathbf{a}$ to scalar index $a$ as $a=R_\mathbf{a}^{\{ N,\mathbf{n}\} }$. For example, in a two-qubit system, for which $\mathbf{n}=(2,2)$, \Eq{U.1} maps the vector index $(2,1)$ to the scalar index $3$.

In other situations, we know a scalar-index value $a$ and want to know to which vector-index $\mathbf{a}$ it corresponds.  For that, we use the \textit{inverse indical register function} \cite[]{HedE},
\begin{Equation}                       {U.2}
\mathbf{a}_a^{\{ N,\mathbf{n}\} }  = (a_1 , \ldots a_N );\;\;\left\{ {\begin{array}{*{20}l}
   {a_m } &\!\! { = \text{floor}(\frac{v_{m - 1}  - 1}{d_{m}})+1}  \\
   {v_m } &\!\! { = v_{m - 1}  - (a_m  - 1)d_m ,}  \\
\end{array}} \right.
\end{Equation}
for $m\in 1,\ldots,N$, where $v_0  \equiv a$, and \smash{$d_m  \equiv \Pi_{q=m+1}^{N}n_{q}$}. Thus \Eq{U.2} maps scalar index $a$ to vector index $\mathbf{a}$ as $\mathbf{a}=\mathbf{a}_a^{\{ N,\mathbf{n}\} }$.  For example, in a two-qubit system, \Eq{U.2} maps scalar index $2$ to vector index $(1,2)$.

Note that throughout this paper, although $\mathbf{n}$ is the mode-size vector to use for register counting of the basis functions with \Eq{U.1} and \Eq{U.2}, in the case of register counting the \textit{decomposition indices} $j\equiv(j_{1},\ldots,j_{N})$, we need to use the decomposition-limit vector $\mathbf{D}\equiv(D_{1},\ldots,D_{N})$ [see \App{P}] instead of $\mathbf{n}$ in \Eq{U.1} and \Eq{U.2} to convert between scalar decomposition indices $j$ and vector decomposition indices $(j_{1},\ldots,j_{N})$ [where again all indices start on $1$]. Thus we have \smash{$j=R_\mathbf{j}^{\{ N,\mathbf{D}\} }$} and \smash{$\mathbf{j}=\mathbf{j}_j^{\{ N,\mathbf{D}\} }$}, where $n_m \to D_m$,  $a_m \to j_m$, $v_0 \equiv j$, and we use $\mathbf{j}$ to specifically mean the vector index as a collection of many indices $\mathbf{j}\equiv(j_{1},\ldots,j_{N})$ rather than as a single equivalent number $j$.
\end{appendix}
%

\begin{thebibliography}{63}%
\makeatletter
\providecommand \@ifxundefined [1]{%
 \@ifx{#1\undefined}
}%
\providecommand \@ifnum [1]{%
 \ifnum #1\expandafter \@firstoftwo
 \else \expandafter \@secondoftwo
 \fi
}%
\providecommand \@ifx [1]{%
 \ifx #1\expandafter \@firstoftwo
 \else \expandafter \@secondoftwo
 \fi
}%
\providecommand \natexlab [1]{#1}%
\providecommand \enquote  [1]{``#1''}%
\providecommand \bibnamefont  [1]{#1}%
\providecommand \bibfnamefont [1]{#1}%
\providecommand \citenamefont [1]{#1}%
\providecommand \href@noop [0]{\@secondoftwo}%
\providecommand \href [0]{\begingroup \@sanitize@url \@href}%
\providecommand \@href[1]{\@@startlink{#1}\@@href}%
\providecommand \@@href[1]{\endgroup#1\@@endlink}%
\providecommand \@sanitize@url [0]{\catcode `\\12\catcode `\$12\catcode
  `\&12\catcode `\#12\catcode `\^12\catcode `\_12\catcode `\%12\relax}%
\providecommand \@@startlink[1]{}%
\providecommand \@@endlink[0]{}%
\providecommand \url  [0]{\begingroup\@sanitize@url \@url }%
\providecommand \@url [1]{\endgroup\@href {#1}{\urlprefix }}%
\providecommand \urlprefix  [0]{URL }%
\providecommand \Eprint [0]{\href }%
\providecommand \doibase [0]{http://dx.doi.org/}%
\providecommand \selectlanguage [0]{\@gobble}%
\providecommand \bibinfo  [0]{\@secondoftwo}%
\providecommand \bibfield  [0]{\@secondoftwo}%
\providecommand \translation [1]{[#1]}%
\providecommand \BibitemOpen [0]{}%
\providecommand \bibitemStop [0]{}%
\providecommand \bibitemNoStop [0]{.\EOS\space}%
\providecommand \EOS [0]{\spacefactor3000\relax}%
\providecommand \BibitemShut  [1]{\csname bibitem#1\endcsname}%
\let\auto@bib@innerbib\@empty
\bibitem [{\citenamefont {Feynman}(1986)}]{Feyn}%
  \BibitemOpen
  \bibfield  {author} {\bibinfo {author} {\bibfnamefont {R.~P.}\ \bibnamefont
  {Feynman}},\ }\href@noop {} {\bibfield  {journal} {\bibinfo  {journal}
  {Found. Phys.}\ }\textbf {\bibinfo {volume} {16}},\ \bibinfo {pages} {507}
  (\bibinfo {year} {1986})}\BibitemShut {NoStop}%
\bibitem [{\citenamefont {DiVincenzo}(2000)}]{DiVi}%
  \BibitemOpen
  \bibfield  {author} {\bibinfo {author} {\bibfnamefont {D.~P.}\ \bibnamefont
  {DiVincenzo}}}\href@noop {} { (\bibinfo {year} {2000})},\ \bibinfo
  {note}
  {\href{http://arxiv.org/abs/quant-ph/0002077}{arXiv:quant-ph/0002077}}\BibitemShut
  {NoStop}%
\bibitem [{\citenamefont {Bennett}\ and\ \citenamefont
  {Brassard}(1984)}]{BB84}%
  \BibitemOpen
  \bibfield  {author} {\bibinfo {author} {\bibfnamefont {C.~H.}\ \bibnamefont
  {Bennett}}\ and\ \bibinfo {author} {\bibfnamefont {G.}~\bibnamefont
  {Brassard}},\ }\href@noop {} {\bibfield  {journal} {\bibinfo  {journal}
  {Proc. IEEE Intern. Conf. on Computers, Systems and Signal Processing}\ ,\
  \bibinfo {pages} {175}} (\bibinfo {year} {1984})}\BibitemShut {NoStop}%
\bibitem [{\citenamefont {Bennett}(1992)}]{B92}%
  \BibitemOpen
  \bibfield  {author} {\bibinfo {author} {\bibfnamefont {C.~H.}\ \bibnamefont
  {Bennett}},\ }\href@noop {} {\bibfield  {journal} {\bibinfo  {journal} {Phys.
  Rev. Lett.}\ }\textbf {\bibinfo {volume} {68}},\ \bibinfo {pages} {3121}
  (\bibinfo {year} {1992})}\BibitemShut {NoStop}%
\bibitem [{\citenamefont {Ekert}(1991)}]{Eker}%
  \BibitemOpen
  \bibfield  {author} {\bibinfo {author} {\bibfnamefont {A.~K.}\ \bibnamefont
  {Ekert}},\ }\href@noop {} {\bibfield  {journal} {\bibinfo  {journal} {Phys.
  Rev. Lett.}\ }\textbf {\bibinfo {volume} {67}},\ \bibinfo {pages} {661}
  (\bibinfo {year} {1991})}\BibitemShut {NoStop}%
\bibitem [{\citenamefont {Bennett}\ \emph {et~al.}(1993)\citenamefont
  {Bennett}, \citenamefont {Brassard}, \citenamefont {Cr{\'e}peau},
  \citenamefont {Jozsa}, \citenamefont {Peres},\ and\ \citenamefont
  {Wootters}}]{BBCJ}%
  \BibitemOpen
  \bibfield  {author} {\bibinfo {author} {\bibfnamefont {C.~H.}\ \bibnamefont
  {Bennett}}, \bibinfo {author} {\bibfnamefont {G.}~\bibnamefont {Brassard}},
  \bibinfo {author} {\bibfnamefont {C.}~\bibnamefont {Cr{\'e}peau}}, \bibinfo
  {author} {\bibfnamefont {R.}~\bibnamefont {Jozsa}}, \bibinfo {author}
  {\bibfnamefont {A.}~\bibnamefont {Peres}}, \ and\ \bibinfo {author}
  {\bibfnamefont {W.~K.}\ \bibnamefont {Wootters}},\ }\href@noop {} {\bibfield
  {journal} {\bibinfo  {journal} {Phys. Rev. Lett.}\ }\textbf {\bibinfo
  {volume} {70}},\ \bibinfo {pages} {1895} (\bibinfo {year}
  {1993})}\BibitemShut {NoStop}%
\bibitem [{\citenamefont {Bouwmeester}\ \emph {et~al.}(1997)\citenamefont
  {Bouwmeester}, \citenamefont {Pan}, \citenamefont {Mattle}, \citenamefont
  {Eibl}, \citenamefont {Weinfurter},\ and\ \citenamefont {Zeilinger}}]{BPM1}%
  \BibitemOpen
  \bibfield  {author} {\bibinfo {author} {\bibfnamefont {D.}~\bibnamefont
  {Bouwmeester}}, \bibinfo {author} {\bibfnamefont {J.~W.}\ \bibnamefont
  {Pan}}, \bibinfo {author} {\bibfnamefont {K.}~\bibnamefont {Mattle}},
  \bibinfo {author} {\bibfnamefont {M.}~\bibnamefont {Eibl}}, \bibinfo {author}
  {\bibfnamefont {H.}~\bibnamefont {Weinfurter}}, \ and\ \bibinfo {author}
  {\bibfnamefont {A.}~\bibnamefont {Zeilinger}},\ }\href@noop {} {\bibfield
  {journal} {\bibinfo  {journal} {Nature}\ }\textbf {\bibinfo {volume} {390}},\
  \bibinfo {pages} {575} (\bibinfo {year} {1997})}\BibitemShut {NoStop}%
\bibitem [{\citenamefont {Bouwmeester}\ \emph {et~al.}(1998)\citenamefont
  {Bouwmeester}, \citenamefont {Pan}, \citenamefont {Mattle}, \citenamefont
  {Eibl}, \citenamefont {Weinfurter},\ and\ \citenamefont {Zeilinger}}]{BPM2}%
  \BibitemOpen
  \bibfield  {author} {\bibinfo {author} {\bibfnamefont {D.}~\bibnamefont
  {Bouwmeester}}, \bibinfo {author} {\bibfnamefont {J.~W.}\ \bibnamefont
  {Pan}}, \bibinfo {author} {\bibfnamefont {K.}~\bibnamefont {Mattle}},
  \bibinfo {author} {\bibfnamefont {M.}~\bibnamefont {Eibl}}, \bibinfo {author}
  {\bibfnamefont {H.}~\bibnamefont {Weinfurter}}, \ and\ \bibinfo {author}
  {\bibfnamefont {A.}~\bibnamefont {Zeilinger}},\ }\href@noop {} {\bibfield
  {journal} {\bibinfo  {journal} {Phil. Trans. R. Soc. Lond. A}\ }\textbf
  {\bibinfo {volume} {356}},\ \bibinfo {pages} {1733} (\bibinfo {year}
  {1998})}\BibitemShut {NoStop}%
\bibitem [{\citenamefont {Hedemann}(2016)}]{HedA}%
  \BibitemOpen
  \bibfield  {author} {\bibinfo {author} {\bibfnamefont {S.~R.}\ \bibnamefont
  {Hedemann}} }\href@noop {} {(\bibinfo {year} {2016})},\ \bibinfo {note}
  {\href{https://arxiv.org/abs/1605.09233}{arXiv:1605.09233}}\BibitemShut
  {NoStop}%
\bibitem [{\citenamefont {Nielsen}\ and\ \citenamefont {Chuang}(2010)}]{NiCh}%
  \BibitemOpen
  \bibfield  {author} {\bibinfo {author} {\bibfnamefont {M.~A.}\ \bibnamefont
  {Nielsen}}\ and\ \bibinfo {author} {\bibfnamefont {I.~L.}\ \bibnamefont
  {Chuang}},\ }\href@noop {} {\emph {\bibinfo {title} {Quantum Computation and
  Quantum Information}}}\ (\bibinfo  {publisher} {Cambridge University Press},\
  \bibinfo {year} {2010})\BibitemShut {NoStop}%
\bibitem [{\citenamefont {Ollivier}\ and\ \citenamefont {Zurek}(2001)}]{OlZu}%
  \BibitemOpen
  \bibfield  {author} {\bibinfo {author} {\bibfnamefont {H.}~\bibnamefont
  {Ollivier}}\ and\ \bibinfo {author} {\bibfnamefont {W.~H.}\ \bibnamefont
  {Zurek}},\ }\href@noop {} {\bibfield  {journal} {\bibinfo  {journal} {Phys.
  Rev. Lett.}\ }\textbf {\bibinfo {volume} {88}},\ \bibinfo {pages} {017901}
  (\bibinfo {year} {2001})}\BibitemShut {NoStop}%
\bibitem [{\citenamefont {Henderson}\ and\ \citenamefont
  {Vedral}(2001)}]{HeVe}%
  \BibitemOpen
  \bibfield  {author} {\bibinfo {author} {\bibfnamefont {L.}~\bibnamefont
  {Henderson}}\ and\ \bibinfo {author} {\bibfnamefont {V.}~\bibnamefont
  {Vedral}},\ }\href@noop {} {\bibfield  {journal} {\bibinfo  {journal} {J.
  Phys. A}\ }\textbf {\bibinfo {volume} {34}},\ \bibinfo {pages} {6899}
  (\bibinfo {year} {2001})}\BibitemShut {NoStop}%
\bibitem [{\citenamefont {Ali}\ \emph {et~al.}(2010)\citenamefont {Ali},
  \citenamefont {Rau},\ and\ \citenamefont {Alber}}]{AlRA}%
  \BibitemOpen
  \bibfield  {author} {\bibinfo {author} {\bibfnamefont {M.}~\bibnamefont
  {Ali}}, \bibinfo {author} {\bibfnamefont {A.~R.~P.}\ \bibnamefont {Rau}}, \
  and\ \bibinfo {author} {\bibfnamefont {G.}~\bibnamefont {Alber}},\
  }\href@noop {} {\bibfield  {journal} {\bibinfo  {journal} {Phys. Rev. A}\
  }\textbf {\bibinfo {volume} {81}},\ \bibinfo {pages} {042105} (\bibinfo
  {year} {2010})}\BibitemShut {NoStop}%
\bibitem [{\citenamefont {Bennett}\ \emph {et~al.}(1999)\citenamefont
  {Bennett}, \citenamefont {DiVincenzo}, \citenamefont {Fuchs}, \citenamefont
  {Mor}, \citenamefont {Rains}, \citenamefont {Shor}, \citenamefont {Smolin},\
  and\ \citenamefont {Wootters}}]{BDFM}%
  \BibitemOpen
  \bibfield  {author} {\bibinfo {author} {\bibfnamefont {C.~H.}\ \bibnamefont
  {Bennett}}, \bibinfo {author} {\bibfnamefont {D.~P.}\ \bibnamefont
  {DiVincenzo}}, \bibinfo {author} {\bibfnamefont {C.~A.}\ \bibnamefont
  {Fuchs}}, \bibinfo {author} {\bibfnamefont {T.}~\bibnamefont {Mor}}, \bibinfo
  {author} {\bibfnamefont {E.}~\bibnamefont {Rains}}, \bibinfo {author}
  {\bibfnamefont {P.~W.}\ \bibnamefont {Shor}}, \bibinfo {author}
  {\bibfnamefont {J.~A.}\ \bibnamefont {Smolin}}, \ and\ \bibinfo {author}
  {\bibfnamefont {W.~K.}\ \bibnamefont {Wootters}},\ }\href@noop {} {\bibfield
  {journal} {\bibinfo  {journal} {Phys. Rev. A}\ }\textbf {\bibinfo {volume}
  {59}},\ \bibinfo {pages} {1070} (\bibinfo {year} {1999})}\BibitemShut
  {NoStop}%
\bibitem [{\citenamefont {Horodecki}\ \emph {et~al.}(2005)\citenamefont
  {Horodecki}, \citenamefont {Horodecki}, \citenamefont {Horodecki},
  \citenamefont {Oppenheim}, \citenamefont {Sen}, \citenamefont {Sen},\ and\
  \citenamefont {Synak-Radtke}}]{HHHO}%
  \BibitemOpen
  \bibfield  {author} {\bibinfo {author} {\bibfnamefont {M.}~\bibnamefont
  {Horodecki}}, \bibinfo {author} {\bibfnamefont {P.}~\bibnamefont
  {Horodecki}}, \bibinfo {author} {\bibfnamefont {R.}~\bibnamefont
  {Horodecki}}, \bibinfo {author} {\bibfnamefont {J.}~\bibnamefont
  {Oppenheim}}, \bibinfo {author} {\bibfnamefont {A.}~\bibnamefont {Sen}},
  \bibinfo {author} {\bibfnamefont {U.}~\bibnamefont {Sen}}, \ and\ \bibinfo
  {author} {\bibfnamefont {B.}~\bibnamefont {Synak-Radtke}},\ }\href@noop {}
  {\bibfield  {journal} {\bibinfo  {journal} {Phys. Rev. A}\ }\textbf {\bibinfo
  {volume} {71}},\ \bibinfo {pages} {062307} (\bibinfo {year}
  {2005})}\BibitemShut {NoStop}%
\bibitem [{\citenamefont {Niset}\ and\ \citenamefont {Cerf}(2006)}]{NiCe}%
  \BibitemOpen
  \bibfield  {author} {\bibinfo {author} {\bibfnamefont {J.}~\bibnamefont
  {Niset}}\ and\ \bibinfo {author} {\bibfnamefont {N.~J.}\ \bibnamefont
  {Cerf}},\ }\href@noop {} {\bibfield  {journal} {\bibinfo  {journal} {Phys.
  Rev. A}\ }\textbf {\bibinfo {volume} {74}},\ \bibinfo {pages} {052103}
  (\bibinfo {year} {2006})}\BibitemShut {NoStop}%
\bibitem [{\citenamefont {Hedemann}(2013{\natexlab{a}})}]{HedX}%
  \BibitemOpen
  \bibfield  {author} {\bibinfo {author} {\bibfnamefont {S.~R.}\ \bibnamefont
  {Hedemann}} }\href@noop {} {(\bibinfo {year} {2013}{\natexlab{a}})},\
  \bibinfo {note}
  {\href{https://arxiv.org/abs/1310.7038}{arXiv:1310.7038}}\BibitemShut
  {NoStop}%
\bibitem [{\citenamefont {Hedemann}(2018{\natexlab{a}})}]{HedE}%
  \BibitemOpen
  \bibfield  {author} {\bibinfo {author} {\bibfnamefont {S.~R.}\ \bibnamefont
  {Hedemann}},\ }\href
  {http://www.rintonpress.com/journals/qiconline.html\#v18n56} {\bibfield
  {journal} {\bibinfo  {journal} {Quant. Inf. Comp.}\ }\textbf {\bibinfo
  {volume} {18}},\ \bibinfo {pages} {389} (\bibinfo {year}
  {2018}{\natexlab{a}})},\ \bibinfo {note}
  {\href{https://arxiv.org/abs/1611.03882}{arXiv:1611.03882}}\BibitemShut
  {NoStop}%
\bibitem [{\citenamefont {Galton}(1877)}]{Galt}%
  \BibitemOpen
  \bibfield  {author} {\bibinfo {author} {\bibfnamefont {F.}~\bibnamefont
  {Galton}},\ }\href@noop {} {\bibfield  {journal} {\bibinfo  {journal}
  {Nature}\ }\textbf {\bibinfo {volume} {15}},\ \bibinfo {pages} {512}
  (\bibinfo {year} {1877})}\BibitemShut {NoStop}%
\bibitem [{\citenamefont {Pearson}(1895)}]{Pear}%
  \BibitemOpen
  \bibfield  {author} {\bibinfo {author} {\bibfnamefont {K.}~\bibnamefont
  {Pearson}},\ }\href@noop {} {\bibfield  {journal} {\bibinfo  {journal} {Proc.
  Roy. Soc. Lond.}\ }\textbf {\bibinfo {volume} {58}},\ \bibinfo {pages} {240}
  (\bibinfo {year} {1895})}\BibitemShut {NoStop}%
\bibitem [{\citenamefont {Devore}(2004)}]{Devo}%
  \BibitemOpen
  \bibfield  {author} {\bibinfo {author} {\bibfnamefont {J.~L.}\ \bibnamefont
  {Devore}},\ }\href@noop {} {\emph {\bibinfo {title} {Probability and
  Statistics for Engineering and the Sciences}}},\ \bibinfo {edition} {sixth}\
  ed.\ (\bibinfo  {publisher} {Brooks/Cole},\ \bibinfo {year}
  {2004})\BibitemShut {NoStop}%
\bibitem [{\citenamefont {Hedemann}(2018{\natexlab{b}})}]{HedC}%
  \BibitemOpen
  \bibfield  {author} {\bibinfo {author} {\bibfnamefont {S.~R.}\ \bibnamefont
  {Hedemann}},\ }\href
  {http://www.rintonpress.com/journals/qiconline.html\#v18n56} {\bibfield
  {journal} {\bibinfo  {journal} {Quant. Inf. Comp.}\ }\textbf {\bibinfo
  {volume} {18}},\ \bibinfo {pages} {443} (\bibinfo {year}
  {2018}{\natexlab{b}})},\ \bibinfo {note}
  {\href{http://arxiv.org/abs/1701.03782}{arXiv:1701.03782}}\BibitemShut
  {NoStop}%
\bibitem [{\citenamefont {Dirac}(1927)}]{Dira}%
  \BibitemOpen
  \bibfield  {author} {\bibinfo {author} {\bibfnamefont {P.~A.~M.}\
  \bibnamefont {Dirac}},\ }\href@noop {} {\bibfield  {journal} {\bibinfo
  {journal} {Proc. R. Soc. A}\ }\textbf {\bibinfo {volume} {114}},\ \bibinfo
  {pages} {243} (\bibinfo {year} {1927})}\BibitemShut {NoStop}%
\bibitem [{\citenamefont {Glauber}(1963{\natexlab{a}})}]{Gla1}%
  \BibitemOpen
  \bibfield  {author} {\bibinfo {author} {\bibfnamefont {R.~J.}\ \bibnamefont
  {Glauber}},\ }\href@noop {} {\bibfield  {journal} {\bibinfo  {journal} {Phys.
  Rev.}\ }\textbf {\bibinfo {volume} {130}},\ \bibinfo {pages} {2529} (\bibinfo
  {year} {1963}{\natexlab{a}})}\BibitemShut {NoStop}%
\bibitem [{\citenamefont {Glauber}(1963{\natexlab{b}})}]{Gla2}%
  \BibitemOpen
  \bibfield  {author} {\bibinfo {author} {\bibfnamefont {R.~J.}\ \bibnamefont
  {Glauber}},\ }\href@noop {} {\bibfield  {journal} {\bibinfo  {journal} {Phys.
  Rev.}\ }\textbf {\bibinfo {volume} {131}},\ \bibinfo {pages} {2766} (\bibinfo
  {year} {1963}{\natexlab{b}})}\BibitemShut {NoStop}%
\bibitem [{\citenamefont {Titulaer}\ and\ \citenamefont
  {Glauber}(1965)}]{TiGl}%
  \BibitemOpen
  \bibfield  {author} {\bibinfo {author} {\bibfnamefont {U.~M.}\ \bibnamefont
  {Titulaer}}\ and\ \bibinfo {author} {\bibfnamefont {R.~J.}\ \bibnamefont
  {Glauber}},\ }\href@noop {} {\bibfield  {journal} {\bibinfo  {journal} {Phys.
  Rev.}\ }\textbf {\bibinfo {volume} {140}},\ \bibinfo {pages} {B676} (\bibinfo
  {year} {1965})}\BibitemShut {NoStop}%
\bibitem [{\citenamefont {Gerry}\ and\ \citenamefont {Knight}(2005)}]{GeKn}%
  \BibitemOpen
  \bibfield  {author} {\bibinfo {author} {\bibfnamefont {C.~C.}\ \bibnamefont
  {Gerry}}\ and\ \bibinfo {author} {\bibfnamefont {P.~L.}\ \bibnamefont
  {Knight}},\ }\href@noop {} {\emph {\bibinfo {title} {Introductory Quantum
  Optics}}}\ (\bibinfo  {publisher} {Cambridge University Press},\ \bibinfo
  {year} {2005})\ pp.\ \bibinfo {pages} {120,130}\BibitemShut {NoStop}%
\bibitem [{\citenamefont {Spekkens}(2007)}]{Spek}%
  \BibitemOpen
  \bibfield  {author} {\bibinfo {author} {\bibfnamefont {R.~W.}\ \bibnamefont
  {Spekkens}},\ }\href@noop {} {\bibfield  {journal} {\bibinfo  {journal}
  {Phys. Rev. A}\ }\textbf {\bibinfo {volume} {75}},\ \bibinfo {pages} {032110}
  (\bibinfo {year} {2007})}\BibitemShut {NoStop}%
\bibitem [{\citenamefont {Pusey}\ \emph {et~al.}(2012)\citenamefont {Pusey},
  \citenamefont {Barrett},\ and\ \citenamefont {Rudolph}}]{PuBR}%
  \BibitemOpen
  \bibfield  {author} {\bibinfo {author} {\bibfnamefont {M.~F.}\ \bibnamefont
  {Pusey}}, \bibinfo {author} {\bibfnamefont {J.}~\bibnamefont {Barrett}}, \
  and\ \bibinfo {author} {\bibfnamefont {T.}~\bibnamefont {Rudolph}},\
  }\href@noop {} {\bibfield  {journal} {\bibinfo  {journal} {Nature}\ }\textbf
  {\bibinfo {volume} {8}},\ \bibinfo {pages} {475} (\bibinfo {year}
  {2012})}\BibitemShut {NoStop}%
\bibitem [{\citenamefont {Harrigan}\ and\ \citenamefont
  {Spekkens}(2010)}]{HaSp}%
  \BibitemOpen
  \bibfield  {author} {\bibinfo {author} {\bibfnamefont {N.}~\bibnamefont
  {Harrigan}}\ and\ \bibinfo {author} {\bibfnamefont {R.~W.}\ \bibnamefont
  {Spekkens}},\ }\href@noop {} {\bibfield  {journal} {\bibinfo  {journal}
  {Found. Phys.}\ }\textbf {\bibinfo {volume} {40}},\ \bibinfo {pages} {125}
  (\bibinfo {year} {2010})}\BibitemShut {NoStop}%
\bibitem [{\citenamefont {Hedemann}(2013{\natexlab{b}})}]{HedU}%
  \BibitemOpen
  \bibfield  {author} {\bibinfo {author} {\bibfnamefont {S.~R.}\ \bibnamefont
  {Hedemann}} }\href@noop {} {(\bibinfo {year} {2013}{\natexlab{b}})},\
  \bibinfo {note}
  {\href{https://arxiv.org/abs/1303.5904}{arXiv:1303.5904}}\BibitemShut
  {NoStop}%
\bibitem [{\citenamefont {Werner}(1989)}]{Wern}%
  \BibitemOpen
  \bibfield  {author} {\bibinfo {author} {\bibfnamefont {R.~F.}\ \bibnamefont
  {Werner}},\ }\href@noop {} {\bibfield  {journal} {\bibinfo  {journal} {Phys.
  Rev. A}\ }\textbf {\bibinfo {volume} {40}},\ \bibinfo {pages} {4277}
  (\bibinfo {year} {1989})}\BibitemShut {NoStop}%
\bibitem [{\citenamefont {Bacciagaluppi}\ and\ \citenamefont
  {Valentini}(2009)}]{BaVa}%
  \BibitemOpen
  \bibfield  {author} {\bibinfo {author} {\bibfnamefont {G.}~\bibnamefont
  {Bacciagaluppi}}\ and\ \bibinfo {author} {\bibfnamefont {A.}~\bibnamefont
  {Valentini}},\ }\href@noop {} {\emph {\bibinfo {title} {Quantum Theory at the
  Crossroads: Reconsidering the 1927 Solvay Conference}}}\ (\bibinfo
  {publisher} {Cambridge University Press},\ \bibinfo {year} {2009})\ p.\
  \bibinfo {pages} {175}\BibitemShut {NoStop}%
\bibitem [{\citenamefont {Einstein}\ \emph {et~al.}(1935)\citenamefont
  {Einstein}, \citenamefont {Podolsky},\ and\ \citenamefont {Rosen}}]{EPR}%
  \BibitemOpen
  \bibfield  {author} {\bibinfo {author} {\bibfnamefont {A.}~\bibnamefont
  {Einstein}}, \bibinfo {author} {\bibfnamefont {B.}~\bibnamefont {Podolsky}},
  \ and\ \bibinfo {author} {\bibfnamefont {N.}~\bibnamefont {Rosen}},\
  }\href@noop {} {\bibfield  {journal} {\bibinfo  {journal} {Phys. Rev.}\
  }\textbf {\bibinfo {volume} {47}},\ \bibinfo {pages} {777} (\bibinfo {year}
  {1935})}\BibitemShut {NoStop}%
\bibitem [{\citenamefont {Einstein}(1936)}]{Eins}%
  \BibitemOpen
  \bibfield  {author} {\bibinfo {author} {\bibfnamefont {A.}~\bibnamefont
  {Einstein}},\ }\href@noop {} {\bibfield  {journal} {\bibinfo  {journal} {J.
  of the Franklin Institute}\ }\textbf {\bibinfo {volume} {221}},\ \bibinfo
  {pages} {313, 349} (\bibinfo {year} {1936})}\BibitemShut {NoStop}%
\bibitem [{\citenamefont {Bell}(1964)}]{Bell}%
  \BibitemOpen
  \bibfield  {author} {\bibinfo {author} {\bibfnamefont {J.~S.}\ \bibnamefont
  {Bell}},\ }\href@noop {} {\bibfield  {journal} {\bibinfo  {journal}
  {Physics}\ }\textbf {\bibinfo {volume} {1}},\ \bibinfo {pages} {195}
  (\bibinfo {year} {1964})}\BibitemShut {NoStop}%
\bibitem [{\citenamefont {Clauser}\ \emph {et~al.}(1969)\citenamefont
  {Clauser}, \citenamefont {Horne}, \citenamefont {Shimony},\ and\
  \citenamefont {Holt}}]{CHSH}%
  \BibitemOpen
  \bibfield  {author} {\bibinfo {author} {\bibfnamefont {J.~F.}\ \bibnamefont
  {Clauser}}, \bibinfo {author} {\bibfnamefont {M.}~\bibnamefont {Horne}},
  \bibinfo {author} {\bibfnamefont {A.}~\bibnamefont {Shimony}}, \ and\
  \bibinfo {author} {\bibfnamefont {R.~A.}\ \bibnamefont {Holt}},\ }\href@noop
  {} {\bibfield  {journal} {\bibinfo  {journal} {Phys. Rev. Lett.}\ }\textbf
  {\bibinfo {volume} {23}},\ \bibinfo {pages} {880} (\bibinfo {year}
  {1969})}\BibitemShut {NoStop}%
\bibitem [{\citenamefont {Aspect}\ \emph {et~al.}(1982)\citenamefont {Aspect},
  \citenamefont {Dalibard},\ and\ \citenamefont {Roger}}]{AsDR}%
  \BibitemOpen
  \bibfield  {author} {\bibinfo {author} {\bibfnamefont {A.}~\bibnamefont
  {Aspect}}, \bibinfo {author} {\bibfnamefont {J.}~\bibnamefont {Dalibard}}, \
  and\ \bibinfo {author} {\bibfnamefont {G.}~\bibnamefont {Roger}},\
  }\href@noop {} {\bibfield  {journal} {\bibinfo  {journal} {Phys. Rev. Lett.}\
  }\textbf {\bibinfo {volume} {49}},\ \bibinfo {pages} {1804} (\bibinfo {year}
  {1982})}\BibitemShut {NoStop}%
\bibitem [{\citenamefont {Greenberger}\ \emph {et~al.}(1989)\citenamefont
  {Greenberger}, \citenamefont {Horne},\ and\ \citenamefont {Zeilinger}}]{GHZ}%
  \BibitemOpen
  \bibfield  {author} {\bibinfo {author} {\bibfnamefont {D.~M.}\ \bibnamefont
  {Greenberger}}, \bibinfo {author} {\bibfnamefont {M.}~\bibnamefont {Horne}},
  \ and\ \bibinfo {author} {\bibfnamefont {A.}~\bibnamefont {Zeilinger}},\
  }\href@noop {} {\emph {\bibinfo {title} {Bell's Theorem, Quantum Theory, and
  Conceptions of the Universe}}},\ edited by\ \bibinfo {editor} {\bibfnamefont
  {M.}~\bibnamefont {Kafatos}}\ (\bibinfo  {publisher} {Kluwer, Dordrecht},\
  \bibinfo {year} {1989})\BibitemShut {NoStop}%
\bibitem [{\citenamefont {Greenberger}\ \emph {et~al.}(1990)\citenamefont
  {Greenberger}, \citenamefont {Horne}, \citenamefont {Shimony},\ and\
  \citenamefont {Zeilinger}}]{GHSZ}%
  \BibitemOpen
  \bibfield  {author} {\bibinfo {author} {\bibfnamefont {D.~M.}\ \bibnamefont
  {Greenberger}}, \bibinfo {author} {\bibfnamefont {M.}~\bibnamefont {Horne}},
  \bibinfo {author} {\bibfnamefont {A.}~\bibnamefont {Shimony}}, \ and\
  \bibinfo {author} {\bibfnamefont {A.}~\bibnamefont {Zeilinger}},\ }\href@noop
  {} {\bibfield  {journal} {\bibinfo  {journal} {Am. J. Phys.}\ }\textbf
  {\bibinfo {volume} {58}},\ \bibinfo {pages} {1131} (\bibinfo {year}
  {1990})}\BibitemShut {NoStop}%
\bibitem [{\citenamefont {Mermin}(1990)}]{Merm}%
  \BibitemOpen
  \bibfield  {author} {\bibinfo {author} {\bibfnamefont {N.~D.}\ \bibnamefont
  {Mermin}},\ }\href@noop {} {\bibfield  {journal} {\bibinfo  {journal} {Phys.
  Today}\ }\textbf {\bibinfo {volume} {43}},\ \bibinfo {pages} {9} (\bibinfo
  {year} {1990})}\BibitemShut {NoStop}%
\bibitem [{\citenamefont {Hill}\ and\ \citenamefont {Wootters}(1997)}]{HiWo}%
  \BibitemOpen
  \bibfield  {author} {\bibinfo {author} {\bibfnamefont {S.}~\bibnamefont
  {Hill}}\ and\ \bibinfo {author} {\bibfnamefont {W.~K.}\ \bibnamefont
  {Wootters}},\ }\href@noop {} {\bibfield  {journal} {\bibinfo  {journal}
  {Phys. Rev. Lett.}\ }\textbf {\bibinfo {volume} {78}},\ \bibinfo {pages}
  {5022} (\bibinfo {year} {1997})}\BibitemShut {NoStop}%
\bibitem [{\citenamefont {Wootters}(1998)}]{Woot}%
  \BibitemOpen
  \bibfield  {author} {\bibinfo {author} {\bibfnamefont {W.~K.}\ \bibnamefont
  {Wootters}},\ }\href@noop {} {\bibfield  {journal} {\bibinfo  {journal}
  {Phys. Rev. Lett.}\ }\textbf {\bibinfo {volume} {80}},\ \bibinfo {pages}
  {2245} (\bibinfo {year} {1998})}\BibitemShut {NoStop}%
\bibitem [{\citenamefont {Horodecki}(1997)}]{Hor1}%
  \BibitemOpen
  \bibfield  {author} {\bibinfo {author} {\bibfnamefont {P.}~\bibnamefont
  {Horodecki}},\ }\href@noop {} {\bibfield  {journal} {\bibinfo  {journal}
  {Phys. Lett. A}\ }\textbf {\bibinfo {volume} {232}},\ \bibinfo {pages} {333}
  (\bibinfo {year} {1997})}\BibitemShut {NoStop}%
\bibitem [{\citenamefont {Horodecki}\ \emph {et~al.}(1999)\citenamefont
  {Horodecki}, \citenamefont {Horodecki},\ and\ \citenamefont
  {Horodecki}}]{Hor3}%
  \BibitemOpen
  \bibfield  {author} {\bibinfo {author} {\bibfnamefont {P.}~\bibnamefont
  {Horodecki}}, \bibinfo {author} {\bibfnamefont {M.}~\bibnamefont
  {Horodecki}}, \ and\ \bibinfo {author} {\bibfnamefont {R.}~\bibnamefont
  {Horodecki}},\ }\href@noop {} {\bibfield  {journal} {\bibinfo  {journal}
  {Phys. Rev. Lett.}\ }\textbf {\bibinfo {volume} {82}},\ \bibinfo {pages}
  {1056} (\bibinfo {year} {1999})}\BibitemShut {NoStop}%
\bibitem [{\citenamefont {Horodecki}\ \emph {et~al.}(1998)\citenamefont
  {Horodecki}, \citenamefont {Horodecki},\ and\ \citenamefont
  {Horodecki}}]{Hor4}%
  \BibitemOpen
  \bibfield  {author} {\bibinfo {author} {\bibfnamefont {M.}~\bibnamefont
  {Horodecki}}, \bibinfo {author} {\bibfnamefont {P.}~\bibnamefont
  {Horodecki}}, \ and\ \bibinfo {author} {\bibfnamefont {R.}~\bibnamefont
  {Horodecki}},\ }\href@noop {} {\bibfield  {journal} {\bibinfo  {journal}
  {Phys. Rev. Lett.}\ }\textbf {\bibinfo {volume} {80}},\ \bibinfo {pages}
  {5239} (\bibinfo {year} {1998})}\BibitemShut {NoStop}%
\bibitem [{\citenamefont {Hedemann}(2014)}]{HedD}%
  \BibitemOpen
  \bibfield  {author} {\bibinfo {author} {\bibfnamefont {S.~R.}\ \bibnamefont
  {Hedemann}},\ }\emph {\bibinfo {title} {Hyperspherical Bloch Vectors with
  Applications to Entanglement and Quantum State Tomography}},\ \href@noop {}
  {Ph.D. thesis},\ \bibinfo  {school} {Stevens Institute of Technology}
  (\bibinfo {year} {2014})\BibitemShut {NoStop}%
\bibitem [{\citenamefont {Mendon{\c c}a}\ \emph {et~al.}(2014)\citenamefont
  {Mendon{\c c}a}, \citenamefont {Marchiolli},\ and\ \citenamefont
  {Galetti}}]{MeMG}%
  \BibitemOpen
  \bibfield  {author} {\bibinfo {author} {\bibfnamefont {P.~E. M.~F.}\
  \bibnamefont {Mendon{\c c}a}}, \bibinfo {author} {\bibfnamefont {M.~A.}\
  \bibnamefont {Marchiolli}}, \ and\ \bibinfo {author} {\bibfnamefont
  {D.}~\bibnamefont {Galetti}},\ }\href@noop {} {\bibfield  {journal} {\bibinfo
   {journal} {Ann. Phys.}\ }\textbf {\bibinfo {volume} {351}},\ \bibinfo
  {pages} {79} (\bibinfo {year} {2014})}\BibitemShut {NoStop}%
\bibitem [{\citenamefont {Mendon{\c c}a}\ \emph {et~al.}(2017)\citenamefont
  {Mendon{\c c}a}, \citenamefont {Marchiolli},\ and\ \citenamefont
  {Hedemann}}]{MeMH}%
  \BibitemOpen
  \bibfield  {author} {\bibinfo {author} {\bibfnamefont {P.~E. M.~F.}\
  \bibnamefont {Mendon{\c c}a}}, \bibinfo {author} {\bibfnamefont {M.~A.}\
  \bibnamefont {Marchiolli}}, \ and\ \bibinfo {author} {\bibfnamefont {S.~R.}\
  \bibnamefont {Hedemann}},\ }\href@noop {} {\bibfield  {journal} {\bibinfo
  {journal} {Phys. Rev. A}\ }\textbf {\bibinfo {volume} {95}},\ \bibinfo
  {pages} {022324} (\bibinfo {year} {2017})}\BibitemShut {NoStop}%
\bibitem [{\citenamefont {Hedemann}(2018{\natexlab{c}})}]{HeXU}%
  \BibitemOpen
  \bibfield  {author} {\bibinfo {author} {\bibfnamefont {S.~R.}\ \bibnamefont
  {Hedemann}},\ }\href {https://doi.org/10.1007/s11128-018-2061-0} {\bibfield
  {journal} {\bibinfo  {journal} {Quant. Inf. Process.}\ }\textbf {\bibinfo
  {volume} {17}},\ \bibinfo {pages} {293} (\bibinfo {year}
  {2018}{\natexlab{c}})},\ \bibinfo {note}
  {\href{http://arxiv.org/abs/1802.03038}{arXiv:1802.03038}}\BibitemShut
  {NoStop}%
\bibitem [{\citenamefont {Streltsov}\ \emph {et~al.}(2010)\citenamefont
  {Streltsov}, \citenamefont {Kampermann},\ and\ \citenamefont {Bru}}]{StKB}%
  \BibitemOpen
  \bibfield  {author} {\bibinfo {author} {\bibfnamefont {A.}~\bibnamefont
  {Streltsov}}, \bibinfo {author} {\bibfnamefont {H.}~\bibnamefont
  {Kampermann}}, \ and\ \bibinfo {author} {\bibfnamefont {D.}~\bibnamefont
  {Bru}},\ }\href@noop {} {\bibfield  {journal} {\bibinfo  {journal} {New J.
  Phys.}\ }\textbf {\bibinfo {volume} {12}},\ \bibinfo {pages} {123004}
  (\bibinfo {year} {2010})}\BibitemShut {NoStop}%
\bibitem [{\citenamefont {Bloch}(1946)}]{Blch}%
  \BibitemOpen
  \bibfield  {author} {\bibinfo {author} {\bibfnamefont {F.}~\bibnamefont
  {Bloch}},\ }\href@noop {} {\bibfield  {journal} {\bibinfo  {journal} {Phys.
  Rev.}\ }\textbf {\bibinfo {volume} {70}},\ \bibinfo {pages} {460} (\bibinfo
  {year} {1946})}\BibitemShut {NoStop}%
\bibitem [{\citenamefont {Stokes}(1852)}]{Stok}%
  \BibitemOpen
  \bibfield  {author} {\bibinfo {author} {\bibfnamefont {G.~G.}\ \bibnamefont
  {Stokes}},\ }\href@noop {} {\bibfield  {journal} {\bibinfo  {journal} {Trans.
  Cambridge Philos. Soc.}\ }\textbf {\bibinfo {volume} {9}},\ \bibinfo {pages}
  {399} (\bibinfo {year} {1852})}\BibitemShut {NoStop}%
\bibitem [{\citenamefont {von Neumann}(1927)}]{VNeu}%
  \BibitemOpen
  \bibfield  {author} {\bibinfo {author} {\bibfnamefont {J.}~\bibnamefont
  {von Neumann}},\ }\href@noop {} {\bibfield  {journal} {\bibinfo  {journal} {G{\"o}ttinger Nachrichten}\ }\textbf {\bibinfo {volume} {1}},\ \bibinfo {pages} {245} (\bibinfo {year} {1927})}\BibitemShut {NoStop}%
\bibitem [{\citenamefont {Ne'eman}(1961)}]{Neem}%
  \BibitemOpen
  \bibfield  {author} {\bibinfo {author} {\bibfnamefont {Y.}~\bibnamefont
  {Ne'eman}},\ }\href@noop {} {\bibfield  {journal} {\bibinfo  {journal} {Nucl.
  Phys.}\ }\textbf {\bibinfo {volume} {26}},\ \bibinfo {pages} {222} (\bibinfo
  {year} {1961})}\BibitemShut {NoStop}%
\bibitem [{\citenamefont {Gell-Mann}(1962)}]{Gell}%
  \BibitemOpen
  \bibfield  {author} {\bibinfo {author} {\bibfnamefont {M.}~\bibnamefont
  {Gell-Mann}},\ }\href@noop {} {\bibfield  {journal} {\bibinfo  {journal}
  {Phys. Rev.}\ }\textbf {\bibinfo {volume} {125}},\ \bibinfo {pages} {1067}
  (\bibinfo {year} {1962})}\BibitemShut {NoStop}%
\bibitem [{\citenamefont {Hioe}\ and\ \citenamefont {Eberly}(1981)}]{HiEb}%
  \BibitemOpen
  \bibfield  {author} {\bibinfo {author} {\bibfnamefont {F.~T.}\ \bibnamefont
  {Hioe}}\ and\ \bibinfo {author} {\bibfnamefont {J.~H.}\ \bibnamefont
  {Eberly}},\ }\href@noop {} {\bibfield  {journal} {\bibinfo  {journal} {Phys.
  Rev. Lett.}\ }\textbf {\bibinfo {volume} {47}},\ \bibinfo {pages} {838}
  (\bibinfo {year} {1981})}\BibitemShut {NoStop}%
\bibitem [{\citenamefont {Sakurai}(1994)}]{Saku}%
  \BibitemOpen
  \bibfield  {author} {\bibinfo {author} {\bibfnamefont {J.~J.}\ \bibnamefont
  {Sakurai}},\ }\href@noop {} {\emph {\bibinfo {title} {Modern Quantum
  Mechanics Revised Edition}}},\ edited by\ \bibinfo {editor} {\bibfnamefont
  {S.~F.}\ \bibnamefont {Tuan}}\ (\bibinfo  {publisher} {Addison-Wesley
  Publishing Company, Inc.},\ \bibinfo {year} {1994})\ p.\ \bibinfo {pages}
  {362}\BibitemShut {NoStop}%
\bibitem [{\citenamefont {Reif}(2009)}]{Reif}%
  \BibitemOpen
  \bibfield  {author} {\bibinfo {author} {\bibfnamefont {F.}~\bibnamefont
  {Reif}},\ }\href@noop {} {\emph {\bibinfo {title} {Fundamentals of
  Statistical and Thermal Physics}}}\ (\bibinfo  {publisher} {Waveland Press,
  Inc.},\ \bibinfo {year} {2009})\ pp.\ \bibinfo {pages} {331--333}\BibitemShut
  {NoStop}%
\bibitem [{\citenamefont {Ehrenfest}(1927)}]{Ehre}%
  \BibitemOpen
  \bibfield  {author} {\bibinfo {author} {\bibfnamefont {P.}~\bibnamefont
  {Ehrenfest}},\ }\href@noop {} {\bibfield  {journal} {\bibinfo  {journal}
  {Zeitschrift f{\"u}r Physik}\ }\textbf {\bibinfo {volume} {45}},\ \bibinfo
  {pages} {455} (\bibinfo {year} {1927})}\BibitemShut {NoStop}%
\bibitem [{\citenamefont {Griffiths}(2005)}]{Grif}%
  \BibitemOpen
  \bibfield  {author} {\bibinfo {author} {\bibfnamefont {D.~J.}\ \bibnamefont
  {Griffiths}},\ }\href@noop {} {\emph {\bibinfo {title} {Introduction to
  Quantum Mechanics}}},\ \bibinfo {edition} {2nd}\ ed.\ (\bibinfo  {publisher}
  {Pearson Education, Inc.},\ \bibinfo {year} {2005})\ pp.\ \bibinfo {pages}
  {18,115}\BibitemShut {NoStop}%
\bibitem [{\citenamefont {Carath{\' e}odory}(1907)}]{Cara}%
  \BibitemOpen
  \bibfield  {author} {\bibinfo {author} {\bibfnamefont {C.}~\bibnamefont
  {Carath{\' e}odory}},\ }\href@noop {} {\bibfield  {journal} {\bibinfo
  {journal} {Mathematische Annalen}\ }\textbf {\bibinfo {volume} {64}},\
  \bibinfo {pages} {95} (\bibinfo {year} {1907})}\BibitemShut {NoStop}%
\bibitem [{\citenamefont {Horodecki}\ \emph {et~al.}(2003)\citenamefont
  {Horodecki}, \citenamefont {Smolin}, \citenamefont {Terhal},\ and\
  \citenamefont {Thapliyal}}]{Hor6}%
  \BibitemOpen
  \bibfield  {author} {\bibinfo {author} {\bibfnamefont {P.}~\bibnamefont
  {Horodecki}}, \bibinfo {author} {\bibfnamefont {J.~A.}\ \bibnamefont
  {Smolin}}, \bibinfo {author} {\bibfnamefont {B.~M.}\ \bibnamefont {Terhal}},
  \ and\ \bibinfo {author} {\bibfnamefont {A.~V.}\ \bibnamefont {Thapliyal}},\
  }\href@noop {} {\bibfield  {journal} {\bibinfo  {journal} {Theor. Comp.
  Sci.}\ }\textbf {\bibinfo {volume} {292}},\ \bibinfo {pages} {589} (\bibinfo
  {year} {2003})}\BibitemShut {NoStop}%
\end{thebibliography}
\end{document}